\documentclass[journal]{IEEEtran}

\usepackage{blindtext}
\usepackage{graphicx}
\usepackage{color,soul, xcolor}

\usepackage{booktabs}
\usepackage{mathtools}
\DeclarePairedDelimiter\abs{\lvert}{\rvert}

\usepackage{pdfpages}
\usepackage{bm}
\usepackage{tabularx} 
\usepackage{algorithm, algorithmic}
\usepackage{mathtools}

\usepackage{cite}
\usepackage{comment}
\usepackage{tabu}
\usepackage{amssymb}
\usepackage{multirow}
\usepackage{hyperref}
\usepackage{xr-hyper}
\usepackage[flushleft]{threeparttable}
\usepackage{breqn}
\usepackage{cancel}
\usepackage[symbol]{footmisc}

\setulcolor{red}

\DeclareMathOperator{\logit}{logit}
\makeatletter
 \def\@textbottom{\vskip \z@ \@plus 1pt}
 \let\@texttop\relax
\makeatother

\usepackage{cite}
\usepackage{array}

\usepackage{eqparbox}

\usepackage[tight,footnotesize]{subfigure}
\begin{document}
\title{
Confidence Calibration and Predictive Uncertainty Estimation for Deep Medical Image Segmentation
}
%
%
%

\author{Alireza~Mehrtash,
        William M.~Wells III,
        Clare~M.~Tempany,
        Purang~Abolmaesumi,
        and~Tina~Kapur
\thanks{
This work was supported by the US National Institutes of Health grants P41EB015898, Natural Sciences and Engineering Research Council (NSERC) of Canada, and the Canadian Institutes of Health Research (CIHR). 

A. Mehrtash is with the Department of Electrical and Computer Engineering, The University of British Columbia, Vancouver, BC, V6T 1Z4, Canada, and also with the Department of Radiology, Brigham and Women's Hospital, Harvard Medical School, Boston, 02115, USA. 

P. Abolmaesumi is with the Department of Electrical and Computer Engineering, The University of British Columbia Vancouver, BC, V5T 1Z4, Canada.

C. M. Tempany, W. M. Wells, and T. Kapur are with the Department of Radiology, Brigham and Women's Hospital, Harvard Medical School, Boston, 02115, USA (e-mail: tkapur@bwh.harvard.edu.)
}
}

\maketitle
\begin{abstract}
Fully convolutional neural networks (FCNs), and in particular U-Nets, have achieved state-of-the-art results in semantic segmentation for numerous medical imaging applications.
Moreover, batch normalization and Dice loss have been used successfully to stabilize and accelerate training.
However, these networks are poorly calibrated i.e. they tend to produce overconfident predictions for both correct and erroneous classifications, making them unreliable and hard to interpret.
In this paper, we study predictive uncertainty estimation
in FCNs for medical image segmentation.
We make the following contributions: 
1) We systematically compare cross-entropy loss with Dice loss in terms of segmentation quality and uncertainty estimation of FCNs;
2) We propose model ensembling for confidence calibration of the FCNs trained with batch normalization and Dice loss;
3) We assess the ability of calibrated FCNs to predict segmentation quality of structures and detect out-of-distribution test examples.
We conduct extensive experiments across three medical image segmentation applications
of the brain, the heart, and the prostate to evaluate our contributions.
The results of this study offer considerable insight into the predictive uncertainty estimation and out-of-distribution detection in medical image segmentation and provide practical recipes for confidence calibration. Moreover, we consistently demonstrate that model ensembling improves confidence calibration.
\end{abstract}




%
\IEEEpeerreviewmaketitle


\newcommand{\cf}{\S}
\newcommand\eg{\textit{e.g.~}}
\newcommand\etal{\textit{et~al.~}}
\newcommand\ie{\textit{i.e.~}}
\newcommand\viz{\textit{viz.~}}

\newcommand\ba{\mathbf{a}}
\newcommand\bb{\mathbf{b}}
\newcommand\bB{\mathbf{B}}
\newcommand\bc{\mathbf{c}}
\newcommand\bC{\mathbf{C}}
\newcommand\bd{\mathbf{d}}
\newcommand\bD{\mathbf{D}}
\newcommand\be{\mathbf{e}}
\newcommand\bff{\mathbf{f}}
\newcommand\bg{\mathbf{g}}
\newcommand\bK{\mathbf{K}}
\newcommand\bn{\mathbf{n}}
\newcommand\bp{\mathbf{p}}
\newcommand\br{\mathbf{r}}
\newcommand\bs{\mathbf{s}}
\newcommand\bt{\mathbf{t}}

\newcommand\bu{\mathbf{u}}
\newcommand\bU{\mathbf{U}}
\newcommand\bv{\mathbf{v}}
\newcommand\bw{\mathbf{w}}
\newcommand\bx{\mathbf{x}}
\newcommand\by{\mathbf{y}}
\newcommand\bz{\mathbf{z}}
\newcommand\balpha{\boldsymbol{\alpha}}
\newcommand\bbeta{\boldsymbol{\beta}}
\newcommand\bepsilon{\boldsymbol{\epsilon}}
\newcommand\blambda{{\boldsymbol{\lambda}}}
\newcommand\bLambda{\boldsymbol\Lambda}
\newcommand\bmu{\boldsymbol{\mu}}
\newcommand\bsigma{\boldsymbol{\sigma}}
\newcommand\bSigma{\boldsymbol{\Sigma}}
\newcommand\btheta{\boldsymbol{\theta}}

\newcommand\bbE{\mathbb{E}}
\newcommand\bbG{\mathbb{G}}
\newcommand\bbR{\mathbb{R}}
\newcommand\bbV{\mathbb{V}}
\newcommand\bbZ{\mathbb{Z}}

\newcommand\cA{\mathcal{A}}
\newcommand\cB{\mathcal{B}}
\newcommand\cD{\mathcal{D}}
\newcommand\cE{\mathcal{E}}
\newcommand\cF{\mathcal{F}}
\newcommand\cG{\mathcal{G}}
\newcommand\cI{\mathcal{I}}
\newcommand\cJ{\mathcal{J}}
\newcommand\cL{\mathcal{L}}
\newcommand\cM{\mathcal{M}}
\newcommand\cN{\mathcal{N}}
\newcommand\cO{\mathcal{O}}
\newcommand\cP{\mathcal{P}}
\newcommand\cT{\mathcal{T}}
\newcommand\cV{\mathcal{V}}
\newcommand\cU{\mathcal{U}}
\newcommand\cX{\mathcal{X}}
\newcommand\cZ{\mathcal{Z}}

\newcommand\dsE{\mathds{E}}
\newcommand\dsG{\mathds{G}}
\newcommand\dsH{\mathds{H}}
\newcommand\dsV{\mathds{V}}

\newcommand\rmA{\mathrm{A}}
\newcommand\rmB{\mathrm{B}}
\newcommand\rmd{\mathrm{d}}
\newcommand\rmD{\mathrm{D}}
\newcommand\rmE{\mathrm{E}}
\newcommand\rmF{\mathrm{F}}
\newcommand\rmH{\mathrm{H}}
\newcommand\rmI{\mathrm{I}}
\newcommand\rmJ{\mathrm{J}}
\newcommand\rmK{\mathrm{K}}
\newcommand\rmM{\mathrm{M}}
\newcommand\rmT{\mathrm{T}}
\newcommand\rmU{\mathrm{U}}
\newcommand\rmV{\mathrm{V}}
\newcommand\rmW{\mathrm{W}}
\newcommand\rmy{\mathrm{y}}
\newcommand\rmz{\mathrm{z}}
\newcommand\rmZ{\mathrm{Z}}

\newcommand{\argmax}{\operatornamewithlimits{argmax}}
\newcommand{\argmin}{\operatornamewithlimits{argmin}}
\newcommand{\distributionequal}{\operatornamewithlimits{\sim}}
\newcommand{\diag}{\textrm{diag}}
\newcommand{\const}{\textrm{const}}
\newcommand{\Cov}{\mathds{C}\text{ov}}
\newcommand{\Corr}{\mathds{C}\text{orr}}
\newcommand{\E}[1]{{\dsE\left\{#1\right\}}}
\newcommand{\Ent}[1]{{\dsH\left\{#1\right\}}}
\newcommand{\Exp}{\dsE}
\newcommand{\half}{\frac{1}{2}}
\newcommand\Hes{\nabla\nabla\tr}
\newcommand{\inv}{^{-1}}
\newcommand{\IP}[1]{{\left\langle{#1}\right\rangle}} 
\newcommand\I{\rmI}
\newcommand\ind{\mathds{1}}
\newcommand\KL{\mathds{KL}}
\newcommand\one{\mathds{1}}
\newcommand{\p}[1]{{p\left(#1\right)}}
\newcommand{\pinv}{^{-}}
\newcommand{\Prec}{\bLambda}
\newcommand{\Prr}{\mathrm{Pr}}
\newcommand\R{\bbR}
\newcommand\rank{{\mathrm{Rank}}}
\newcommand{\tr}{^\top}
\newcommand{\Trace}{\mathrm{Tr}}
\newcommand\Var[1]{{\mathds{V}\mathrm{ar}\left\{#1\right\}}}
\newcommand\vol{\mathrm{vol}}


\newcommand{\TV}{\mathrm{TV}}
\newcommand{\img}{\mathrm{y}}
\renewcommand{\ij}{{i,j}}
\newcommand{\ioj}{{i+1,j}}
\newcommand{\ijo}{{i,j+1}}
\newcommand{\iojo}{{i+1,j+1}}
\newcommand{\x}{{x}}
\newcommand{\y}{{\rmy}}
\newcommand{\z}{{\rmz}}
\newcommand{\q}{{q}}
\newcommand{\Q}{{Q}}
\newcommand{\Py}{{P}}
\renewcommand{\E}[1]{{\dsE_\Q\left\{#1\right\}}}

\newcommand{\KLD}[2]{{\mathds{KL}}[#1 || #2]}
\newcommand{\EV}[2]{{\dsE}_{#1}\left[#2\right]}
\newcommand{\PD}{~.}
\newcommand{\CM}{~,}
\newcommand{\zvar}{\rmz}
\newcommand{\area}{}
\newcommand{\wrt}{wrt.~}
\newcommand{\ase}{\overset{a.s.}{=}}
\newcommand\where{\text{where }}
\newcommand\thrfr{\text{therefore }}

\newcommand{\indx}{i\in{\mathbb I}_{\cX}}

\renewcommand{\Ent}[1]{{\dsH\left[#1\right]}}

\section{Introduction}
\IEEEPARstart{F}{ully} convolutional neural networks (FCNs), and in particular the U-Net \cite{ronneberger2015u}, have become a de facto standard for semantic segmentation in general and in medical image segmentation tasks in particular.
The U-Net architecture has been used for segmentation of both normal organs and lesions and achieved top ranking results in several international segmentation challenges  \cite{kuijf2019standardized, kaggle_salt, mrbrains18}.
Despite numerous applications of U-Nets, very few works have studied  the capability of these networks in capturing predictive uncertainty.
Predictive uncertainty or prediction confidence is described as the ability of a
decision-making system to provide an expectation of success (i.e. correct classification) or failure for the test examples at inference time.
Using a frequentist interpretation of uncertainty, predictions (i.e. class probabilities) of  a \textit{well-calibrated} model should match the probability of success of those inferences in the long run \cite{guo2017calibration}.
For instance, if a well-calibrated brain tumor segmentation model classifies 100 pixels each with the probability of 0.7 as cancer,  we expect 70 of those pixels to be correctly classified as cancer. However, a poorly calibrated model with similar classification probabilities is expected to result in many more or less correctly classified pixels.
Miscalibration frequently occurs in many modern neural networks 
(NNs) that are trained with advanced optimization methods\cite{guo2017calibration}.
Poorly-calibrated NNs are often highly confident in misclassification \cite{amodei2016concrete}.
In some applications, for example, medical image analysis, or automated driving, overconfidence can be dangerous.


The soft Dice loss function \cite{milletari2016v}, 
also known as Dice loss, is a generalized measure where the probabilistic output of a segmenter is compared to the training data,  set memberships are augmented with label probability,
and a smoothing factor is added to the denominator to make the loss function differentiable.
With the Dice loss, the model parameter set is chosen to minimize the negative of weighted Dice of different structures.
Dice loss is robust to class imbalance and has been successfully applied in many segmentation problems \cite{sudre2017generalised}.
Furthermore, Batch Normalization (BN) effectively stabilizes convergence and also improves performance of networks for natural image classification tasks \cite{ioffe2015batch}.
BN and Dice loss have made FCN optimization seamless.
The addition of BN to the U-Net has improved optimization and segmentation quality \cite{cciccek20163d}.
However, it has been reported that both BN and Dice loss have adverse effects on calibration quality \cite{guo2017calibration, sander2019towards, bertels2019optimization}.
Consequently, FCNs trained with BN and Dice loss do not produce well-calibrated probabilities leading to poor uncertainty estimation.
In contrast to Dice loss, cross-entropy loss provides better calibrated predictions and uncertainty estimates,
as it is a strictly proper scoring rule \cite{gneiting2007strictly}.
Yet, the use of cross-entropy as the loss function for training FCNs can be challenging in situations where there is a high class imbalance, e.g., where most of an image is considered background \cite{sudre2017generalised}.
Hence, it is of great significance and interest to study methods for confidence calibration of FCNs trained with BN and Dice loss.

\begin{figure}[t]
\centering
\setlength{\tabcolsep}{1pt}
\begin{tabular}{lcccc}
&
&\multicolumn{1}{c}{\scriptsize uncalibrated}
&\multicolumn{1}{c}{\scriptsize calibrated}&\\
\parbox[t]{3mm}{\rotatebox[origin=l]{90}{\footnotesize{$~~~~~~$(a)}}}&
\includegraphics[height=18mm]{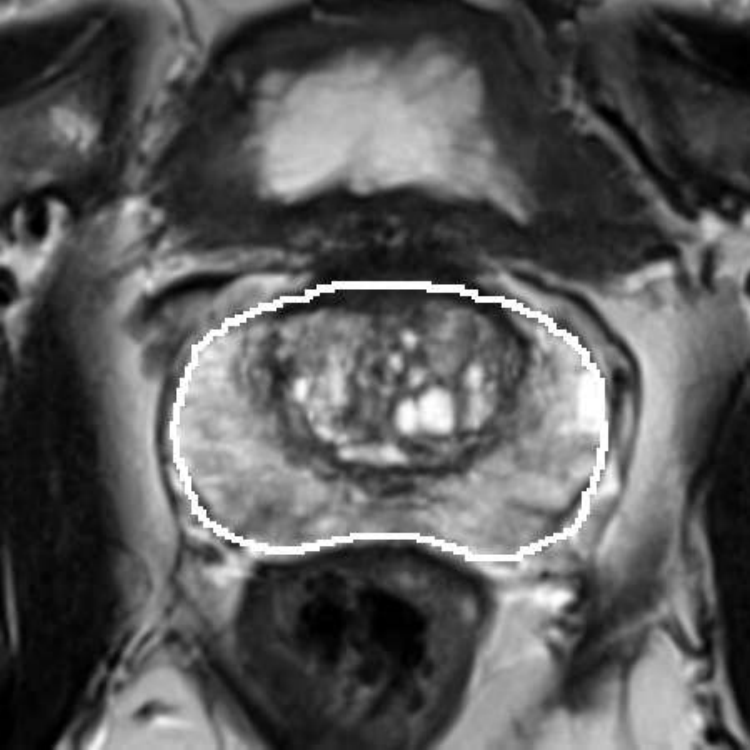}&
\includegraphics[height=18mm]{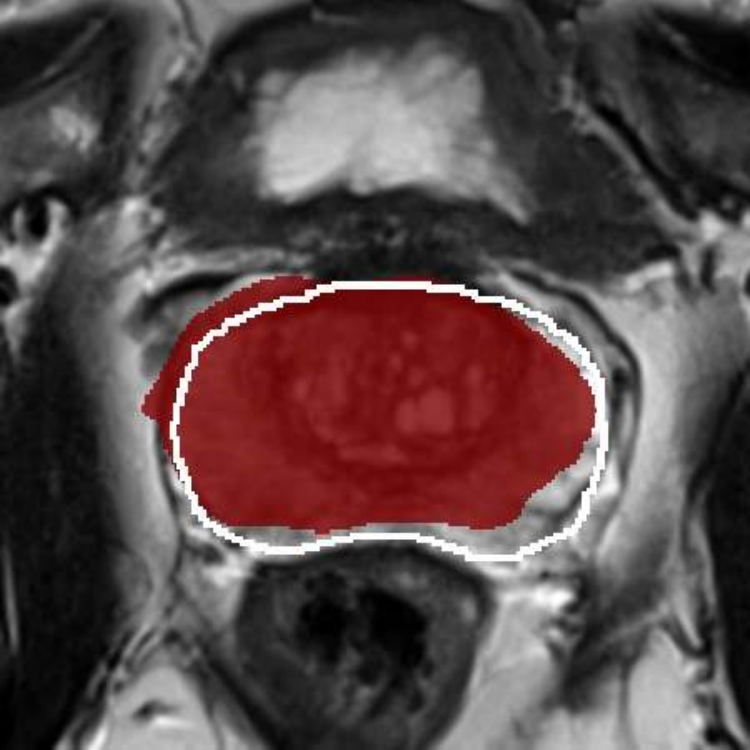}&
\includegraphics[height=18mm]{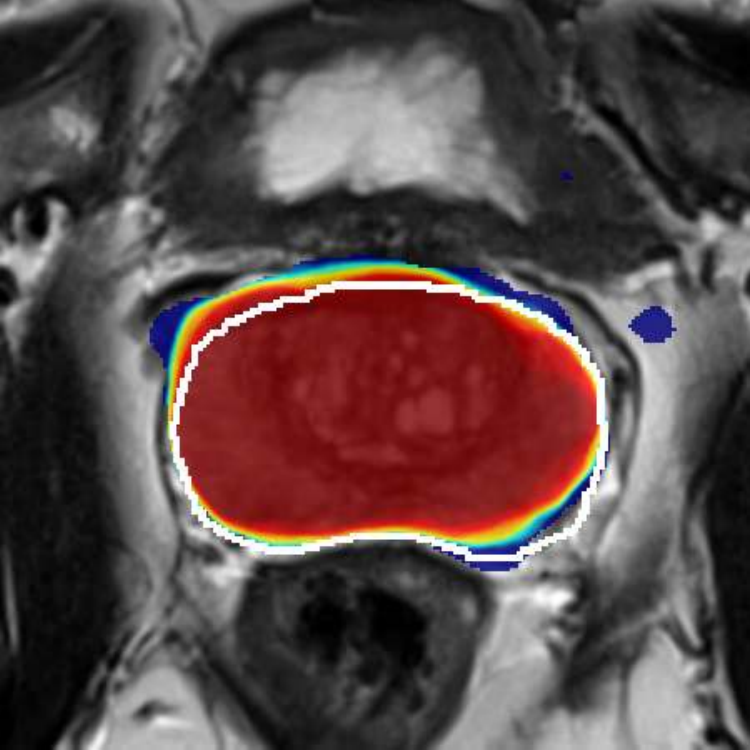}&
\includegraphics[height=18mm]{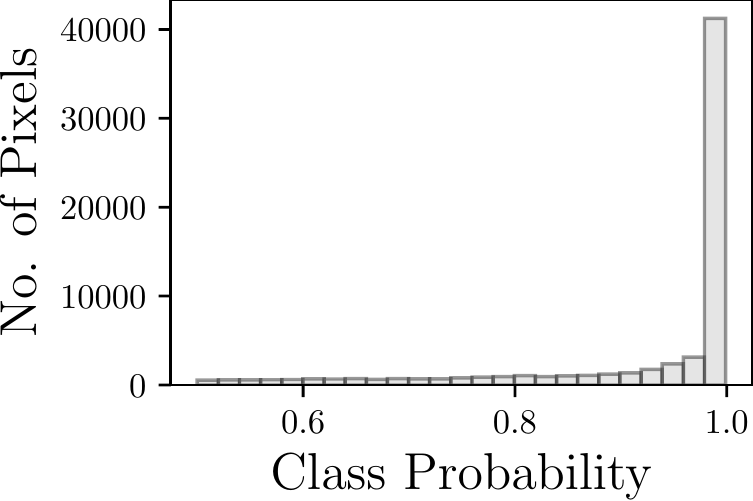}\\
\parbox[t]{3mm}{\rotatebox[origin=l]{90}{\footnotesize{$~~~~~~~$(b)}}}&
\includegraphics[height=18mm]{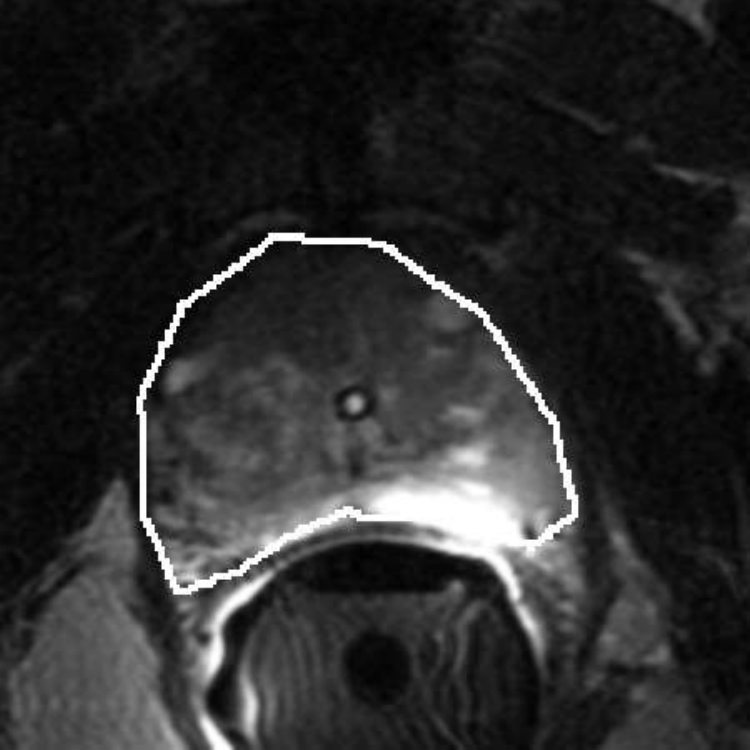}&
\includegraphics[height=18mm]{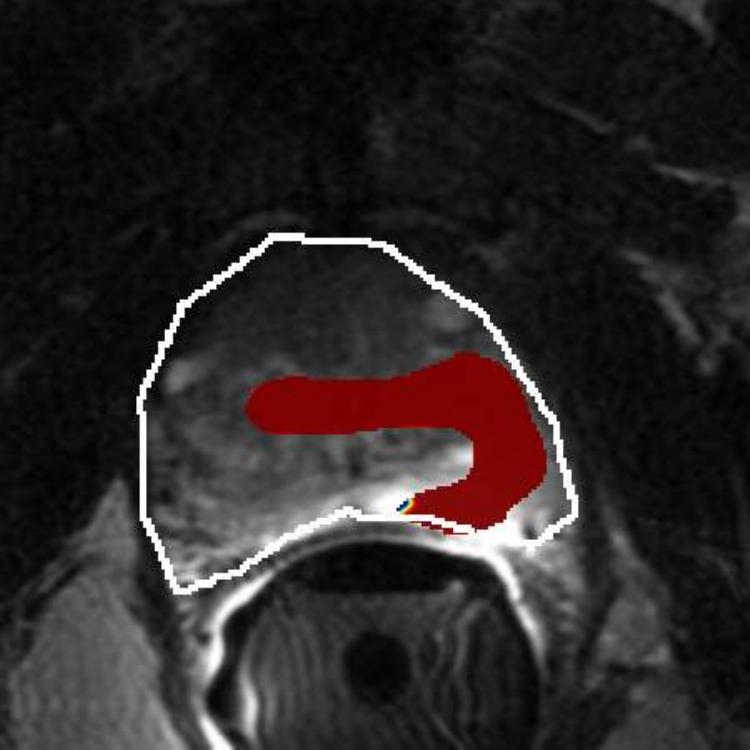}&
\includegraphics[height=18mm]{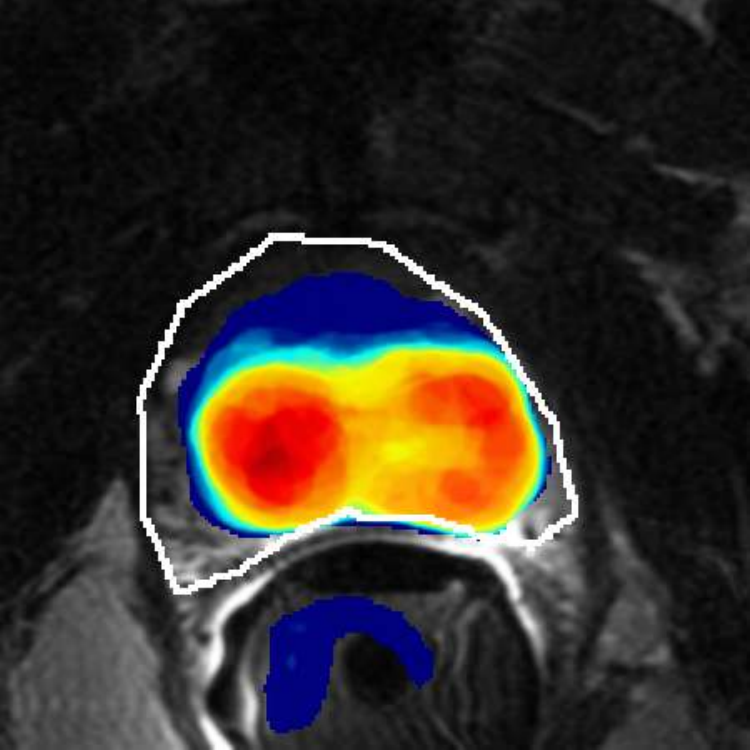}&
\includegraphics[height=18mm]{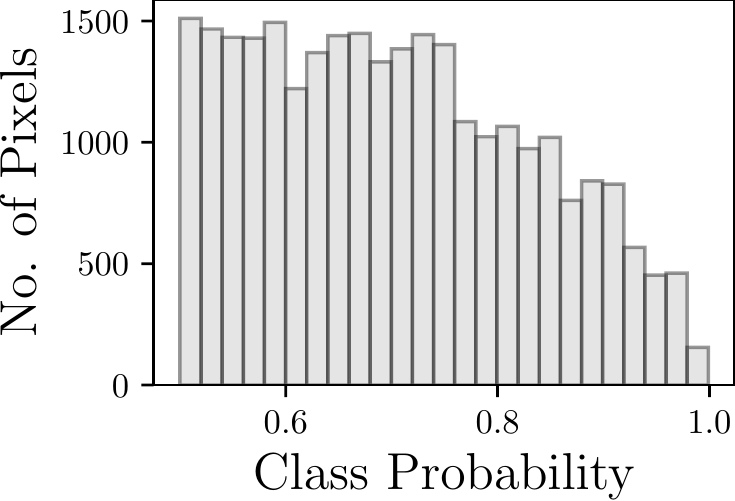}\\

\multicolumn{5}{c}{\includegraphics[width=50mm]{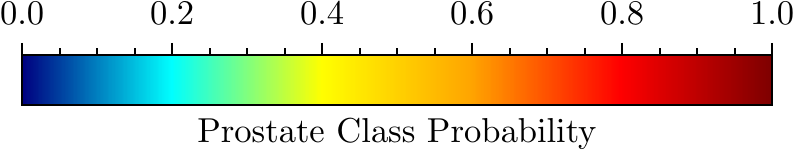}}
\end{tabular}
\caption{
\textbf{Calibration and out-of-distribution detection}.
Models for prostate gland segmentation were trained with T2-weighted MR images acquired using phased-array coils.
The results of inference are shown for two test examples imaged with: (a) phased-array coil (in-distribution example), and (b) endorectal coil (out-of-distribution example).
The first column shows T2-weighted MRI images with the prostate gland boundary drawn by an expert (white line).
The second column shows the MRI overlaid with uncalibrated segmentation predictions of an FCN trained with Dice loss.
The third column shows the calibrated segmentation predictions of an ensemble of FCNs trained with Dice loss.
The fourth column shows the histogram of the calibrated class probabilities over the predicted prostate segment of the whole volume. 
Note that the bottom row has a much wider distribution compared to the top row, indicating that this is an out of distribution example. 
In the middle column, prediction prostate class probabilities $\leq0.001$ has been masked out.
}
\label{fig:introduction}
\end{figure}

Another important aspect of uncertainty estimation 
is the ability of a predictive model to distinguish \textit{in-distribution} test examples (i.e. those similar to the training data) from \textit{out-of-distribution} test examples (i.e. those that do not fit the distribution of the training data) \cite{hendrycks2016baseline}.
The ability of the models to detect out-of-distribution inputs is specifically important for medical imaging applications as deep networks are sensitive to \textit{domain shift}, which is a recurring situation in medical imaging \cite{ghafoorian2017transfer}.
For instance, networks trained on one MRI protocol often do not perform satisfactorily on images obtained with slightly different parameters or out-of-distribution test images.
Hence, in the face of an out-of-distribution sample, an ideal model knows
and announces \textit{``I do not know''} and seeks human intervention -- if possible -- instead of a silent failure.
Figure \ref{fig:introduction} shows 
an example of out-of-distribution detection
from a U-Net model that was trained with BN and Dice loss for prostate gland segmentation before and after confidence calibration.

\section{Related Works}
There has been a recent growing interest in uncertainty estimation and confidence measurement with deep NNs.
Although most studies on uncertainty estimation have been done through Bayesian modeling of the NN, there has been some recent interest in using non-Bayesian approaches such as ensembling methods.
Here, we first briefly review Bayesian and non-Bayesian methods and then review the recent literature for uncertainty estimation for semantic segmentation applications.

In the Bayesian approach, the deterministic parameters of the NN are replaced by 
prior probability distributions.
Using Bayesian inference, given the data samples, a posterior probability distribution over the parameters is calculated. 
At inference time, instead of single scalar probability, the Bayesian NN gives  probability distributions over the output label probabilities \cite{mackay1992practical}, which models NN predictive uncertainty. 
Gal and Ghahramani  \cite{gal2015dropout} proposed to use dropout \cite{srivastava2014dropout} as a Bayesian approximation.
They proposed \textit{Monte Carlo dropout} (MC dropout) in which dropout layers are applied before every weight together with non-linearities.
The probabilistic Gaussian process is approximated at inference time by running the model several times with active dropout layers.
Implementing MC dropout is straightforward and has been applied in several application domains including medical imaging \cite{leibig2017leveraging}.
In a similar Bayesian approach, Teye et al. \cite{teye2018MCBN} showed that training NNs with BN \cite{ioffe2015batch} can be used to approximate inference of Bayesian NNs.
For networks with BN and without dropout, \textit{Monte Carlo Batch Normalization} (MCBN) can be considered an alternative to MC dropout. 
In another Bayesian work, Heo et al. \cite{heo2018uncertainty} proposed a method that allows the attention model to leverage uncertainty.
By learning the \textit{Uncertainty-aware Attention} (UA) with variational inference, they improved both model calibration and performance in attention models.
Seo et al. \cite{seo2019learning} proposed a variance-weighted loss function that enables learning single-shot calibration scores.
In combination with stochastic depth and dropout, their method can improve confidence calibration and classification accuracy.
Recently, Liao et al. \cite{liao2019modelling} proposed a method for modeling such uncertainty in intra-observer variability of 2D echocardiography  using the proposed cumulative density function probability method.

Non-Bayesian approaches have been proposed for probability calibration and uncertainty estimation.
Guo et al. \cite{guo2017calibration} studied the problem of confidence calibration in deep NNs.
Through experiments, they analyzed different parameters such as depth, width, weight decay, and BN and their effect on calibration.
They also used temperature scaling to easily calibrate trained models.
Ensembling has been used as an effective tool to improve classification performance
of deep NNs in several applications including medical image segmentation 
\cite{kamnitsas2017ensembles, mehrtash2018automatic}.
Following the success of ensembling methods \cite{dietterich2000ensemble} in improving baseline performance, Lakshminarayanan  proposed {\em Deep Ensembles}  in which model averaging was used to estimate predictive uncertainty \cite{lakshminarayanan2017simple}. 
By training collections of models with random initialization of parameters and adversarial training, they provided a simple approach to assess uncertainty. This observation motivated some of the experimental design in our work.
Unlike MC dropout, using Deep Ensembles does not require network architecture modification. In \cite{lakshminarayanan2017simple} authors showed that Deep Ensembles outperforms MC dropout on two image classification problems.
On the downside, Deep Ensembles requires retraining a model from scratch, which is computationally expensive for large datasets and complex models.

Predictive uncertainty estimation has been studied specifically for the problem of semantic segmentation with deep NNs.
Bayesian SegNet  \cite{kendall2015bayesian} was among the first that addressed uncertainty estimation in FCNs by using MC dropout.
They applied MC dropout by adding dropout layers after the pooling and upsampling blocks of the three innermost layers of the encoder and decoder sections of the SegNet architecture.
Using similar approaches for uncertainty estimation, Kwon et al. \cite{kwon2018uncertainty} and Sedai et al. \cite{sedai2018joint} 
used Bayesian NNs for uncertainty quantification in 
segmentation of ischemic stroke lesions and visualization of retinal layers, respectively.
Sander et al. \cite{sander2019towards} applied MC dropout to capture instance segmentation uncertainty in ambiguous regions and compared different loss functions in terms of the resultant miscalibration.
Kohl et al. \cite{kohl2018probabilistic} proposed a \textit{Probabilistic U-Net} that combined an FCN with a conditional variance autoencoder to provide multiple segmentation hypotheses for ambiguous images. 
In similar work, Hu et al. \cite{hu2019supervised} studied uncertainty quantification in the presence of multiple annotations as a result of inter-observer disagreement.
They used a probabilistic U-Net to quantify uncertainty in the segmentation of lung abnormalities.
Baumgartner et al. \cite{baumgartner2019phiseg} 
presented a probabilistic hierarchical 
model where separate latent variable are used for different resolutions and 
variational autoencoder is used for inference.
Rottmann and Schubert \cite{rottmann2019uncertainty} proposed a prediction quality rating method for segmentation of nested multi-resolution street scene images by measuring both pixel-wise and segment-wise measures of uncertainty as predictive metrics for segmentation quality.
Recently, Karimi et al. \cite{karimi2019accurate} used ensembling for uncertainty estimation of difficult to segment regions and used this information to improve clinical target volume estimation in prostate ultrasound images.
In another recent work, Jungo and Reyes \cite{jungo2019assessing} studied uncertainty estimation for brain tumor and skin lesion segmentation tasks.

In conjunction with uncertainty estimation and confidence calibration, several works have studied out-of-distribution detection \cite{hendrycks2016baseline, liang2017enhancing, lee2017training, devries2018learning, shalev2018out}.
In a non-Bayesian approach, Hendrycks and Gimpel \cite{hendrycks2016baseline} used softmax prediction probability baseline to effectively predict misclassificaiton and out-of-distribution in test examples. 
Liang et al. \cite{liang2017enhancing} used temperature scaling and input perturbations to enhance the baseline method of Hendrycks and Gimpel \cite{hendrycks2016baseline}. 
In the context of a generative NN scheme, Lee et al.\cite{lee2017training} used a loss function that encourages confidence calibration and this resulted in improvements in out-of-distribution detection. 
Similarly, DeVries and Taylor \cite{devries2018learning} proposed a hybrid with a confidence term to improve out-of-distribution detection.
Shalev et al. \cite{shalev2018out} used multiple semantic dense representations of the target labels to detect misclassified and adversarial examples.

\section{Contributions}
In this work, we study predictive uncertainty estimation for semantic segmentation with FCNs and propose ensembling for confidence calibration and reliable predictive uncertainty estimation of segmented structures.
In summary, we make the following contributions:
\begin{itemize}
    \item We analyze the choice of loss function for semantic segmentation in FCNs.
    We compare the two most commonly used loss functions in training FCNs for semantic segmentation: cross-entropy loss and Dice loss.
    We train models with these loss functions and compare the resulting segmentation quality and predictive uncertainty estimation.
    We observe that FCNs trained with Dice loss perform significantly better segmentation compared to those trained with cross-entropy but at the cost of poor calibration.\\
    \item 
    We propose model ensembling \cite{lakshminarayanan2017simple} for confidence calibration of FCNs trained with Dice loss and batch normalization.
    By training collections of FCNs with random initialization of parameters and random shuffling of training data, we create an ensemble that improves both segmentation quality and uncertainty estimation.
    We also compare ensembling with MC dropout \cite{gal2015dropout,kendall2015bayesian}.
    We empirically quantify the effect of the number of models on calibration and segmentation quality.
    \item 
    We propose to use average entropy over the predicted segmented object as a metric to predict segmentation quality of foreground structures, which can be further used to detect out-of-distribution test inputs. 
    Our results demonstrate that object segmentation quality 
    correlates inversely with the average entropy over the segmented object
    and can be used effectively for detecting out-of-distribution inputs.
    \item We demonstrate our method for uncertainty estimation and confidence calibration on three different segmentation tasks from MRI images of the brain, the heart, and the prostate.
    Where appropriate, we report the statistical significance of our findings.
\end{itemize}

\section{Applications \& Data}
Table \ref{datasets} shows the number of patient images in each dataset and how we split these into training, validation, and test sets.
In the following subsections, we briefly describe each segmentation task, data characteristics, and pre-processing.

\begin{table}[h]
\centering
\caption{Number of patients for training, validation, and test sets used in this study.
\label{datasets}
}
\begin{threeparttable}
\begin{tabular}{l|cc|c|cc}
\toprule
Application& \multicolumn{2}{c|}{Brain} & Heart &\multicolumn{2}{c}{Prostate}\\
\midrule
Dataset & CBICA & TCIA & ACDC 
& PROSTATEx 
& PROMISE12{$^\dagger$}\\
\midrule
\# Training & 66 & $-$ & 40 & 16 & $-$ \\
\midrule
\# Validation & 22 & $-$ & 10 & 4 & $-$\\
\midrule
\# Test & $-$ & 102 & 50 & 20 & 35 \\
\bottomrule

\end{tabular}
\begin{tablenotes}
    \item[$\dagger$] Used only for out-of-distribution detection experiments. 
\end{tablenotes}
\end{threeparttable}
\end{table}

\subsection{Brain Tumor Segmentation Task}
For brain tumor segmentation, data from the MICCAI 2017 BraTS challenge \cite{bakas2017advancing,menze2015multimodal} was used. 
This is a four-class segmentation task; multiparametric MRI of brain tumor patients are to be segmented into enhancing tumor, non-enhancing tumor, edema, and background.
The training dataset consists of 190 multiparametric MRI (T1-weighted, contrast-enhanced T1-weighted, T2-weighted, and FLAIR sequences) from brain tumor patients.
The dataset is further subdivided into two sets: CBICA and TCIA.
The images in CBICA set were acquired at the Center for Biomedical Image Computing and Analytics (CBICA) at the University of Pennsylvania \cite{bakas2017advancing}. 
The images in the TCIA set were acquired across multiple institutions and hosted by the National Cancer Institute, The Cancer Imaging Archive (TCIA).
The CBICA subset was used for training and validation and the TCIA subset was reserved as the test set.

\subsection{Ventricular Segmentation Task}
For heart ventricle segmentation, data from the MICCAI 2017 ACDC challenge for automated cardiac diagnosis was used \cite{wolterink2017automatic}. 
This is a four-class segmentation task; cine MR images (CMRI) of patients are to be segmented into the left ventricle, the myocardium, the right ventricle, and the background.
This dataset consists of end-diastole (ED) and end-systole (ES) images of 100 patients. 
We used only the ED images in our study. 

\subsection{Prostate Segmentation Task}
For prostate segmentation, the public datasets, PROSTATEx \cite{Litjens2014-prostatex} and PROMISE12 \cite{litjens2014evaluation} were used. This is a two-class segmentation task; Axial T2-weighted images of men suspected of having prostate cancer are to be segmented into the prostate gland and the background. 
For PROSTATEx dataset, 40 images with annotations from Meyer et al. \cite{Meyer2018-lo} were used.  All these images were acquired at the same institution.
PROSTATEx dataset was used for both training and testing purposes, and PROMISE12 dataset was set aside for test only.
PROMISE12 dataset is a heterogeneous multi-institutional dataset acquired using different MR scanners and acquisition parameters.
We used the 50 training images for which ground truth is available.

\subsection{Data Pre-processing} 
Prostate and cardiac images were resampled to the common in-plane resolution of $0.5 \times 0.5$ mm and $2 \times 2$~mm, respectively. 
Brain images were resampled to the resolution of $1 \times 1 \times 2$~mm. 
All axial slices were then cropped at the center to create images of size $224 \times 224$ pixels as the input size of the FCN.
Image intensities were normalized to be within the range of [0,1]. 

\section{Methods}
\label{sec:Methods}
\subsection{Model}
\label{sec:Methods:Model}
Semantic segmentation can be formulated as a pixel-level classification problem, which can be solved by convolutional neural networks \cite{litjens2017survey}.
The pixels in the training image and label pairs can be considered as N i.i.d data points $\mathcal{D}=\{\bm{x}_n, y_n\}_{n=1}^N$, where $\bm{x}\in\mathbb{R}^M$ is the M-dimensional input and $y$ can be one and only one of the $k$ possible classes $k \in \{1, ..., K\}$.
The use of FCNs for image segmentation allows for end-to-end learning, with each pixel of the input image being mapped by the FCN to the output segmentation map.
Compared to FCNs, patch-based  NNs are much slower at inference time as they require sliding window mechanisms for predicting each pixel \cite{long2015fully}.
Moreover, it is more straightforward to implement segment-level loss functions such as Dice loss in FCN architectures.
FCNs for segmentation usually consist of an encoder (contracting) path and a decoder (expanding) path \cite{long2015fully,ronneberger2015u}.
FCNs with skip-connections are able to combine high level abstract features with low-level high-resolution features, which has been shown to be successful in segmentation tasks \cite{ronneberger2015u, cciccek20163d}.
NNs can be formulated as parametric conditional probability models, $p(y_j|x_j, \theta)$, and the parameter set $\theta$ is chosen to minimize a loss function.
Both cross-entropy (CE) and negative of Dice Similarity Coefficient (DSC), known as Dice loss, have been used as loss functions for training FCNs.
Class weights are used for optimization convergence and dealing with the class imbalance issue.
With CE loss, parameter set is chosen to maximize the average log-likelihood over training data: 

\begin{equation}
   \mathcal{L}_{CE} = - \frac{1}{N} \sum_{i=1}^N \sum_{k=1}^K
    \omega_{k} \ln{(p(\hat{y}_i= k| x_i, \theta))} \cdot \left(y_i = k \right),
\end{equation}
where $p(\hat{y}_i= k| x_i, \theta)$ is the probability of pixel $i$ belonging to class $k$,
$(y_i = k)$ is the binary indicator which denotes if the class label k is the correct class of $i$th pixel, $\omega_k$ is the weight for class $k$, and $N$ is the number of pixels that are used in each mini-batch. 

With the Dice loss, the parameter set is chosen to minimize the negative of weighted Dice of different structures:
\begin{equation}
   \mathcal{L}_{DSC} = -2\sum_{k=1}^{K}\frac{\omega_{k}{\sum_{i=1}^N \left[ p(\hat{y}_i=k| x_i, \theta) \cdot \left( y_i = k \right) \right] }}
   {\sum_{i=1}^{N} \left[ p(\hat{y}_i=k| x_i, \theta) +  \left( y_i = k \right) \right]  + \epsilon},
\end{equation}
where $p(\hat{y}_i= k| x_i, \theta)$ is the probability of pixel belonging to class $k$,
$(y_i = k)$ is the binary indicator which denotes if the class label k is the correct class of $i$th pixel, $\omega_k$ is the weight for class $k$, $N$ is the number of pixels that are used in each mini-batch, and $\epsilon$ is the smoothing factor to make the loss function differentiable. Subsequently, $p(y_i|x_i, \theta^{*})$ is used for inference, where $\theta^{*}$ is the optimized parameter set. 

\subsection{Calibration Metrics}
The output of an FCN for each input pixel is a class prediction $\hat{y_j}$ and its associated class probability $p(y_j|x_j,\theta)$.
The class probability can be considered the model confidence or probability of correctness and can be used as a measure for predictive uncertainty  at the pixel level.
Strictly proper scoring rules are used to assess the calibration quality of predictive models \cite{gneiting2007strictly}.
In general, scoring rules assess the quality of uncertainty estimation in models by awarding well-calibrated probabilistic forecasts.
Negative log-likelihood (NLL), and Brier score \cite{brier1950verification}, are both strictly proper scoring rules that have been previously used in several studies for evaluating predictive uncertainty \cite{guo2017calibration, lakshminarayanan2017simple, gal2015dropout}.
In a segmentation problem, for a collection of $N$ pixels, NLL is calculated as:

\begin{equation}
   NLL = - \frac{1}{N} \sum_{i=1}^N 
   \sum_{k=1}^K
    \ln{(p(\hat{y}_i=y_k| x_i, \theta))} \cdot \left( \hat{y}_i = y_k \right)
\end{equation}

Brier score (Br) measures the accuracy of probabilistic predictions:
\begin{equation}
   Br =  \frac{1}{N} \sum_{i=1}^N 
   \frac{1}{K} \sum_{k=1}^K 
    \left[p(\hat{y}_i=y_k| x_i, \theta) - \left( \hat{y}_i = y_k \right)\right]^2
\end{equation}

In addition to NLL and Brier score, we directly assess the predictive power of a model by analyzing test examples confidence values versus their measured expected accuracy values.
To do so, we use reliability diagrams as visual representations of model calibration and Expected Calibration Error (ECE) as summary statistics for calibration \cite{guo2017calibration,naeini2015obtaining}.
Reliability diagrams plot expected accuracy as a function of class probability (confidence).
The reliability diagram of a perfectly calibrated model is the identity function.
For expected accuracy measurement, the samples are binned into N groups and the accuracy and confidence for each group are computed. Assuming $D_m$ to be indices of samples whose confidence predictions are in the range of $ \left( \frac{m-1}{M}, \frac{m}{M} \right]$, the expected accuracy of the $D_m$ is ${Acc}(D_m) = \frac{1}{|D_m|}\sum_{i \in D_m} \bm{1}(\hat{y}_i = y_i)$.
The average confidence on bin $D_{m}$ is calculated as
$\overline{P}(D_m) = \frac{1}{|D_m|}\sum_{i \in D_m}p(\hat{y}_i=y_i| x_i, \theta)$.
ECE is calculated by summing up the weighted average 
of the differences between accuracy and the average confidence over the bins: 

\begin{equation}
    \text{ECE} = \sum_{m=1}^{M}\frac{|D_m|}{N} \abs*{ACC(D_m)-\overline{P}(D_m)},
\end{equation}
where $N$ is the total number of samples. In other words, ECE is the average of gaps on the reliability diagram.

\subsection{Confidence Calibration with Ensembling}

We propose to empirically determine whether ensembling  \cite{dietterich2000ensemble} results in confidence calibration of otherwise poorly calibrated FCNs trained with Dice loss. 
To this end, similar to the Deep Ensembles method \cite{lakshminarayanan2017simple}, we train $M$ FCNs with random initialization of the network parameters and random shuffling of the training dataset in mini-batch stochastic gradient descent. 
However, unlike the Deep Ensemble method, we do not use any form of adversarial training.
We train each of the $M$ models in the ensemble from scratch and then compute the probability of the ensemble $p_E$ as the average of the baseline probabilities as follows: 

 \begin{equation}
\label{eq:ensembling}
p_E(y_j=k|x_j) = \frac{1}{M} \sum_{m=1}^{M} p(y_j=k| x_j, \theta^{*}_m), 
\end{equation}
where $p(y_i=k|x_i, \theta^{*}_m)$ are the individual probabilities.

\subsection{Segment-level Predictive Uncertainty Estimation}
For segmentation applications, besides the pixel-level confidence metric, it is desirable to have a confidence metric that captures model uncertainty at the segment-level.
Such a metric would be very useful in clinical applications for decision making.
For a well-calibrated system, we anticipate that a segment-level confidence metric can predict the segmentation quality in the absence of ground truth. 
The metric can be used to detect out-of-distribution samples and hard or ambiguous cases.
Such metrics have been previously proposed for street scene segmentation \cite{rottmann2019uncertainty}.
Given the pixel-level class predictions $\hat{y}_i$ and their associated ground truth class $y_i$ for a predicted segment $\hat{\mathcal{S}}_k = \{s \in (x_i, \hat{y}_i) | \hat{y}_i=k \}$, we propose to use the average of pixel-wise entropy values over the predicted foreground \footnote{Following the convention in the semantic segmentation literature, we are using foreground and background labels regardless  of  the  fact  that  the  problem  is  binary  or  k-class  segmentation \cite{long2015fully}.} segment $\hat{\mathcal{S}}_k$ as a scalar metric for volume-level confidence of that segment as:

\begin{dmath}
\label{eq:avg_entr}
\overline{\mathcal{H}(\hat{\mathcal{S}}_k)} = - \frac{1}{\left|\hat{\mathcal{S}}_k\right|} \sum_{i\in \hat{\mathcal{S}}_k} [p(\hat{y}_i=k| x_i,\theta)\cdot \ln{\left(p(\hat{y}_i=k|x_i,\theta)\right)} + 
     \left(1- p(\hat{y}_i=k| x_i,\theta) \right) \cdot \ln{\left(1-p(\hat{y}_i=k|x_i,\theta)\right)}].
\end{dmath} 

In calculating the average entropy of $\hat{\mathcal{S}}_k$, we assumed binary classification: the probability of belonging to class $k$, $p(\hat{y}_i=k| x_i, \theta)$ and the probability of belonging to other classes $1 - p(\hat{y}_i=k| x_i, \theta)$.

\section{Experiments}
\subsection{Training Baselines}
For all of the experiments, we used a baseline FCN model similar to the two-dimensional U-Net architecture \cite{ronneberger2015u} but with fewer kernel filters at each layer.
The input and output of the FCN have a size of $224 \times 224$ pixels. 
Except for the brain tumor segmentation that used a three-channel input (T1CE, T2, FLAIR), for the rest of the problems the input was a single channel.
The network has the same number of layers as the original U-Net but with fewer kernels.
The number of kernels for the encoder section of U-Net were 8, 8, 16, 16, 32, 32, 64, 64, 128, and 128.
The parameters of the convolutional layers were initialized randomly from a Gaussian distribution \cite{he2015delving}.
For each of the three segmentation problems, the model was trained 100 times with CE and 100 times with Dice loss, each with random weight initialization and random shuffling of the training data.
For the models that were trained with Dice loss, the softmax activation function of the last layer was substituted with sigmoid function as it improved the convergence substantially.
For CE loss, class weights $\omega_k$, were  calculated  as  inverse  frequencies of  the  class  labels  for  the  combined  pixels  in  training  and validation  data.  
For Dice loss, uniform  class  weights, $\omega_k$, were used for all the foreground segments, except for the myocardium  class  in  heart  segmentation  where  the  class weight  was  three  times  the  other  two  foreground  classes.
For optimization, stochastic gradient descent with the Adam update rule \cite{kingma2014adam} was used.  
During the training, we used a mini-batch of 16 examples for prostate segmentation and 32 examples for brain tumor and cardiac segmentation tasks.
The initial learning rate was set to $0.005$ and it was reduced by a factor of $0.5-0.8$ if the average validation Dice score did not improve by $0.001$ in 10 epochs. 
We used 1000 epochs for training the models with an early stopping policy.
For each run, the model checkpoint was saved at the epoch where the validation DSC was the highest.

\subsection{Cross-entropy vs. Dice}
CE loss aims to minimize the average negative log-likelihood over the pixels, while Dice loss improves segmentation quality in terms of the Dice coefficient directly. 
As a result, we expect to observe models trained with CE to achieve a lower NLL and models trained with Dice loss to achieve better Dice coefficients.
Here, our main focuses are to observe the segmentation quality of a model that is trained with CE in terms of Dice loss and the calibration quality of a model that was trained with Dice loss.
We compare models trained with CE with those trained with Dice on three segmentation tasks.

\subsection{MC dropout}
MC dropout was implemented by modifying the baseline network as it was done in  Bayesian SegNet \cite{kendall2015bayesian}.
Dropout layers were added to the three inner-most encoder and decoder layers with a dropout probability of 0.5. 
At inference time, Monte Carlo sampling was done with 50 samples and the mean of the samples was used as the final prediction.

\subsection{Confidence Calibration}
We used ensembling (Equation \ref{eq:ensembling}) to calibrate batch normalized FCNs trained with Dice loss.
For the three segmentation problems, we made ensemble predictions and compared them with baselines in terms of calibration and segmentation quality.
For calibration quality, we compared NLL, Brier score, and ECE\%.
For segmentation quality, we compared dice and $95^{th}$ percentile Hausdorff distance.
Moreover, for calibration quality assessment we calculated the metrics on two sets of samples from the held-out test datasets: 1) the whole test dataset (all pixels of the test volumes) 2) pixels belonging to dilated bounding boxes around the foreground segments. 
The foreground segments and the adjacent background around them usually have the highest uncertainty and difficulty.
At the same time, background pixels far from foreground segments show less uncertainty but outnumber the foreground pixels. 
Using bounding boxes removes most of the correct (certain) background predictions from the statistics and will lead to a better highlighting of the differences among models.
For all three problems, we constructed bounding boxes of the foreground structures.
The boxes are then dilated by 8 mm in each direction of the in-plane axes and 2 slices (which translates to 4mm to 20mm) in each direction of the out-of-plane axis.

We also measured the effect of ensembles by calculating $p_{E}(y|x)$ (Equation \ref{eq:ensembling}) for ensembles with number of models ($M$) of 1, 2, 5, 10, 25, and 50.
To provide better statistics and reduce the effect of chance in reporting the performance, for each ensemble, we sampled the 100 baseline models $n$ times and reported the averages and 95\% CI of the NLL and Dice.
For example, for M=50, instead of reporting the means of NLL and Dice on a single set of 50 models (out of the 100 trained models), we sampled $n$ sets of 50 models and reported the averages and 95\% CI of the NLL and Dice.
For prostate and heart segmentation tasks $n$ was set to 50 and for brain tumor segmentation $n$ was set to 10.

\subsection{Segment-level Predictive Uncertainty}
For each of the segmentation problems, we calculated volume-level confidence for each of the foreground labels and $\overline{\mathcal{H(\hat{\mathcal{S})}}}$ (Equation \ref{eq:avg_entr}) vs. Dice.
For prostate segmentation, we are also interested in observing the difference between the two datasets of PROSTATEx test set (which is the same as the source domain) and PROMISE-12 set (which can be considered as a target set).

Finally, in all the experiments, for statistical tests and calculating 95\% confidence intervals (CI), we used bootstrapping (n=100). 
P-values of less than 0.01 were regarded as statistically significant.
In all the presented tables, boldfaced text indicates the best results for each instance and shows that the differences are statistically significant.

\begin{table*}[t]
\centering
\caption{
Calibration quality and segmentation performance for baselines trained with cross-entropy ($\mathcal{L}_{CE}$)  are compared with those that were trained with Dice loss ($\mathcal{L}_{DSC}$) and those that were calibrated with ensembling (M=50) and MC dropout. 
Boldfaced font indicates the best results for each application (model) and shows that the differences are statistically significant.
} 
\begin{threeparttable}
\begin{tabular}{l|ccc|ccc}
\toprule
& \multicolumn{3}{c}{Calibration Quality $^{\dagger}$}
& \multicolumn{3}{c}{Segmentation Performance (Average Dice Score (95\% CI)) $^{\ddagger}$}\\

\midrule
Application (Model) & NLL (95\% CI) & Brier (95\% CI)  & ECE\%  (95\% CI) 
& Segment I$^*$ & Segment II$^*$ & Segment III$^*$ \\
\midrule
Brain (${\mathcal{L}_{CE}}$) & 
0.52 (0.16$-$1.66) & 
0.23 (0.08$-$0.62) & 
8.11 (1.54$-$26.23) & 
0.37 (0.00$-$0.84)  & 
0.47 (0.07$-$0.82)  & 
0.58 (0.03$-$0.87)  
\\

Brain (MCDO ${\mathcal{L}_{CE}}$) & 
0.81 (0.16$-$2.62) & 
0.36 (0.08$-$0.92) & 
13.41 (0.80$-$43.26)  &
0.34 (0.00$-$0.81)  & 
0.34 (0.03$-$0.76)  & 
0.54 (0.02$-$0.86) \\ 

Brain (EN ${\mathcal{L}_{CE}}$) & 
\textbf{0.29 (0.11$-$0.71)} & 
0.15 (0.05$-$0.40) & 
\textbf{3.28 (0.52$-$10.06)} & 
0.49 (0.00$-$0.92)  & 
0.59 (0.11$-$0.86)  & 
0.68 (0.04$-$0.91)   
\\
Brain ($\mathcal{L}_{DSC}$)& 
0.62 (0.17$-$2.70) & 
0.23 (0.06$-$0.55) & 
13.20 (2.60$-$33.55) & 
0.45 (0.00$-$0.89)  & 
0.60 (0.10$-$0.90)  & 
0.67 (0.07$-$0.91)   
\\
Brain (MCDO ${\mathcal{L}_{DSC}}$)&
1.14 (0.28$-$4.04) & 
0.18 (0.06$-$0.49) & 
8.96 (2.41$-$23.87) & 
0.43 (0.00$-$0.88)  & 
0.58 (0.08$-$0.89)  & 
0.64 (0.03$-$0.91)  
\\

Brain (EN ${\mathcal{L}_{DSC}}$) & 
0.31 (0.16$-$0.78) & 
\textbf{0.14 (0.08$-$0.35)} & 
3.71 (0.94$-$15.27) & 
\textbf{0.51 (0.00$-$0.93)}  & 
\textbf{0.66 (0.11$-$0.91)}  & 
\textbf{0.74 (0.16$-$0.92)} 
\\

\midrule
Heart (${\mathcal{L}_{CE}}$)& 
0.36 (0.16$-$1.18) & 
0.17 (0.09$-$0.41) & 
5.75 (1.42$-$17.99) & 
0.77 (0.17$-$0.91)  & 
0.73 (0.45$-$0.86)  & 
0.91 (0.63$-$0.97)  
\\

Heart (MCDO ${\mathcal{L}_{CE}}$)& 
0.36 (0.17$-$1.10) & 
0.17 (0.09$-$0.41) & 
5.70 (1.39$-$17.93) & 
0.78 (0.27$-$0.90)  & 
0.73 (0.47$-$0.86)  & 
0.92 (0.64$-$0.97)  \\ 

Heart (EN ${\mathcal{L}_{CE}}$) & 
\textbf{0.23 (0.13$-$0.58)} & 
\textbf{0.13 (0.07$-$0.30)} & 
\textbf{2.51 (0.58$-$10.15)} & 
0.81 (0.18$-$0.93)  & 
0.77 (0.56$-$0.88)  & 
\textbf{0.93 (0.79$-$0.97)}  
\\

Heart ($\mathcal{L}_{DSC}$) & 
0.62 (0.17$-$2.70) & 
0.23 (0.06$-$0.55) & 
13.20 (2.60$-$33.55) & 
0.84 (0.14$-$0.96)  & 
0.81 (0.49$-$0.90)  & 
0.92 (0.64$-$0.97)   
\\

Heart (MCDO ${\mathcal{L}_{DSC}}$) &
0.41 (0.17$-$1.51) & 
0.45 (0.11$-$0.81) & 
36.79 (6.17$-$70.58) &  
0.84 (0.12$-$0.96)  & 
0.78 (0.04$-$0.89)  & 
0.91 (0.61$-$0.97)   
\\

Heart (EN ${L}_{DSC}$) & 
0.31 (0.16$-$0.78) & 
0.14 (0.08$-$0.35) & 
3.71 (0.94$-$15.27) & 
\textbf{0.87 (0.12$-$0.96)}  & 
\textbf{0.83 (0.59$-$0.91)}  & 
\textbf{0.93 (0.71$-$0.98)}  
\\

\midrule
Prostate ($\mathcal{L}_{CE}$)& 
0.40 (0.22$-$0.79) & 
0.25 (0.13$-$0.47) & 
8.08 (1.60$-$25.50) & 
%
0.83 (0.62$-$0.91)  & 
$-$ & $-$
\\

Prostate ( MCDO $\mathcal{L}_{CE}$)& 
0.30 (0.14$-$0.69) & 
0.16 (0.08$-$0.30) & 
5.23 (0.70$-$12.75) & 
0.77 (0.49$-$0.89) & 
$-$ & $-$ 
\\

Prostate (EN ${\mathcal{L}_{CE}}$) & 
0.16 (0.13$-$0.25) & 
0.09 (0.06$-$0.16) & 
4.12 (1.92$-$7.04) & 
0.87 (0.68$-$0.92)  & 
$-$ & $-$ 
\\
Prostate (${\mathcal{L}_{DSC}}$)& 
0.74 (0.31$-$1.60) & 
0.11 (0.06$-$0.27) & 
5.72 (3.20$-$12.57) & 
0.88 (0.72$-$0.93)  & 
$-$ & $-$ 
\\

Prostate (MCDO ${\mathcal{L}_{DSC}}$) &
0.48 (0.22$-$1.03) & 
0.11 (0.07$-$0.25) & 
5.23 (2.75$-$11.60) & 
0.86 (0.67$-$0.93)  & 
$-$ & $-$ 
\\

Prostate (EN ${\mathcal{L}_{DSC}}$)& 
\textbf{0.15 (0.07$-$0.25)} & 
\textbf{0.07 (0.04$-$0.14)} & 
\textbf{2.02 (0.48$-$3.89)} & 
\textbf{0.90 (0.76$-$0.95)} & 
$-$ & $-$ \\
\bottomrule
\end{tabular}
\begin{tablenotes}
		\item[$\dagger$] 
		The presented calibration quality metrics are calculated for bounding boxes. For whole volume results see Table 
		I
		of the Supplementary Material.
		\item[$\ddagger$] 
		Comparison between Hausdorff distance of different models is provided in Table 
		II
		of the Supplementary Material.
    \item[*] 
    For brain application segments, I, II, and III correspond to non-enhancing tumor, edema, and enhancing tumor, respectively.
    For heart application segments, I, II, and III correspond to the right ventricle, the myocardium, and the left ventricle, respectively.
    For prostate application segment I corresponds to the prostate gland.
\end{tablenotes}
\end{threeparttable}
\label{tab:confidence_calibration_and_segmentation}
\end{table*}

\section{Results}

Table \ref{tab:confidence_calibration_and_segmentation} compares 
the calibration quality and segmentation performance of baselines and ensembles (M=50) trained with CE loss with those that were trained with Dice loss and those that were calibrated with MC dropout.
The averages and 95\% CI values for NLL, Brier score, and ECE\% for the bounding boxes around the segments are provided.
Table \ref{tab:confidence_calibration_and_segmentation} also compares the averages and 95\% CI values of Dice coefficients of foreground segments for baselines trained with cross-entropy loss, Dice loss, and baselines calibrated with ensembling (M=50) for the whole volume.
Calibration quality results for whole volumes
and segmentation quality results in terms of Hausdorff distances
are provided in Tables 
I
and 
II
of the Supplementary Material, respectively.
For all tasks across all segments, in terms of segmentation performance, baselines trained with Dice loss outperform those trained with CE loss and ensembles of models trained with Dice loss outperform all the other models.
For all three segmentation tasks, calibration quality was significantly better in terms of NLL and ECE\% for baseline (single) models trained with CE comparing to those that were trained with Dice loss.
However, the direction of change for Brier score was not consistent among models trained with CE vs models trained with Dice loss. 
For bounding boxes of brain tumor and prostate segmentation, the Brier scores were significantly better for models trained with Dice loss compared to those trained with CE, while in the case of the heart segmentation was the opposite.
The ensemble models show significantly better calibration qualities for all metrics across all tasks.
In all cases ensembling outperformed baselines and  MC dropout models in terms of calibration quality.
We observe that ensembling improves the calibration quality of the models trained with Dice loss significantly.
MC dropout consistently improves the calibration quality of the models trained with Dice loss. 
However, for models trained with CE loss, MC dropout only improves the calibration quality of prostate application models and not brain and heart applications.

\begin{figure}
	\centering
	\begin{tabular}{lll}
	\includegraphics[height=28mm]{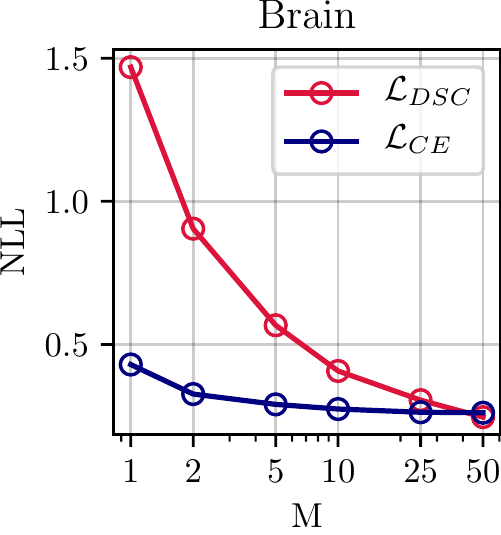}&
	\includegraphics[height=28mm]{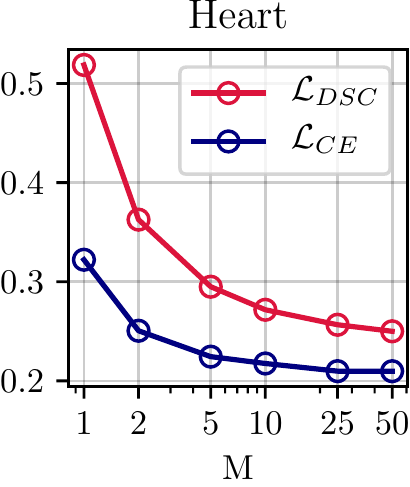}&
	\includegraphics[height=28mm]{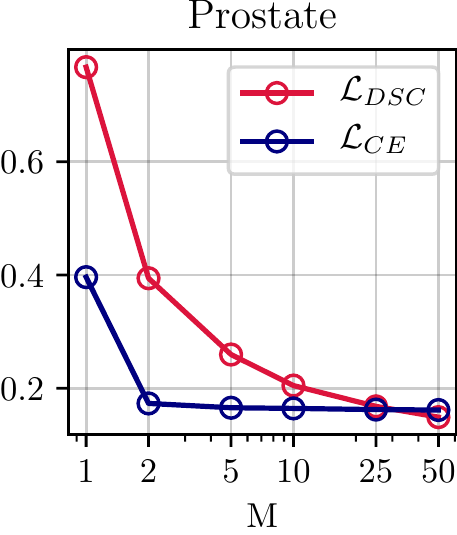}\\
	\end{tabular}
\caption{Improvements in calibration as a function of the number of models in the ensemble for baselines trained with cross-entropy and Dice loss functions.
Calibration quality in terms of NLL improves as number of models $M$ increases for prostate, heart, and brain tumor segmentation.
For each task an ensemble of size M=10 trained with Dice loss outperforms the baseline model (M=1) trained with cross-entropy in terms of NLL.
Same plot with 95\% CIs and for both whole volume and bounding box measurements are given in Figure
4
of the Supplementary Material.
}
\label{fig:n_ensembles}
\end{figure}

The graphs in Figure \ref{fig:n_ensembles} show the quantitative improvement in the calibration and segmentation as a function of the number of models in the ensemble, for each of the three segmentation applications of the prostate, the heart, and the brain tumors.
As we see, for the prostate, the heart, and the brain tumor segmentation, using even five ensembles (M=5) of baselines trained with Dice loss can reduce the NLL by about $66\%$, $44\%$, and $62\%$, respectively.
Qualitative examples for improvement as a function of number of models in ensemble are provided in the Supplementary Material Figures 
5
and 
6.

\begin{figure*}[t]
\centering
\setlength{\tabcolsep}{1pt}
  \begin{tabular}{ccc}
  \small{~~~(A) Prostate Segmentation} &
  \small{~~~~(B) Brain Tumor Segmentation} &
  \small{~~~(C) Cardiac Segmentation} \\
  \includegraphics[width=58mm]{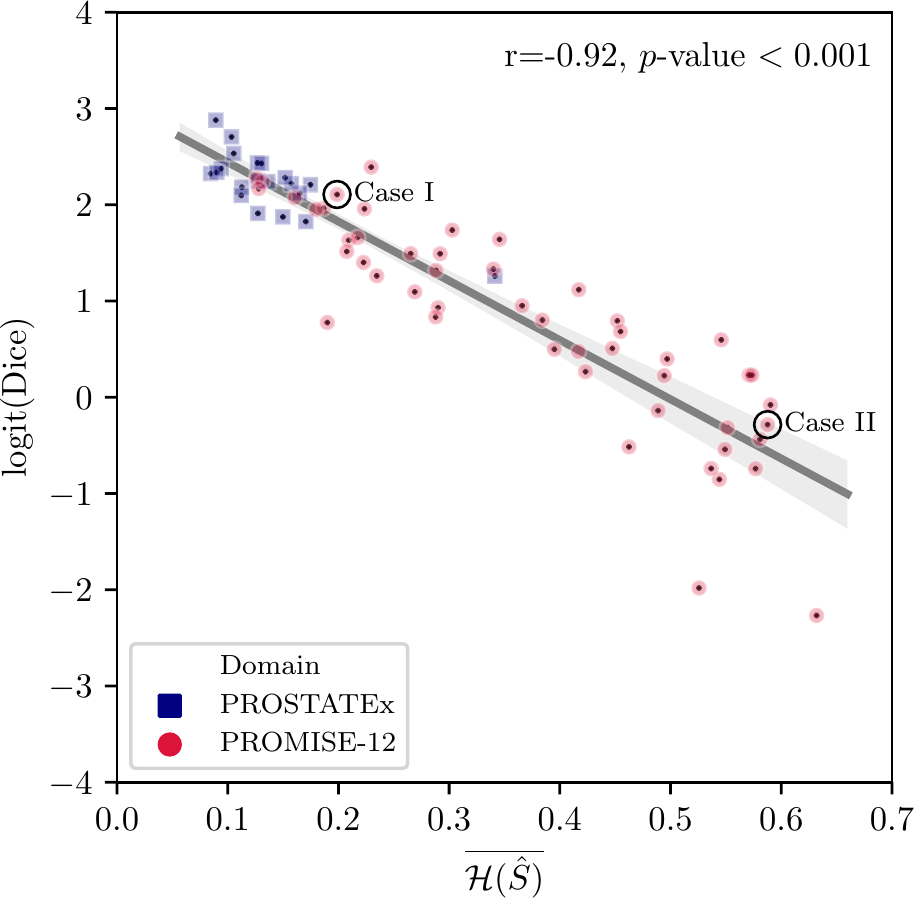} &
  \includegraphics[width=58mm]{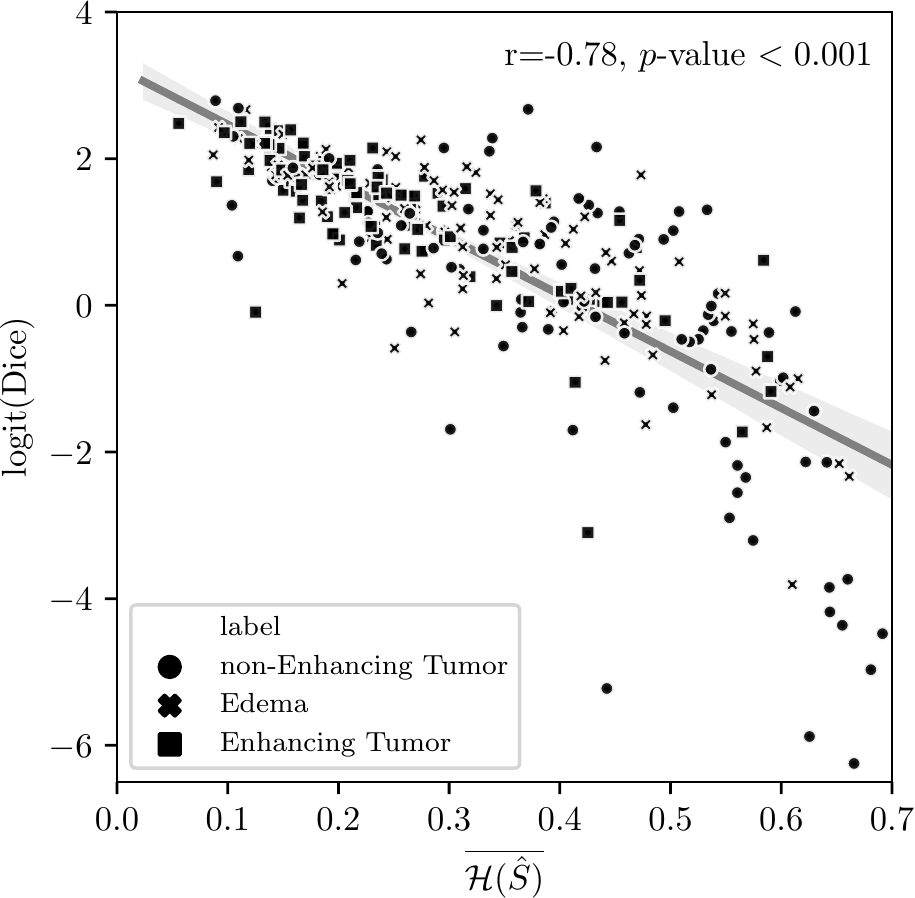} &
  \includegraphics[width=58mm]{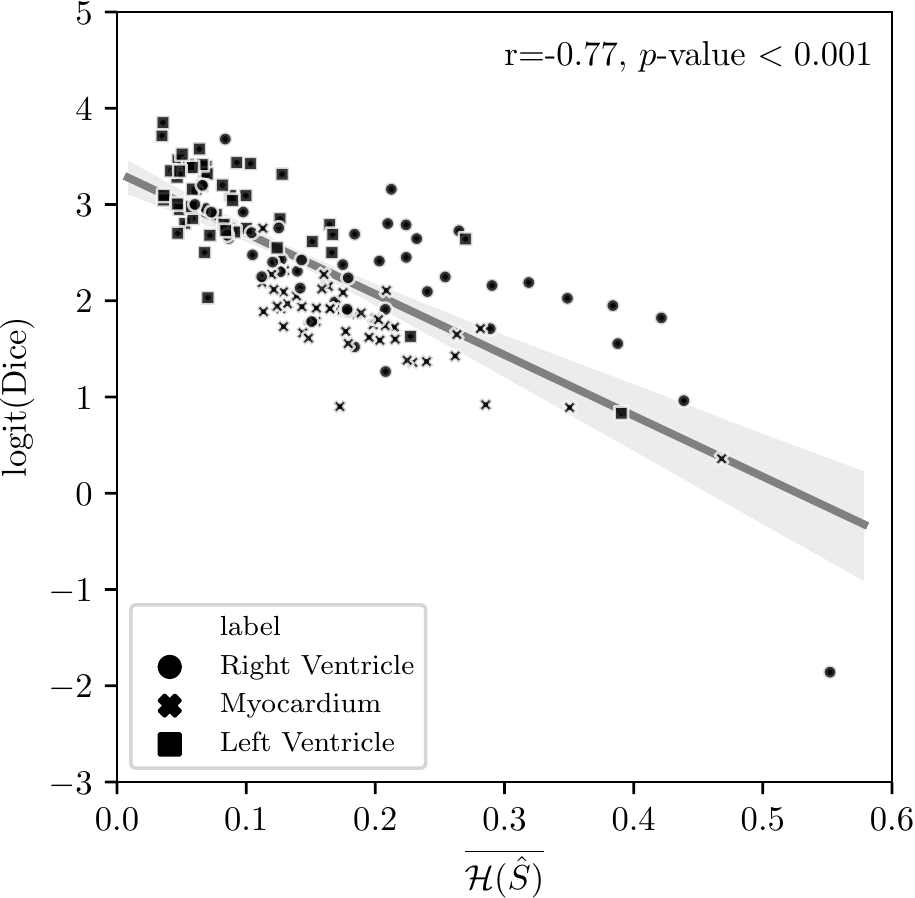} \\
  \end{tabular}
  \begin{tabular}{ccccccccc}
  \includegraphics[width=18mm,height=17.8mm]{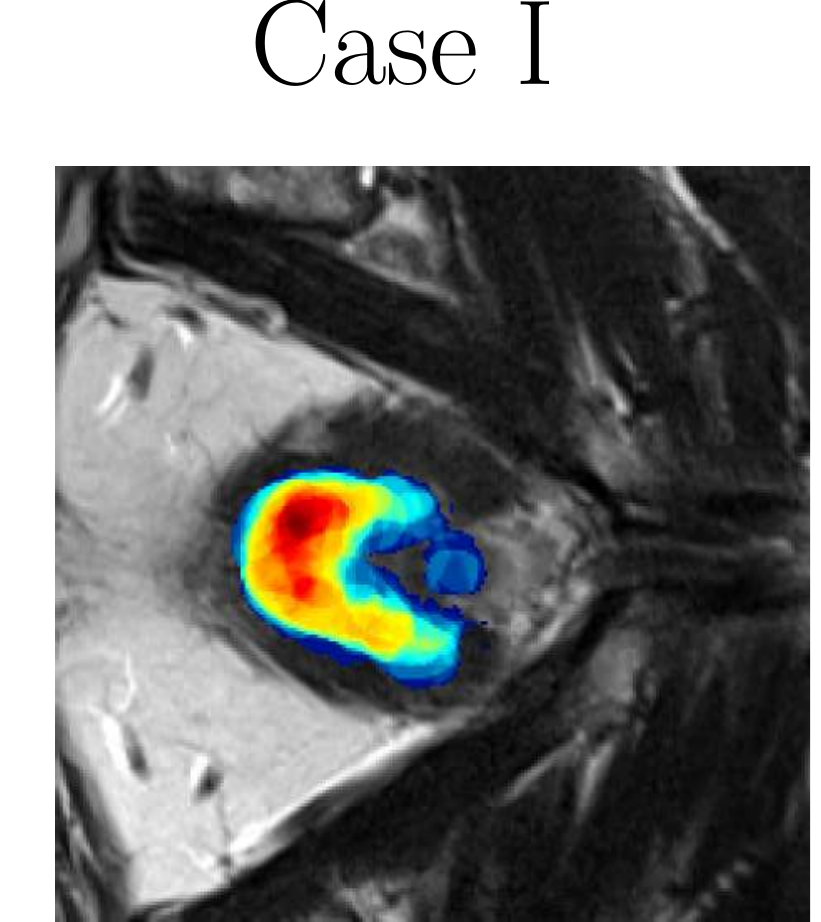} &
  \includegraphics[width=18mm]{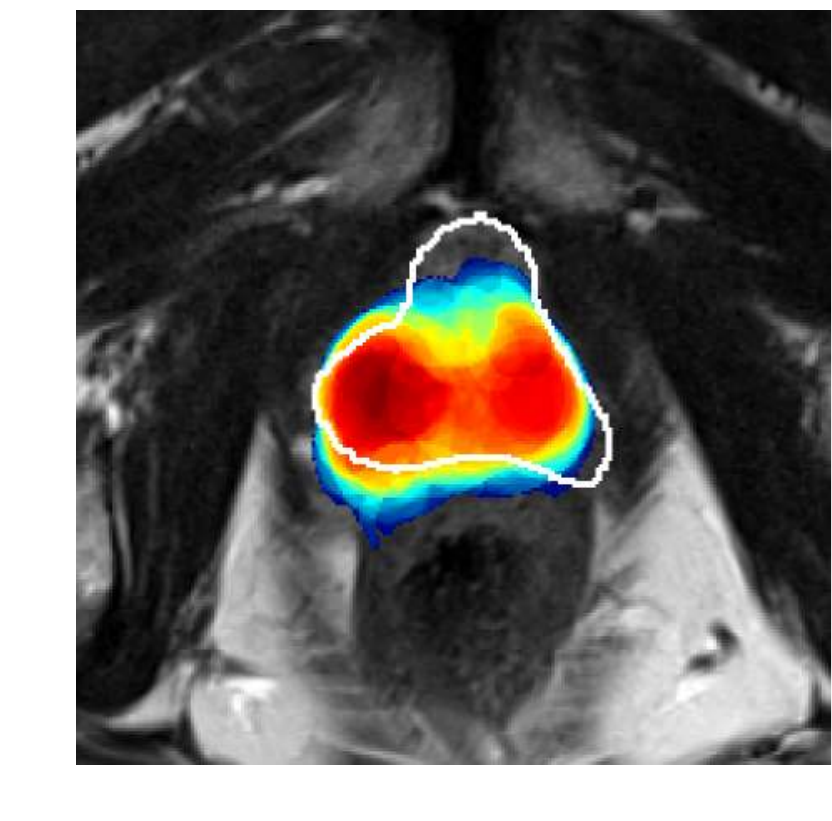} &
  \includegraphics[width=18mm]{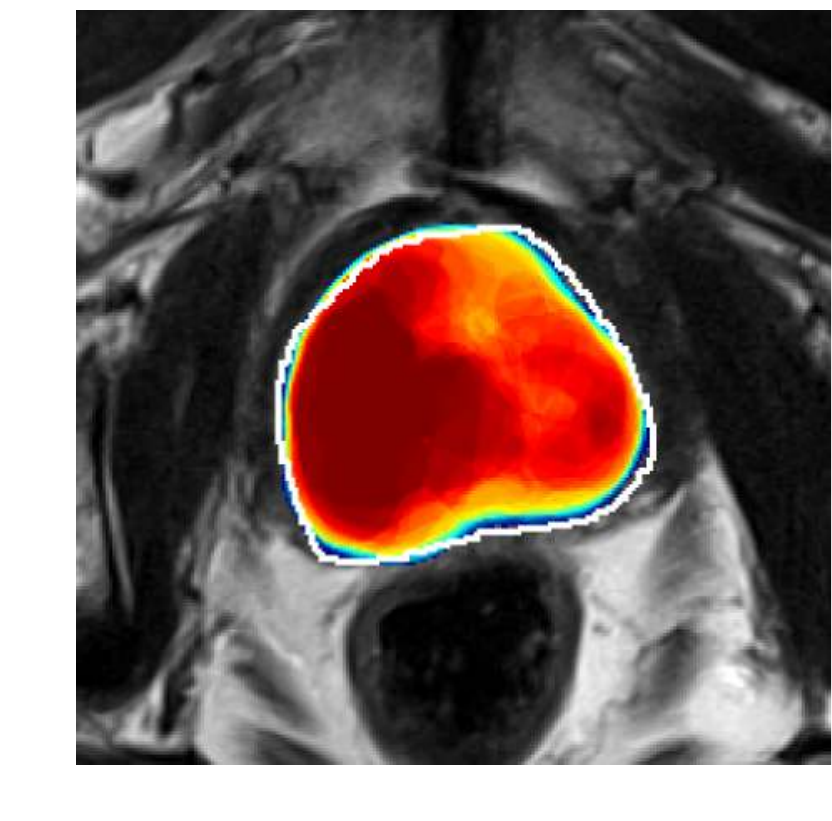} &
  \includegraphics[width=18mm]{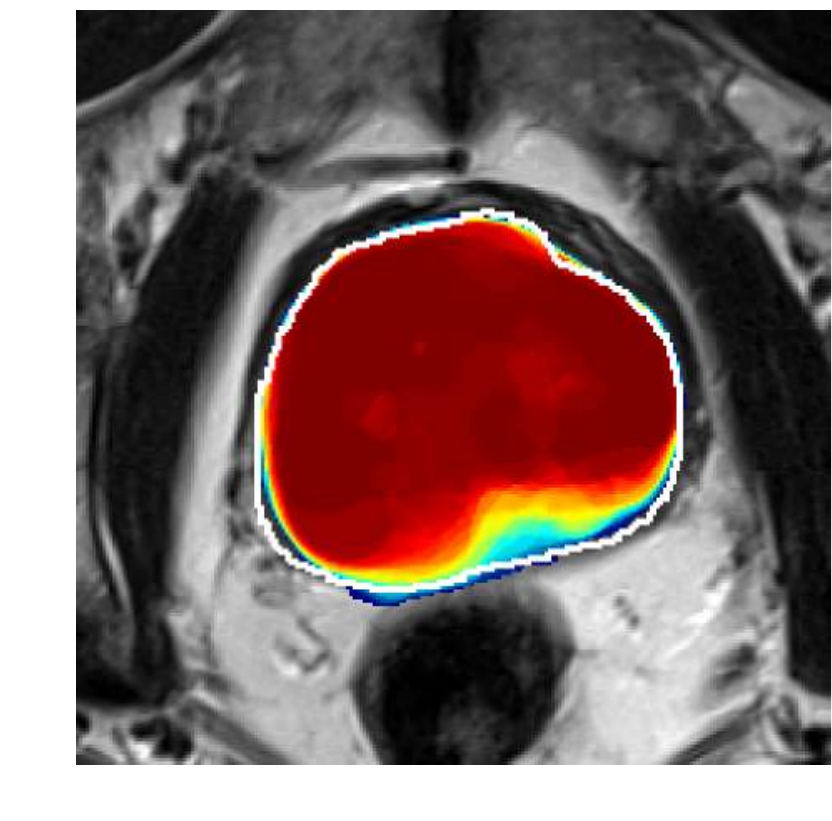} &
  \includegraphics[width=18mm]{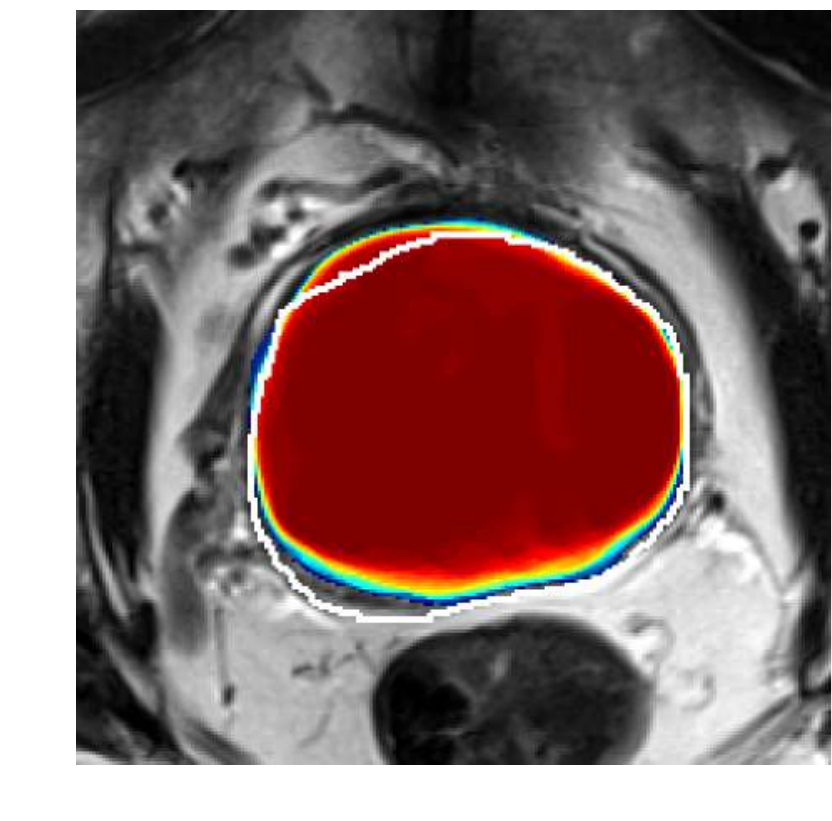} &
  \includegraphics[width=18mm]{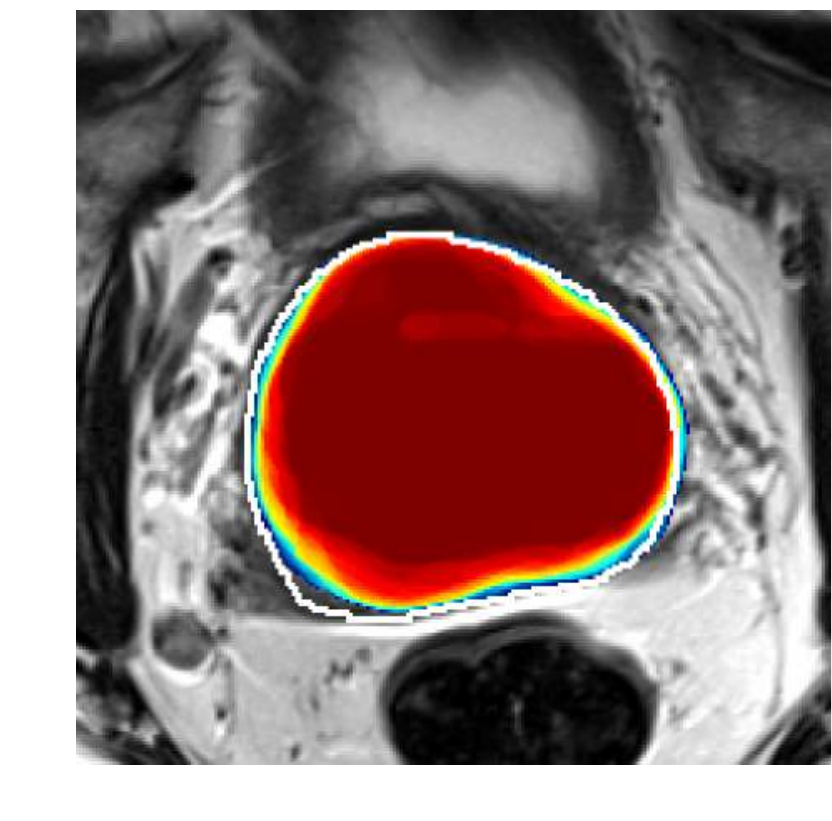} &
  \includegraphics[width=18mm]{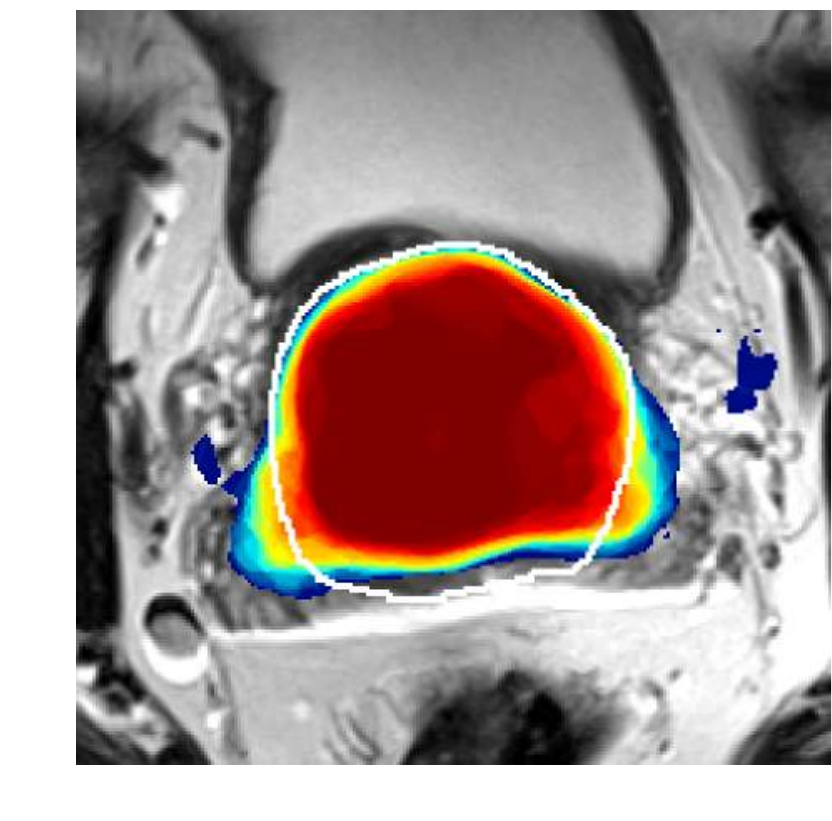} &
  \includegraphics[width=18mm]{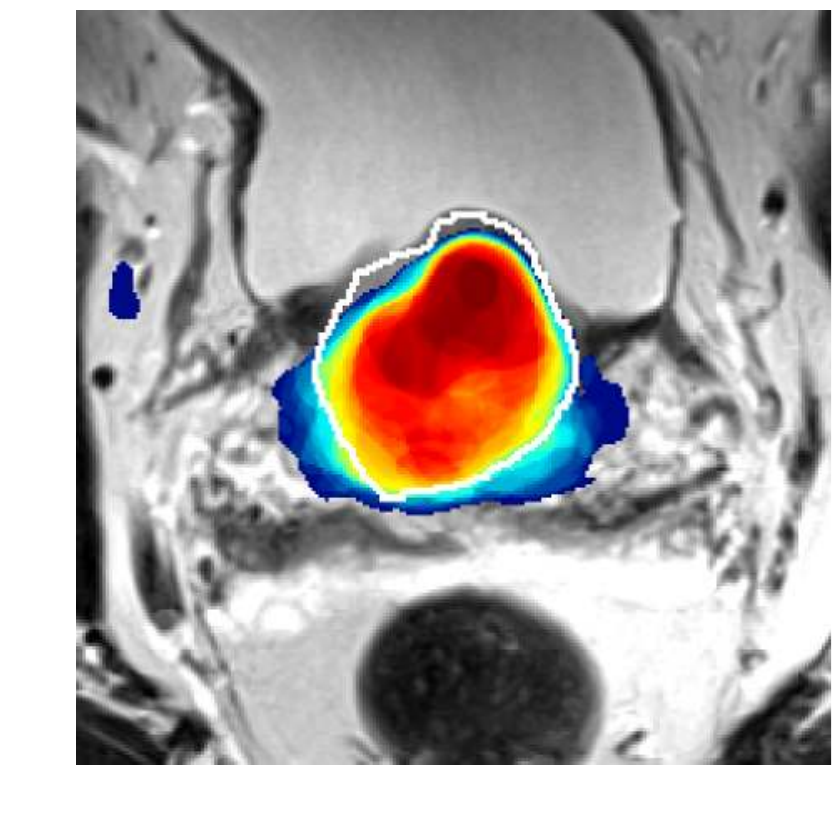} &
  \includegraphics[width=27mm, height=18mm]{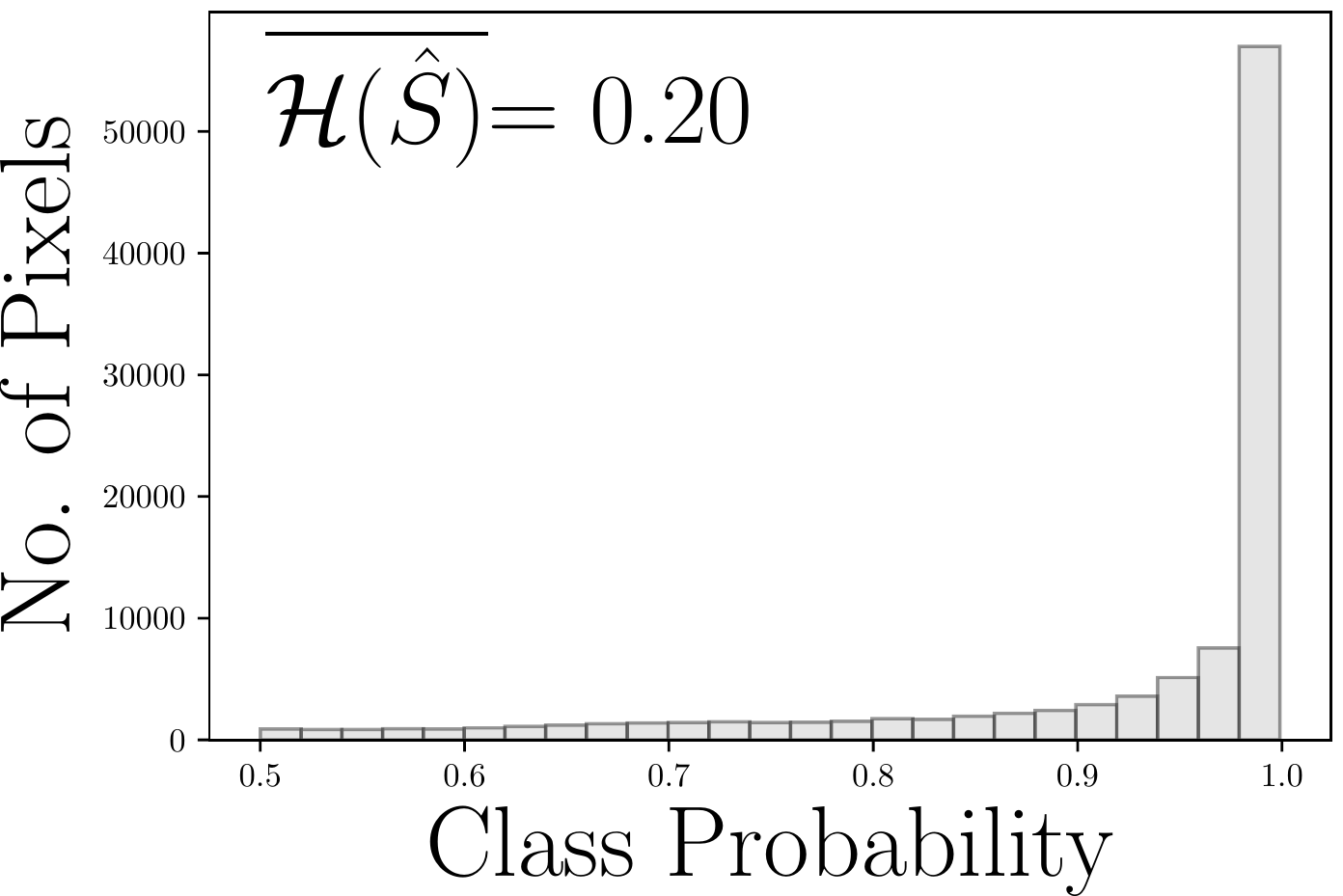} \\
  \includegraphics[width=18mm,height=17.8mm]{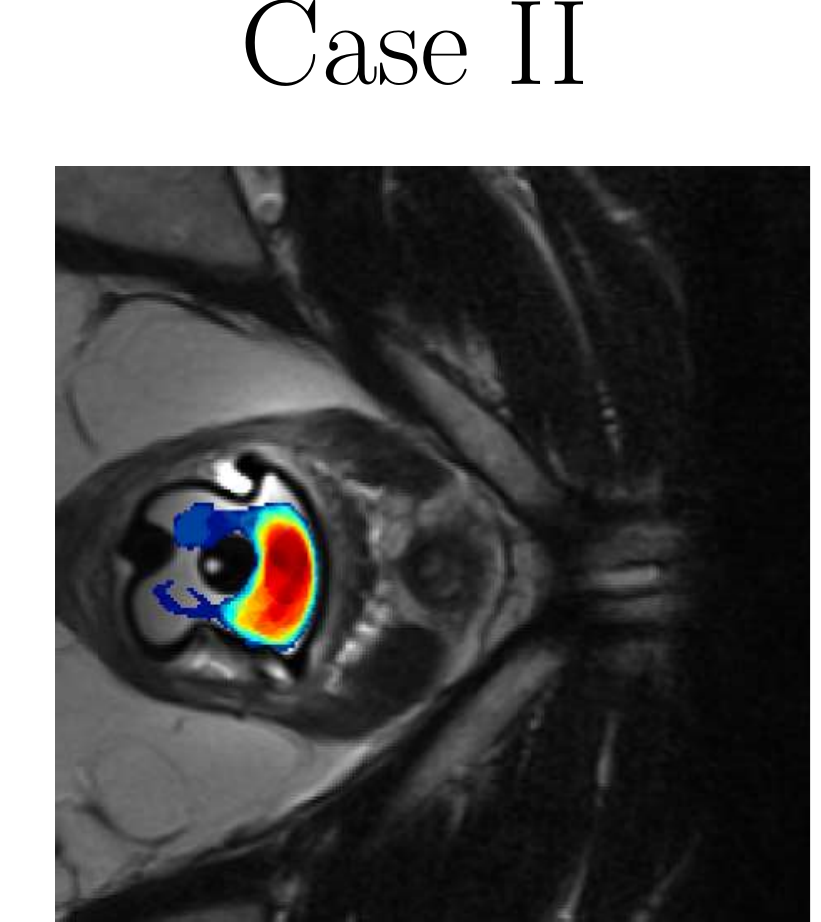} &
  \includegraphics[width=18mm]{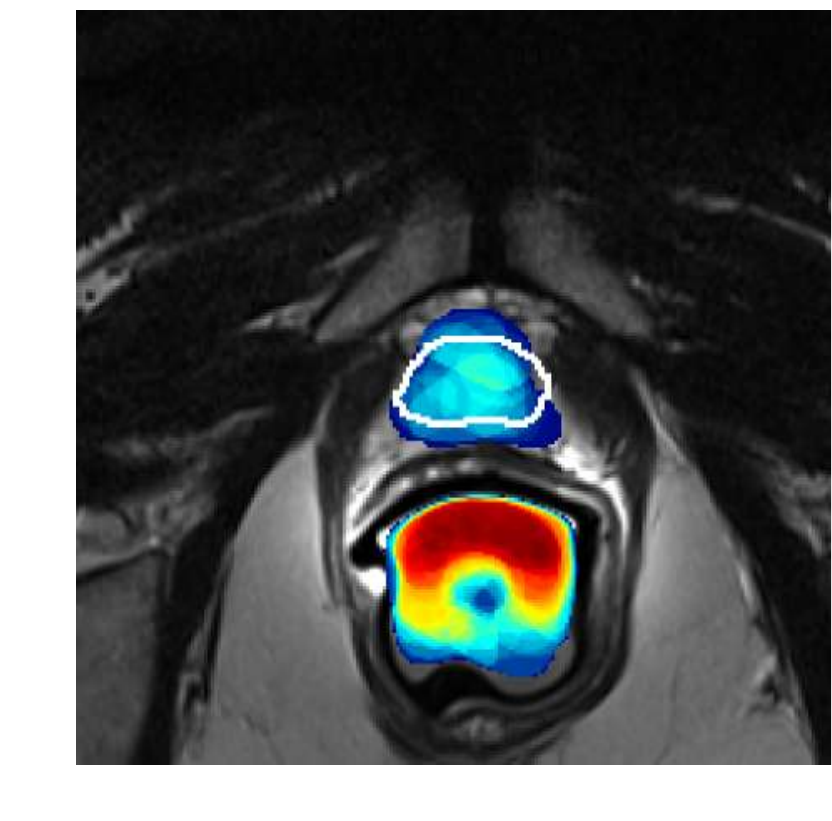} &
  \includegraphics[width=18mm]{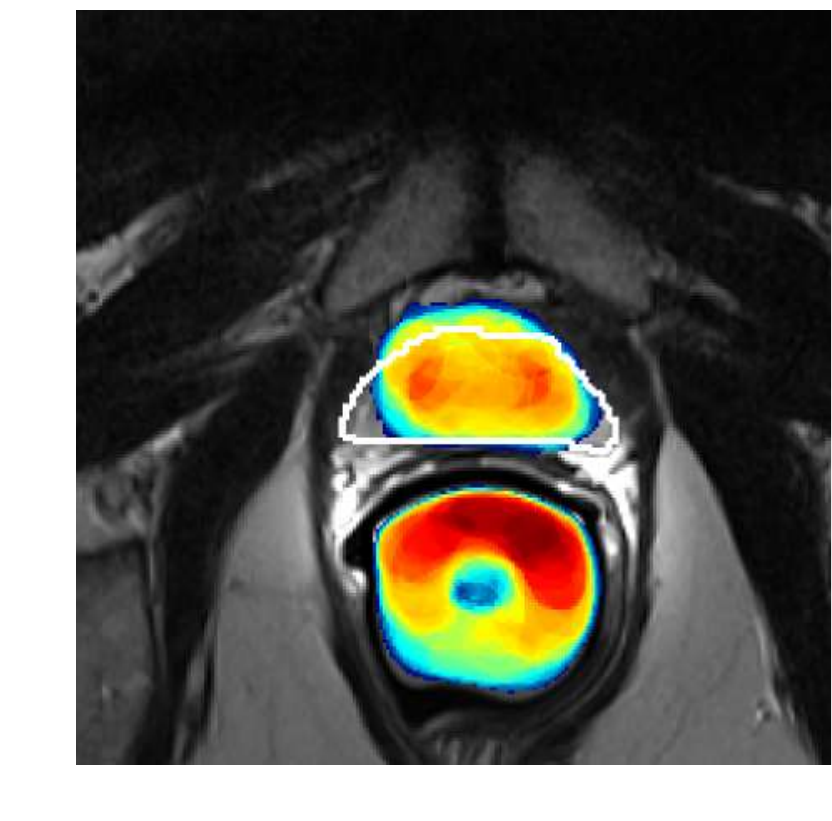} &
  \includegraphics[width=18mm]{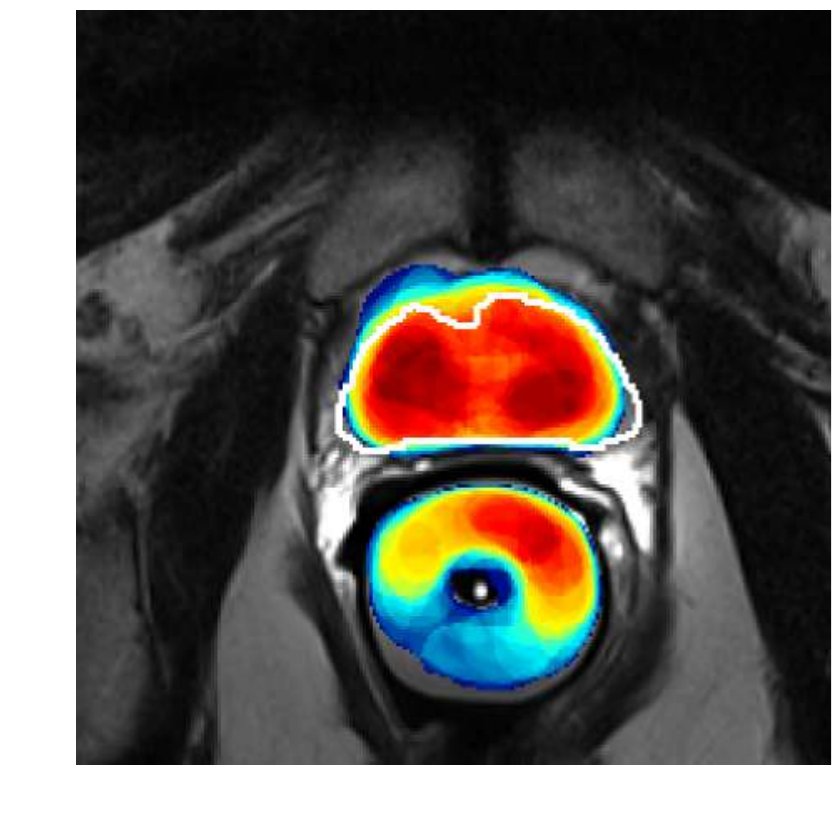} &
  \includegraphics[width=18mm]{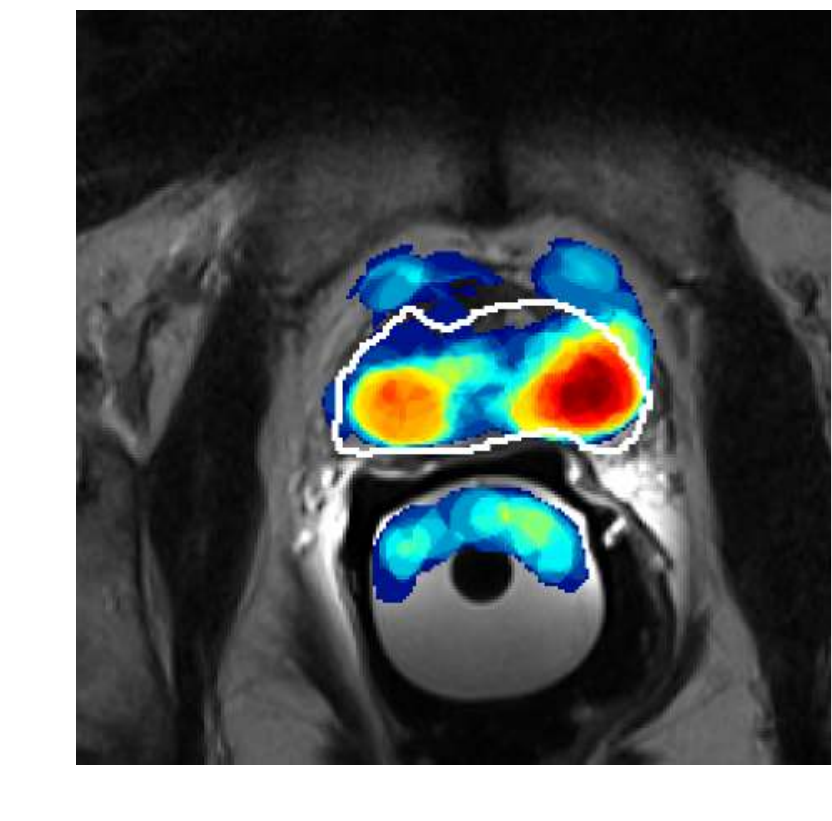} &
  \includegraphics[width=18mm]{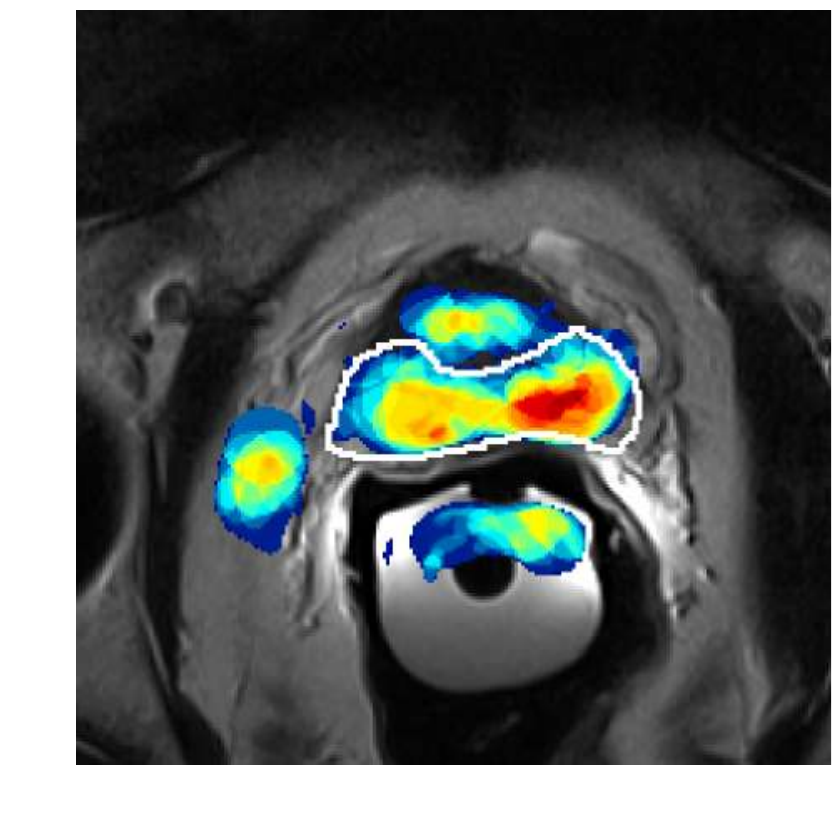} &
  \includegraphics[width=18mm]{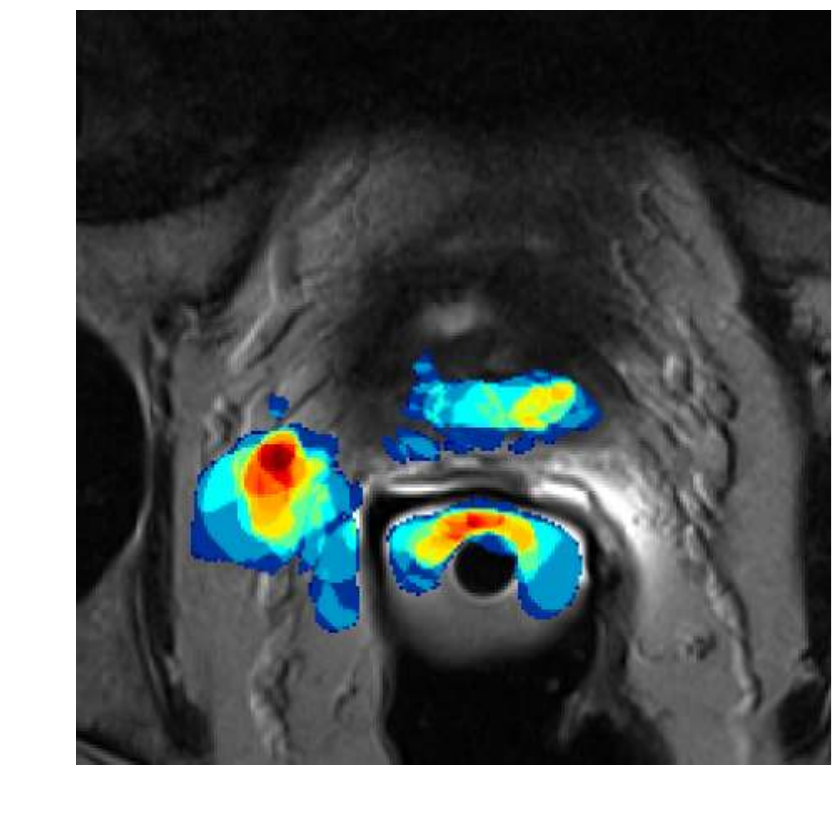} &
  \includegraphics[width=18mm]{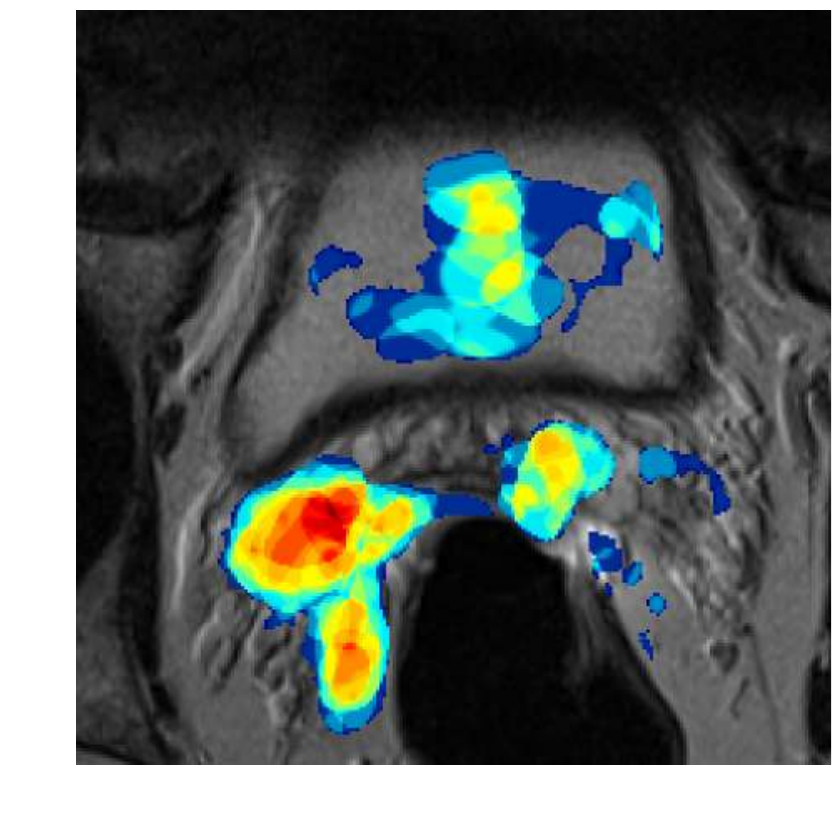} &
  \includegraphics[width=27mm, height=18mm]{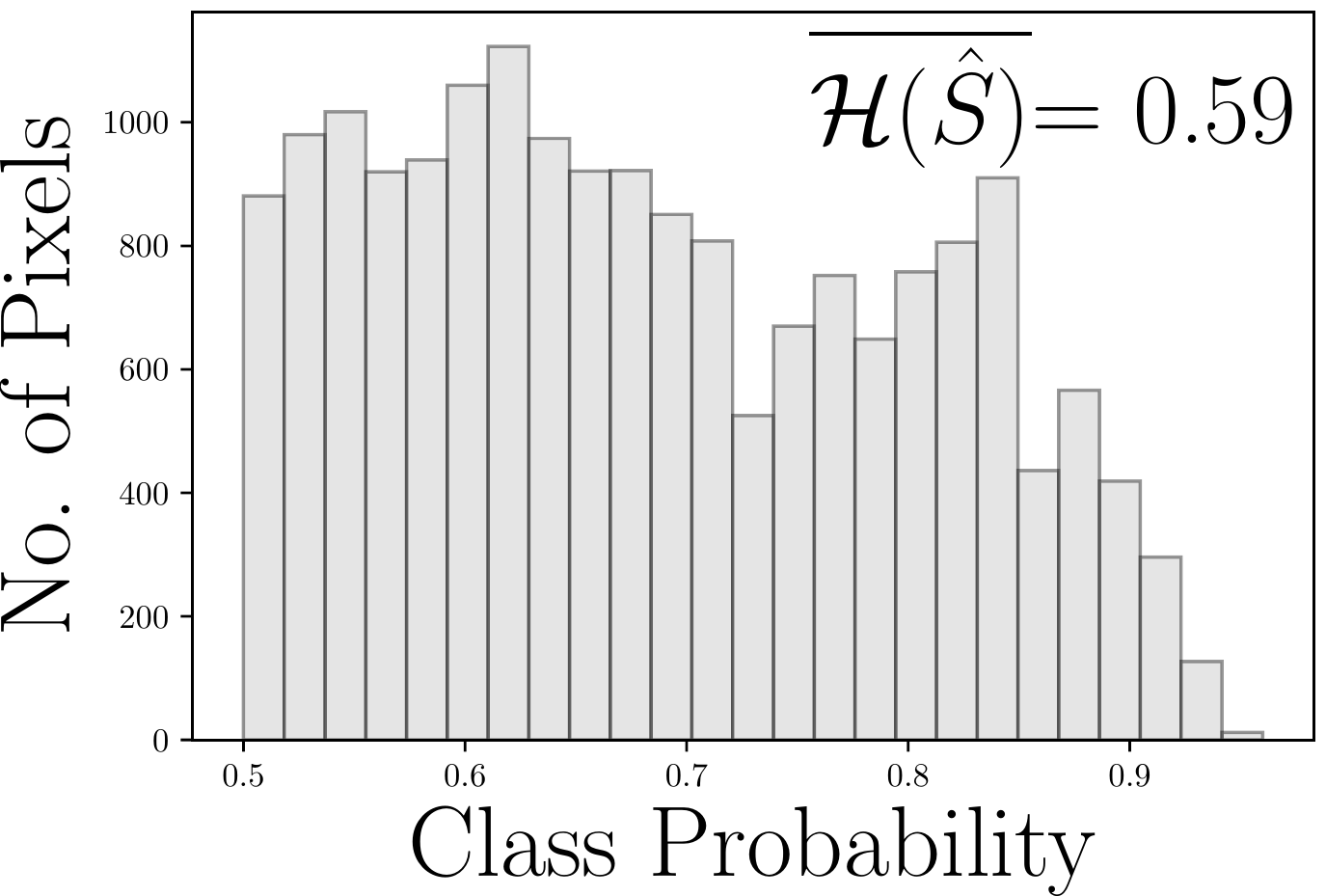} \\
   \end{tabular}
\caption{
Segment-level predictive uncertainty estimation: Top row: Scatter plots and linear regression between Dice coefficient and average of entropy over the predicted segment 
$\overline{\mathcal{H(\hat{\mathcal{S}})}}$.
For each of the regression plots, Pearson's correlation coefficient ($r$) and 2-tailed p-value for testing non-correlation are provided.
Dice coefficients are logit transformed before plotting and regression analysis.
For the majority of the cases in all three segmentation tasks, the average entropy correlates well with Dice coefficient, meaning that it can be used as a reliable metric for predicting the segmentation quality of the predictions at test-time.
Higher entropy means less confidence in predictions and more inaccurate classifications leading to poorer Dice coefficients.
However, in all three tasks there are few cases that can be considered outliers.
(A) For prostate segmentation, samples are marked by their domain: PROSTATEx (source domain), and the multi-device multi-institutional PROMISE12 dataset (target domain).
As expected, on average, the source domain performs much better than the target domain, meaning that average entropy can be used to flag out-of-distribution samples.
The two bottom rows correspond to two of the cases from the PROMISE12 dataset are marked in (A): Case I and Case II; These show the prostate T2-weighted MRI at different locations of the same patient with overlaid calibrated class probabilities (confidences) and histograms depicting distribution of probabilities over the segmented regions.  
The white boundary overlay on prostate denotes the ground truth.
The wider probability distribution in Case II associates with a higher average entropy which correlates with a lower Dice score.
Case-I was imaged with phased-array coil (same as the images that was used for training the models), while Case II was imaged with endorectal coil (out-of-distribution case in terms of imaging parameters). 
The samples in scatter plots in (B) and (C) are marked by their associated foreground segments.
The color bar for the class probability values is given in Figure \ref{fig:introduction}.
Qualitative examples for brain and heart applications and scatter plots for models trained with cross-entropy are given in Figures 7 and 8 of the Supplementary Material, respectively.
}
\label{fig:segment_level_uncertainty}
\end{figure*}

Figure \ref{fig:segment_level_uncertainty} provides scatter plots of Dice coefficient vs. the proposed segment-level predictive uncertainty metric,  $\overline{\mathcal{H(\hat{\mathcal{S}})}}$ (Equation \ref{eq:avg_entr}), for models trained with Dice loss and calibrated with ensembling (M=50).
For better visualization, Dice values were logit transformed $\logit(p) = \ln(\frac{p}{1-p})$ as in \cite{niethammer2017active}.
In all three segmentation tasks, we observed a strong correlation ($ 0.77 \leq r \leq 0.92$) between logit of Dice coefficient and average of entropy over the predicted segment.
For the prostate segmentation task, a clustering is obvious among the test set from the source domain (PROSTATEx dataset) and those from the target domain (PROMISE12).
Investigation of individual cases reveals that most of the poorly segmented cases, which were predicted correctly using $\overline{\mathcal{H}(\hat{\mathcal{S}})}$, can be considered out-of-distribution examples as they were imaged with endorectal coils.

\section{Discussion}
Through extensive experiments, we have rigorously assessed uncertainty estimation for medical image segmentation with FCNs.
Furthermore, we proposed ensembling for confidence calibration of FCNs trained with Dice loss.
We have performed these assessments using three common medical image segmentation tasks to ensure the generalizability of the findings.
The results consistently show that for baseline (single) models, cross-entropy loss is better than Dice loss in terms of uncertainty estimation in terms of NLL and ECE\%, but falls short in segmentation quality.
We then showed that ensembling with $M\geq5$ notably calibrates the confidence of models trained with Dice loss and CE loss.
Importantly, we also observed that in addition to NLL reduction, the segmentation accuracy in terms of the Dice coefficient and Hausdorff distance was also improved through ensembling.
We also showed that ensembling outperforms MC dropout 
in estimating the uncertainty of deep image segmenters.
This confirms previous findings in the image classification literature 
\cite{lakshminarayanan2017simple}.
Consistent with the results of previous studies \cite{kuijf2019standardized}, we observed that segmentation quality improved with ensembling.
The results of our experiments for comparing cross-entropy with Dice loss are in line with the achieved results of Sanders et al. \cite{sander2019towards}.

Importantly, we demonstrated the feasibility of constructing metrics that can capture predictive uncertainty of individual segments. 
We showed that the average entropy of segments can predict the quality of the segmentation in terms of Dice coefficient.
Preliminary results suggest that calibrated FCNs have the potential to detect out-of-distribution samples.
Specifically, for prostate segmentation, the ensemble correctly predicted the cases where it failed due to differences in imaging parameters (such as different imaging coils).
However, it should be noted that this is an early attempt to capture the segment-level quality of segmentation and the results thus need to be interpreted with caution. 
The proposed metric can be improved by adding prior knowledge about the labels. 
Furthermore, it should be noted that the proposed metric does not encompass any information on number of samples used in that estimation. 

As introduced in the methods section, some loss functions are ``proper scoring rules'', a desirable quality that promotes well-calibrated probabilistic predictions.
The Deep Ensembles method has a proper scoring rule requirement for the baseline loss function \cite{lakshminarayanan2017simple}.
The question arises: \textit{``Is the Dice loss a proper scoring rule?''}
Here, we argue that there is a fundamental mismatch in the potential
usage of the Dice loss for scoring rules.
Scoring rules are functions that compare a probabilistic prediction with an outcome.  
In the context of binary segmentation, an outcome
corresponds to a binary vector of length $n$, where $n$ is the number of
pixels.
The difficulty with using scoring rules here is that the
corresponding probabilistic prediction is a distribution on binary
vectors. However, the predictions made by deep segmenters are
collections of $n$ label probabilities.  This is in contrast to
distributions on binary vectors, which are more complex and in general characterized by probability mass functions with $2^n$ parameters, one for each of
the $2^n$ possible outcomes (the number of possible binary
segmentations).
The essential problem is that deep segmenters do not predict
distributions on outcomes (binary vectors).
One potential workaround is to say that the network does predict the
required distributions, by constructing them as the product of the
marginal distributions.  This, though, has the problem that
the predicted distributions will not be similar to the
more general data distributions, so in that sense, they
are bound to be  poor predictions.

We used segmentation tasks in the brain, the heart and the prostate to assess uncertainty estimation. 
Although each of these tasks was performed on MRI images, there were subtle differences between them.
The brain segmentation task was performed on three-channel input (T1 contrast-enhanced, FLAIR, and T2) while the other two were performed on single-channel input (T2 for prostate and Cine images for the heart). 
Moreover, the number of training samples, the size of the target segments, and the homogeneity of samples were different in each task.
Only publicly available datasets were used in this study to allow others to easily reproduce these experiments and results.
The ground truth was created by experts and independent test sets were used for all experiments. 
For prostate gland segmentation and brain tumor segmentation tasks, we used multi-scanner, multi-institution test sets.
For all three tasks, the boundaries of the target segments were commonly identified as areas of high uncertainty estimate. 
Compared to the prostate and heart applications, we observed lower segmentation quality in the brain tumor application.
Segmentation of lesions (in this case brain tumors) is generally a harder problem compared to the segmentation of organs (in this case the heart, and the prostate gland). 
This is partly due to the fact that lesions are more heterogeneous. 
However, as shown in Figure \ref{fig:segment_level_uncertainty} the calibrated models successfully predicted the segmentation quality and total failures (where the model failed to predict any meaningful 
structure -- e.g. Dice score $\leq$ 0.05. 

Our focus was not on achieving state-of-the-art results on the three mentioned segmentation tasks, but on using these to understand and improve the uncertainty prediction capabilities of FCNs.
Since we performed several rounds of training with different loss functions, we limited the number of parameters in the models to speed up each training round; we carried out experiments with 2D CNNs (not 3D), used fewer convolutional filters in our baseline compared to the original U-Net, and performed limited (not exhaustive) hyperparameter tuning to allow reasonable convergence. 
2D U-Nets have been used extensively to segment 3D images and we used these to conduct the experiments reported above. 2D vs 3D is one of the many design choices or hyper-parameters of constructing deep networks for semantic segmentation, without a clear-cut answer that 2D U-Nets are always better for 2D images and 3D-UNets are always better for 3D images. In fact, in some applications, 2D networks have outperformed 3D networks \cite{kuijf2019standardized}. However, in the case of confidence calibration using deep ensembles, preliminary experiments (that we have included in Appendix F of the Supplementary Material) indicate no difference between using 3D U-Nets or 2D U-Nets.  A comprehensive empirical study on this topic would be quite interesting.

In this paper, we compared calibration qualities of two commonly used loss functions
and showed that loss function directly affects calibration quality and segmentation performance.  As stated earlier, calibration quality is an important metric that provides information about the quality of the predictions. We think it is important for users of deep networks to be aware of the calibration qualities associated with different loss functions, and to that end, we think that 
it would be interesting to investigate the calibration and segmentation quality of other commonly used loss functions such as combinations of Dice loss and cross-entropy loss, as well as the recently proposed  Lov\'asz-Softmax loss \cite{berman2018lovasz} that we think is promising for medical image segmentation.

For the proposed segment-level predictive uncertainty measure (Equation \ref{eq:avg_entr}),
we assumed binary classification and entropy of the foreground 
class was calculated by considering every other class as background.
However, there are neighborhood relationships between classes and adjacent pixels that could be further integrated using measures such as multi-class entropy  
or similar strategies such as the Wasserstein losses
\cite{fidon2017generalised}.

There remains a need to study calibration methods that, unlike ensembling, do not require training from scratch which is time-consuming.
In this work, we only investigated uncertainty estimation for MR images.
Although parameter changes occur more often in MRI comparing to computed tomography (CT), it would  still be very interesting to study uncertainty estimation in CT images.
Parameter changes in CT can also be a source of failure in CNNs. 
For instance, changes in slice thickness or use of contrast can result in failures in predictions and it is highly desirable to predict such failures through model confidence.

We believe that our research will serve as a base for future studies on uncertainty estimation and confidence calibration for medical image segmentation.
Further study of the sources of uncertainty in medical image segmentation is needed.
Uncertainty has been classified as aleatoric or epistemic in medical applications \cite{indrayan2012medical} and Bayesian modeling \cite{kendall2017uncertainties}.
Aleatoric refers to types of uncertainties that exist due to noise or the stochastic behavior of a system.
In contrast, epistemic uncertainties are rooted in  limitation in knowledge about the model or the data.
In this study, we consistently observed higher levels of uncertainty at specific locations such as boundaries. 
For example in the prostate segmentation task, single and multiple raters often have higher inter- and intra-disagreements in the delineation of the base and apex of the prostate rather than at the mid-gland boundaries \cite{litjens2014evaluation}.
Such disagreements can leave their traces on models that are trained using ground truth labels with such discrepancies.
It has been shown that with enough training data from multiple raters, deep models are able to surpass human agreements on segmentation tasks \cite{litjens2017survey}. 
However, few works have been addressed the correlation of ground truth quality and model uncertainty that results from rater disagreements \cite{sudre2019let, tanno2019learning}.

We conclude that model ensembling can be used successfully for confidence calibration of FCNs trained with Dice Loss.
Also, the proposed average entropy metric can be used as an effective predictive metric for estimating the performance of the model at test-time when the ground-truth is unknown.


\ifCLASSOPTIONcaptionsoff
  \newpage
\fi

%





\bibliography{references}
\bibliographystyle{IEEEtran}

\newpage
\onecolumn
\appendices
\setcounter{figure}{0}    
\setcounter{table}{0}    
\section{Calibration Quality (Whole Volume Results)}

Table \ref{tab:calibration_whole_volume}
compares the calibration quality of different models calculated on the whole volume.
The baselines and ensembles (M=50) trained with CE loss are compared with those that were trained with Dice loss and those that were calibrated with MC dropout.

\begin{table}[h]
	\centering
	\caption{
		Calibration quality for baselines trained with Dice loss ($\mathcal{L}_{DSC}$) are compared with those that trained with cross-entropy ($\mathcal{L}_{CE}$) and those that were calibrated with ensembling (M=$50$) and MC dropout.
        Bold faced indicates best results for each application (model) and shows that the differences are statistically significant.
	} 
	\label{tab:calibration_whole_volume}
		\begin{tabular}{l|lll}
			\toprule
			Organ (Model) & 
			NLL (95\% CI) & Brier (95\% CI)  & ECE\%  (95\% CI)) \\
			\midrule
			Brain (${\mathcal{L}_{CE}}$)& 
			0.08 (0.01$-$0.38) & 
            0.03 (0.01$-$0.14) & 
            1.06 (0.12$-$5.69) \\ 
			
			Brain (MCDO ${\mathcal{L}_{CE}}$) &
			0.16 (0.01$-$0.70) & 
            0.07 (0.01$-$0.24) & 
            2.45 (0.08$-$11.11) \\ 

			Brain (EN ${\mathcal{L}_{CE}}$) &
			0.04 (0.01$-$0.16) & 
            0.02 (0.00$-$0.09) & 
            0.97 (0.29$-$2.50) \\ 
			
			Brain  ($\mathcal{L}_{DSC}$)&
			0.16 (0.03$-$0.56) & 
            0.02 (0.00$-$0.07) & 
            1.16 (0.17$-$3.32) \\ 

			Brain (MCDO ${\mathcal{L}_{DSC}}$) &
			0.13 (0.02$-$0.63) & 
            0.02 (0.00$-$0.11) & 
            1.05 (0.14$-$4.81) \\ 
			
			Brain (EN ${\mathcal{L}_{DSC}}$) & 
			\textbf{0.03 (0.01$-$0.11)} & 
            \textbf{0.01 (0.00$-$0.03)} & 
            \textbf{0.49 (0.03$-$1.58)} \\ 
			
			\midrule
			Heart (${\mathcal{L}_{CE}}$)& 
			0.04 (0.01$-$0.12) & 
            0.02 (0.01$-$0.04) & 
            0.53 (0.14$-$1.55) \\ 
            
			Heart (MCDO ${\mathcal{L}_{CE}}$)& 
            0.04 (0.01$-$0.11) & 
            0.02 (0.01$-$0.04) & 
            0.52 (0.11$-$5.02) \\ 
			
			Heart (EN ${\mathcal{L}_{CE}}$) &
			\textbf{0.03 (0.01$-$0.06)} & 
            \textbf{0.01 (0.01$-$0.03)} & 
            0.51 (0.25$-$0.71) \\ 
			
			Heart ($\mathcal{L}_{DSC}$)& 
			0.07 (0.01$-$0.24) & 
            0.02 (0.00$-$0.05) & 
            1.10 (0.10$-$3.29) \\ 

			Heart (MCDO ${\mathcal{L}_{DSC}}$) &
			0.04 (0.01$-$0.15) & 
            0.34 (0.02$-$0.85) & 
            47.39 (6.24$-$90.55) \\ 
			
			Heart (EN ${\mathcal{L}_{DSC}}$) & 
			0.04 (0.01$-$0.07) & 
            0.01 (0.01$-$0.03) & 
            \textbf{0.36 (0.07$-$1.33)} \\ 
		
			\midrule
			Prostate ($\mathcal{L}_{CE}$)& 
			0.08 (0.04$-$0.16) & 
            0.04 (0.02$-$0.09) & 
            2.15 (0.50$-$7.17) \\ 

			Prostate (MCDO ${\mathcal{L}_{CE}}$) &
            0.11 (0.04$-$0.24) & 
            0.06 (0.03$-$0.11) & 
            1.82 (0.23$-$4.65) \\ 
            
			Prostate (EN ${\mathcal{L}_{CE}}$) &
			0.07 (0.05$-$0.10) & 
            0.03 (0.02$-$0.06) & 
            2.62 (1.65$-$3.87) \\ 
			
			Prostate (${\mathcal{L}_{DSC}}$) & 
			0.26 (0.10$-$0.58) & 
            0.04 (0.02$-$0.08) & 
            1.94 (0.97$-$4.12) \\ 
			
			Prostate (MCDO ${\mathcal{L}_{DSC}}$) &
			0.17 (0.07$-$0.37) & 
            0.04 (0.02$-$0.08) & 
            1.79 (0.80$-$3.99) \\ 
			
			Prostate (EN ${\mathcal{L}_{DSC}}$)&  
			\textbf{0.05 (0.02$-$0.09)} & 
            \textbf{0.02 (0.01$-$0.04)} & 
            \textbf{0.65 (0.13$-$1.26)} \\ 

			\bottomrule
		\end{tabular}
\end{table}

\clearpage
\section{Hausdorff Distance Metric}
Table \ref{tab:segmentation_hd95}
compares the segmentation performance of different models with $95^{th}$ Hausdorff distance (in mm).
The baselines and ensembles (M=50) trained with CE loss are compared with those that were trained with Dice loss and those that were calibrated with MC dropout.

\begin{table}[h]
\centering
\caption{
Segmentation quality of baselines in terms of 95^{th} Hausdorff distances in mm. 
Models trained with Dice loss ($\mathcal{L}_{DSC}$) are compared with those that trained with cross-entropy ($\mathcal{L}_{CE}$) and those that were calibrated with ensembling (M=$50$) and MC dropout.
For brain application segments, I, II, and III correspond to non-enhancing tumor, edema, and enhancing tumor, respectively.
For heart application segments, I, II, and III correspond to the right ventricle, the myocardium, and the left ventricle, respectively.
For prostate application segment I corresponds to the prostate gland.
Bold faced indicates best results for each application (model) and shows that the differences are statistically significant.
} 
\label{tab:segmentation_hd95}
\begin{tabular}{l|lll}
\toprule
Organ (Model) & Segment I* & Segment II* & Segment III* \\

\midrule
Brain (${\mathcal{L}_{CE}}$)& 
59.05 (5.39$-$107.69)  & 
52.93 (7.28$-$83.41)  & 
56.85 (3.00$-$101.92)  \\ 

Brain (MCDO ${\mathcal{L}_{CE}}$)& 
63.95 (5.83$-$110.00)  & 
60.71 (10.20$-$86.34)  & 
62.45 (3.61$-$103.77)  \\ 

Brain (EN ${\mathcal{L}_{CE}}$) &
36.53 (3.46$-$95.12)  & 
35.63 (3.46$-$80.14)  & 
41.24 (2.24$-$100.92)  \\ 

Brain  ($\mathcal{L}_{DSC}$)&
40.05 (4.00$-$102.69)  & 
39.22 (3.00$-$80.06)  & 
40.49 (2.00$-$99.70)  \\ 

Brain (MCDO ${\mathcal{L}_{DSC}}$) &
44.19 (4.12$-$107.24)  & 
44.83 (3.61$-$81.61)  & 
45.80 (2.24$-$100.84)  \\ 

Brain (EN ${\mathcal{L}_{DSC}}$) & 
\textbf{22.55 (2.56$-$93.82)}  & 
\textbf{24.75 (2.24$-$71.03)}  & 
\textbf{30.81 (2.00$-$94.88)}  \\ 

\midrule
Heart (${\mathcal{L}_{CE}}$)& 
26.30 (7.21$-$135.30)  & 
20.55 (4.00$-$136.38)  & 
24.04 (2.00$-$154.95)  \\ 

Heart (MCDO ${\mathcal{L}_{CE}}$) &
30.60 (7.21$-$169.28)  & 
22.15 (4.00$-$146.43)  & 
24.64 (2.00$-$164.93)  \\ 

Heart (EN ${\mathcal{L}_{CE}}$) &
14.42 (5.66$-$30.00)  & 
7.37 (4.00$-$20.69)  & 
6.43 (2.00$-$20.79)  \\ 

Heart ($\mathcal{L}_{DSC}$)& 
15.18 (2.00$-$79.97)  & 
10.47 (2.00$-$88.23)  & 
13.69 (2.00$-$126.91)  \\ 

Heart (MCDO ${\mathcal{L}_{DSC}}$) &
15.53 (2.00$-$79.51)  & 
9.51 (2.00$-$64.18)  & 
11.85 (2.00$-$104.34)  \\ 

Heart (EN ${\mathcal{L}_{DSC}}$) & 
\textbf{9.50 (2.00$-$26.91)}  & 
\textbf{5.90 (2.00$-$14.70)}  & 
\textbf{6.30 (2.00$-$20.98)}  \\ 

\midrule
Prostate ($\mathcal{L}_{CE}$) & 
11.67 (5.00$-$25.07) 
& $-$ & $-$ \\ 

Prostate (MCDO ${\mathcal{L}_{CE}}$) &
14.54 (6.18$-$28.40) 
& $-$ & $-$ \\ 

Prostate (EN ${\mathcal{L}_{CE}}$) &
6.62 (3.54$-$19.91)  
& $-$ & $-$  \\ 

Prostate (${\mathcal{L}_{DSC}}$) & 
8.22 (3.64$-$20.59)  
& $-$ & $-$ 
\\ 

Prostate (MCDO ${\mathcal{L}_{DSC}}$) &
9.84 (4.12$-$23.34)  
& $-$ & $-$ 
\\ 

Prostate (EN ${\mathcal{L}_{DSC}}$)&  
\textbf{5.66 (3.16$-$18.71)}  
& $-$ & $-$ 
\\ 
\bottomrule
\end{tabular}
\end{table}
\clearpage
\section{Quantitative Results}

\begin{figure*}[h]
\centering
\setlength{\tabcolsep}{1pt}
\begin{tabular}{lcccc}
		\parbox[t]{4mm}{\rotatebox[origin=l]{90}{\footnotesize{$~~~~~~~~{\mathcal{L}_{CE}}$}}} &
		\includegraphics[width=25mm]{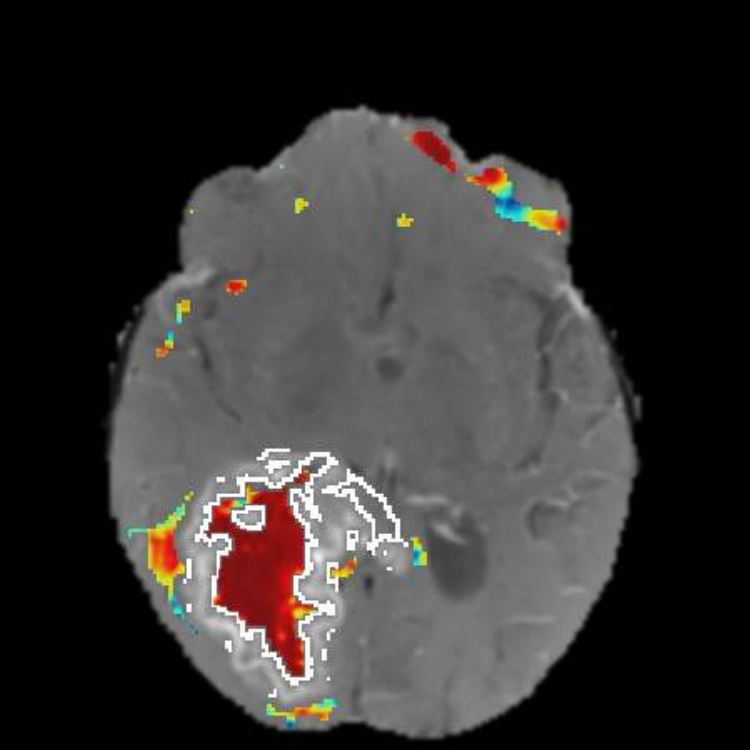} &
		\includegraphics[width=25mm]{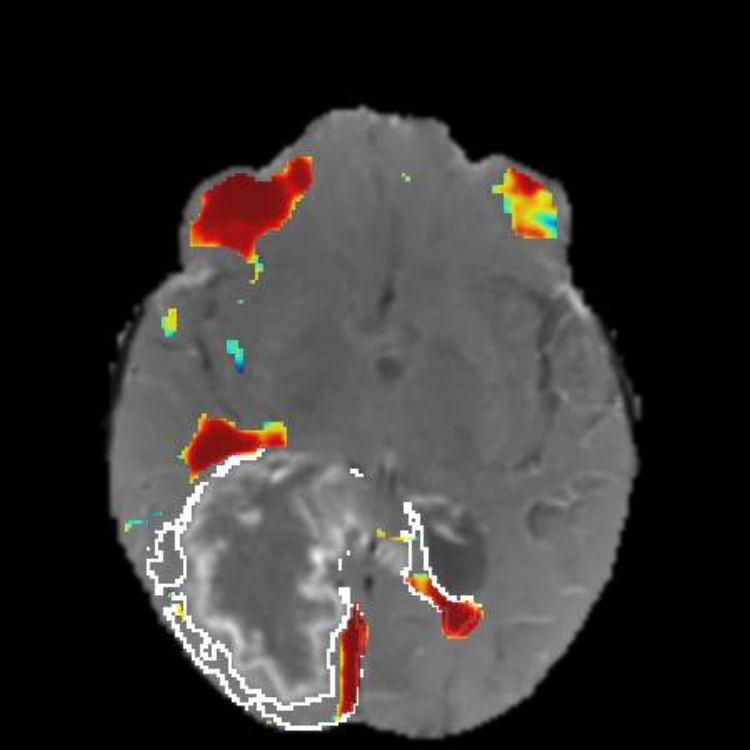} &
		\includegraphics[width=25mm]{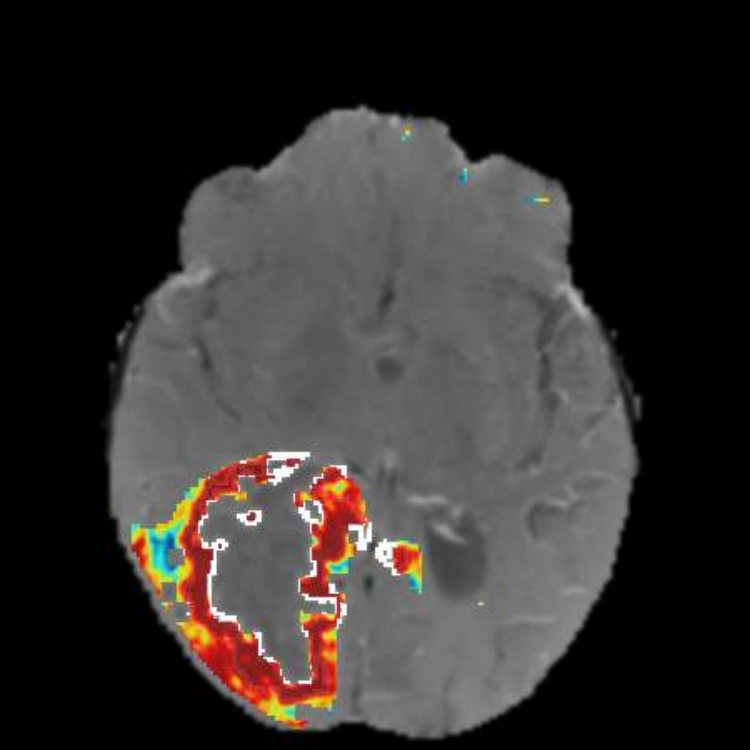} &
		\includegraphics[width=26mm]{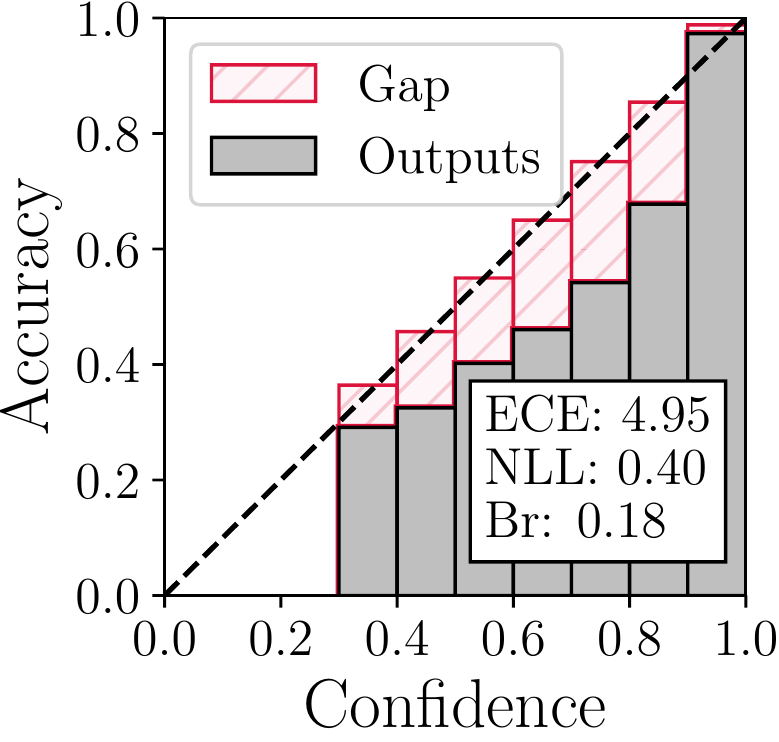} \\
		
		\parbox[t]{4mm}{\rotatebox[origin=l]{90}{\footnotesize{$~~{\mathcal{L}_{CE}}~(MCDO)$}}} &
		\includegraphics[width=25mm]{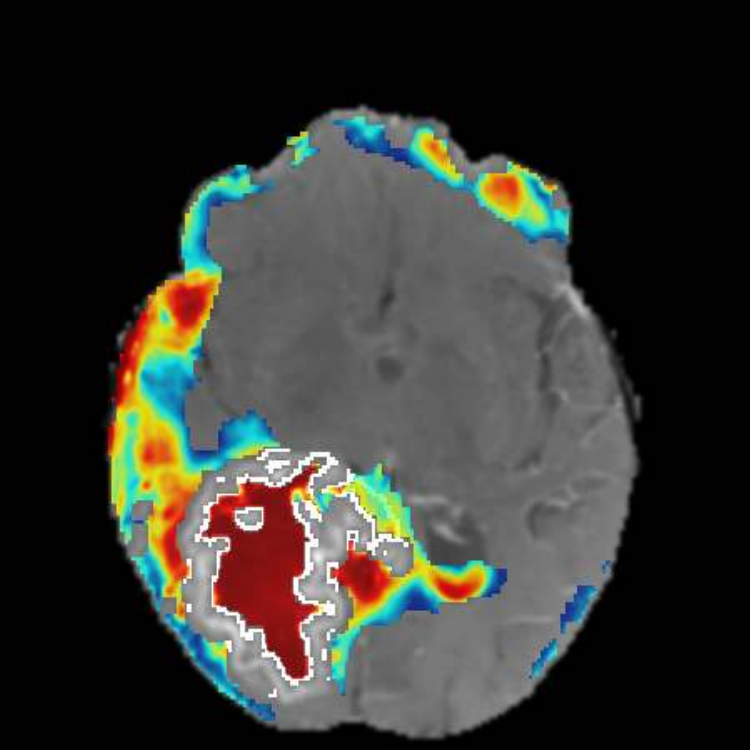} &
		\includegraphics[width=25mm]{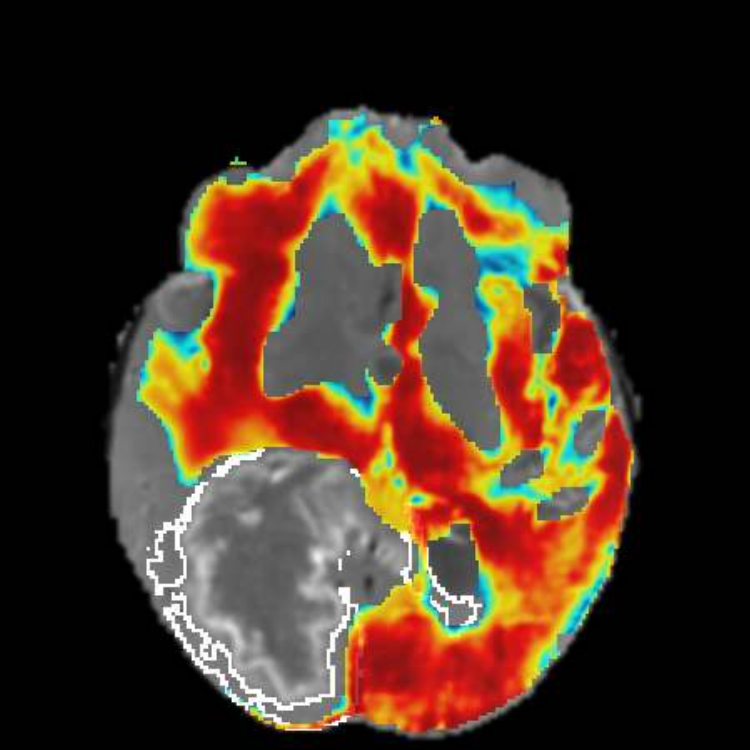} &
		\includegraphics[width=25mm]{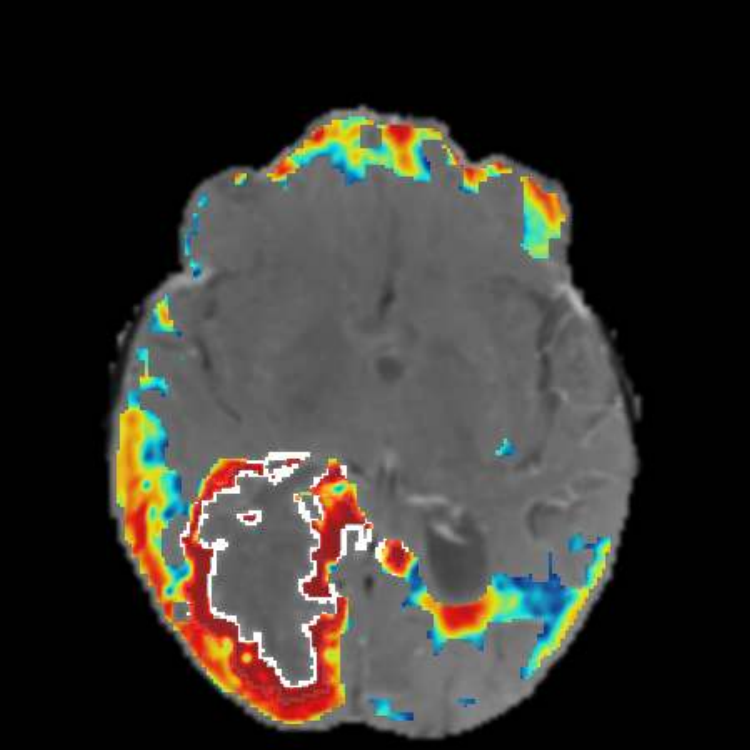} &
		\includegraphics[width=26mm]{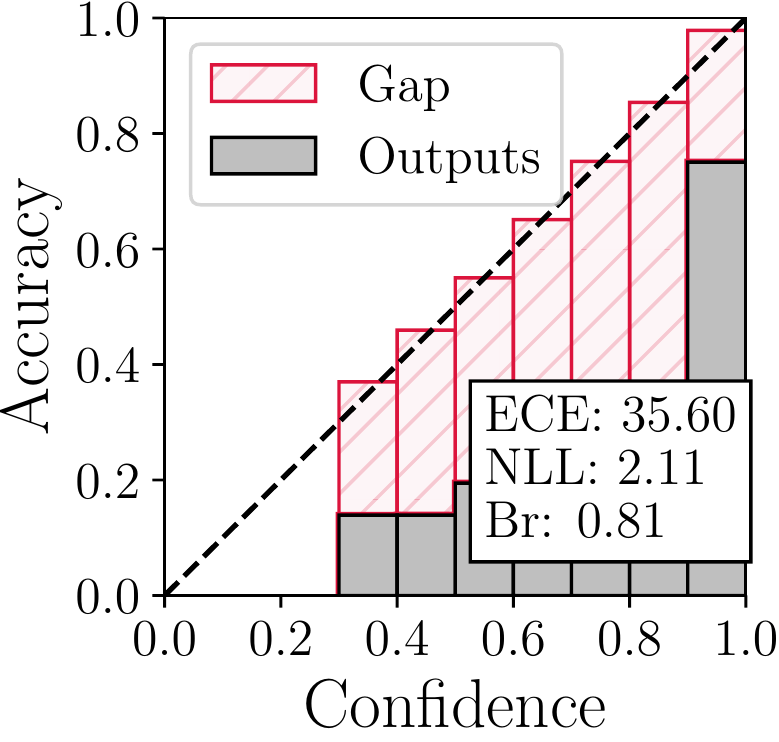} \\

		\parbox[t]{4mm}{\rotatebox[origin=l]{90}{\footnotesize{$~~~~{\mathcal{L}_{CE}~(EN)}$}}} &
		\includegraphics[width=25mm]{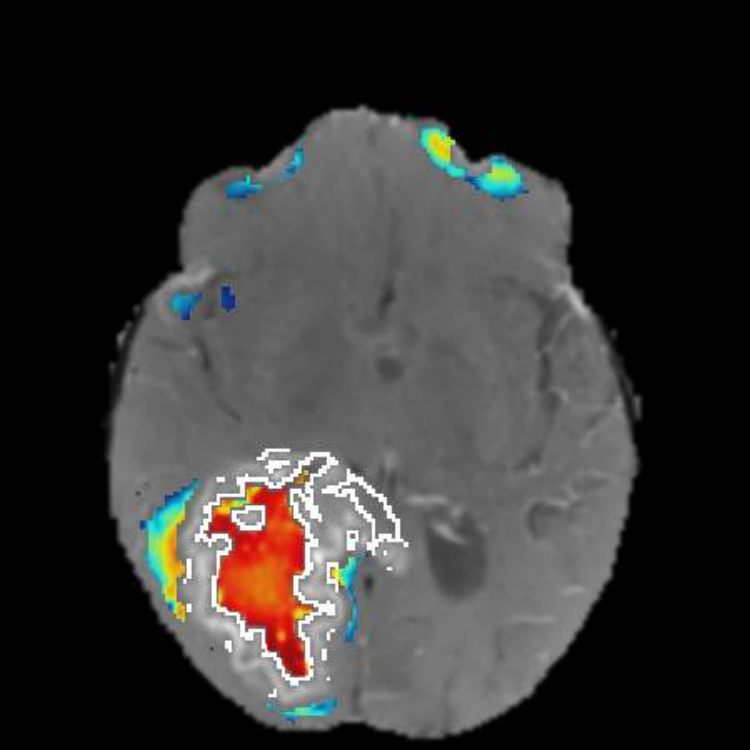} &
		\includegraphics[width=25mm]{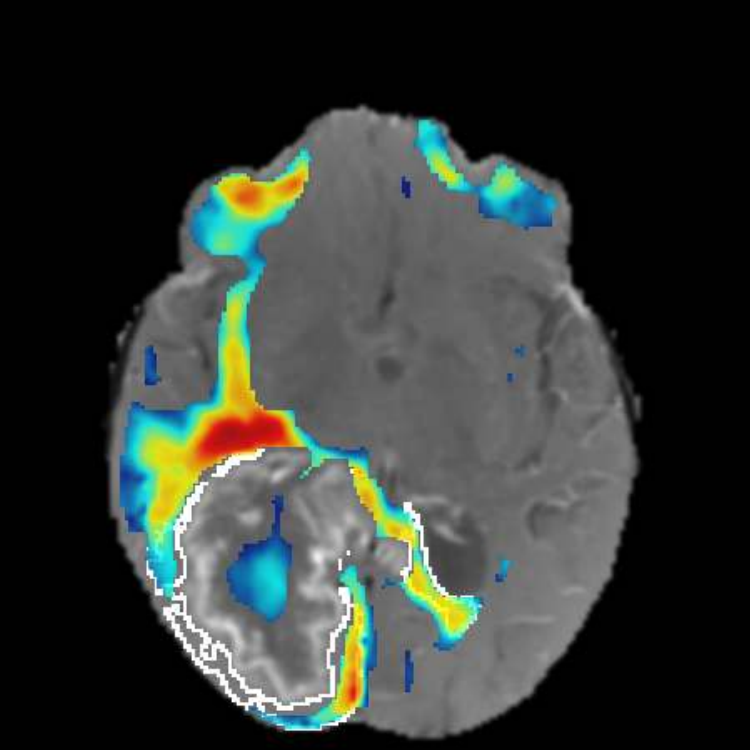} &
		\includegraphics[width=25mm]{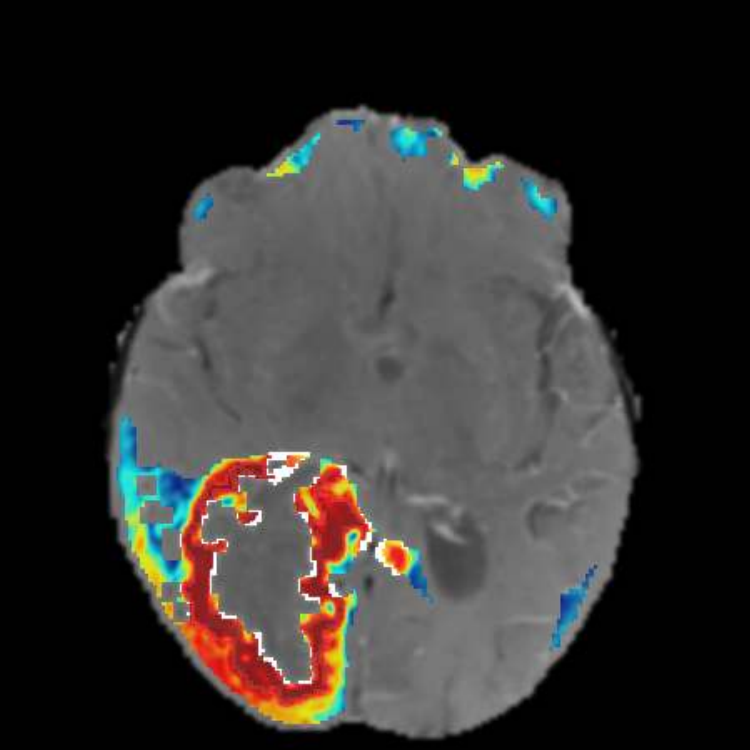} &
		\includegraphics[width=26mm]{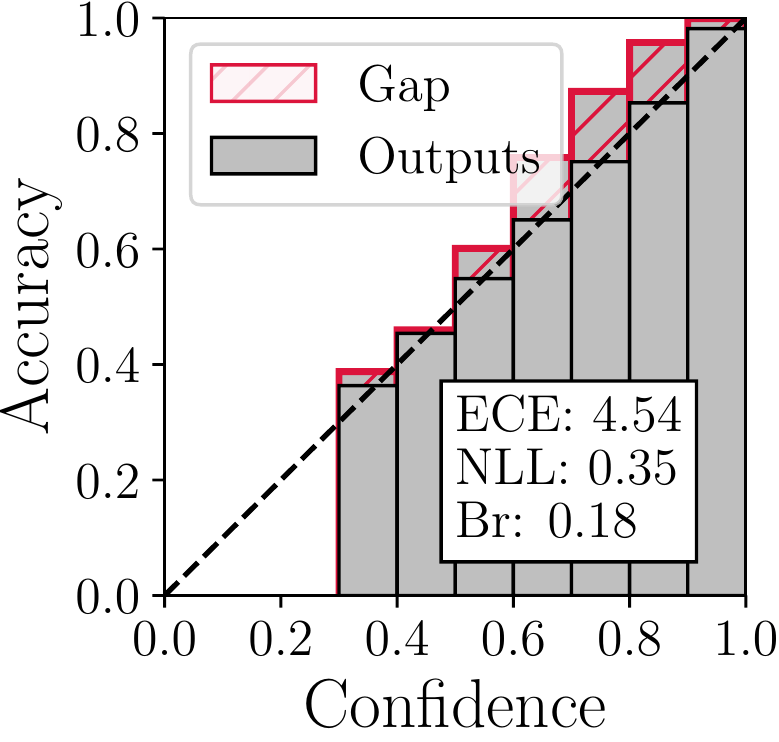} \\
		
		\parbox[t]{4mm}{\rotatebox[origin=l]{90}{\footnotesize{$~~~~~~~~{\mathcal{L}_{DSC}}$}}} &
		\includegraphics[width=25mm]{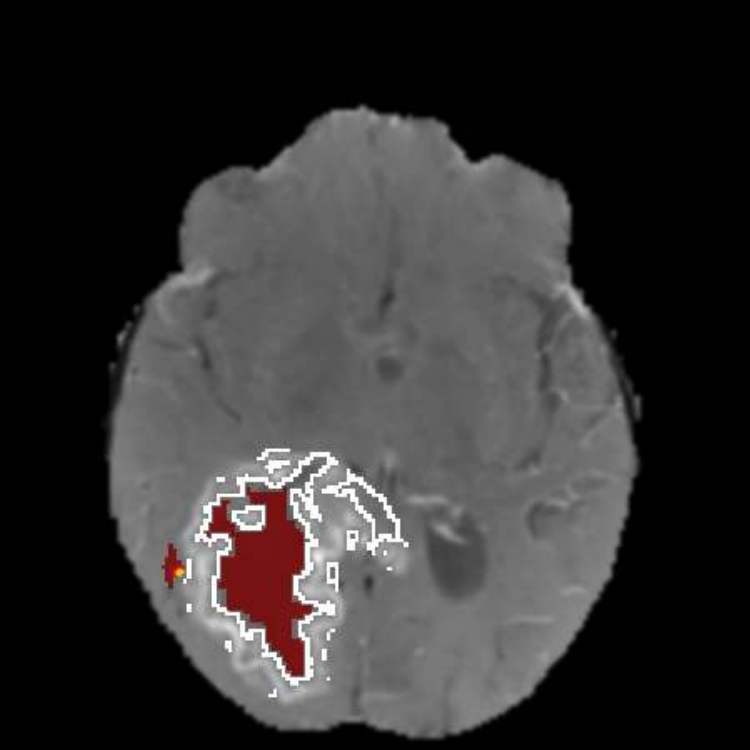} &
		\includegraphics[width=25mm]{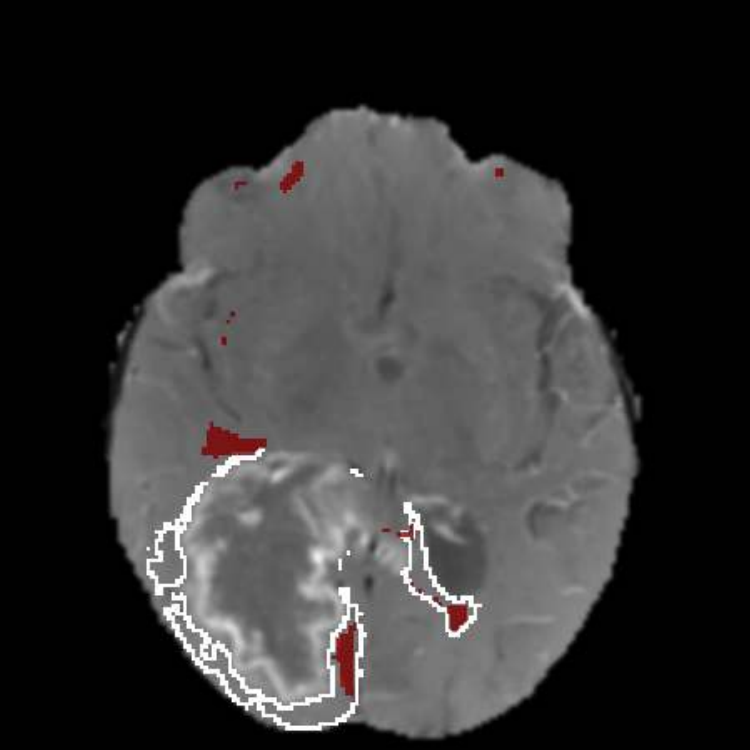} &
		\includegraphics[width=25mm]{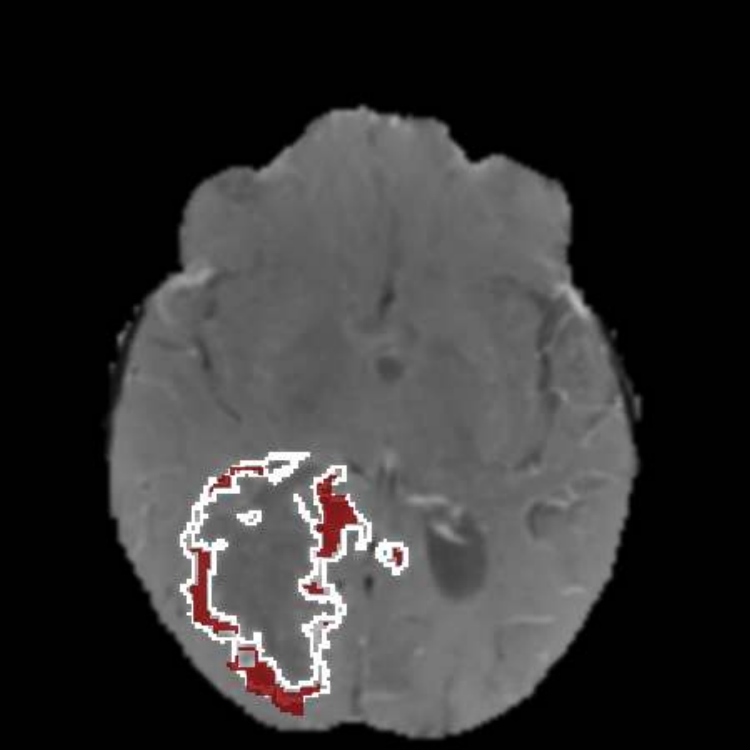} &
		\includegraphics[width=26mm]{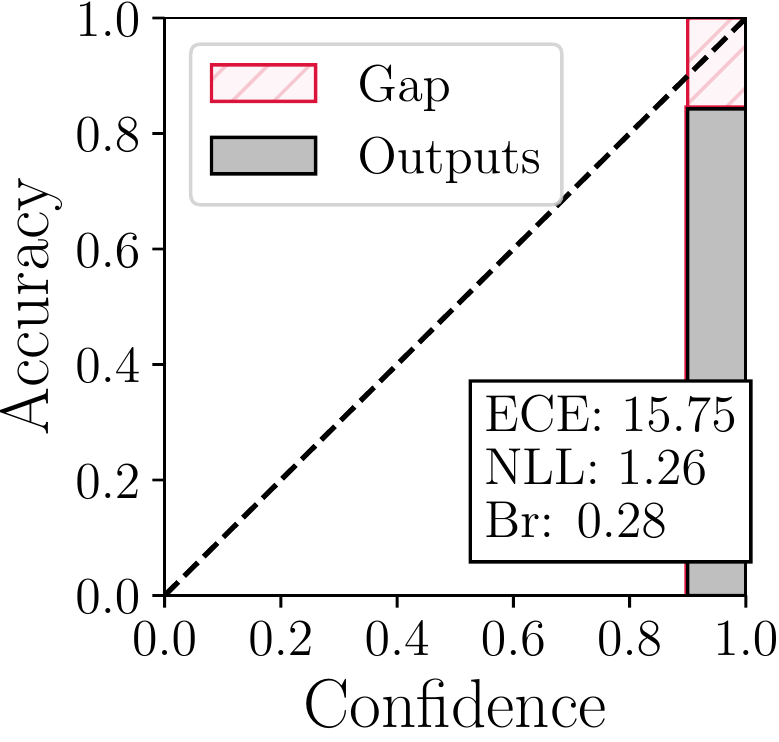} \\
		
		\parbox[t]{4mm}{\rotatebox[origin=l]{90}{\footnotesize{$~~{\mathcal{L}_{DSC}~(MCDO)}$}}} &
		\includegraphics[width=25mm]{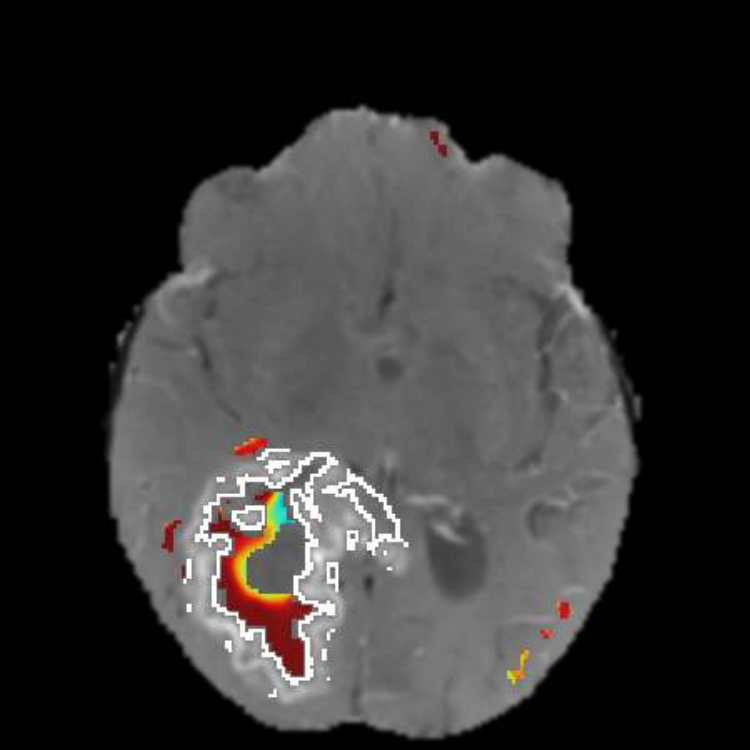} &
		\includegraphics[width=25mm]{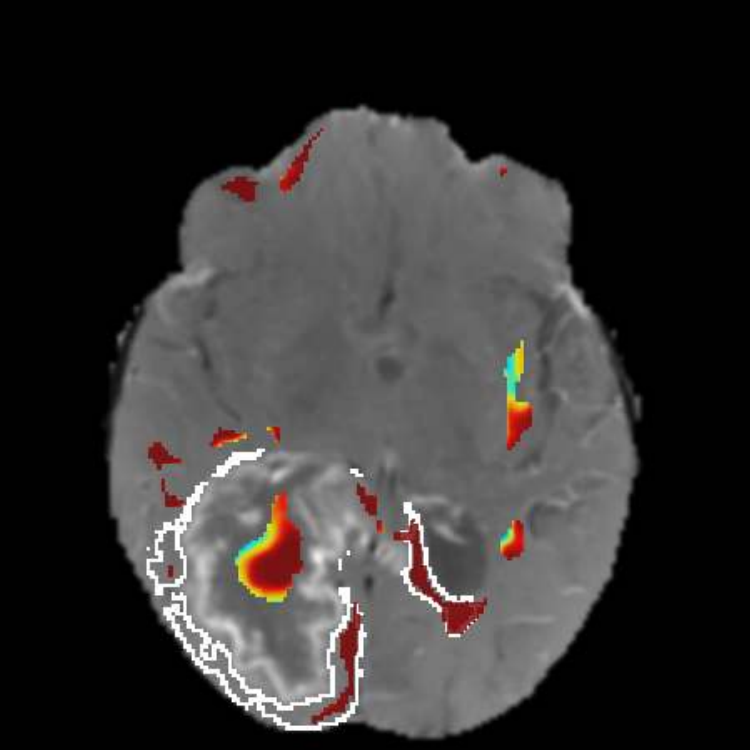} &
		\includegraphics[width=25mm]{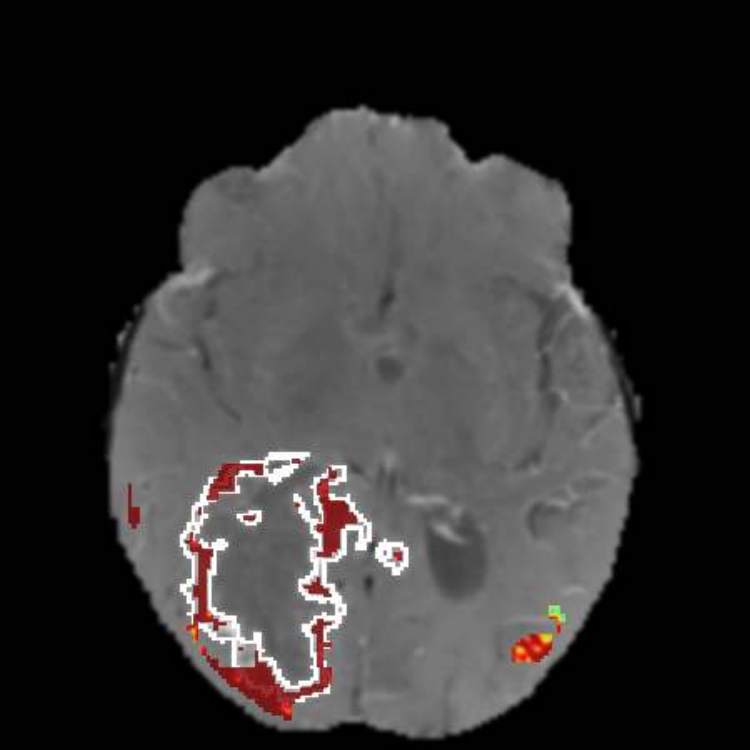} &
		\includegraphics[width=26mm]{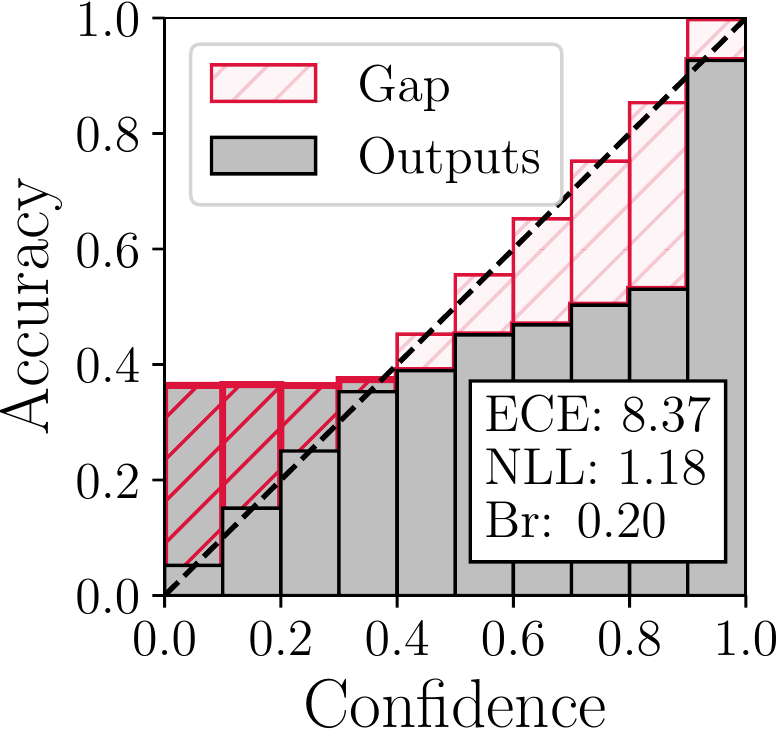} \\
		
		\parbox[t]{4mm}{\rotatebox[origin=l]{90}{\footnotesize{$~~~{\mathcal{L}_{DSC}~(EN)}$}}} &
		\includegraphics[width=25mm]{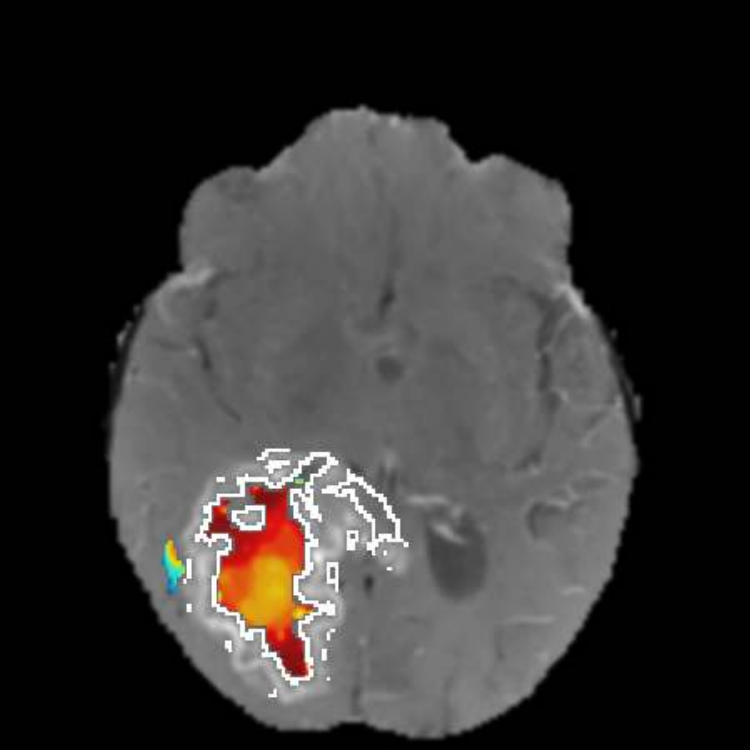} &
		\includegraphics[width=25mm]{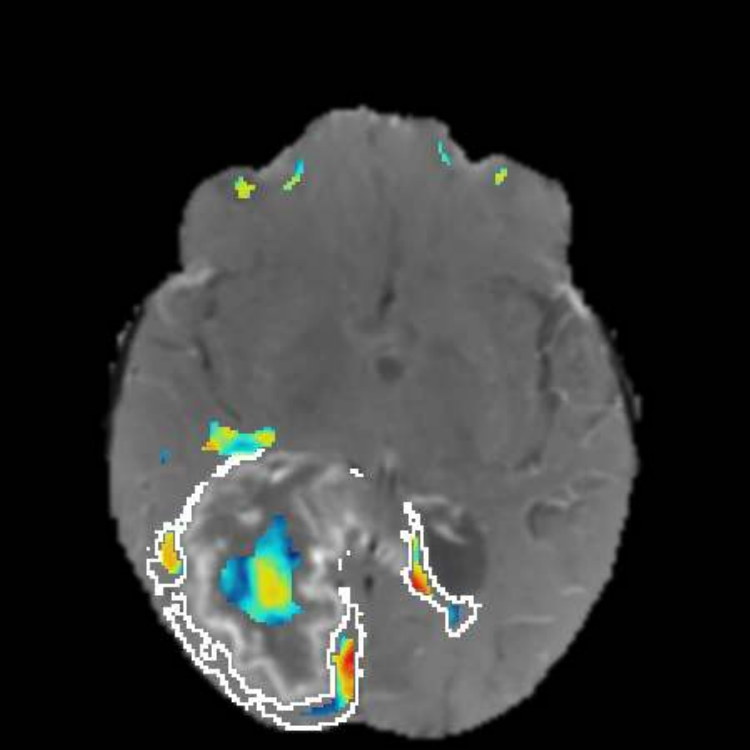} &
		\includegraphics[width=25mm]{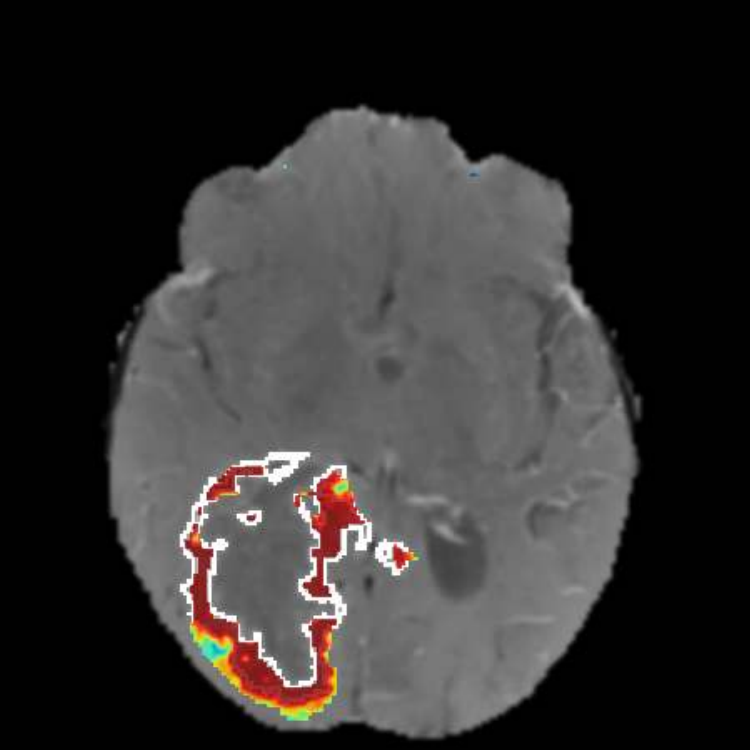} &
		\includegraphics[width=26mm]{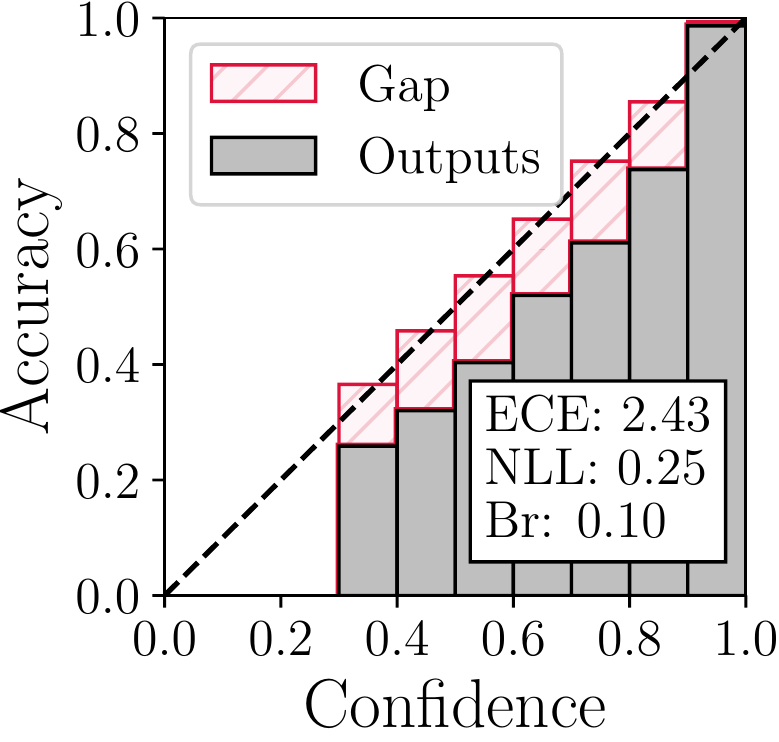} \\
		\multicolumn{5}{r}{\includegraphics[width=50mm]{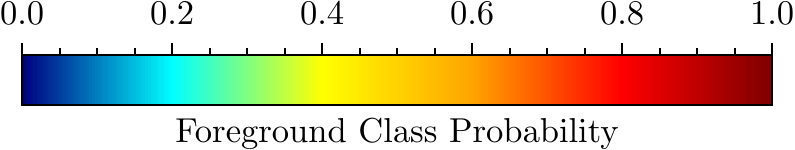}}\\

\end{tabular}
\caption{
Examples of uncertainty estimation quality for brain tumor segmentation using different methods.
MRI images are overlaid with class probabilities, and reliability diagrams (together with ECE\%, NLL,
and Brier score) are given for that specific volume. 
In the reliability diagrams only the bins with greater than 1000 samples are shown. 
}
\label{fig:drawing_brain}
\end{figure*}

\begin{figure*}[h]
	\centering
	\setlength{\tabcolsep}{1pt}
	\begin{tabular}{lcccc}
		\parbox[t]{4mm}{\rotatebox[origin=l]{90}{\footnotesize{$~~~~~~~~{\mathcal{L}_{CE}}$}}} &
		\includegraphics[width=25mm]{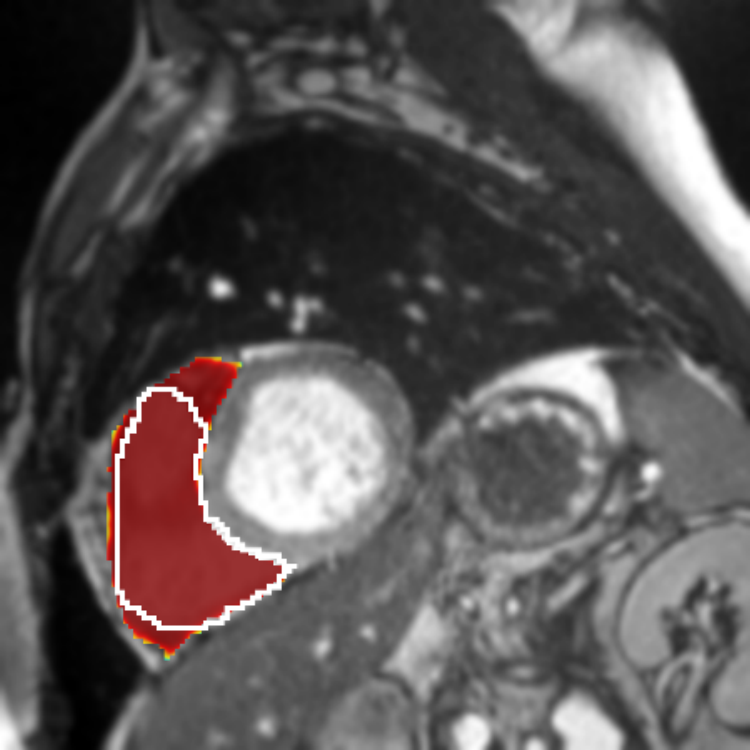} &
		\includegraphics[width=25mm]{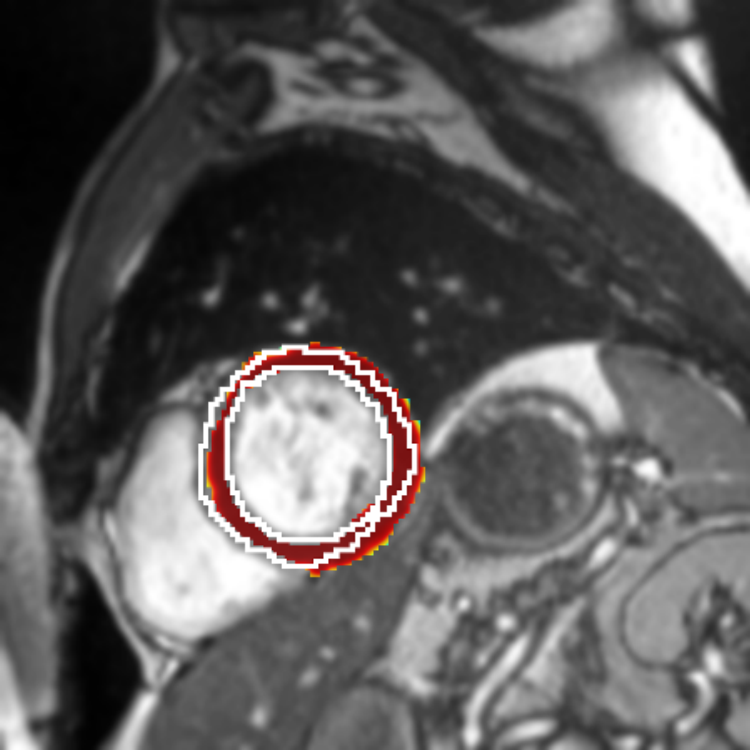} &
		\includegraphics[width=25mm]{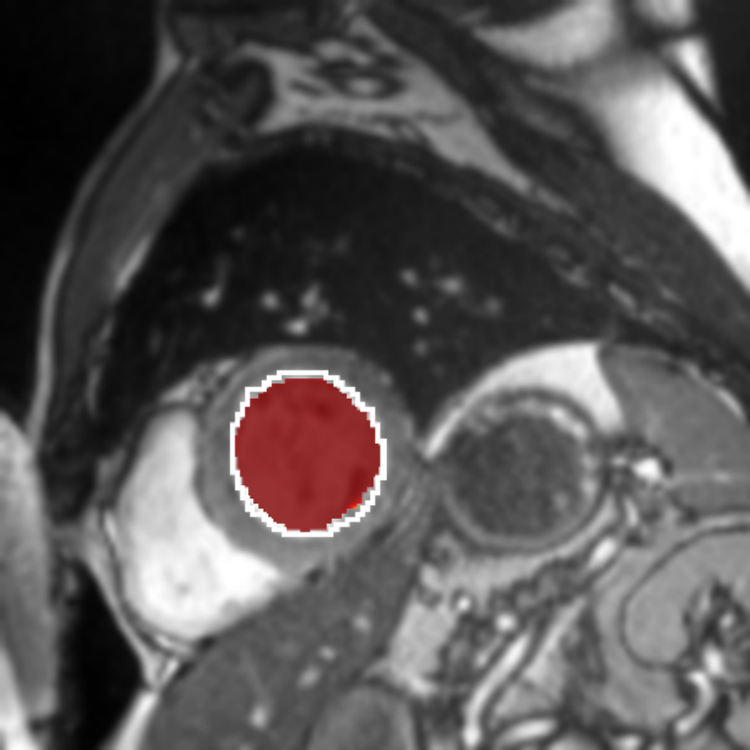} &
		\includegraphics[width=26mm]{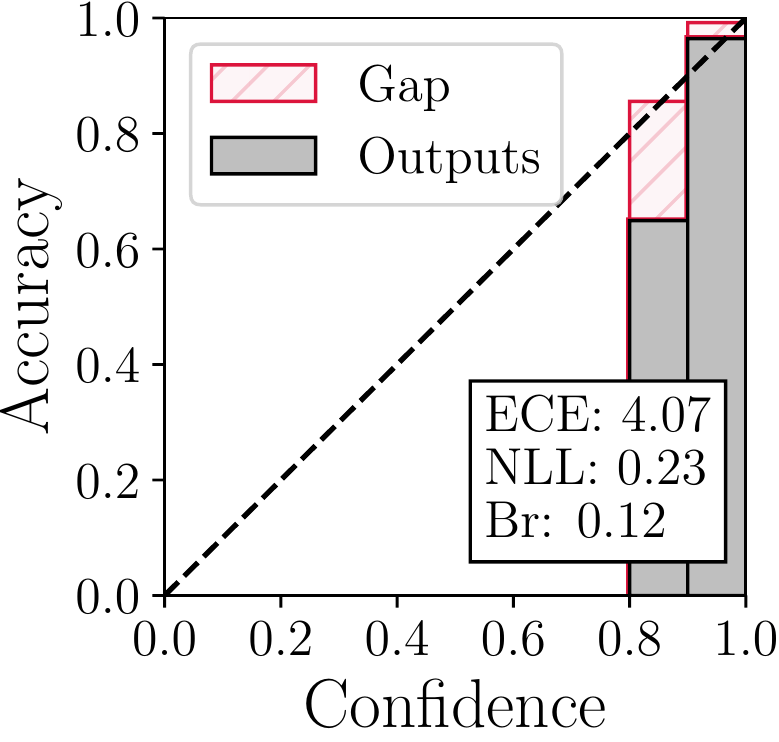} \\
		
		\parbox[t]{4mm}{\rotatebox[origin=l]{90}{\footnotesize{$~~{\mathcal{L}_{CE}}~(MCDO)$}}} &
		\includegraphics[width=25mm]{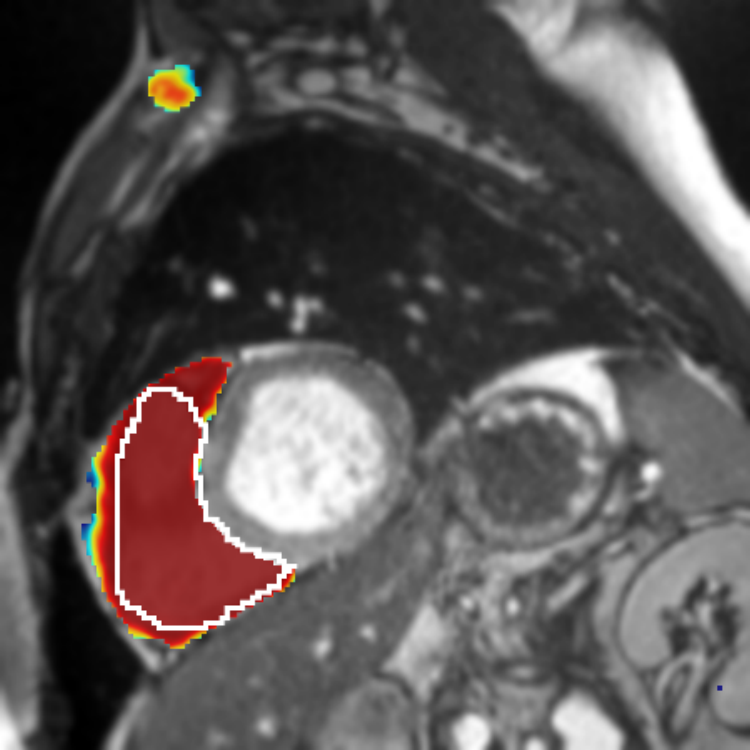} &
		\includegraphics[width=25mm]{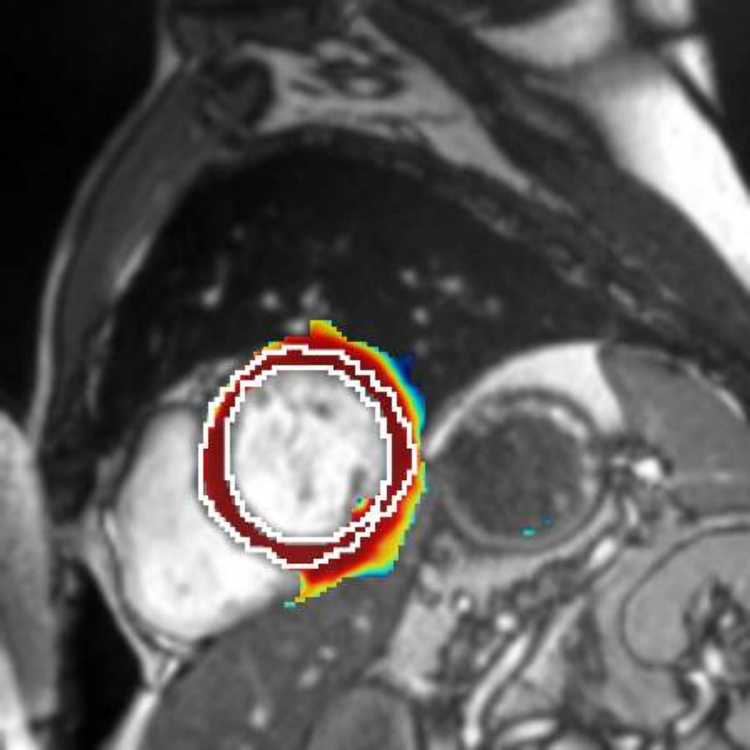} &
		\includegraphics[width=25mm]{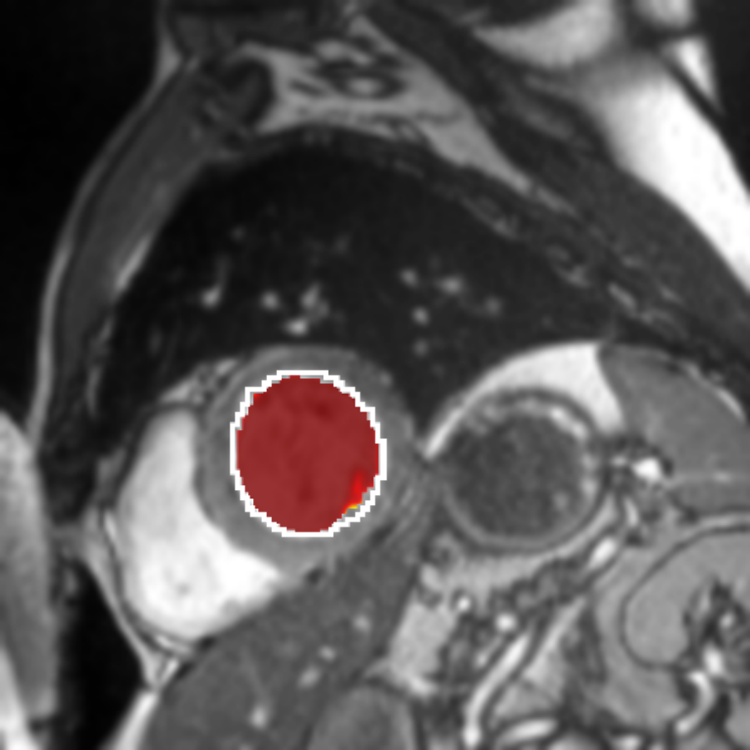} &
		\includegraphics[width=26mm]{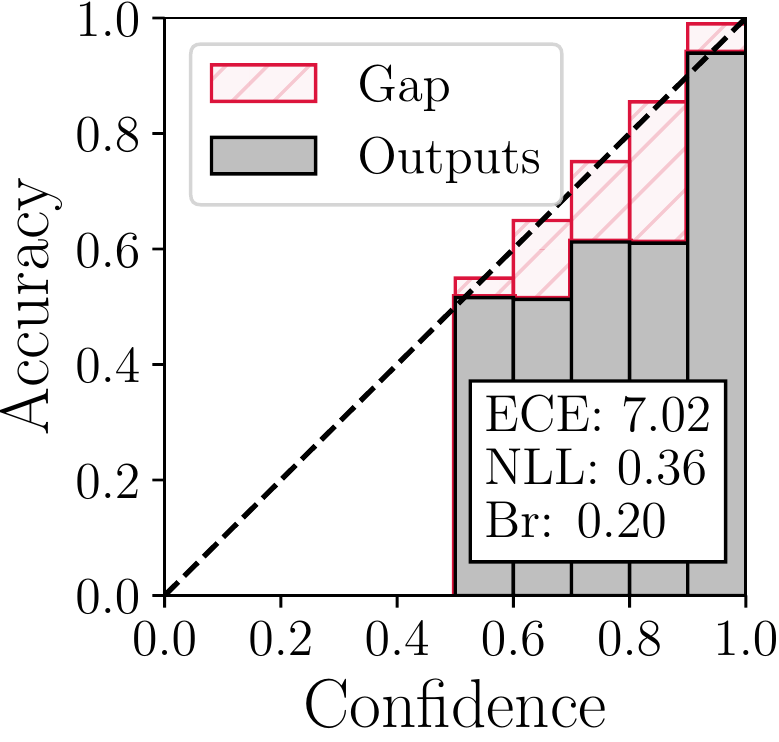} \\

		\parbox[t]{4mm}{\rotatebox[origin=l]{90}{\footnotesize{$~~~~{\mathcal{L}_{CE}~(EN)}$}}} &
		\includegraphics[width=25mm]{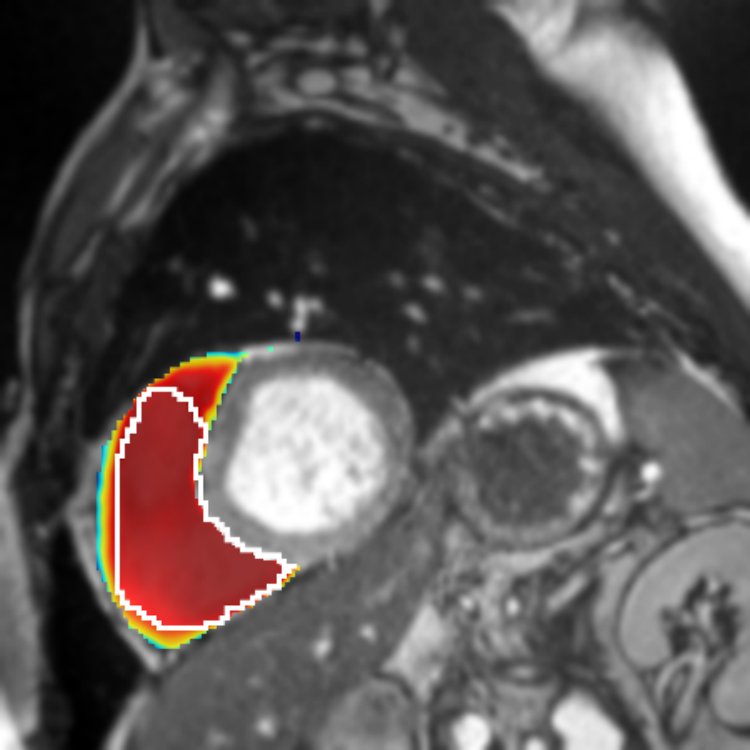} &
		\includegraphics[width=25mm]{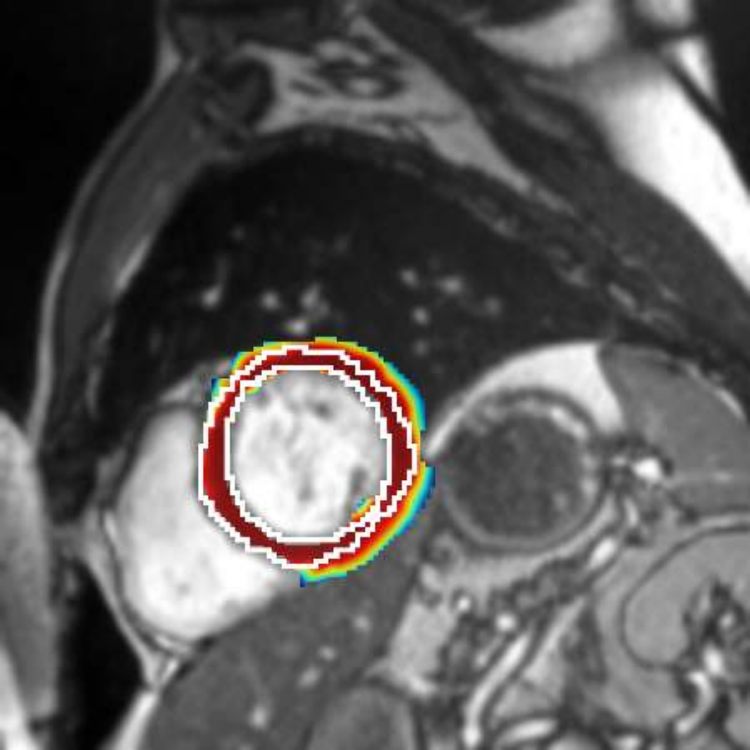} &
		\includegraphics[width=25mm]{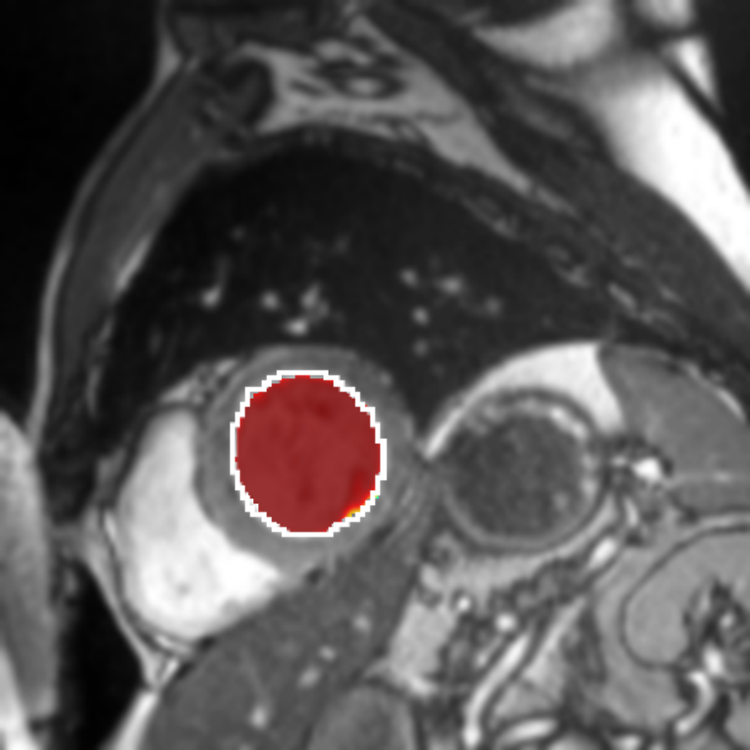} &
		\includegraphics[width=26mm]{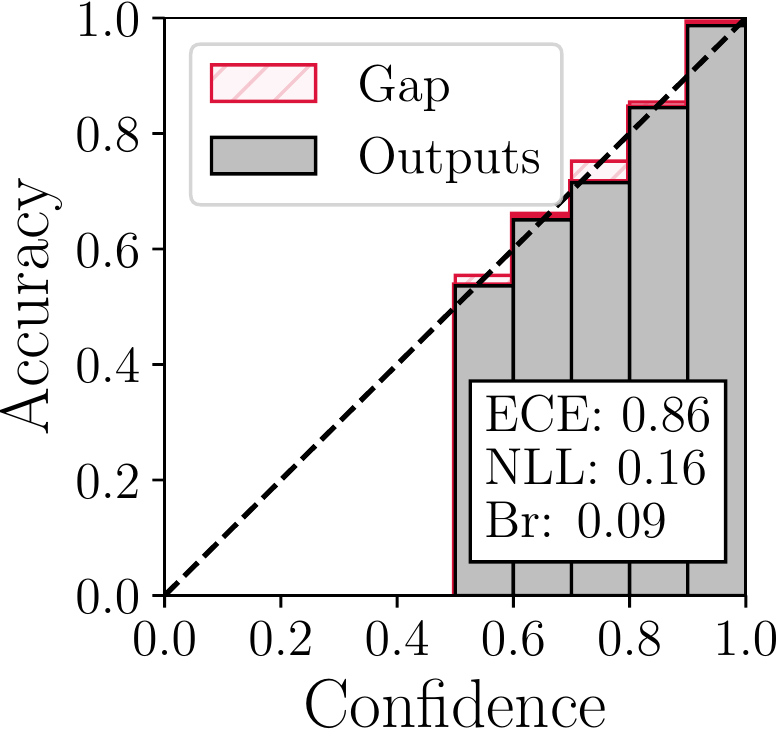} \\
		
		\parbox[t]{4mm}{\rotatebox[origin=l]{90}{\footnotesize{$~~~~~~~~{\mathcal{L}_{DSC}}$}}} &
		\includegraphics[width=25mm]{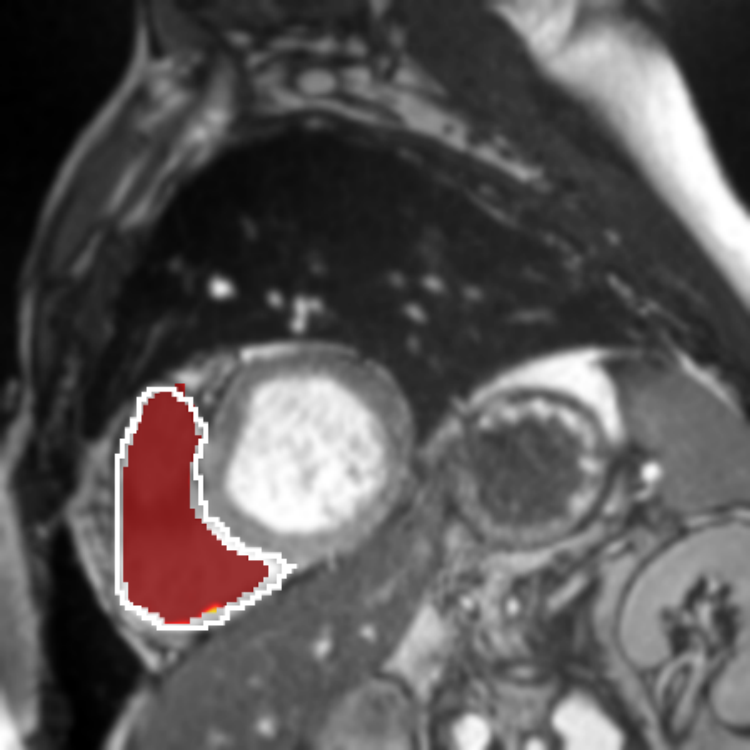} &
		\includegraphics[width=25mm]{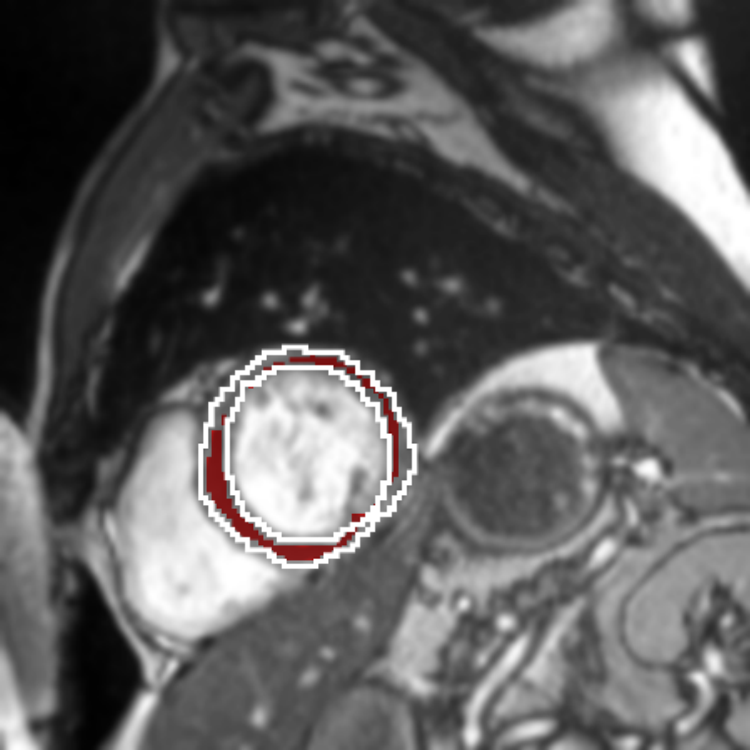} &
		\includegraphics[width=25mm]{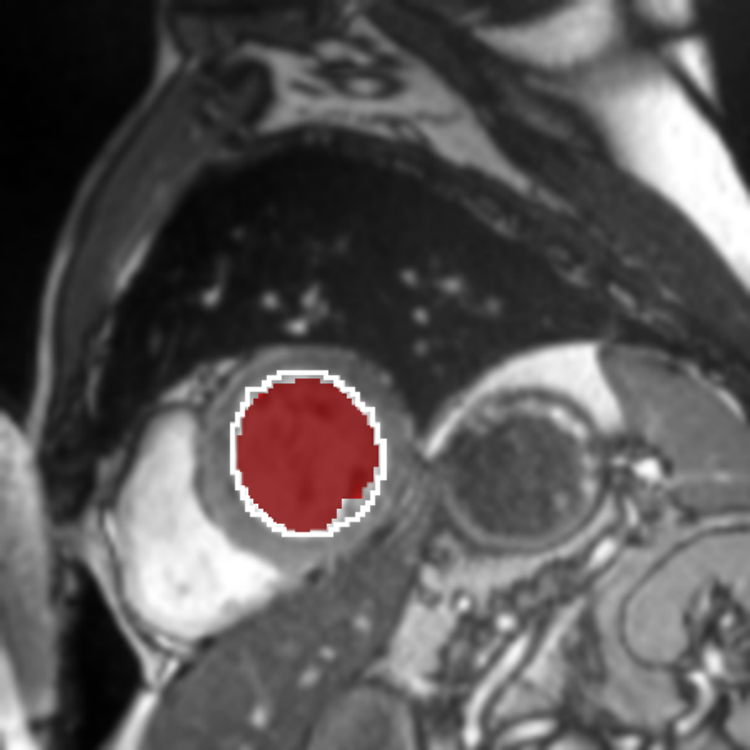} &
		\includegraphics[width=26mm]{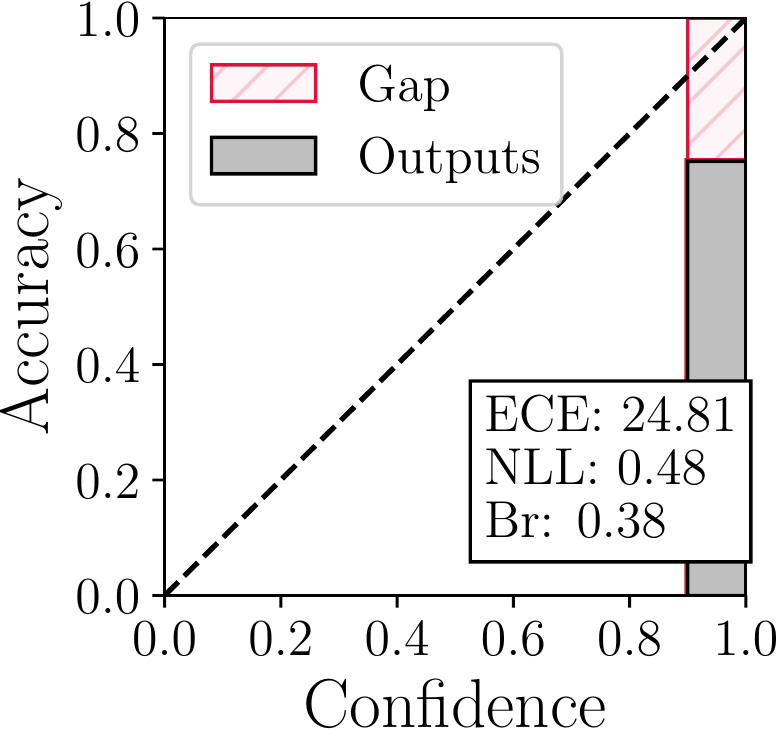} \\
		
		\parbox[t]{4mm}{\rotatebox[origin=l]{90}{\footnotesize{$~~{\mathcal{L}_{DSC}~(MCDO)}$}}} &
		\includegraphics[width=25mm]{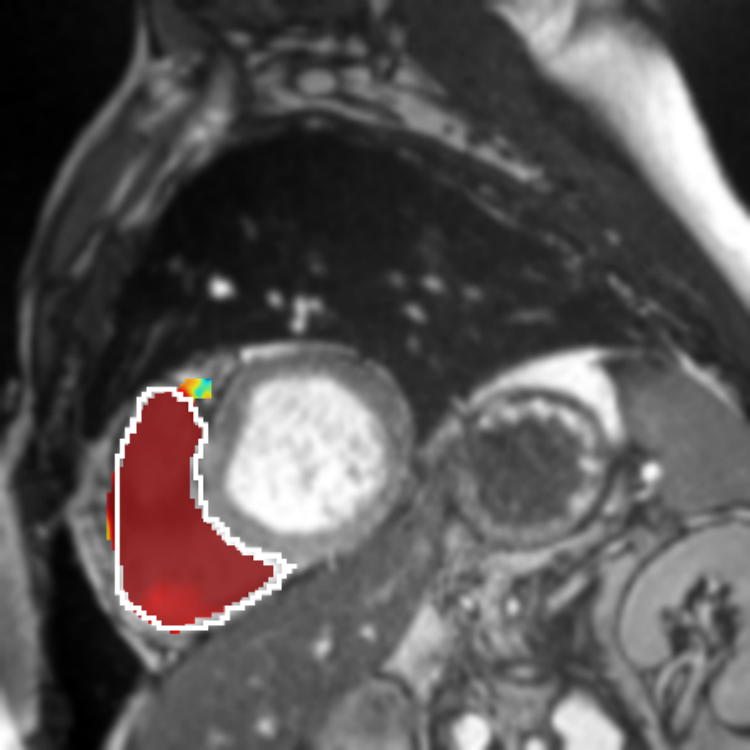} &
		\includegraphics[width=25mm]{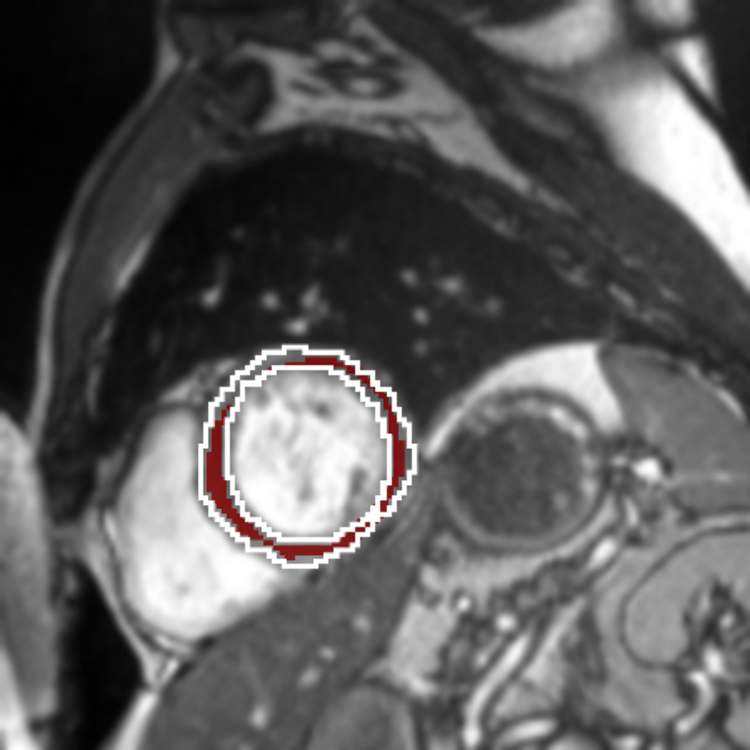} &
		\includegraphics[width=25mm]{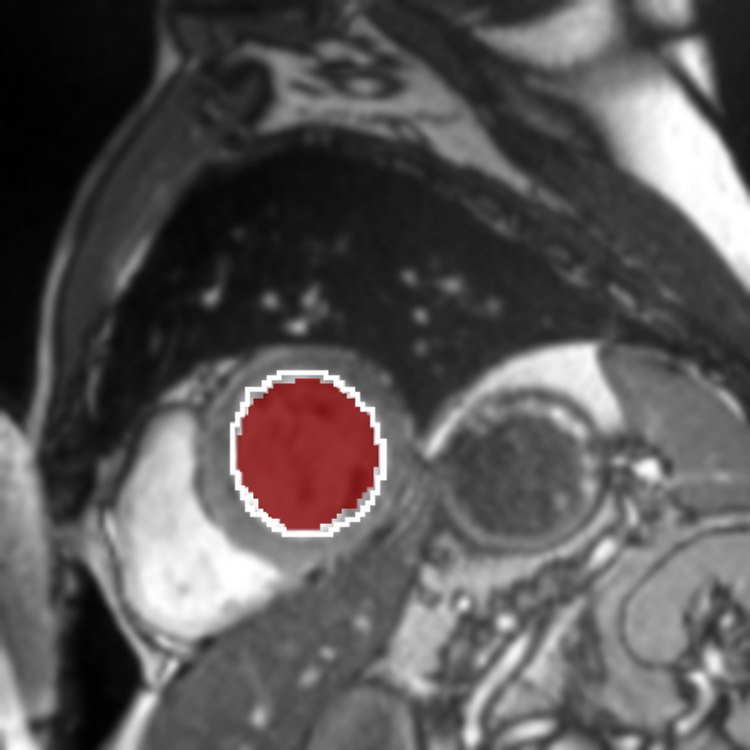} &
		\includegraphics[width=26mm]{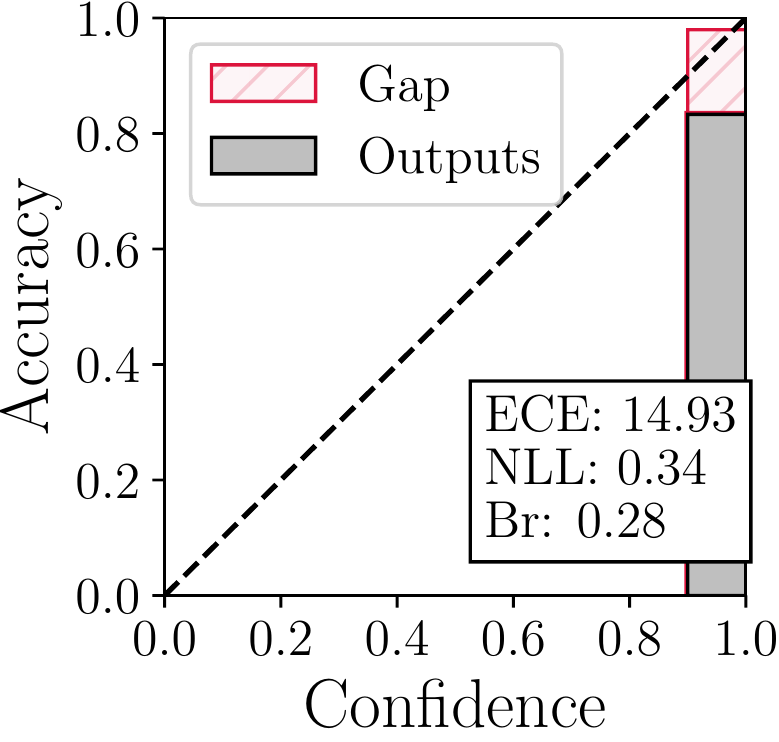} \\
		
		\parbox[t]{4mm}{\rotatebox[origin=l]{90}{\footnotesize{$~~~{\mathcal{L}_{DSC}~(EN)}$}}} &
		\includegraphics[width=25mm]{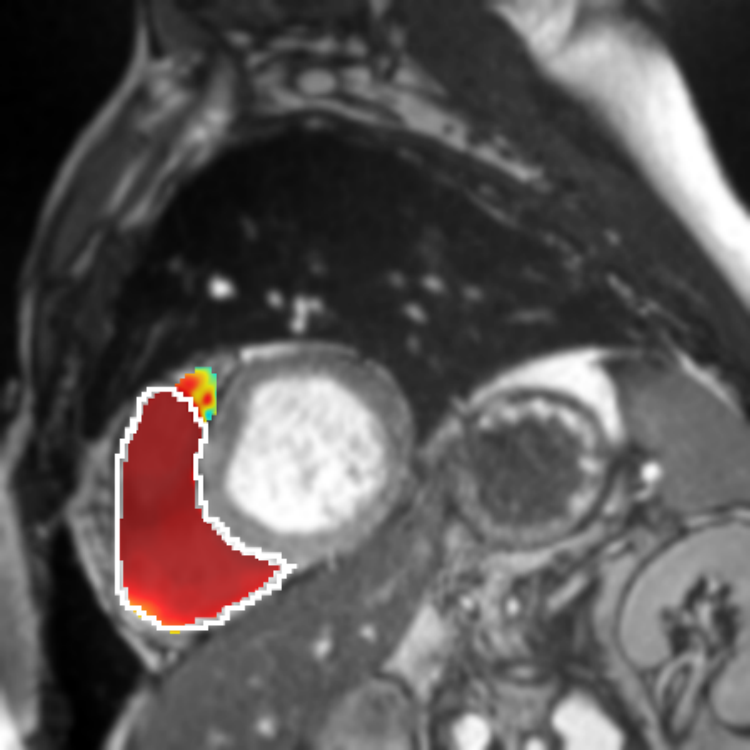} &
		\includegraphics[width=25mm]{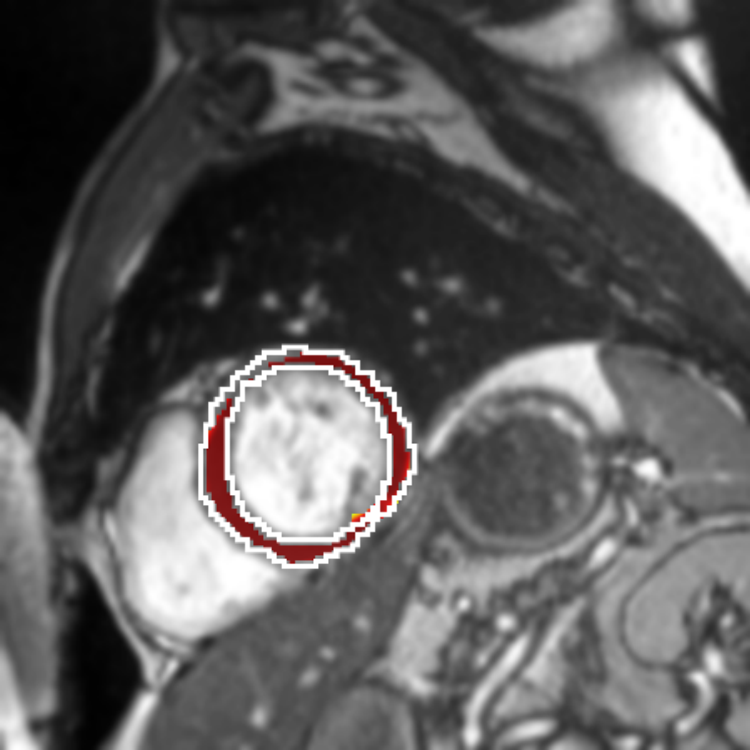} &
		\includegraphics[width=25mm]{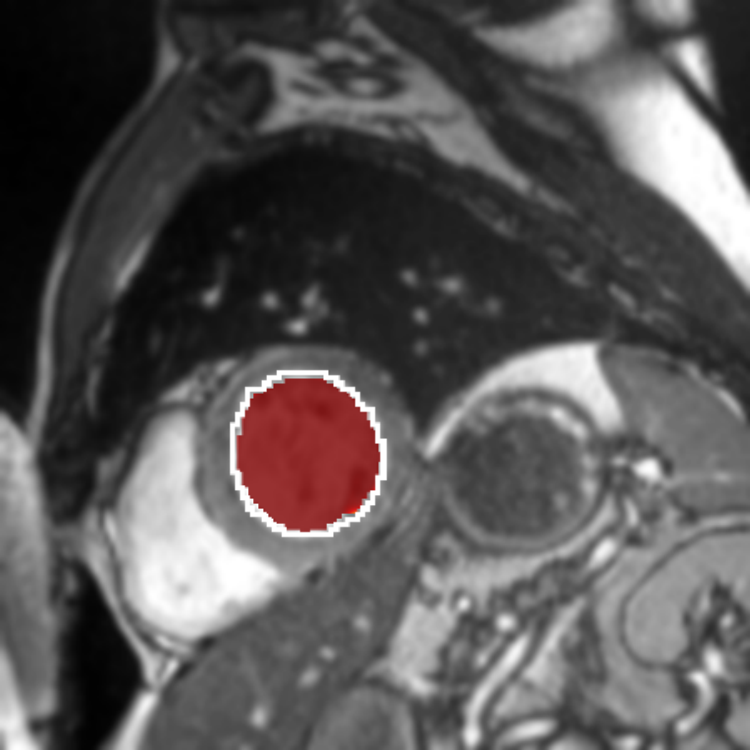} &
		\includegraphics[width=26mm]{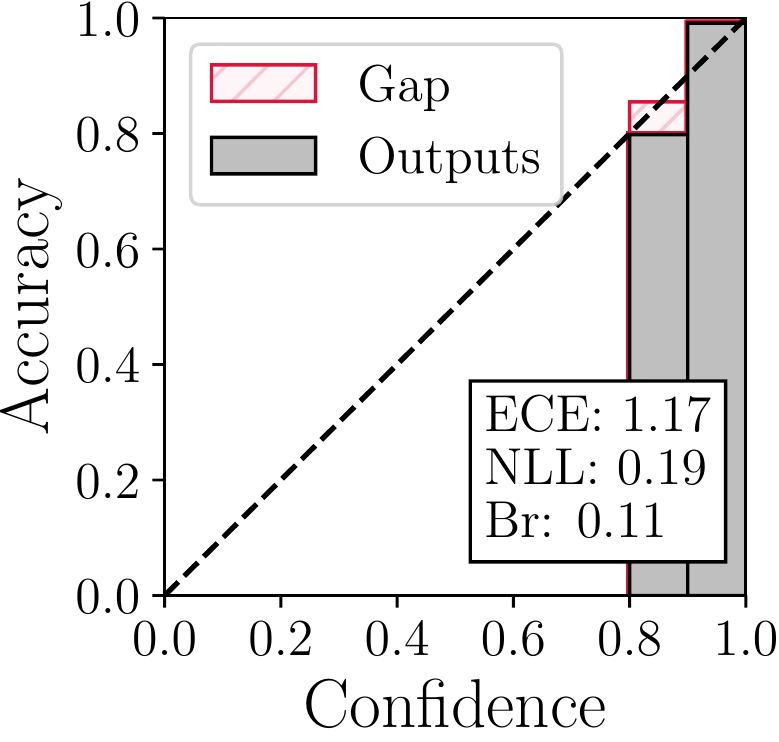} \\
		\multicolumn{5}{r}{\includegraphics[width=50mm]{figures/cb.pdf}}\\
	\end{tabular}
	\caption{
Examples of uncertainty estimation quality for heart segmentation using different methods.
MRI images are overlaid with class probabilities, and reliability diagrams (together with ECE\%, NLL,
and Brier score) are given for that specific volume. 
In the reliability diagrams only the bins with greater than 1000 samples are shown. 
	}
	\label{fig:drawing_heart}
\end{figure*}

\begin{figure*}[h]
	\centering
	\setlength{\tabcolsep}{1pt}
	\begin{tabular}{lcccl}
		\parbox[t]{4mm}{\rotatebox[origin=l]{90}{\footnotesize{$~~~~~~~~{\mathcal{L}_{CE}}$}}} &
		\includegraphics[width=25mm]{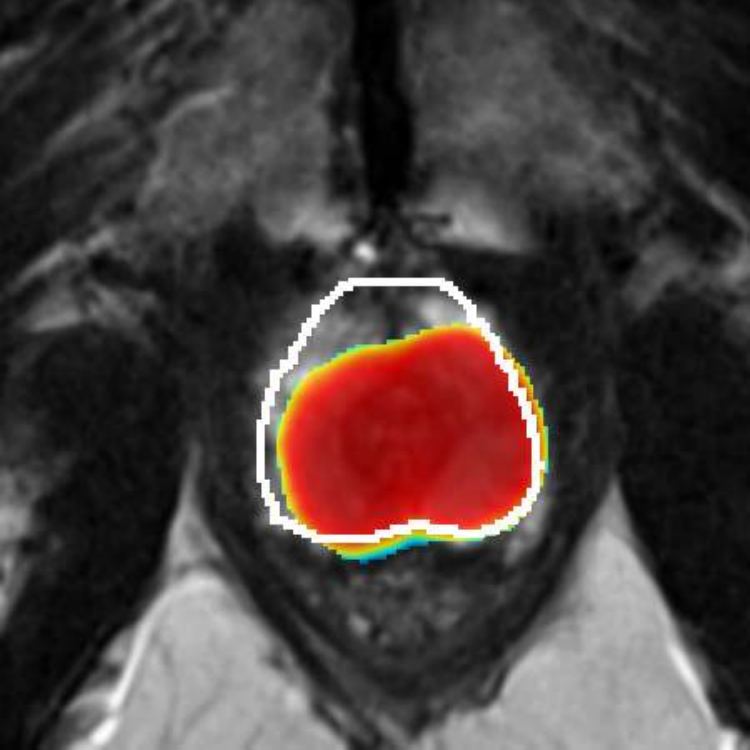} &
		\includegraphics[width=25mm]{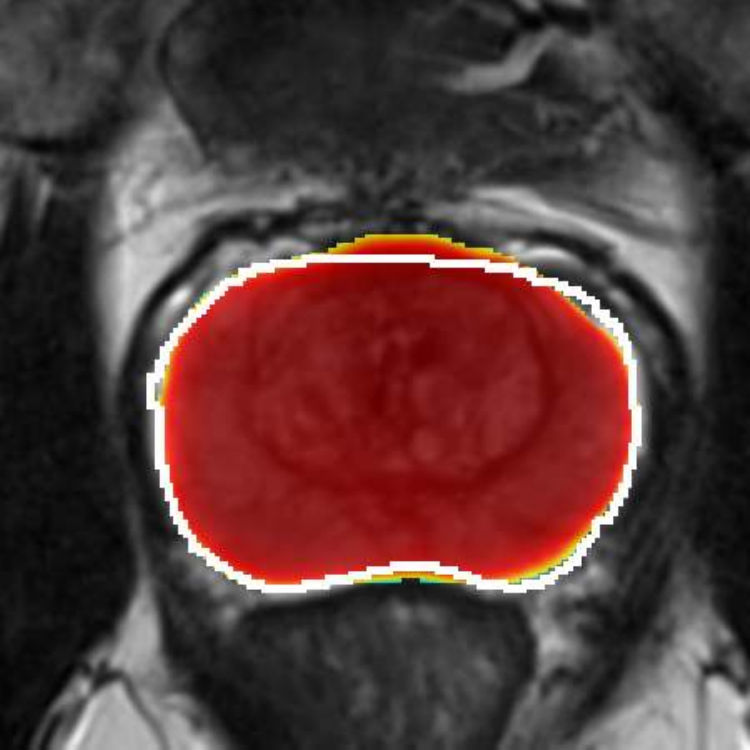} &
		\includegraphics[width=25mm]{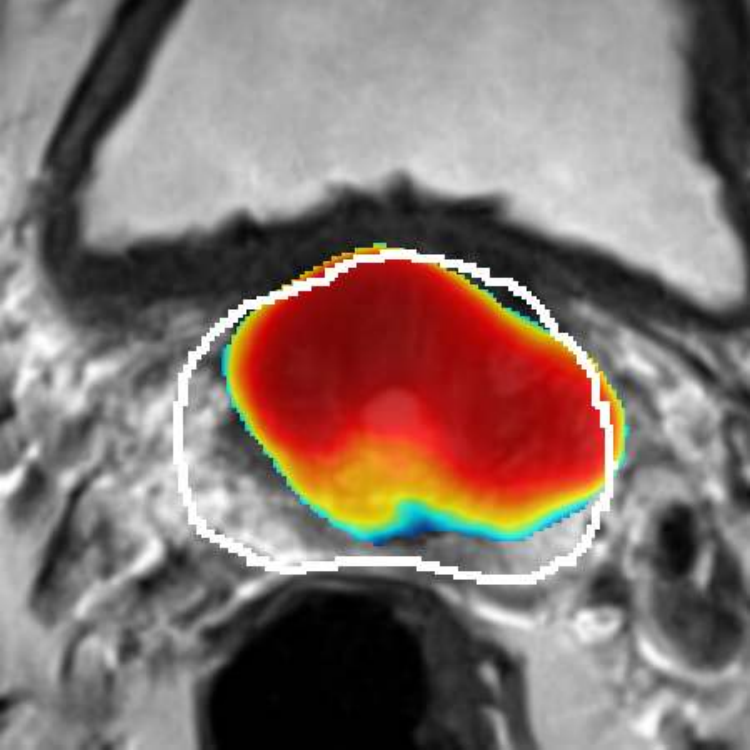} &
		\includegraphics[width=26mm]{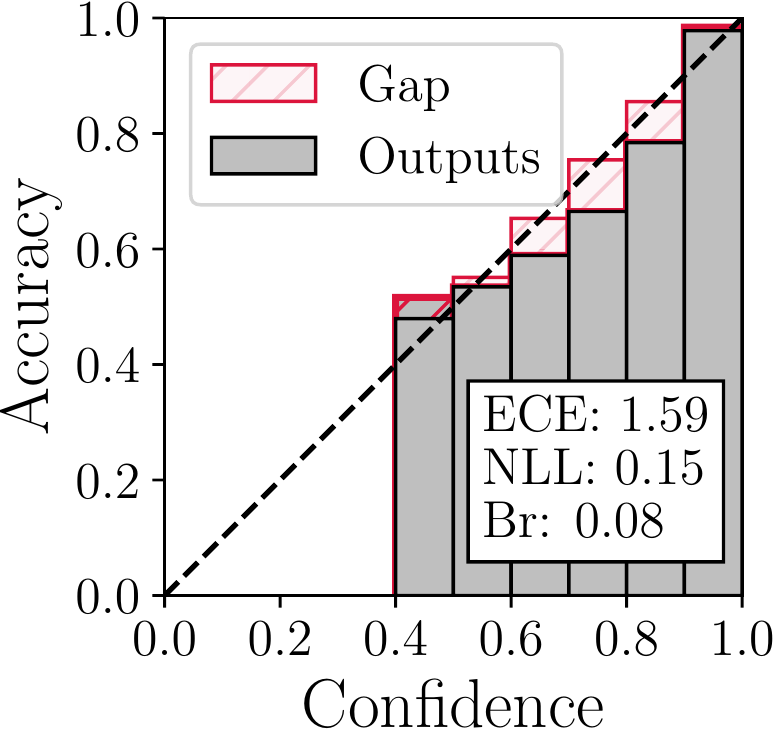} \\
		
		\parbox[t]{4mm}{\rotatebox[origin=l]{90}{\footnotesize{$~{\mathcal{L}_{CE}}~(MCDO)$}}} &
		\includegraphics[width=25mm]{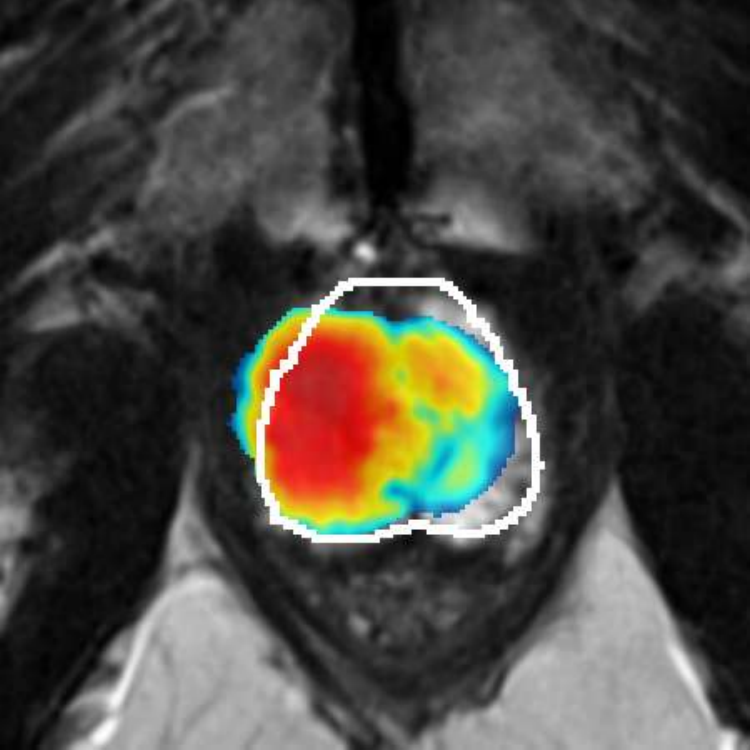} &
		\includegraphics[width=25mm]{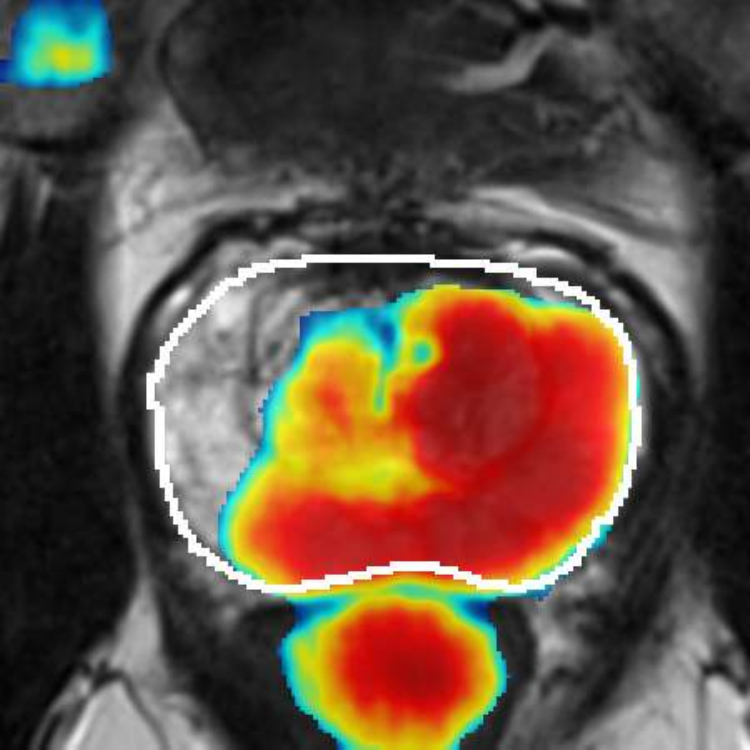} &
		\includegraphics[width=25mm]{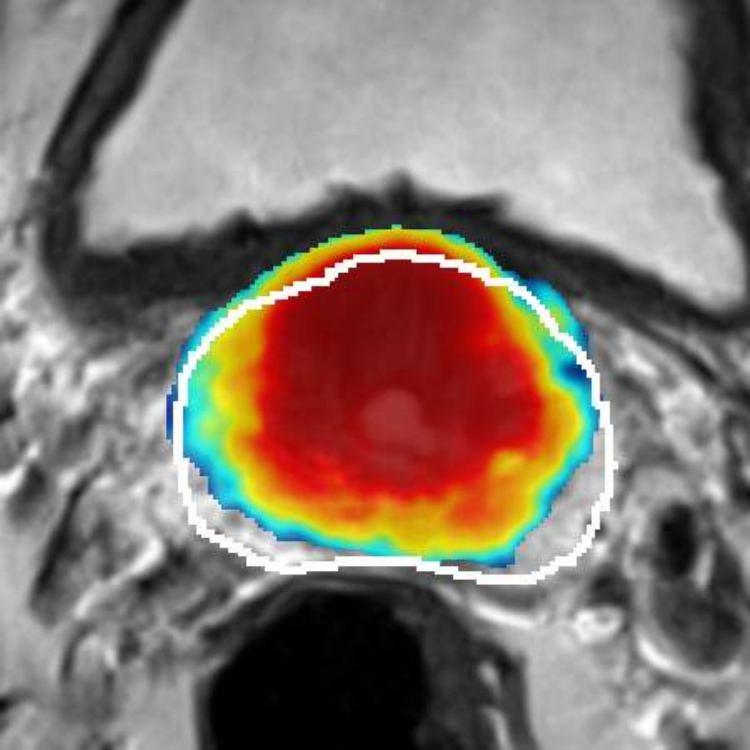} &
		\includegraphics[width=26mm]{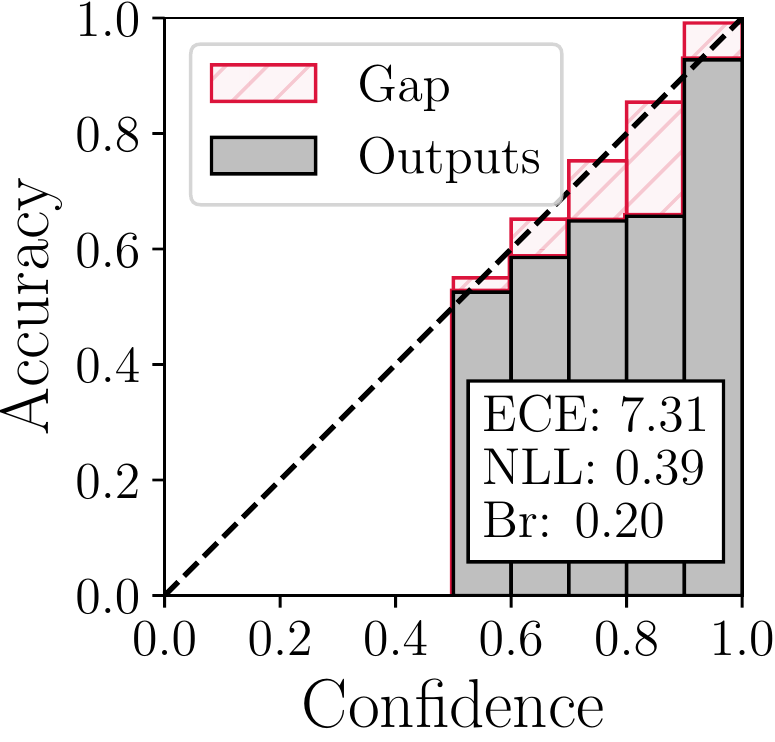} \\
		
		\parbox[t]{4mm}{\rotatebox[origin=l]{90}{\footnotesize{$~{\mathcal{L}_{CE}}~(EN)$}}} &
		\includegraphics[width=25mm]{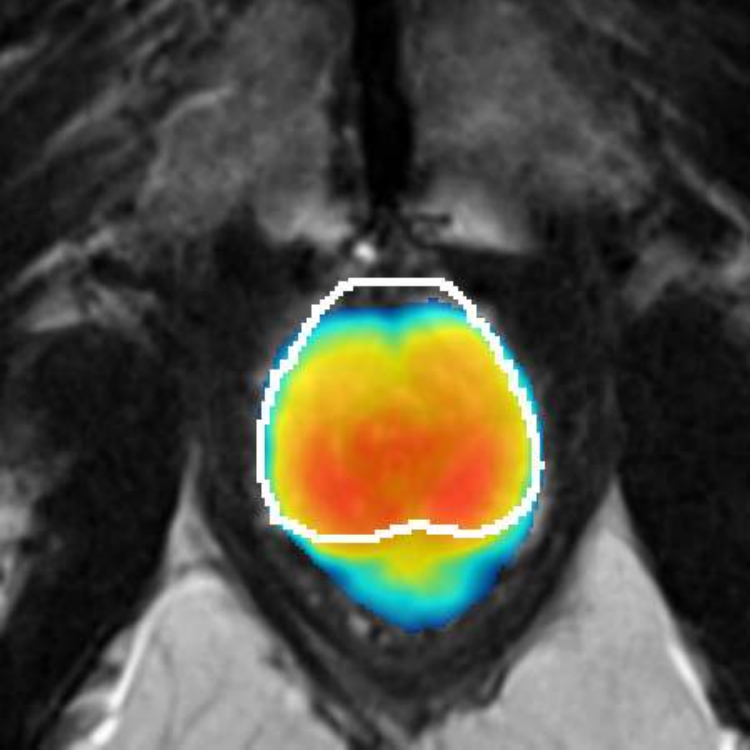} &
		\includegraphics[width=25mm]{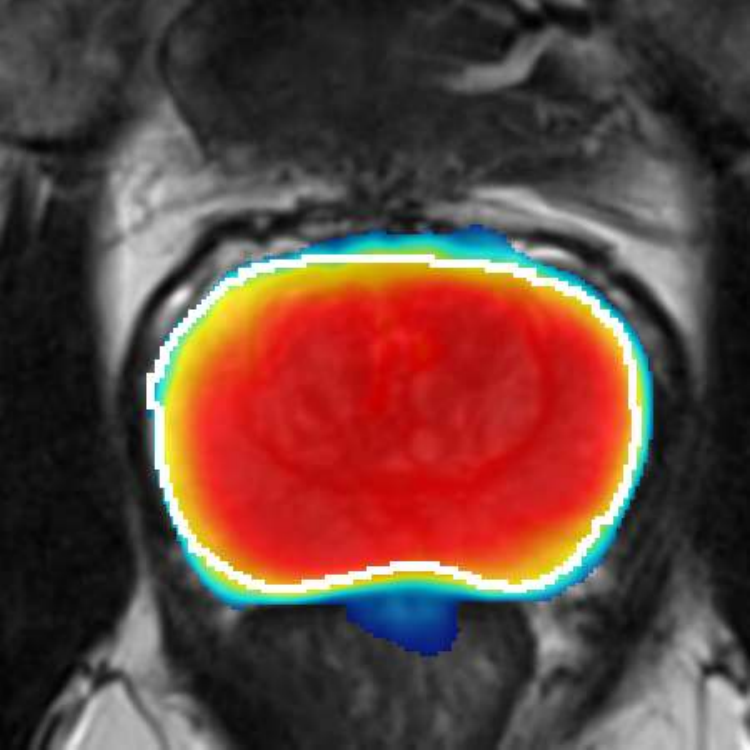} &
		\includegraphics[width=25mm]{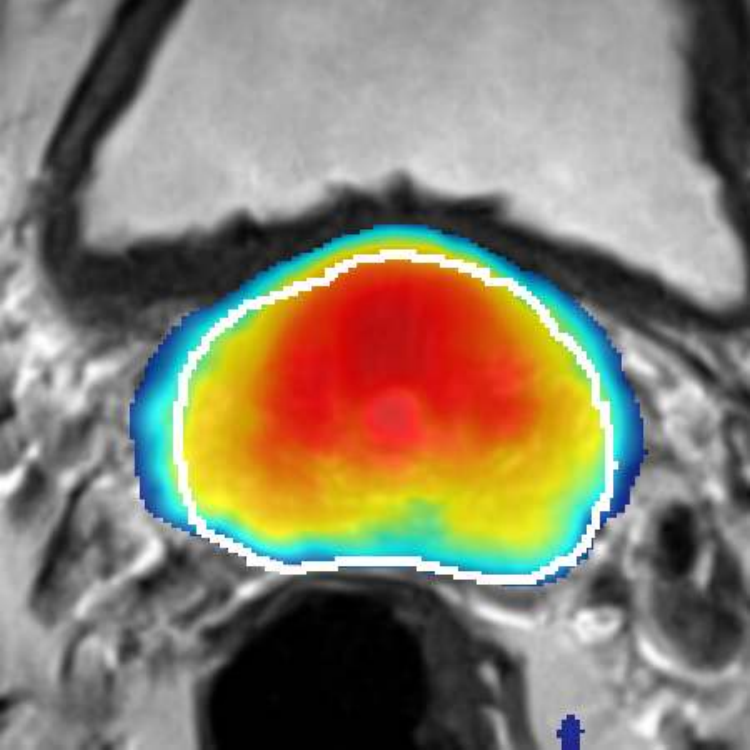} &
		\includegraphics[width=26mm]{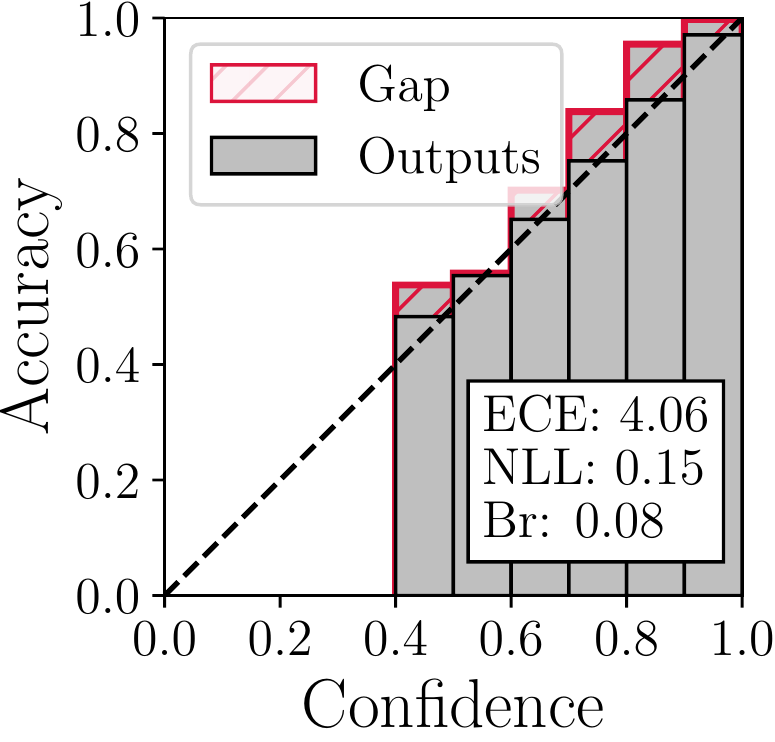} \\
		
		\parbox[t]{4mm}{\rotatebox[origin=l]{90}{\footnotesize{$~~~~~~~~{\mathcal{L}_{DSC}}$}}} &
		\includegraphics[width=25mm]{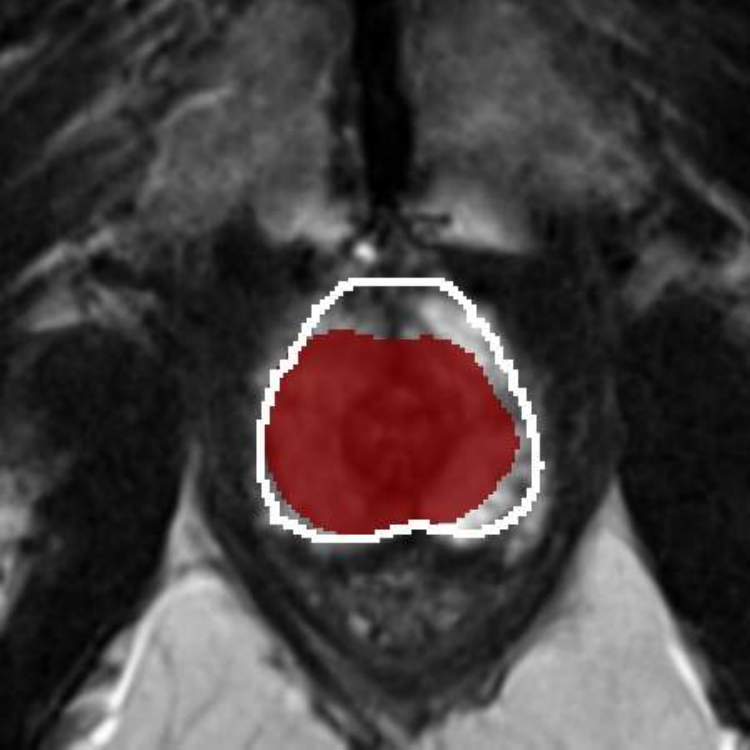} &
		\includegraphics[width=25mm]{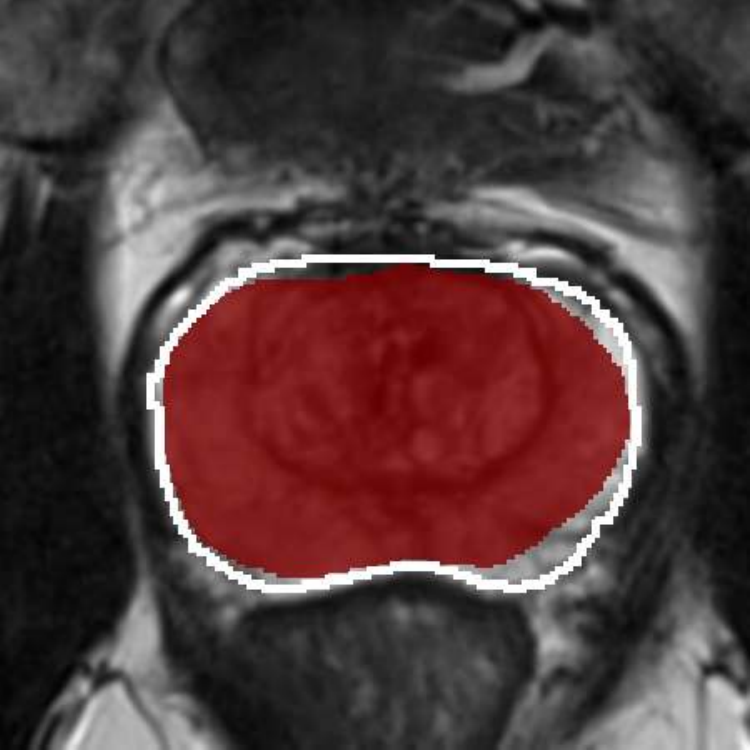} &
		\includegraphics[width=25mm]{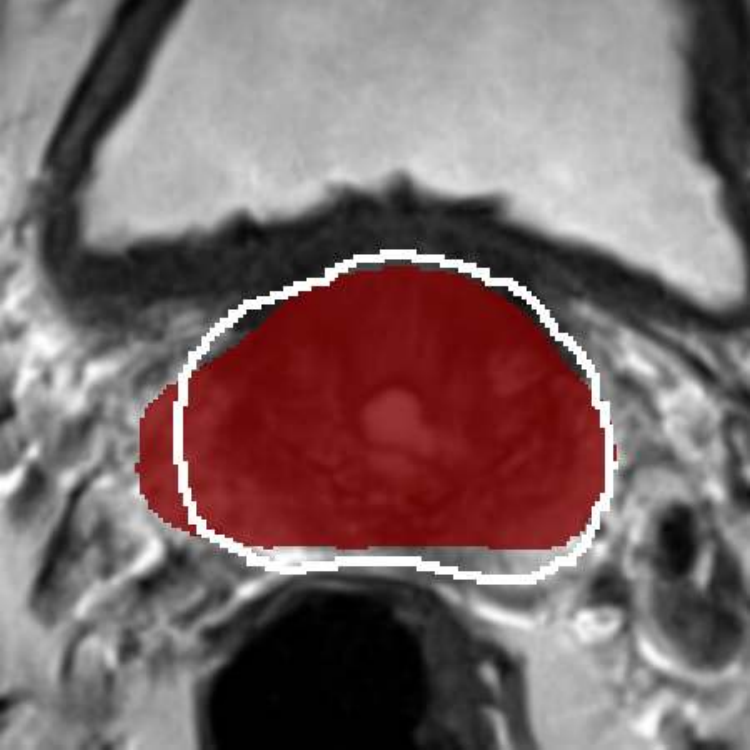} &
		\includegraphics[width=26mm]{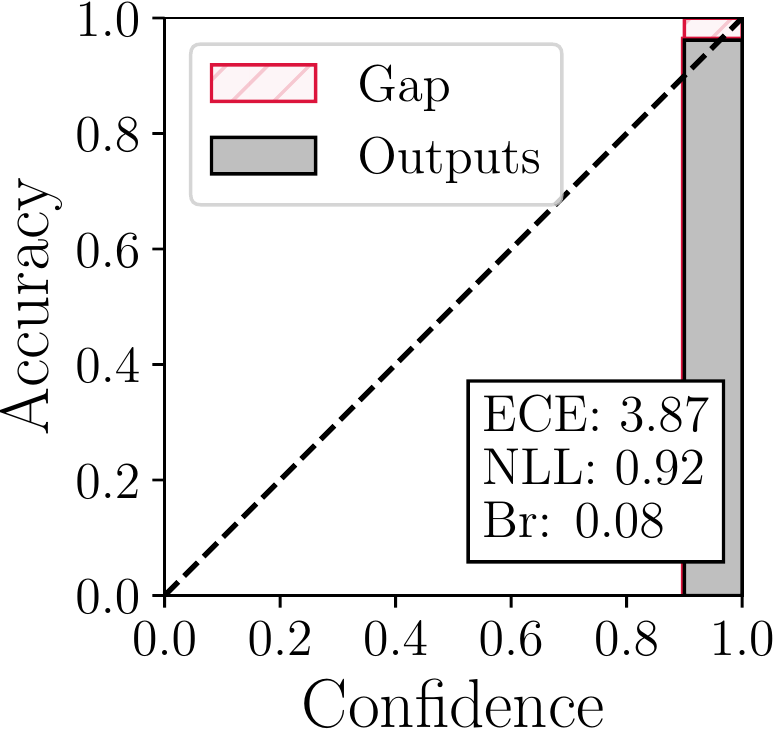} \\
		
		\parbox[t]{4mm}{\rotatebox[origin=l]{90}{\footnotesize{$~~~{\mathcal{L}_{DSC}~(MCDO)}$}}} &
		\includegraphics[width=25mm]{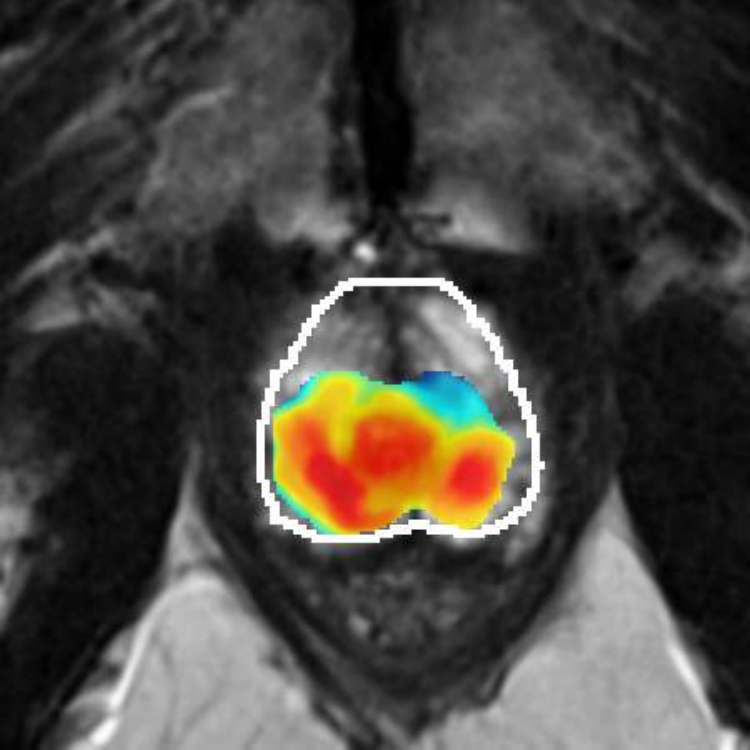} &
		\includegraphics[width=25mm]{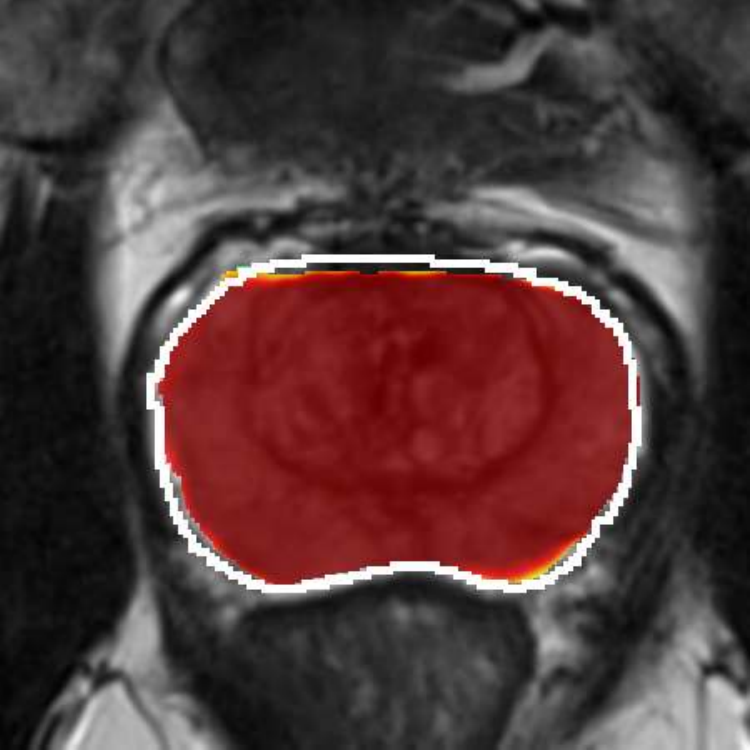} &
		\includegraphics[width=25mm]{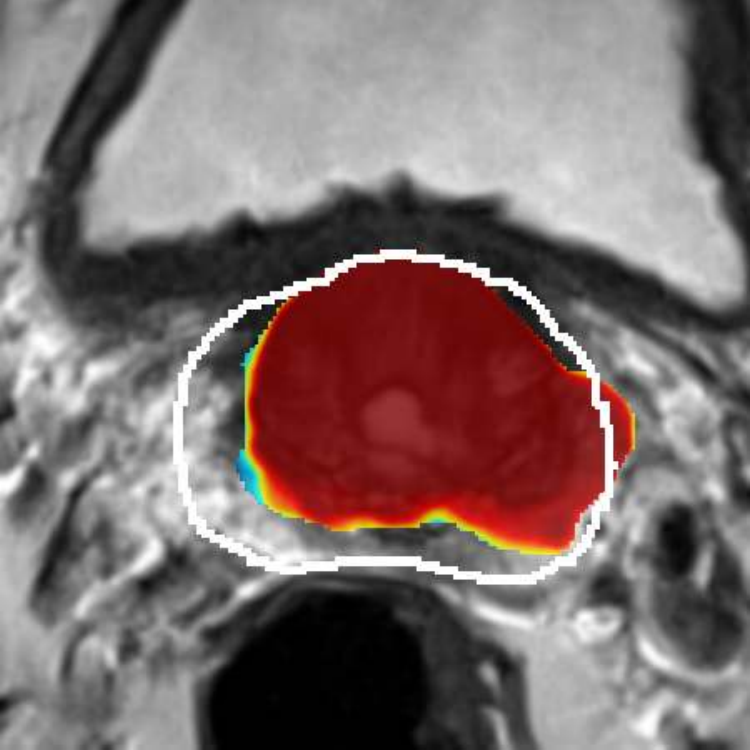} &
		\includegraphics[width=26mm]{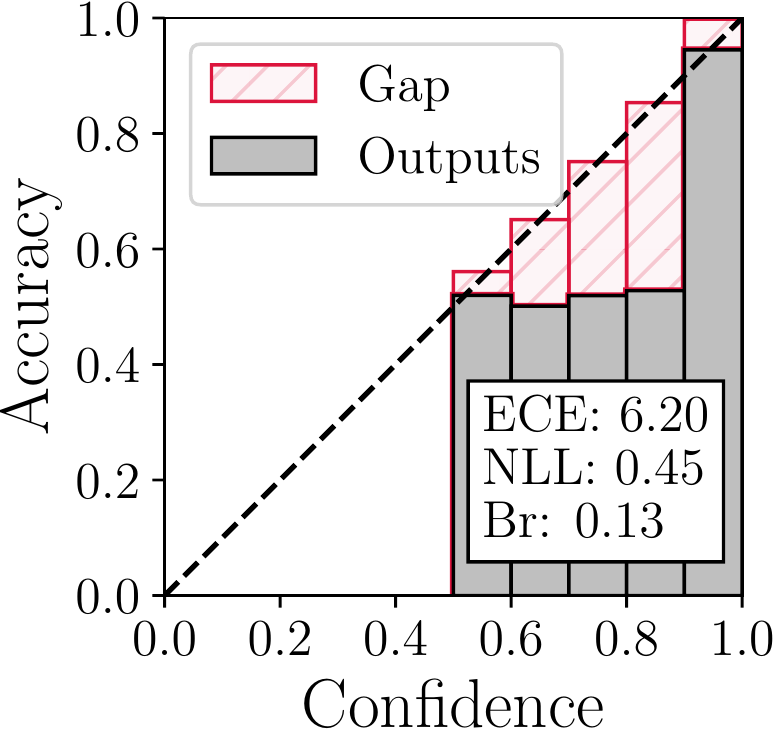} \\

		\parbox[t]{4mm}{\rotatebox[origin=l]{90}{\footnotesize{$~~~{\mathcal{L}_{DSC}~(EN)}$}}} &
		\includegraphics[width=25mm]{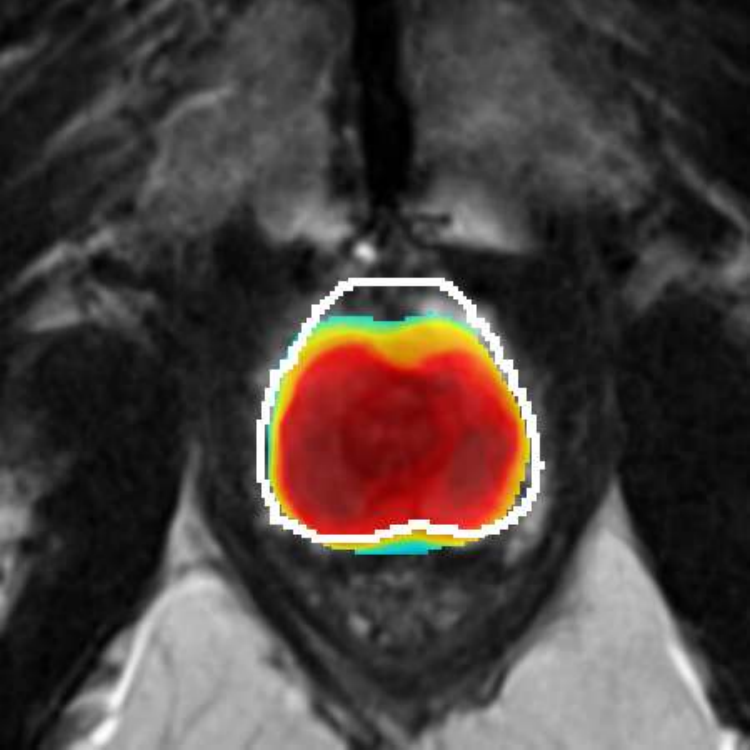} &
		\includegraphics[width=25mm]{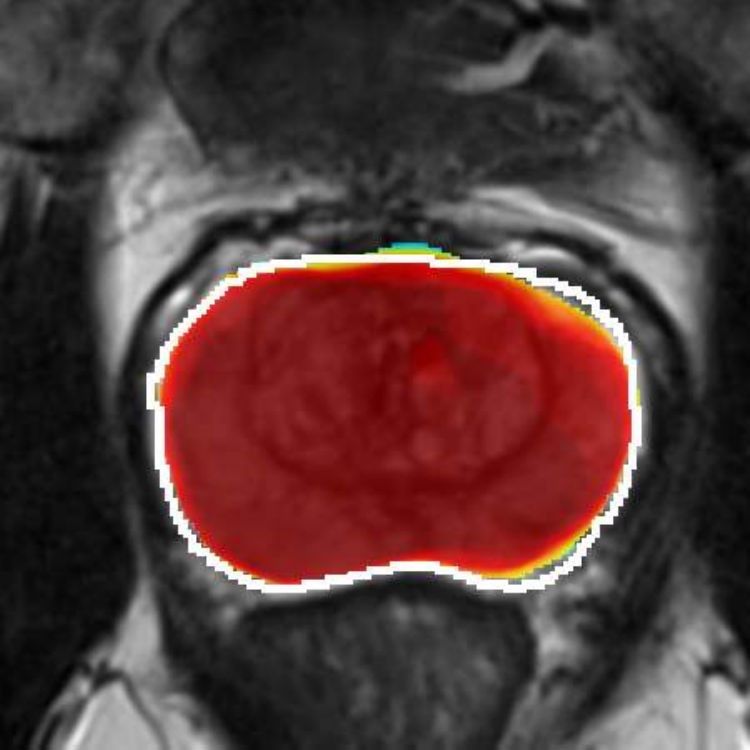} &
		\includegraphics[width=25mm]{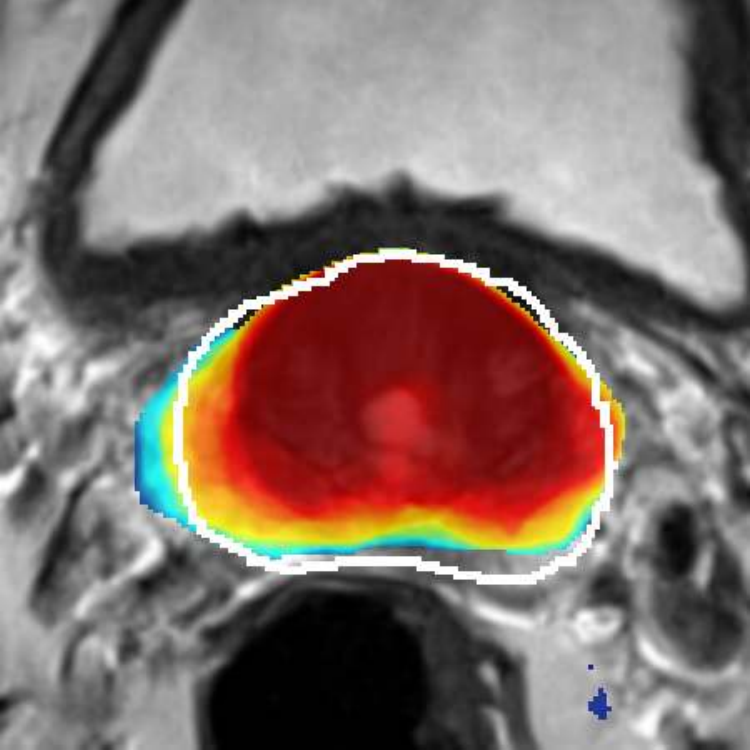} &
		\includegraphics[width=26mm]{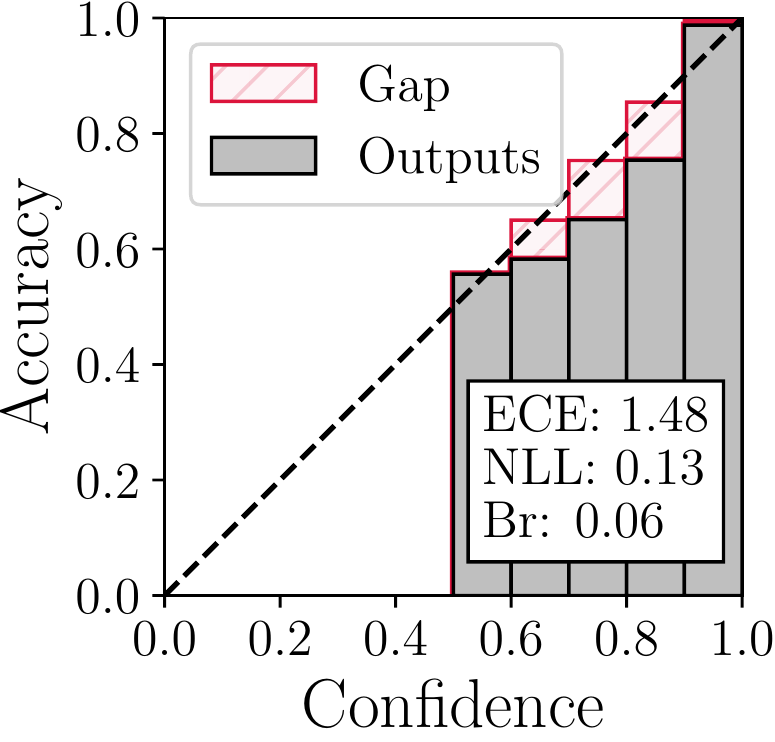} \\
		\multicolumn{5}{r}{\includegraphics[width=50mm]{figures/cb.pdf}}\\
		
	\end{tabular}
	\caption{
Examples of uncertainty estimation quality for prostate segmentation using different methods.
MRI images are overlaid with class probabilities, and reliability diagrams (together with ECE\%, NLL,
and Brier score) are given for that specific volume. 
In the reliability diagrams only the bins with greater than 1000 samples are shown. 
	}
	\label{fig:drawing_prostate}
\end{figure*}

Figure \ref{fig:drawing_brain} visually compares the baselines trained with cross entropy, $\mathcal{L}_{CE}$, Dice loss,  $\mathcal{L}_{DSC}$, with those calibrated with MC dropout and ensembling over the three segmentation tasks.
For each prediction map, a reliability diagram over the whole volume is provided.
In rendering the reliability diagrams only bins with greater than 1000 samples are shown.
Figures \ref{fig:drawing_heart} and \ref{fig:drawing_prostate} show example 
cases for heart and prostate applications.

\clearpage
\section{Number of Models in Ensemble}
\begin{figure}[h]
	\centering
	\begin{tabular}{lll}
		\includegraphics[height=0.25\textwidth]{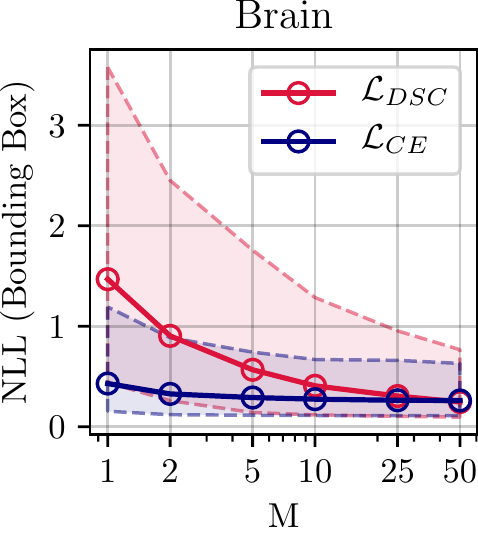}&
		\includegraphics[height=0.25\textwidth]{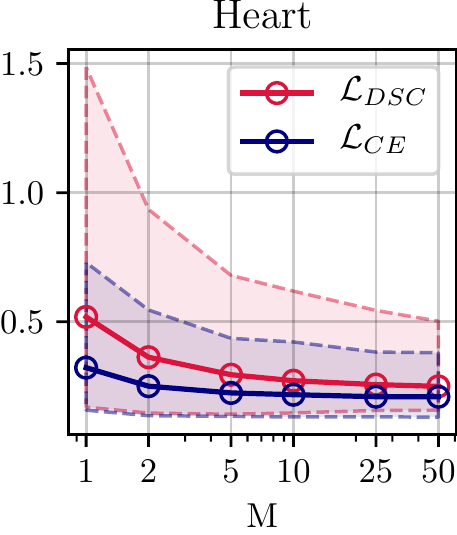}&
		\includegraphics[height=0.25\textwidth]{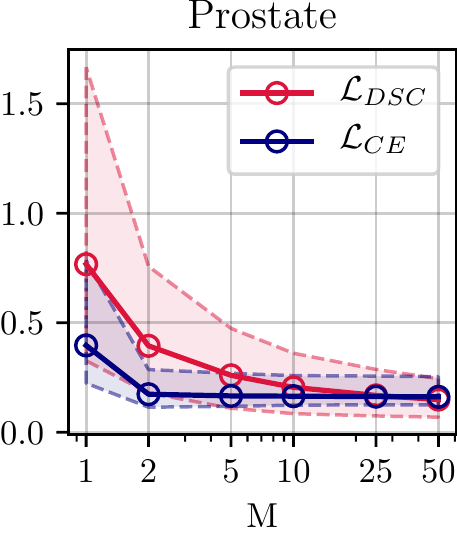}\\
		\includegraphics[height=0.25\textwidth]{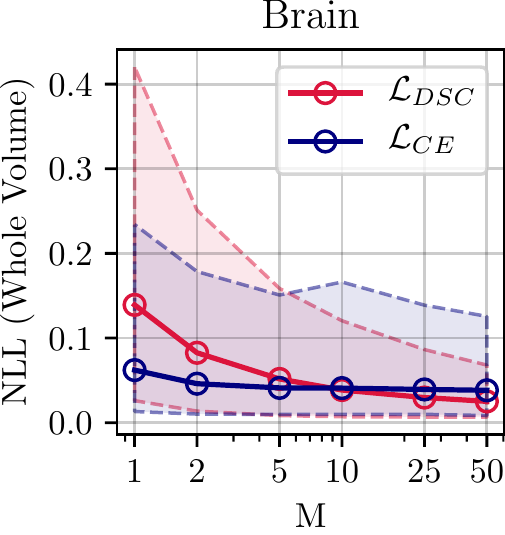}&
		\includegraphics[height=0.25\textwidth]{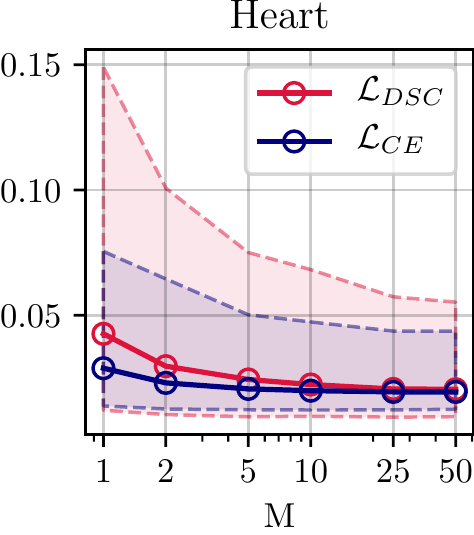}&
		\includegraphics[height=0.25\textwidth]{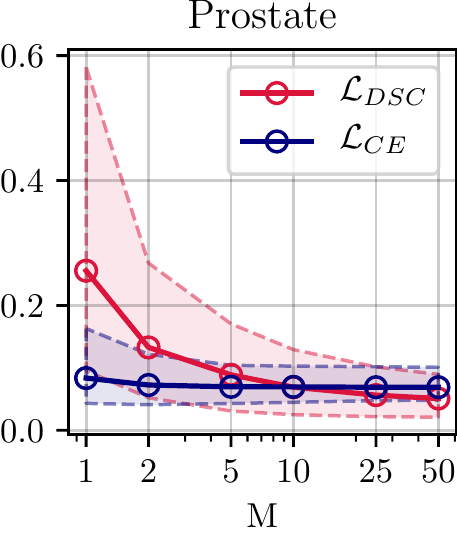}\\
	\end{tabular}
	\caption{
		Improvements in calibration as a function of the number of models in the ensemble.
		Calibration quality in terms of NLL as number of models $M$ increases for prostate, heart, and brain tumor segmentation.
	}
	\label{fig:n_ensembles_with_ci}
\end{figure}

\begin{figure*}[h]
\centering
\setlength{\tabcolsep}{1pt}
\begin{tabular}{llllllll}
	\parbox[t]{3mm}{\rotatebox[origin=l]{90}{\footnotesize{$~~~~~{M=2}$}}} &
	\includegraphics[width=20mm]{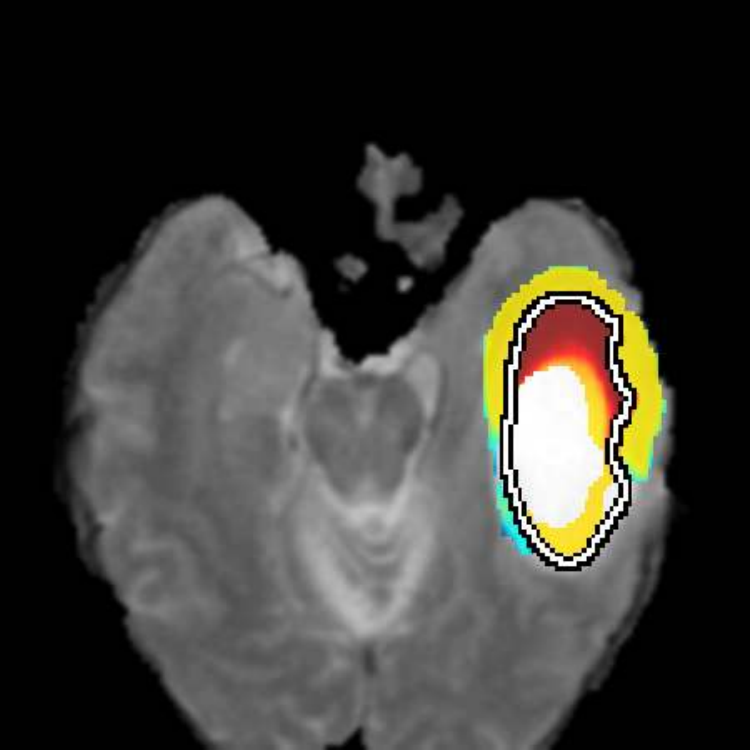}&
	\includegraphics[width=20mm]{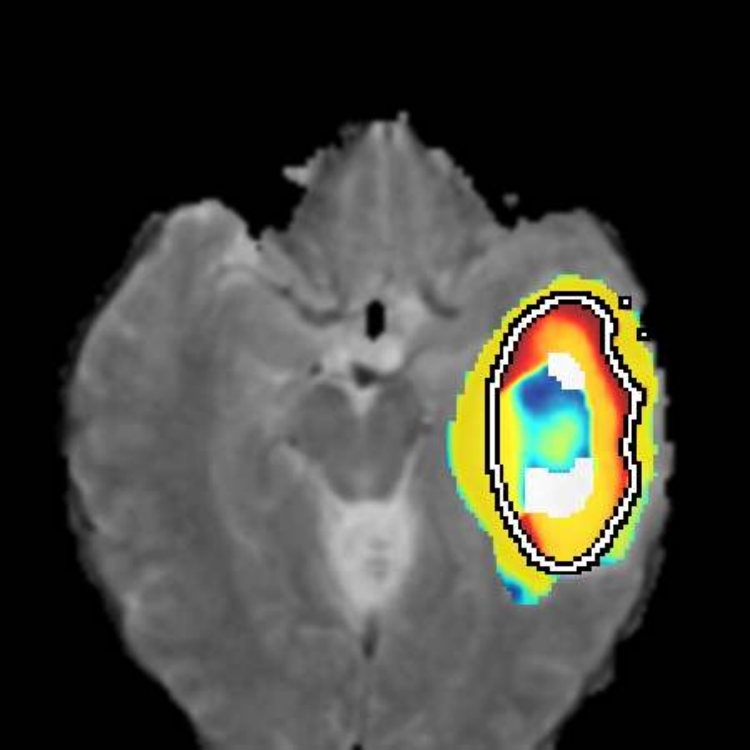}&
	\includegraphics[width=20mm]{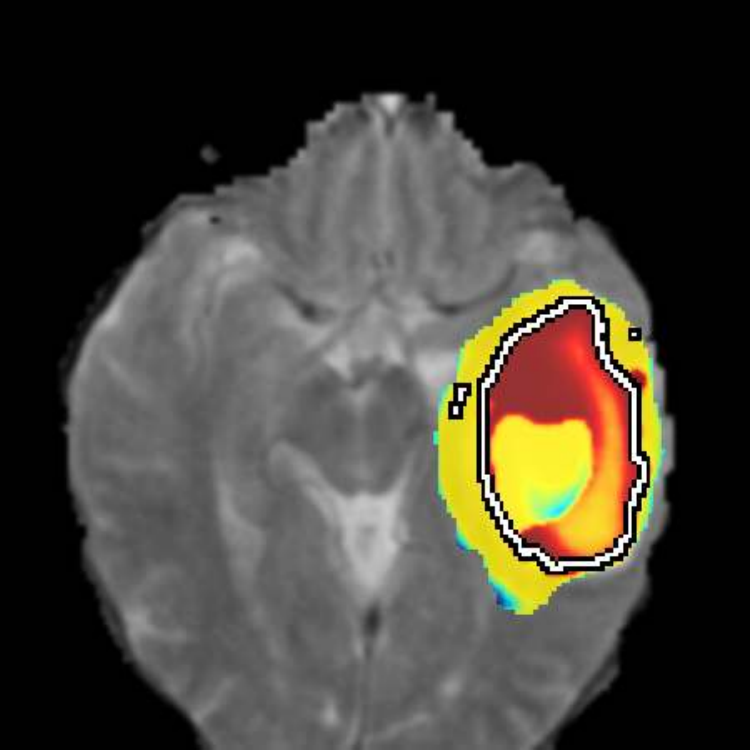}&
	\includegraphics[width=20mm]{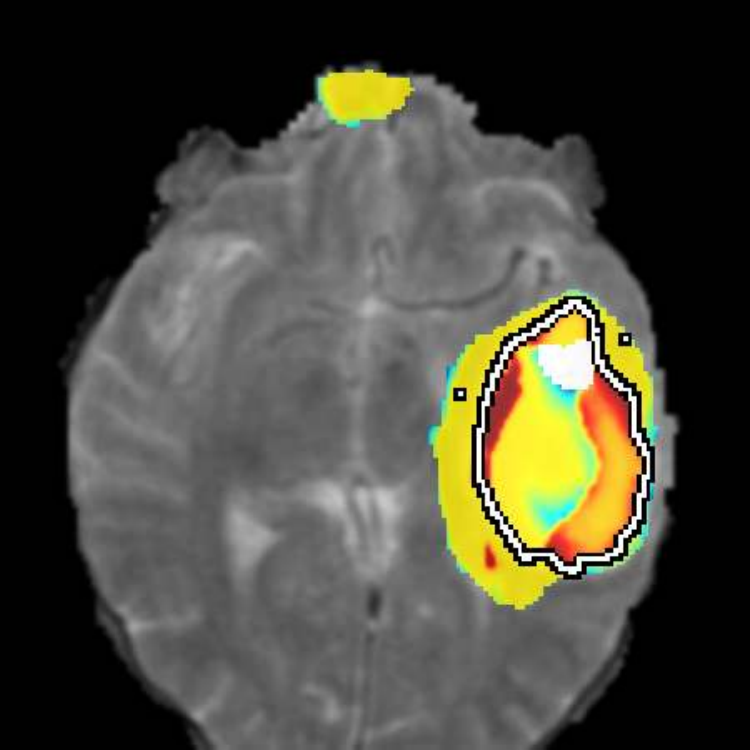}&
	\includegraphics[width=20mm]{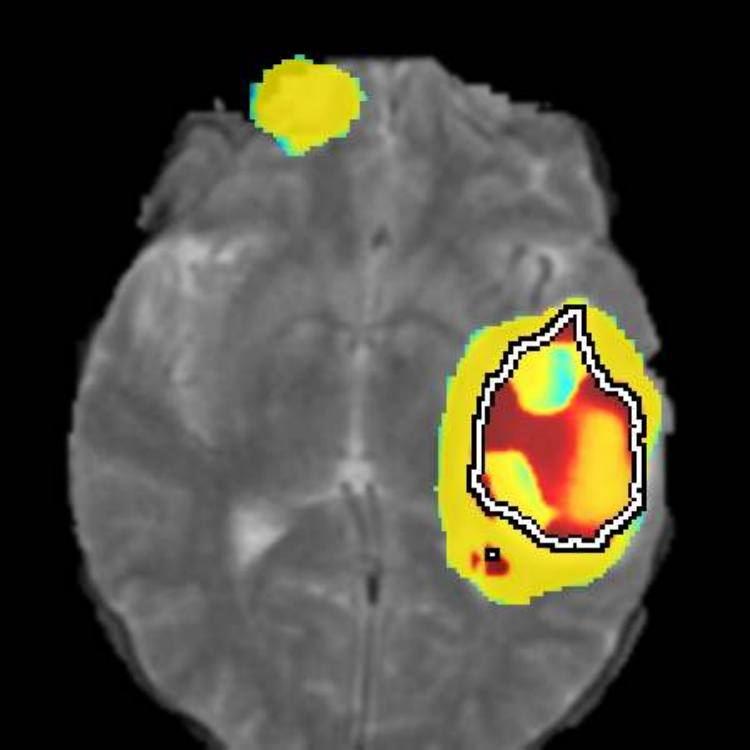}&
	\includegraphics[width=20mm]{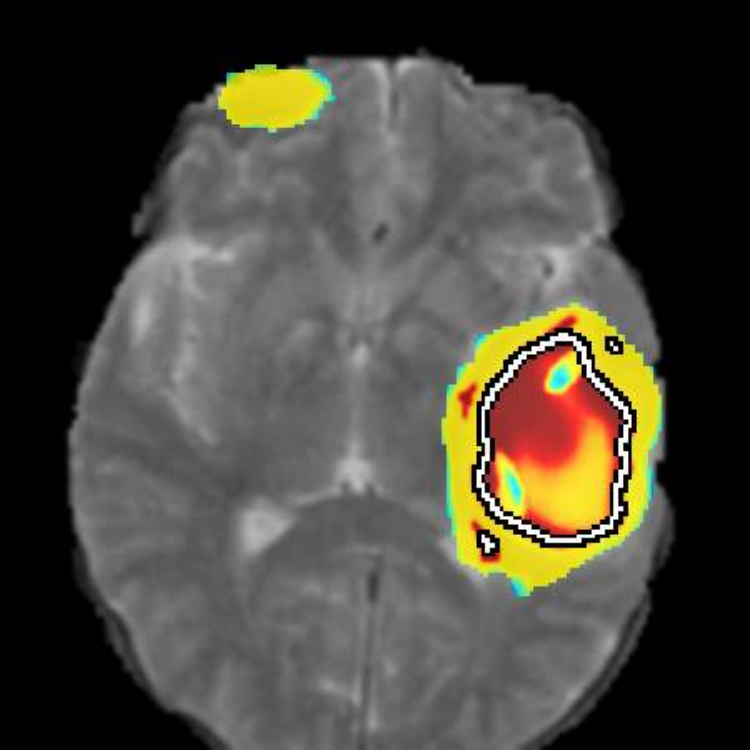}&
	\includegraphics[width=20mm]{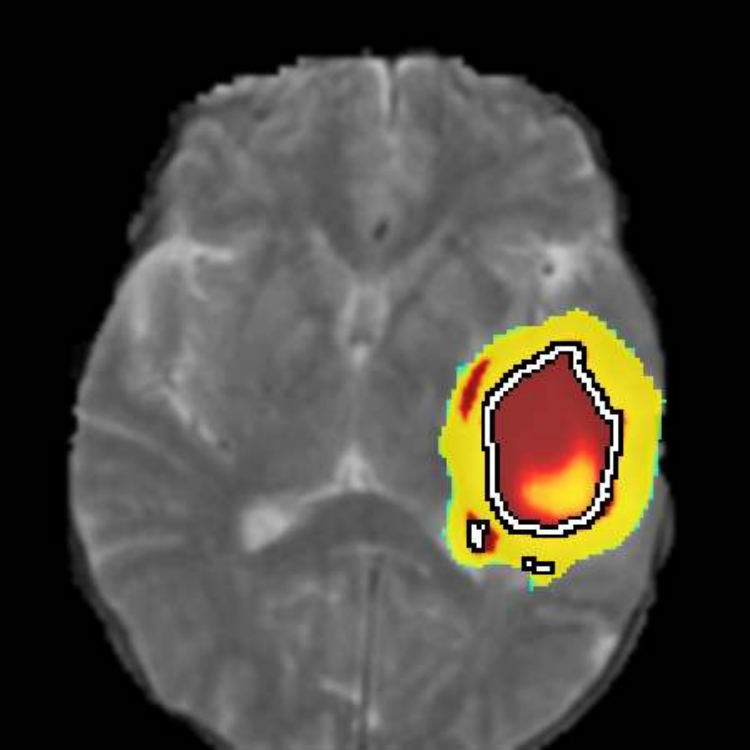}\\
	\parbox[t]{3mm}{\rotatebox[origin=l]{90}{\footnotesize{$~~~~{M=10}$}}} &
	\includegraphics[width=20mm]{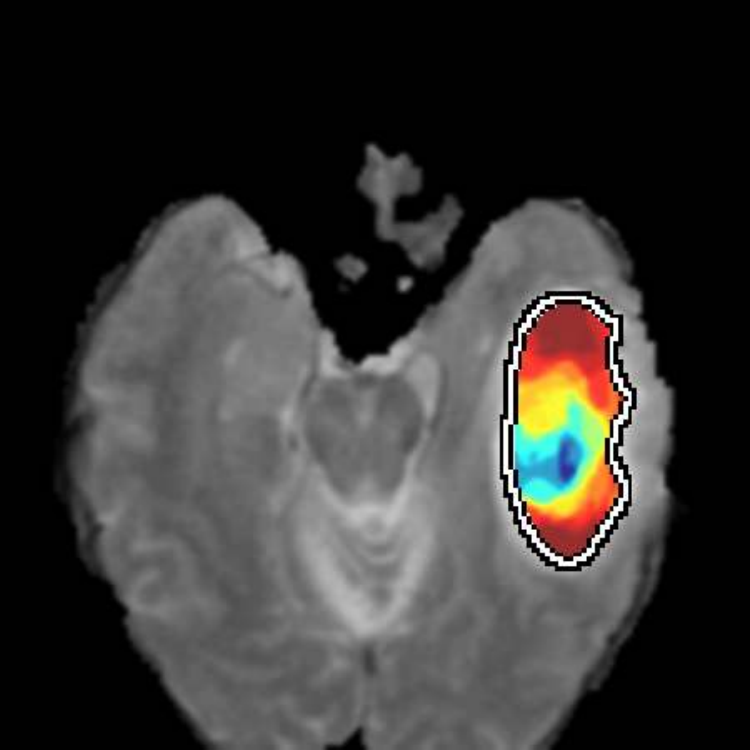}&
	\includegraphics[width=20mm]{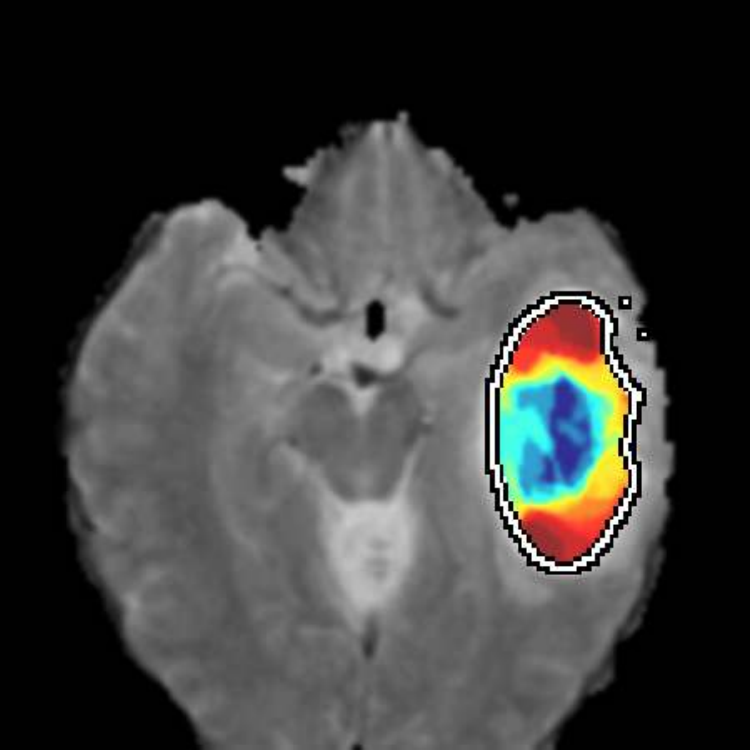}&
	\includegraphics[width=20mm]{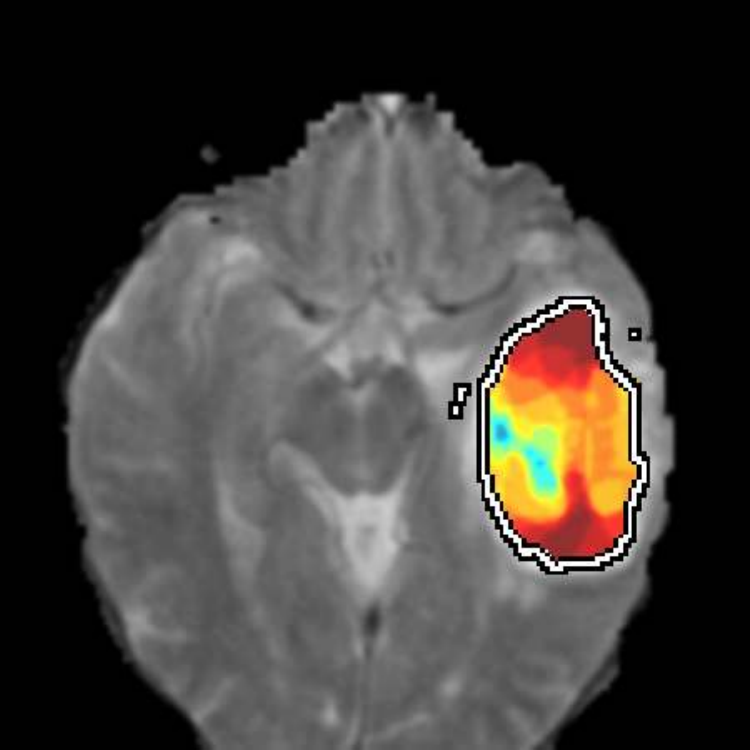}&
	\includegraphics[width=20mm]{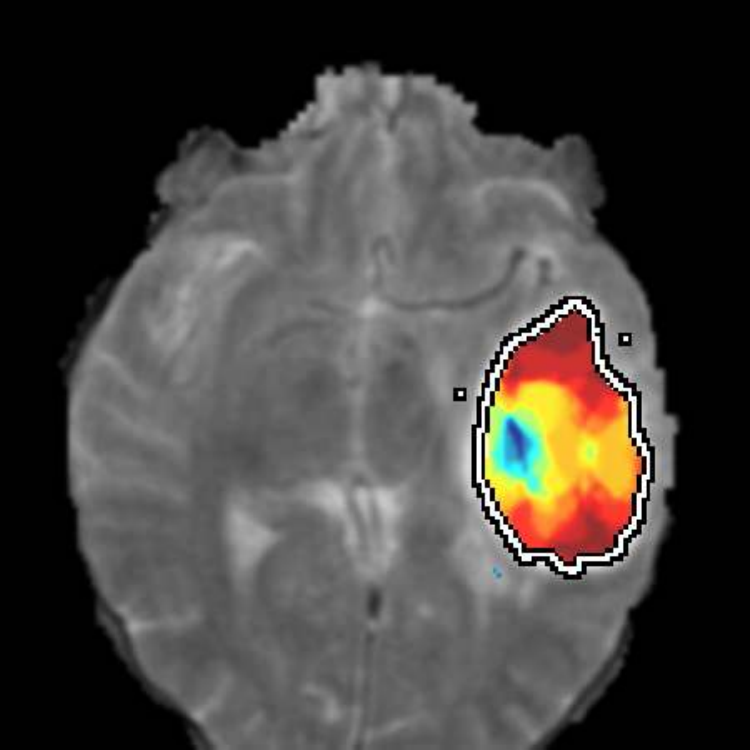}&
	\includegraphics[width=20mm]{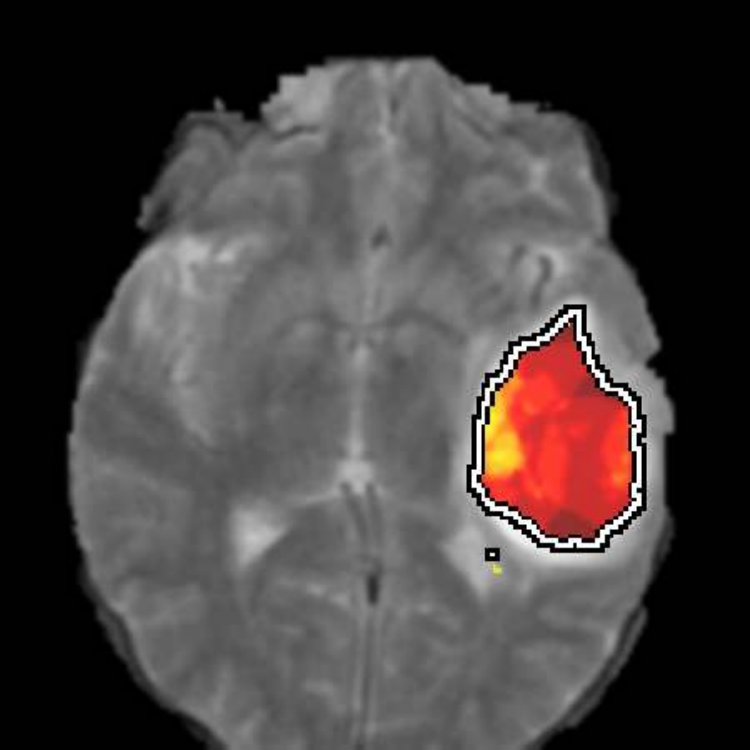}&
	\includegraphics[width=20mm]{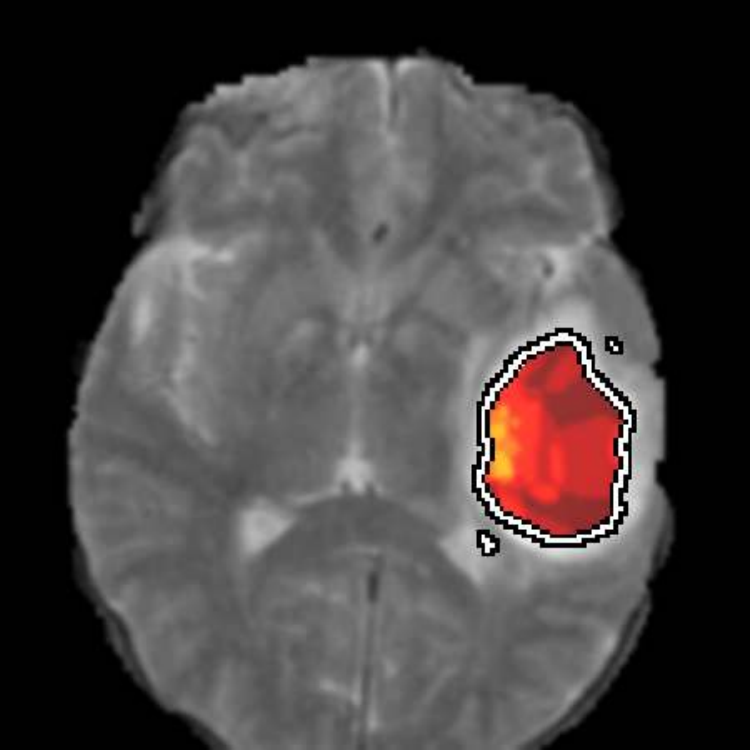}&
	\includegraphics[width=20mm]{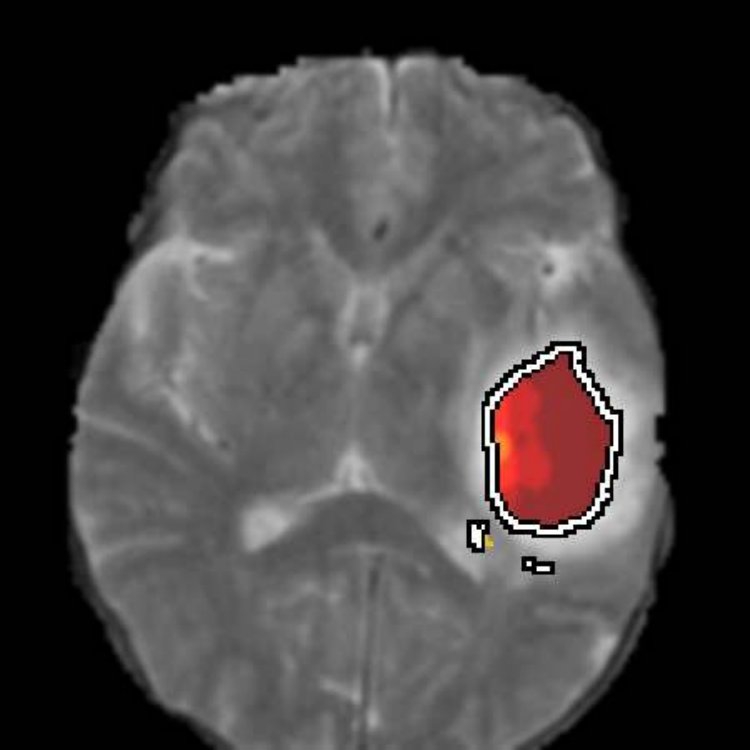}\\
	\parbox[t]{3mm}{\rotatebox[origin=l]{90}{\footnotesize{$~~~~{M=50}$}}} &
	\includegraphics[width=20mm]{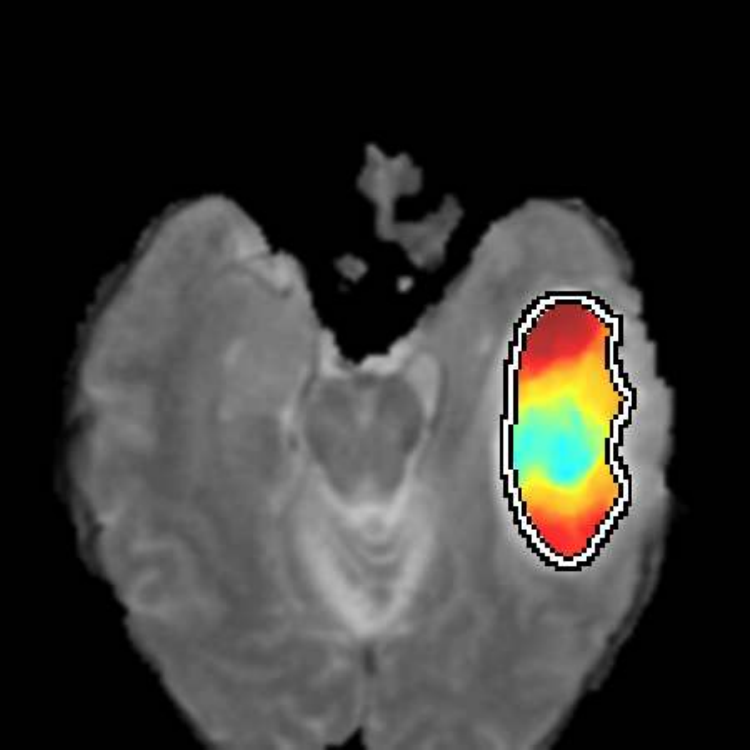}&
	\includegraphics[width=20mm]{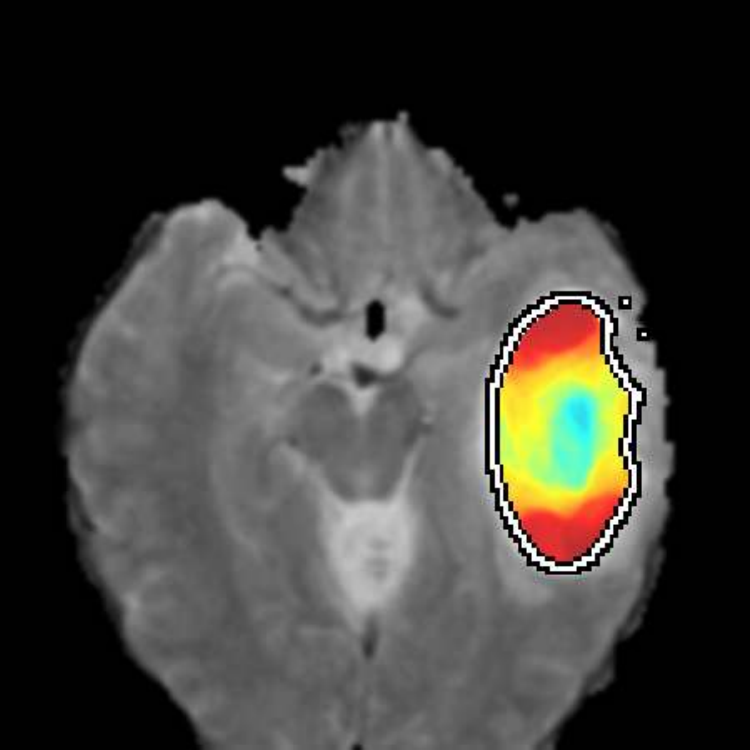}&
	\includegraphics[width=20mm]{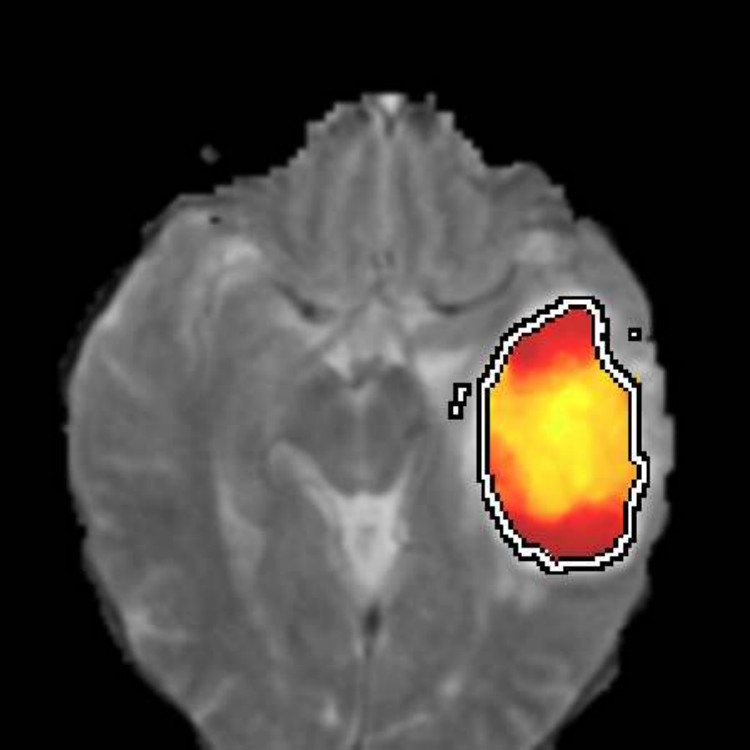}&
	\includegraphics[width=20mm]{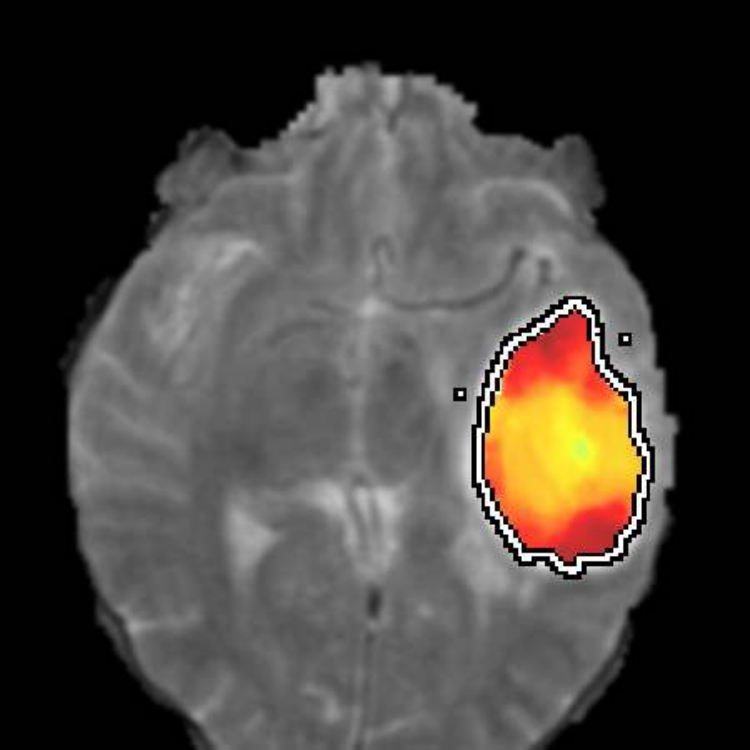}&
	\includegraphics[width=20mm]{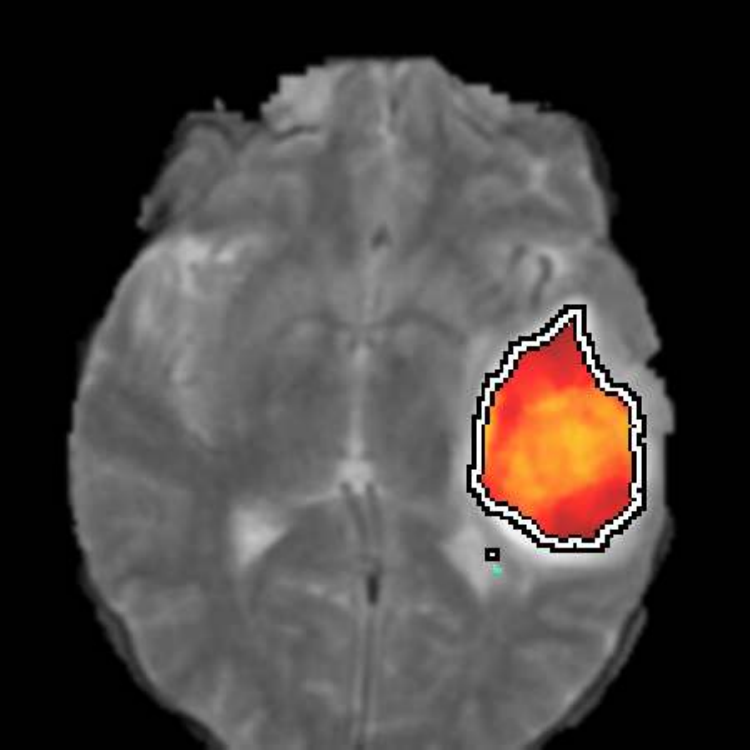}&
	\includegraphics[width=20mm]{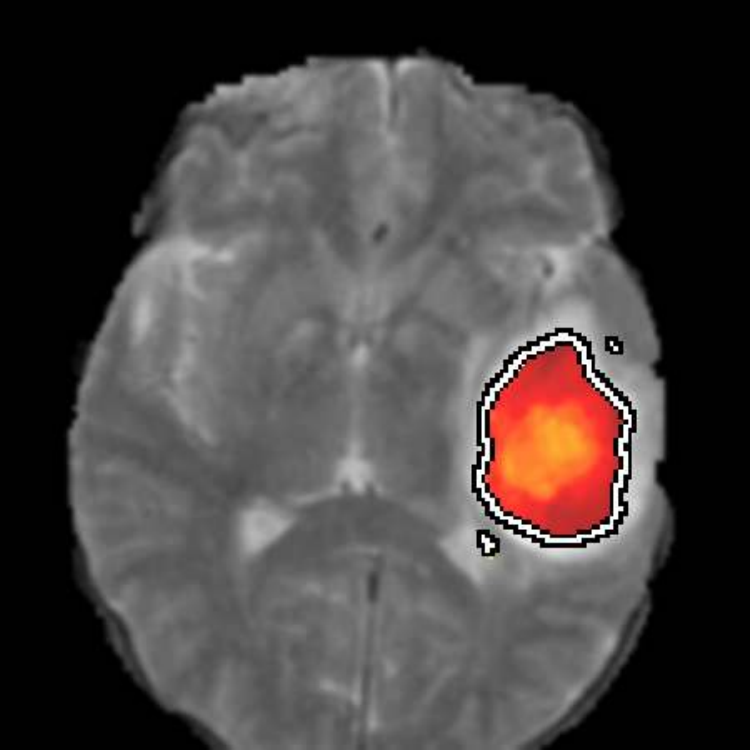}&
	\includegraphics[width=20mm]{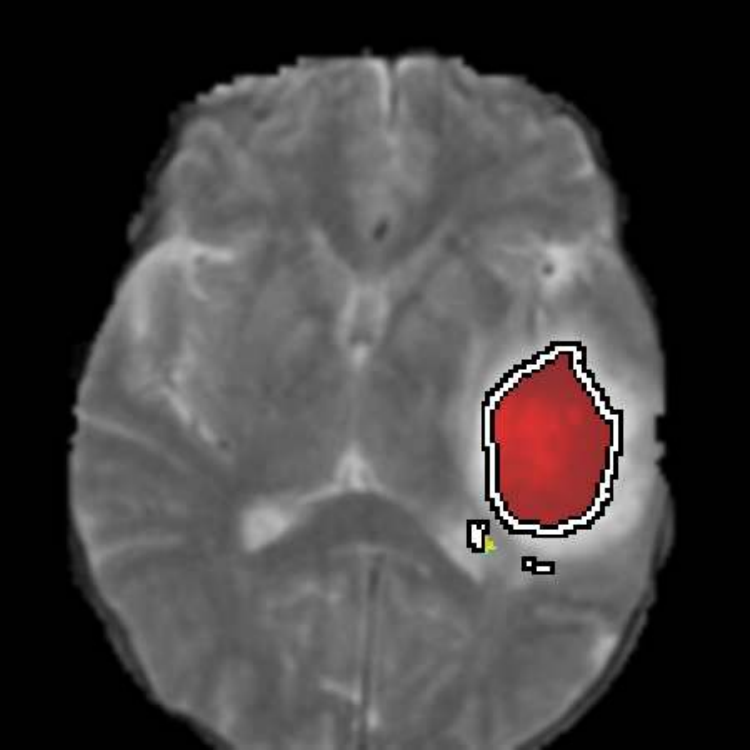}\\
\end{tabular}

\begin{tabular}{llllllll}
	\parbox[t]{3mm}{\rotatebox[origin=l]{90}{\footnotesize{$~~~~~{M=2}$}}} &
	\includegraphics[width=20mm]{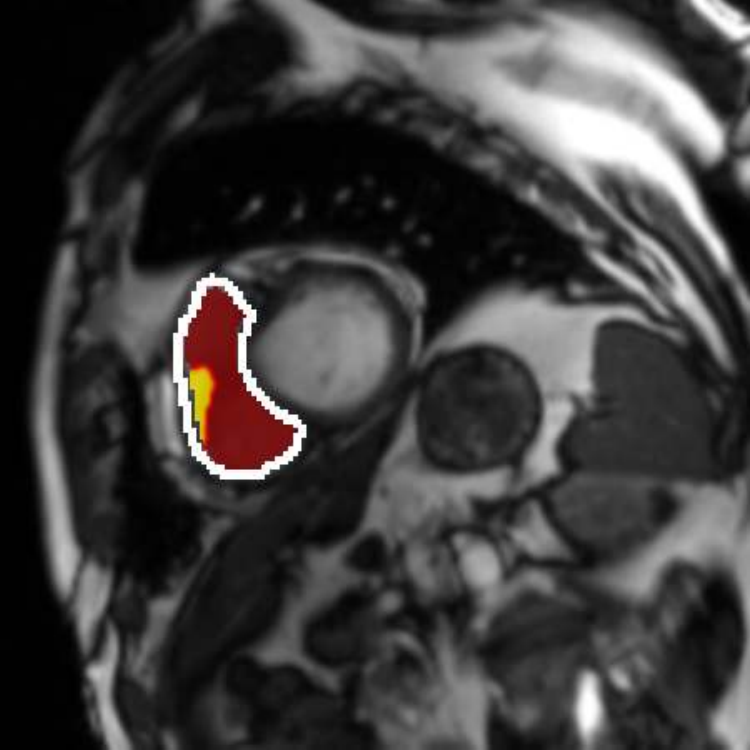}&
	\includegraphics[width=20mm]{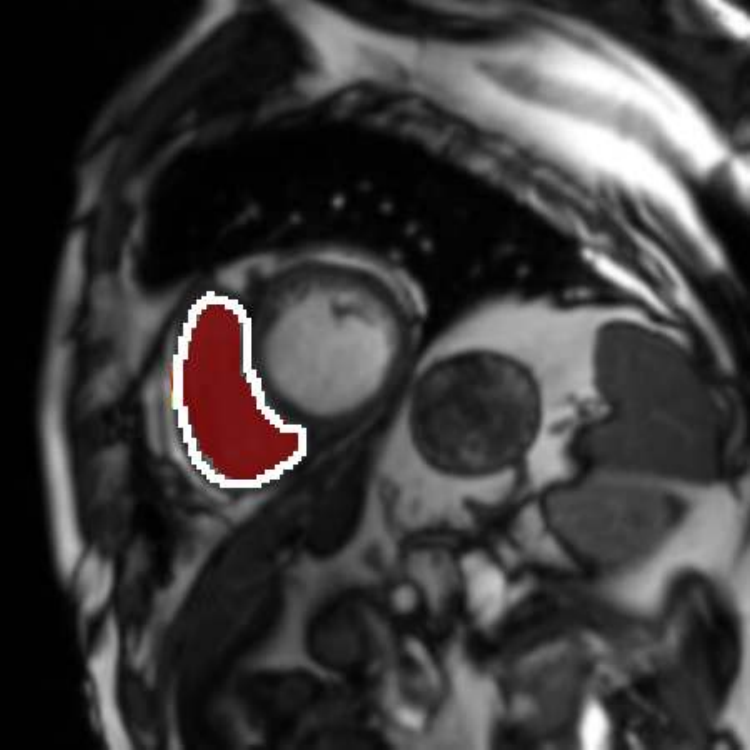}&
	\includegraphics[width=20mm]{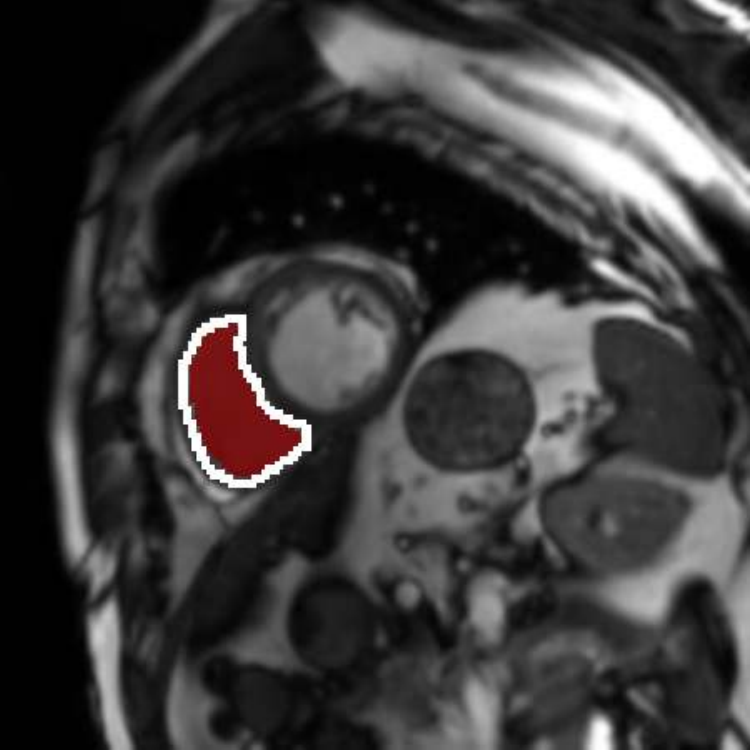}&
	\includegraphics[width=20mm]{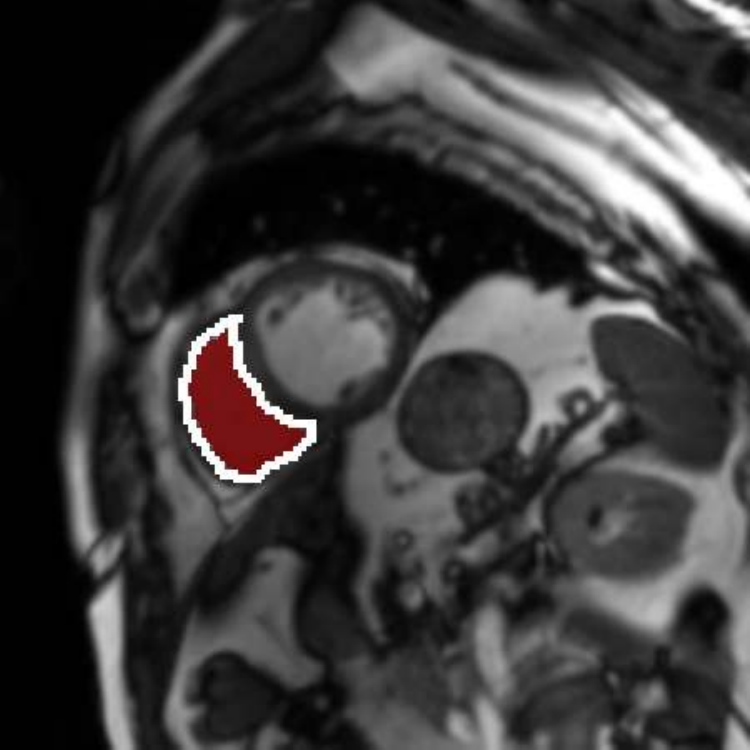}&
	\includegraphics[width=20mm]{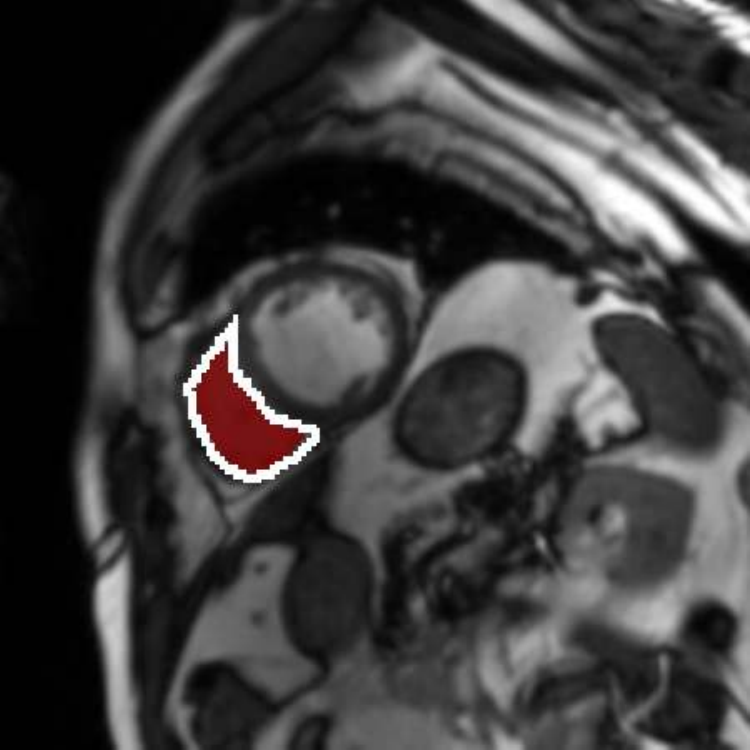}&
	\includegraphics[width=20mm]{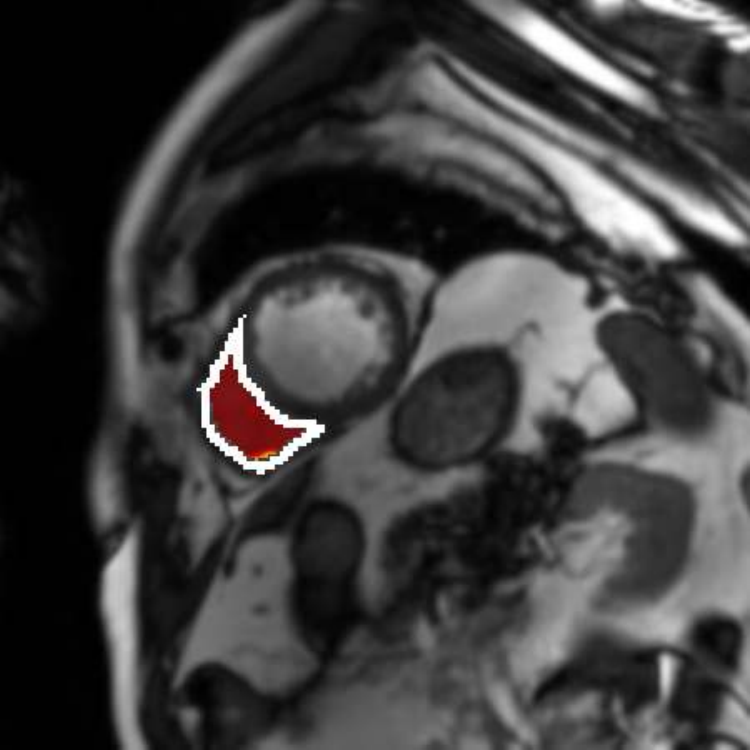}&
	\includegraphics[width=20mm]{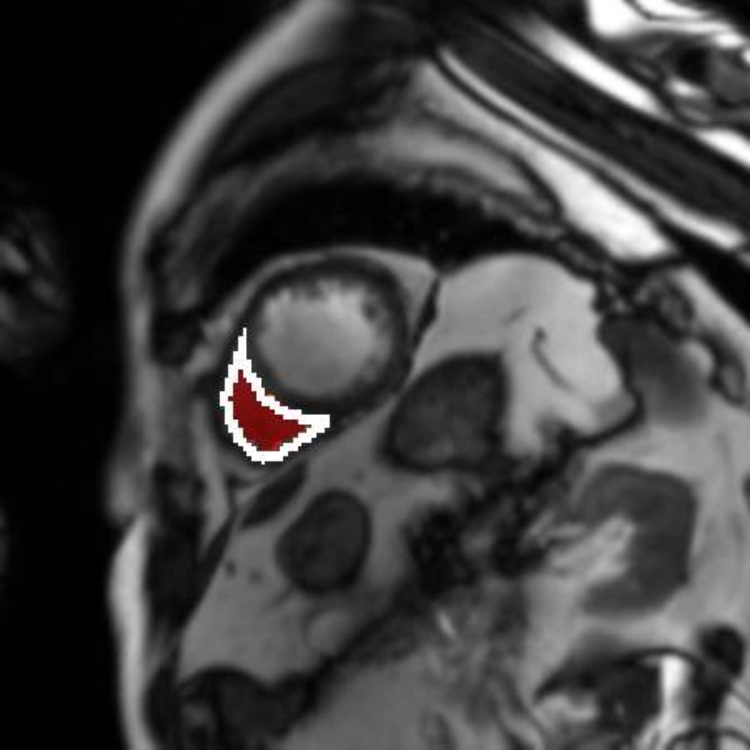}\\
	\parbox[t]{3mm}{\rotatebox[origin=l]{90}{\footnotesize{$~~~~{M=10}$}}} &
	\includegraphics[width=20mm]{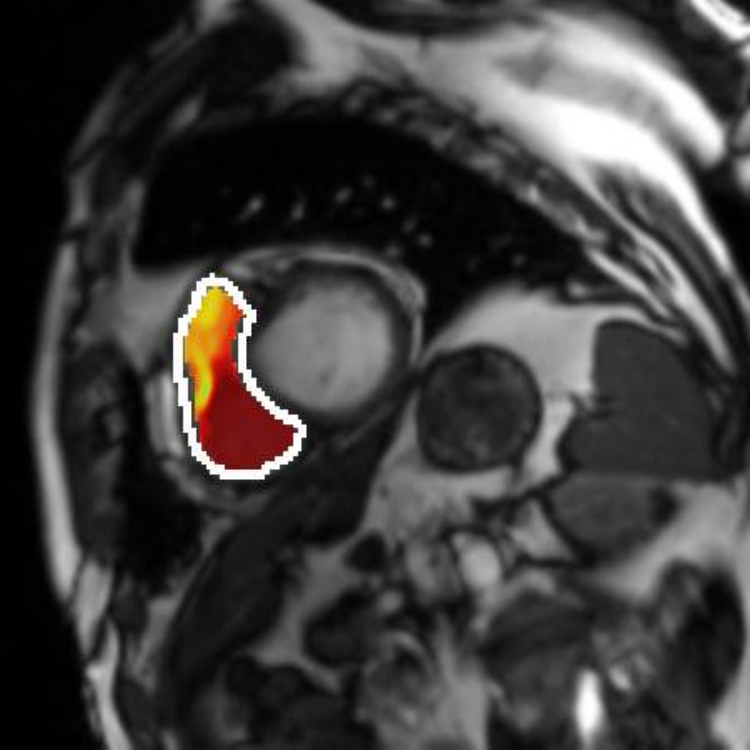}&
	\includegraphics[width=20mm]{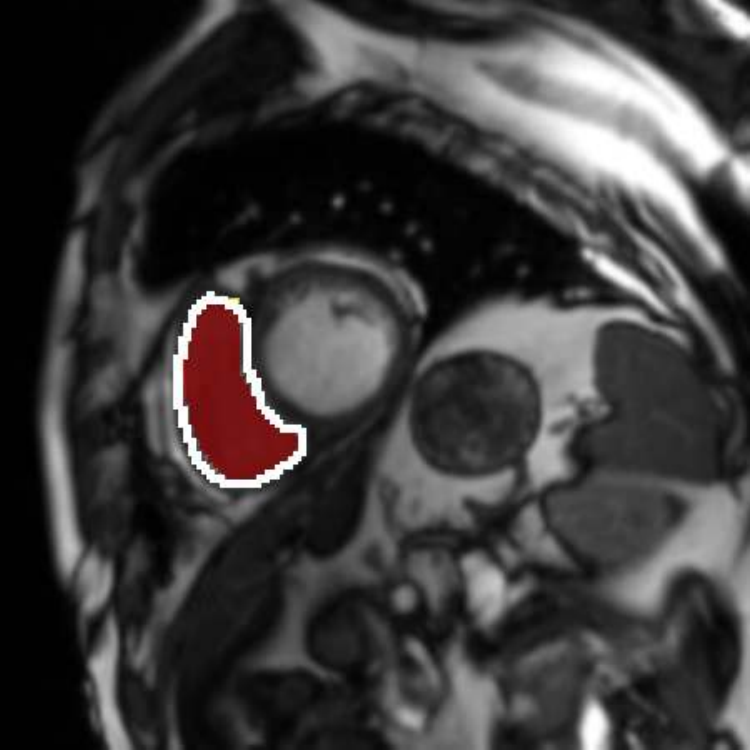}&
	\includegraphics[width=20mm]{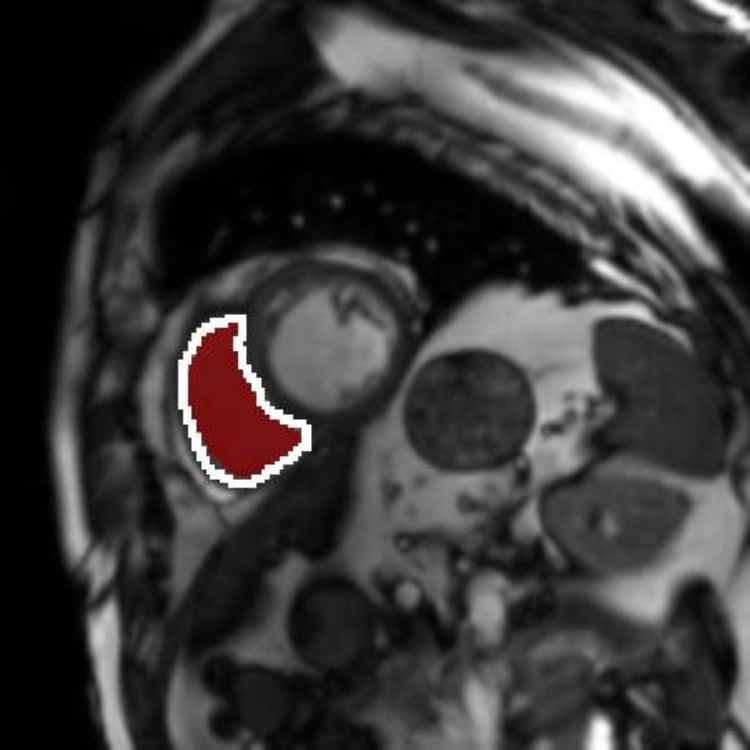}&
	\includegraphics[width=20mm]{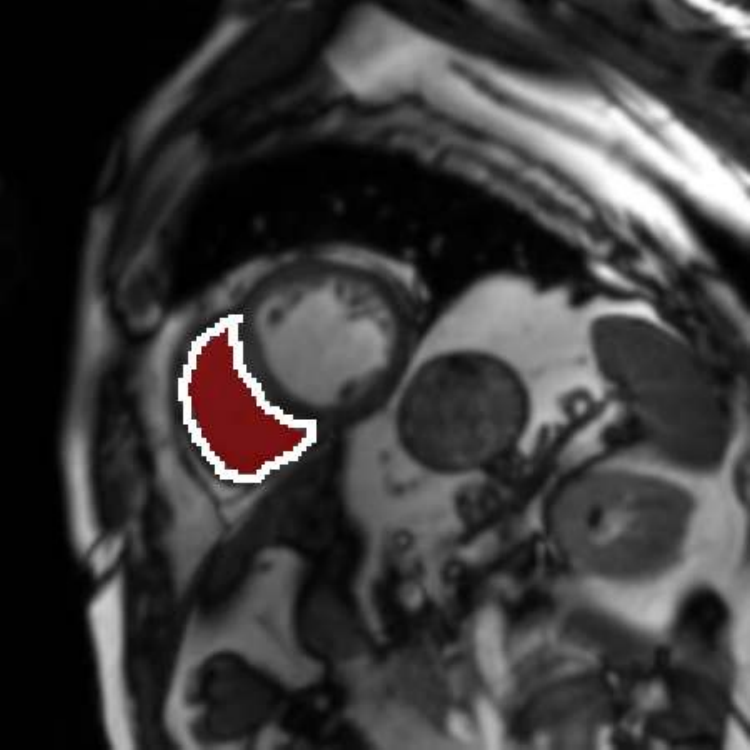}&
	\includegraphics[width=20mm]{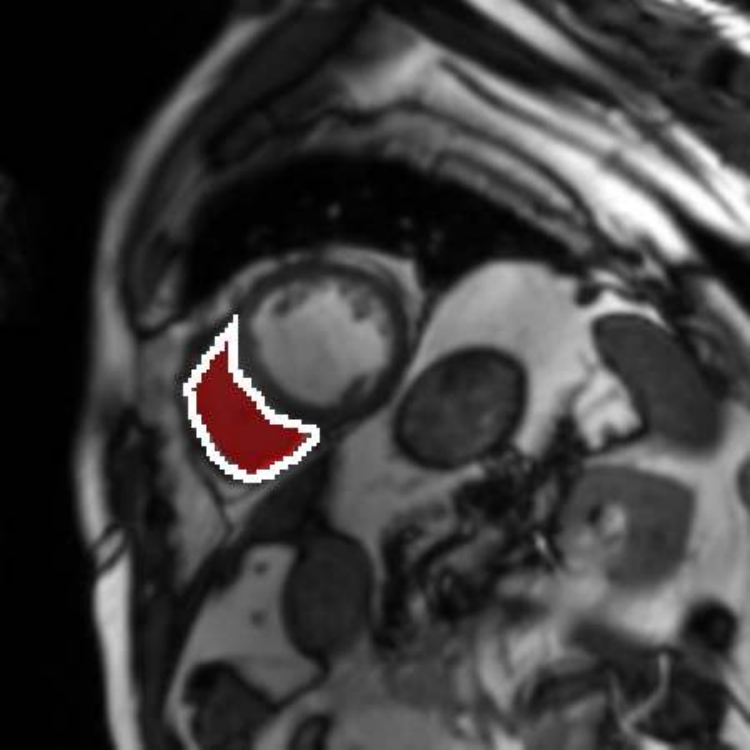}&
	\includegraphics[width=20mm]{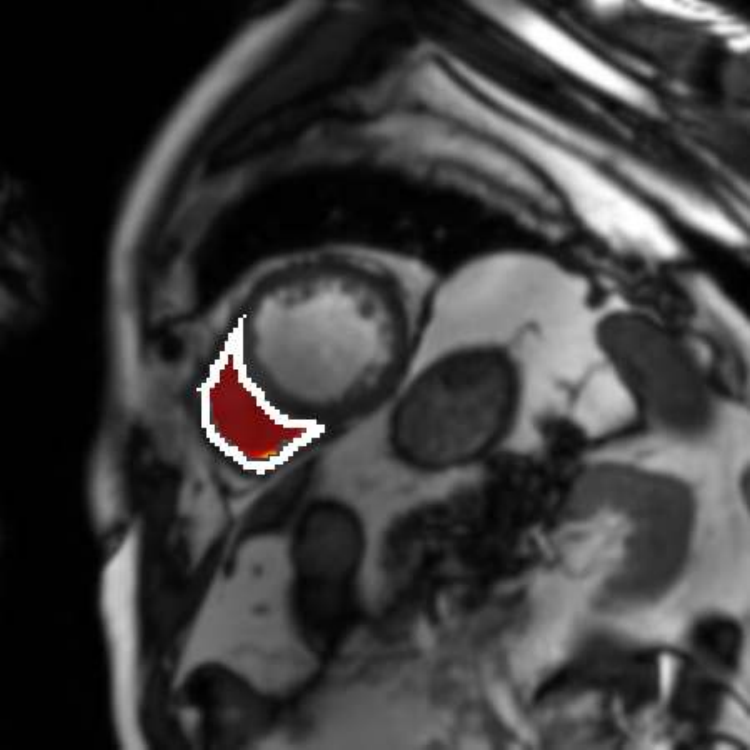}&
	\includegraphics[width=20mm]{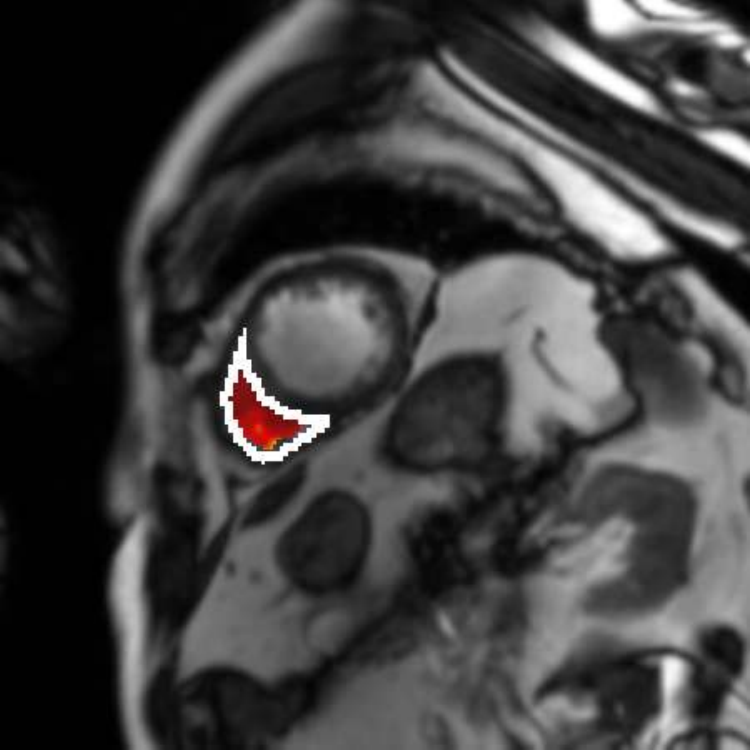}\\
	\parbox[t]{3mm}{\rotatebox[origin=l]{90}{\footnotesize{$~~~~{M=50}$}}} &
	\includegraphics[width=20mm]{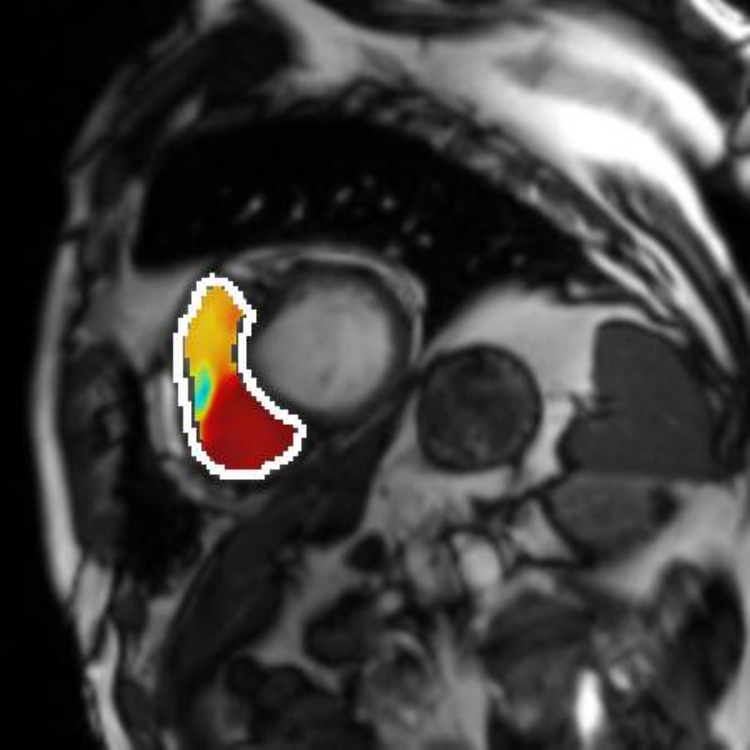}&
	\includegraphics[width=20mm]{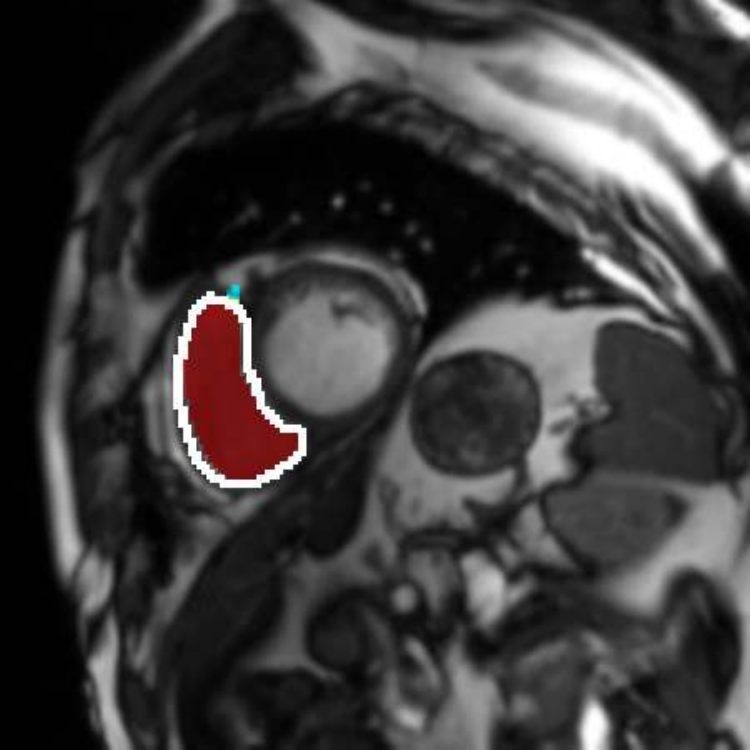}&
	\includegraphics[width=20mm]{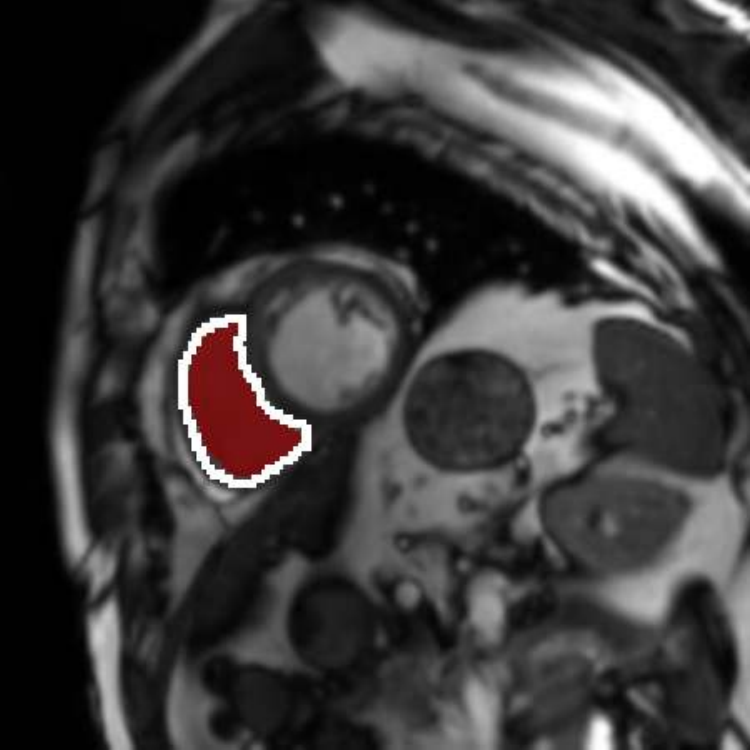}&
	\includegraphics[width=20mm]{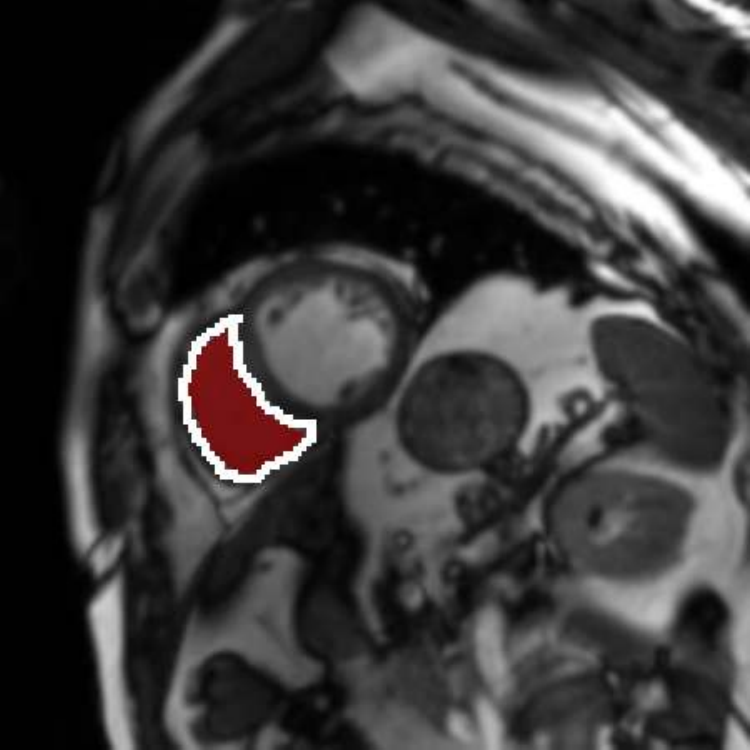}&
	\includegraphics[width=20mm]{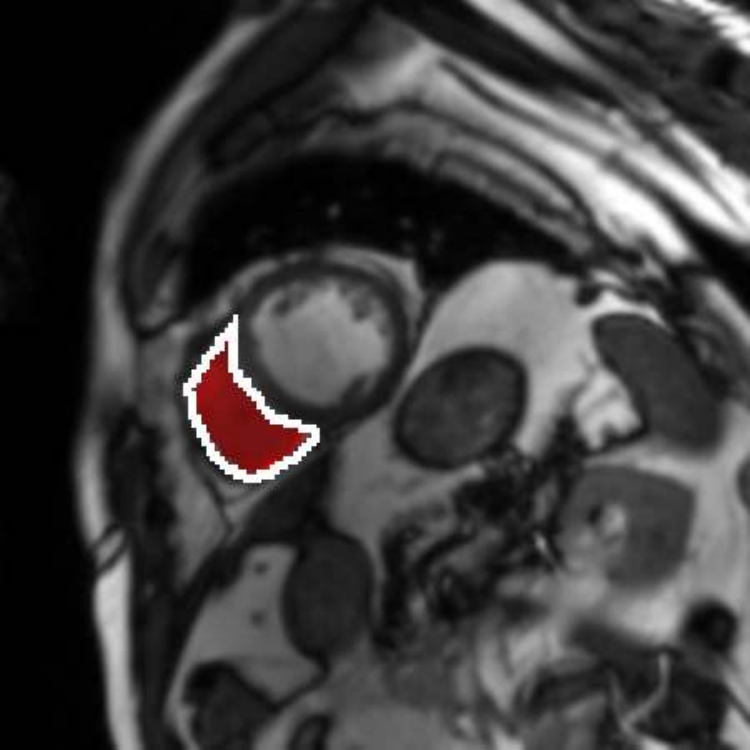}&
	\includegraphics[width=20mm]{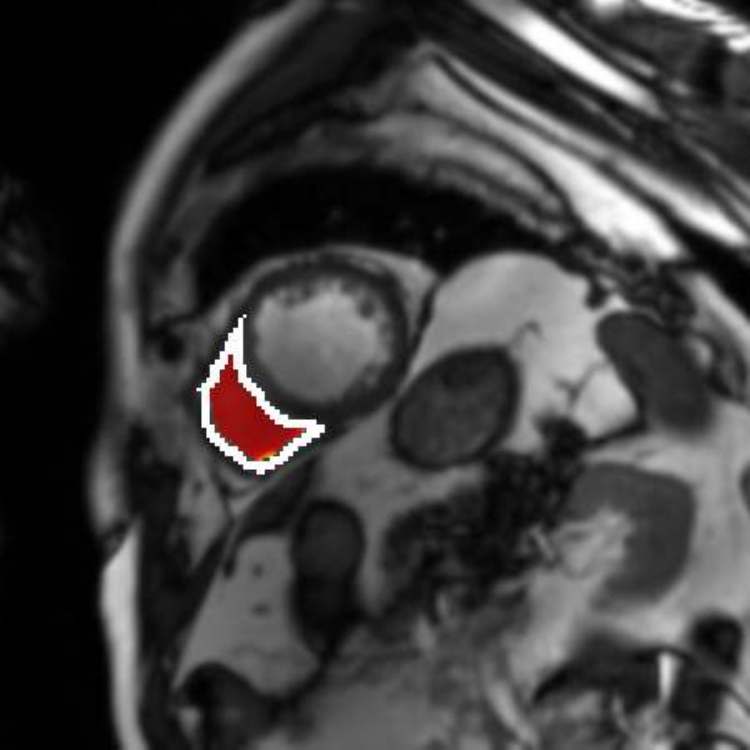}&
	\includegraphics[width=20mm]{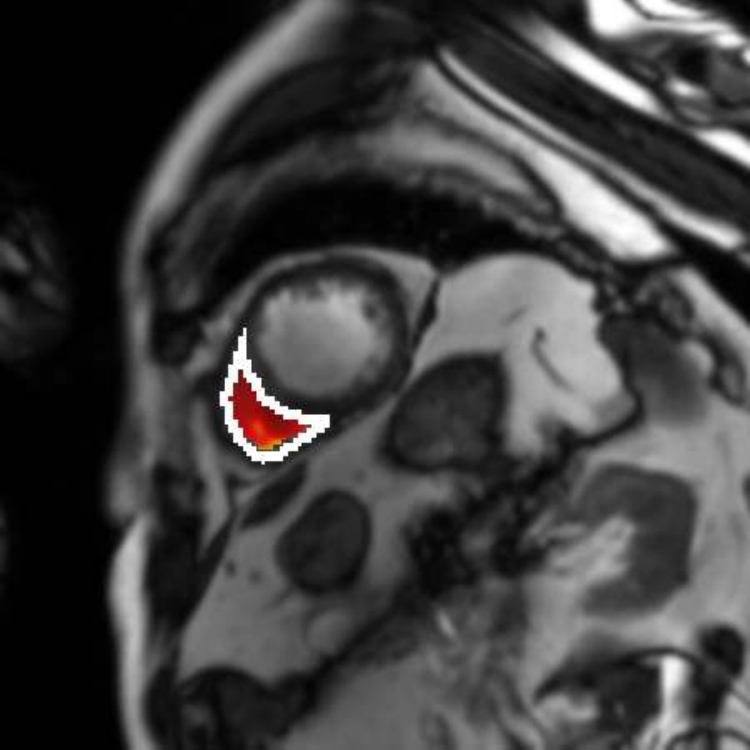}\\
\end{tabular}

\begin{tabular}{rlllllll}
	\parbox[t]{3mm}{\rotatebox[origin=l]{90}{\footnotesize{$~~~~~{M=2}$}}} &
	\includegraphics[width=20mm]{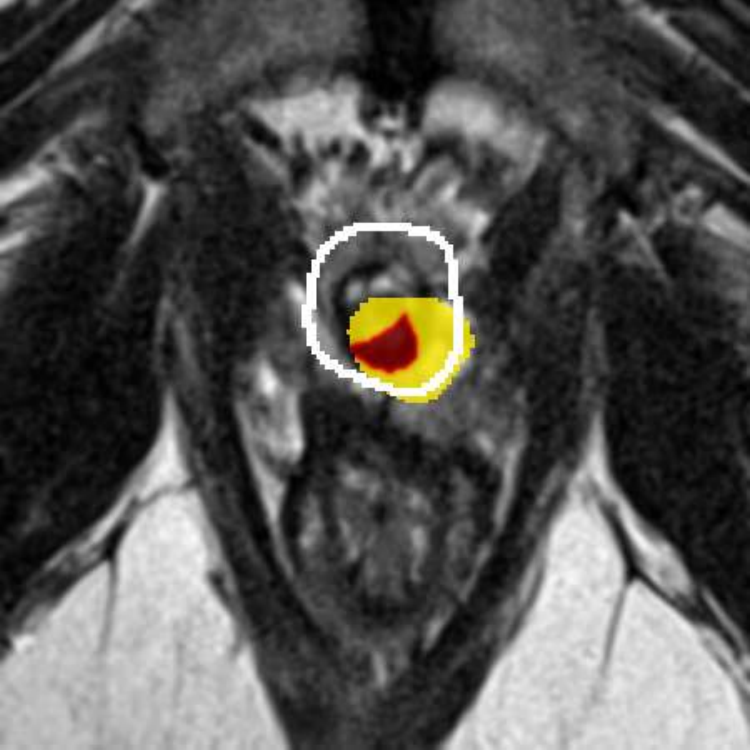}&
	\includegraphics[width=20mm]{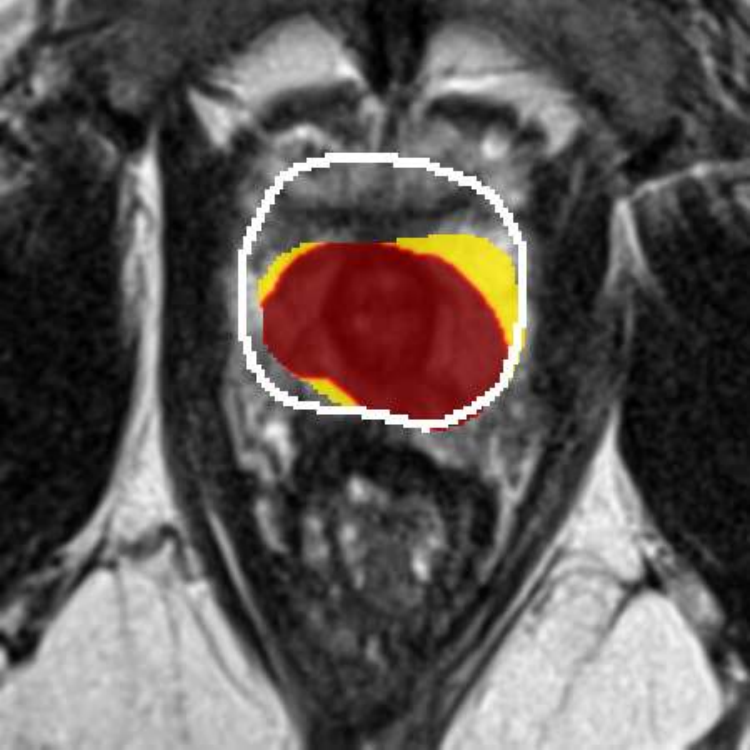}&
	\includegraphics[width=20mm]{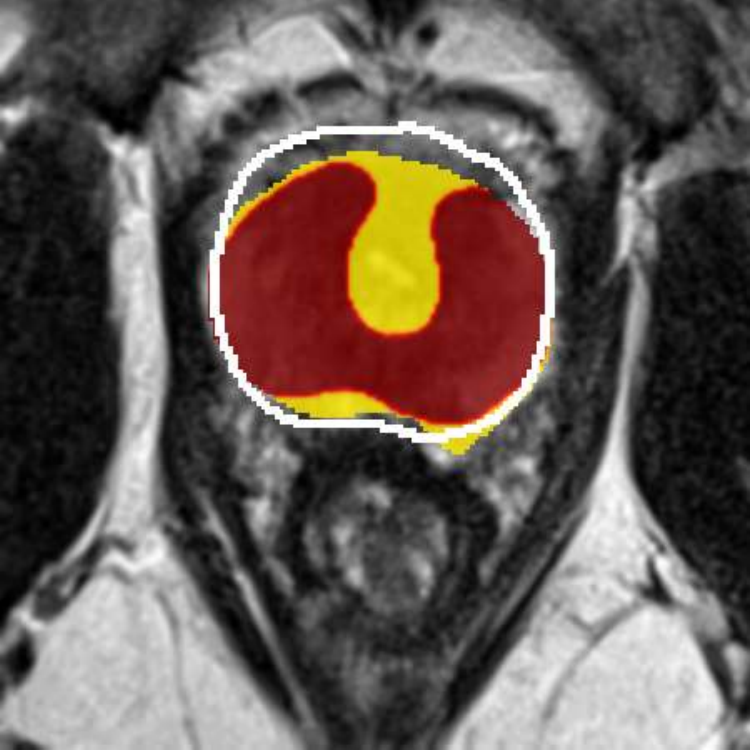}&
	\includegraphics[width=20mm]{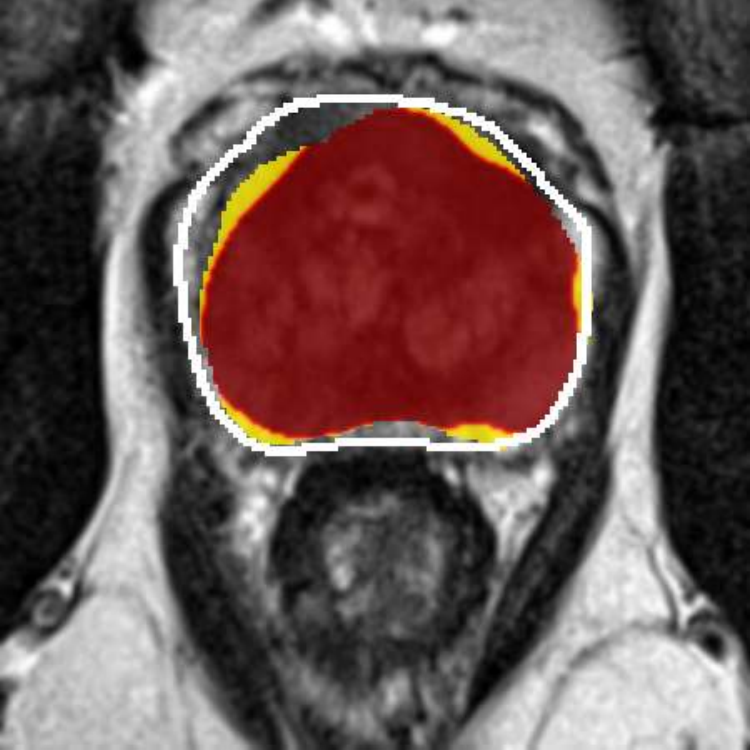}&
	\includegraphics[width=20mm]{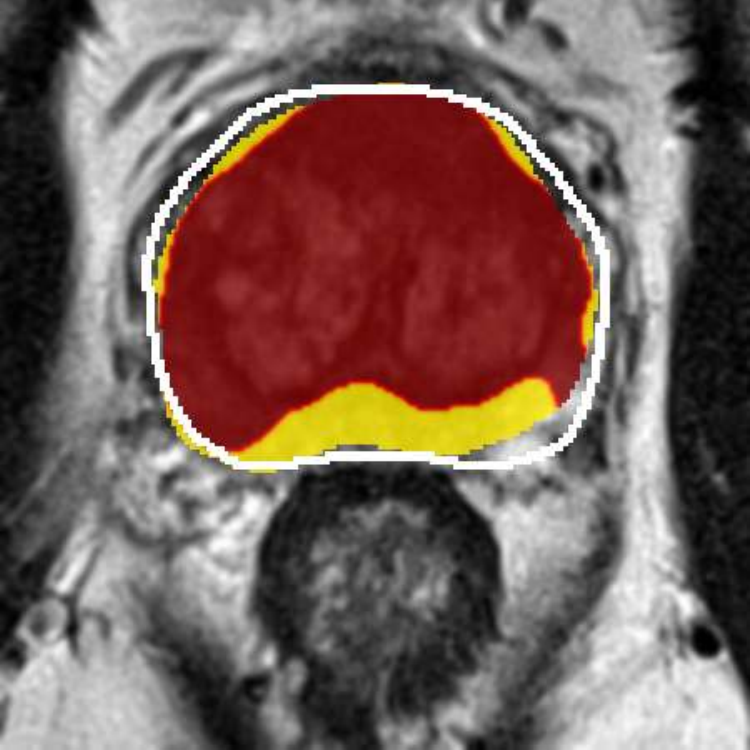}&
	\includegraphics[width=20mm]{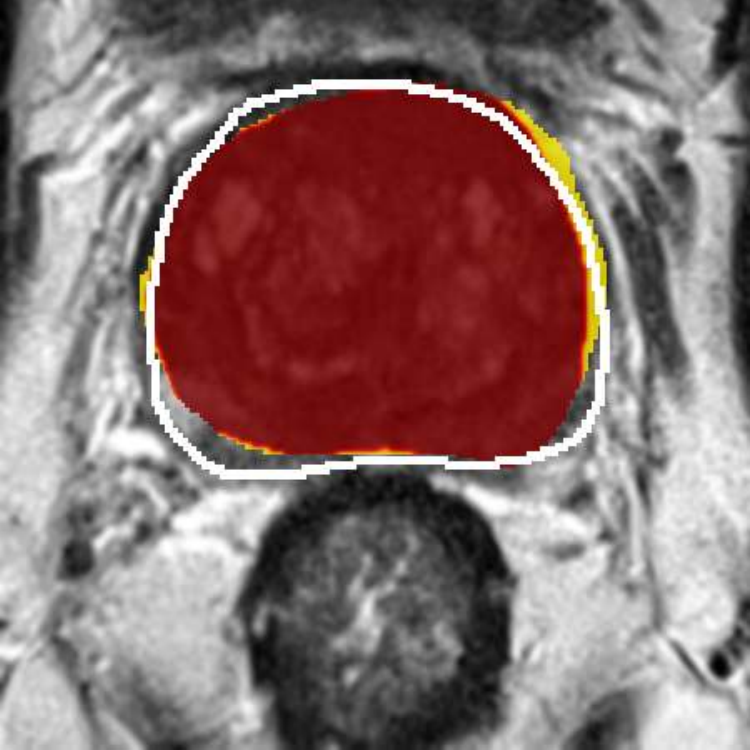}&
	\includegraphics[width=20mm]{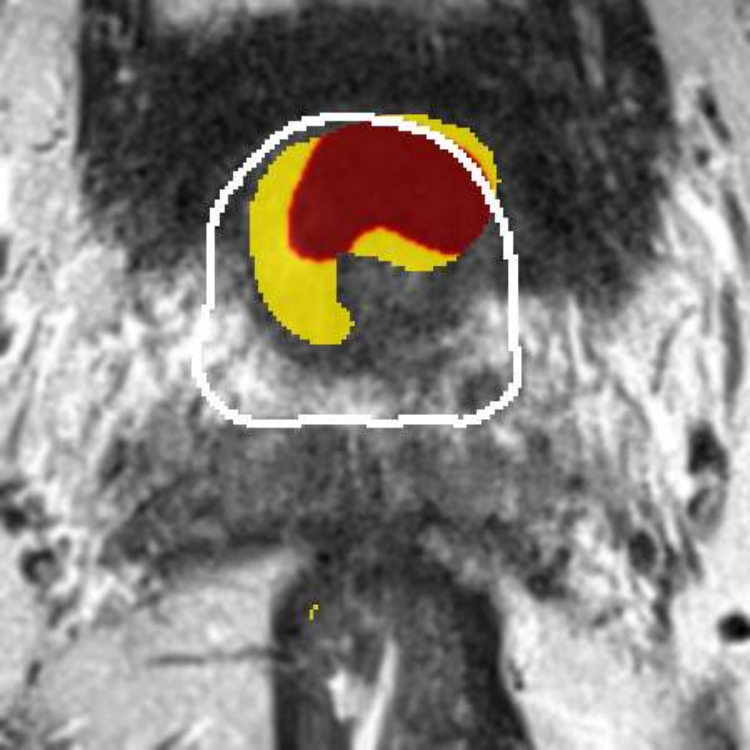}\\
	\parbox[t]{3mm}{\rotatebox[origin=l]{90}{\footnotesize{$~~~~{M=10}$}}} &
	\includegraphics[width=20mm]{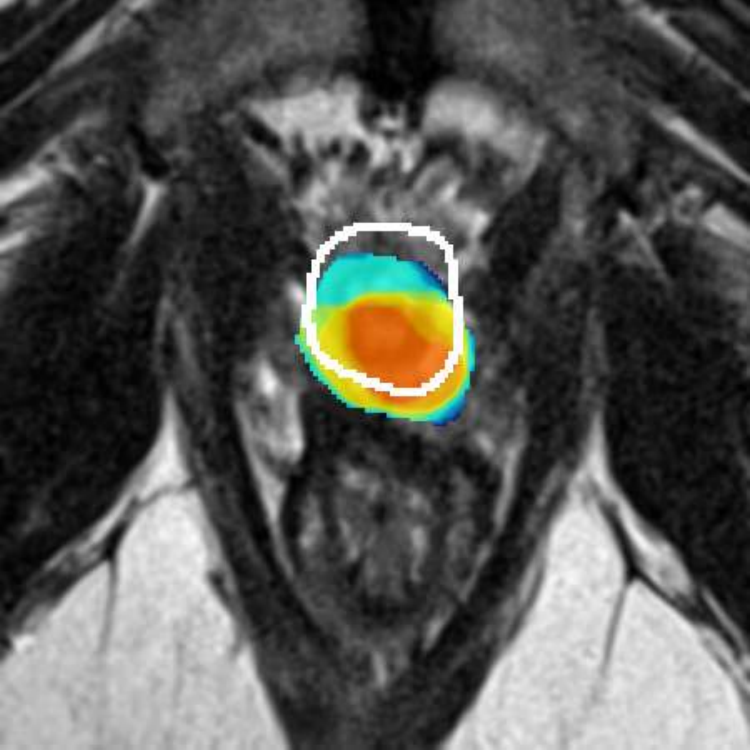}&
	\includegraphics[width=20mm]{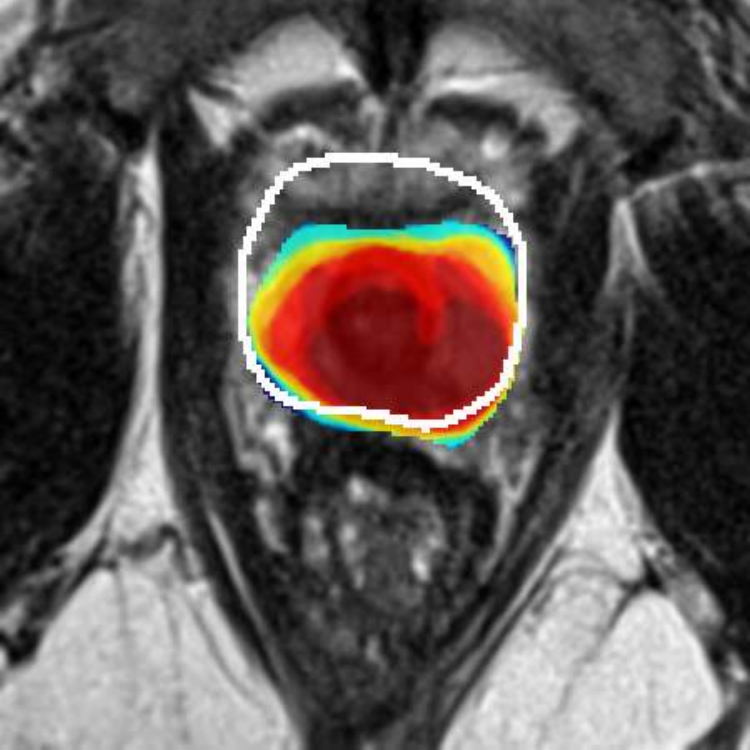}&
	\includegraphics[width=20mm]{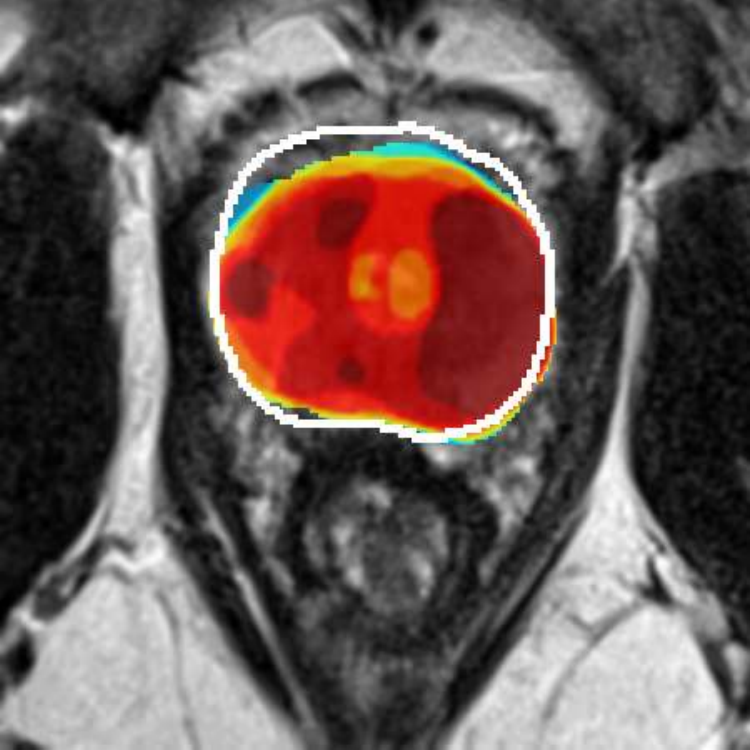}&
	\includegraphics[width=20mm]{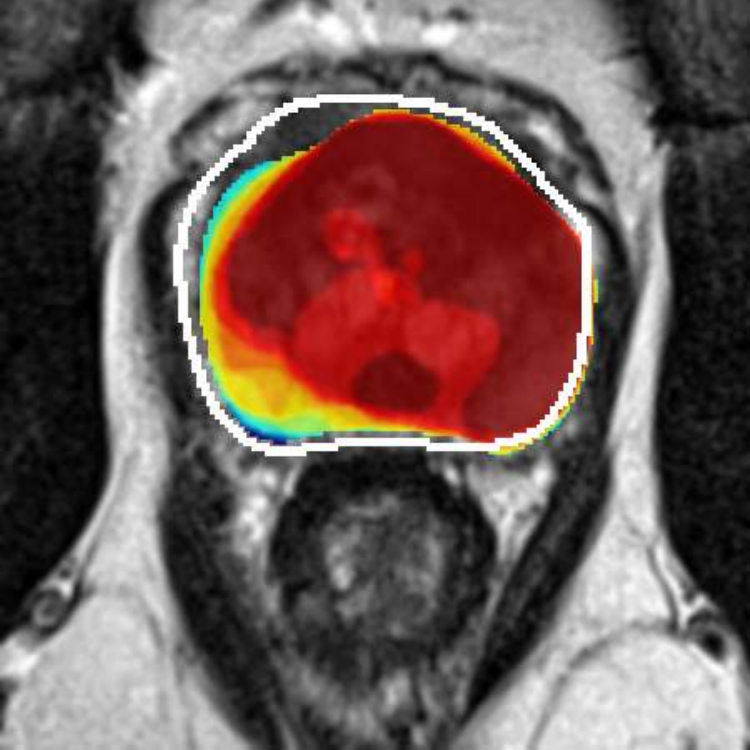}&
	\includegraphics[width=20mm]{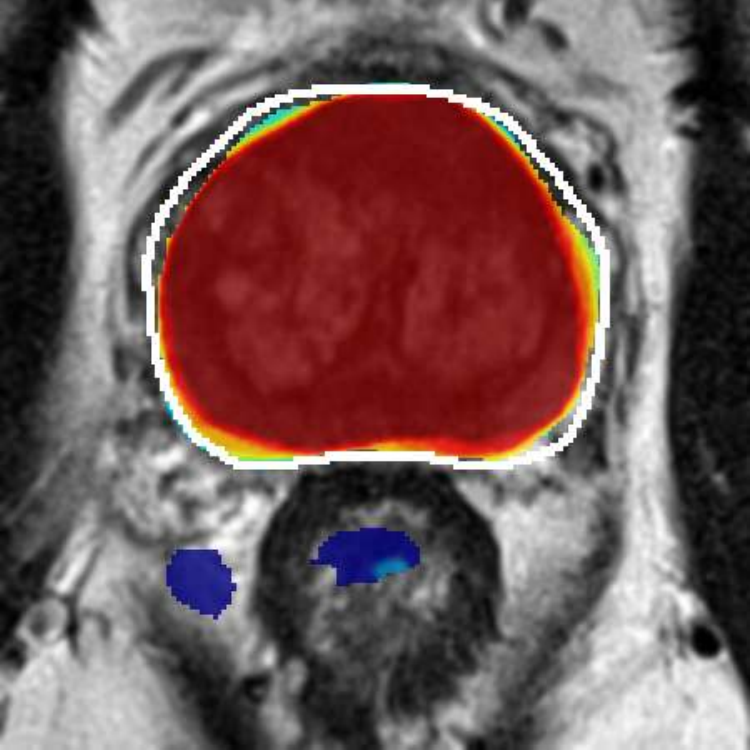}&
	\includegraphics[width=20mm]{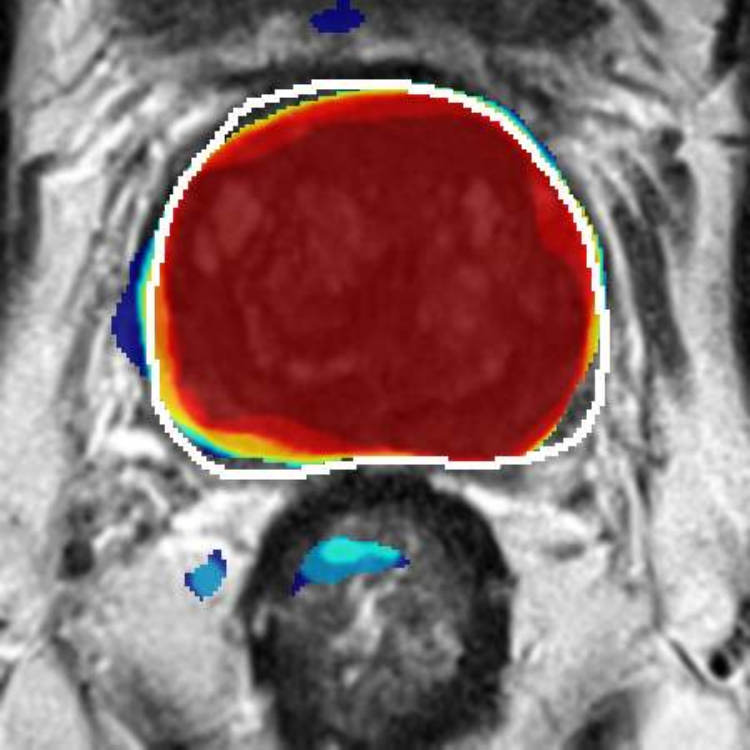}&
	\includegraphics[width=20mm]{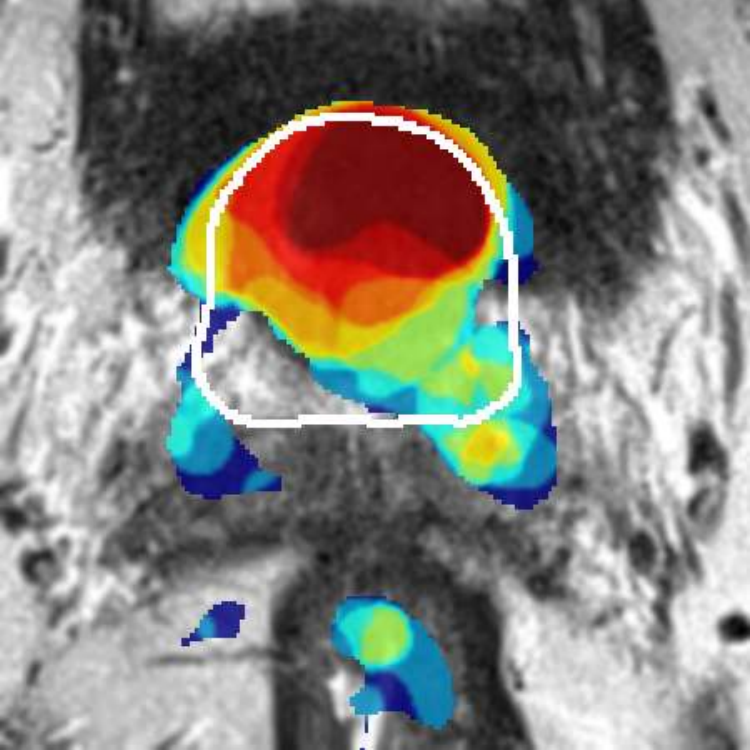}\\
	\parbox[t]{3mm}{\rotatebox[origin=l]{90}{\footnotesize{$~~~~{M=50}$}}} &
	\includegraphics[width=20mm]{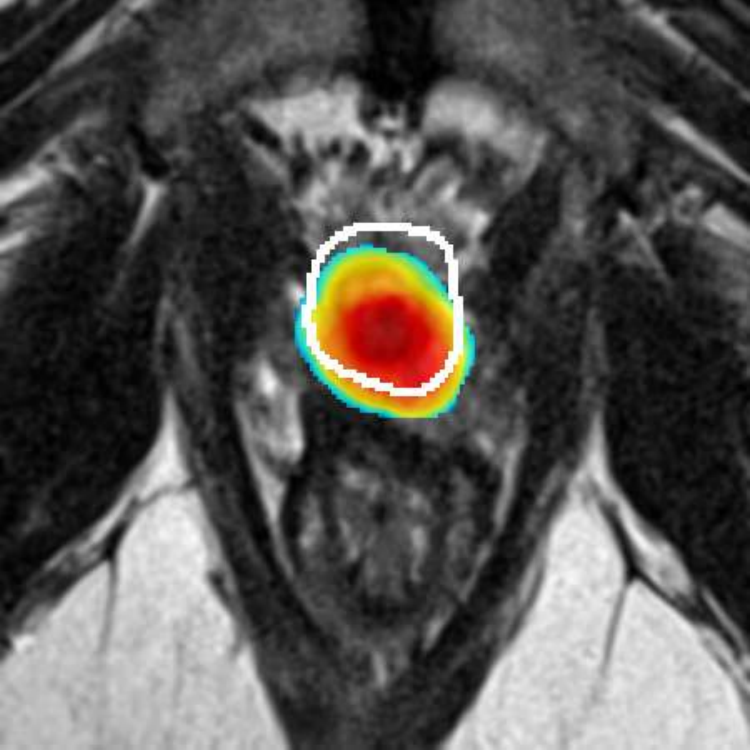}&
	\includegraphics[width=20mm]{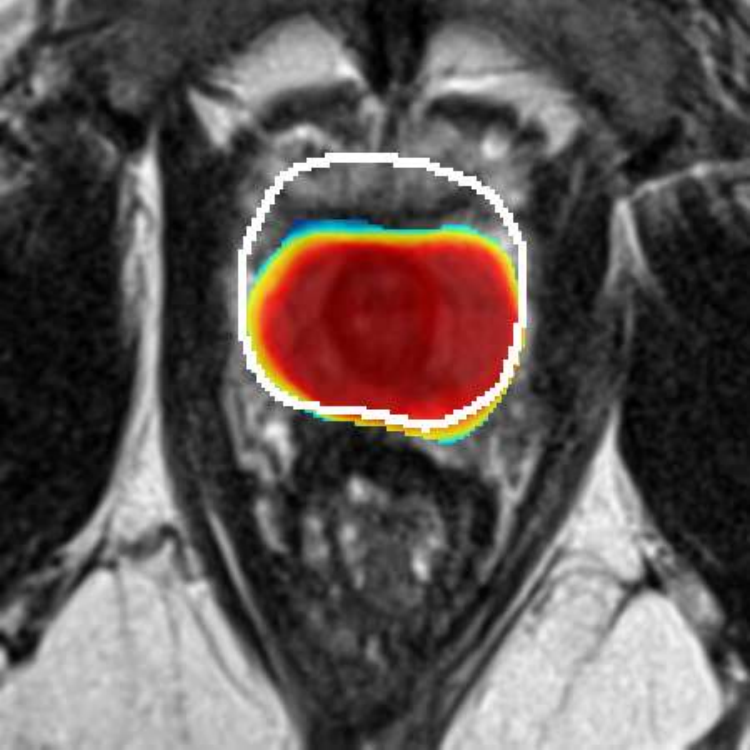}&
	\includegraphics[width=20mm]{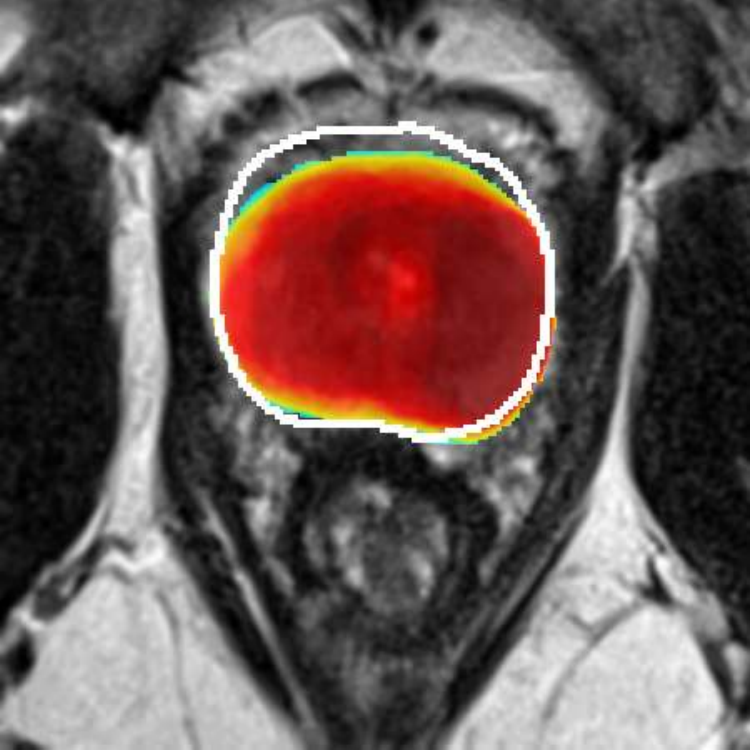}&
	\includegraphics[width=20mm]{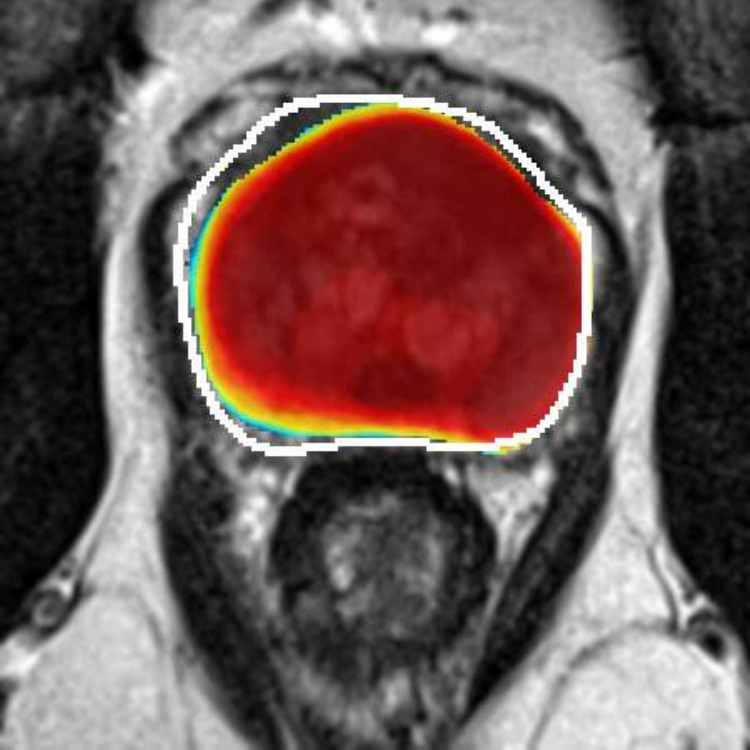}&
	\includegraphics[width=20mm]{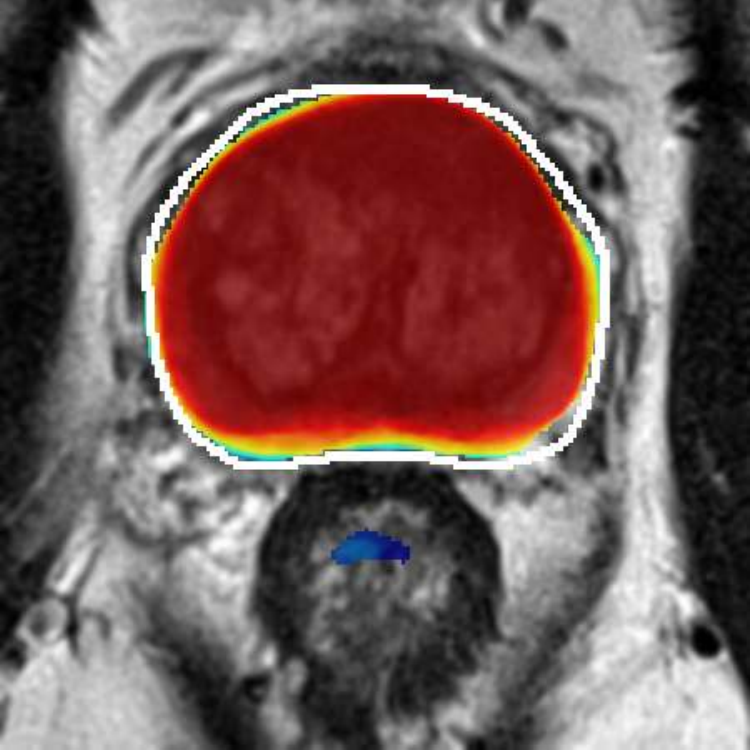}&
	\includegraphics[width=20mm]{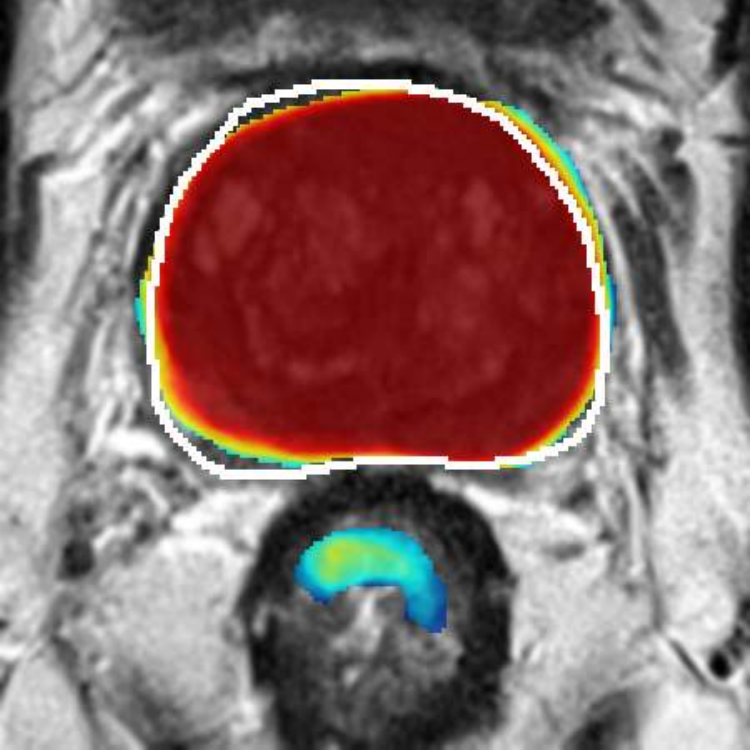}&
	\includegraphics[width=20mm]{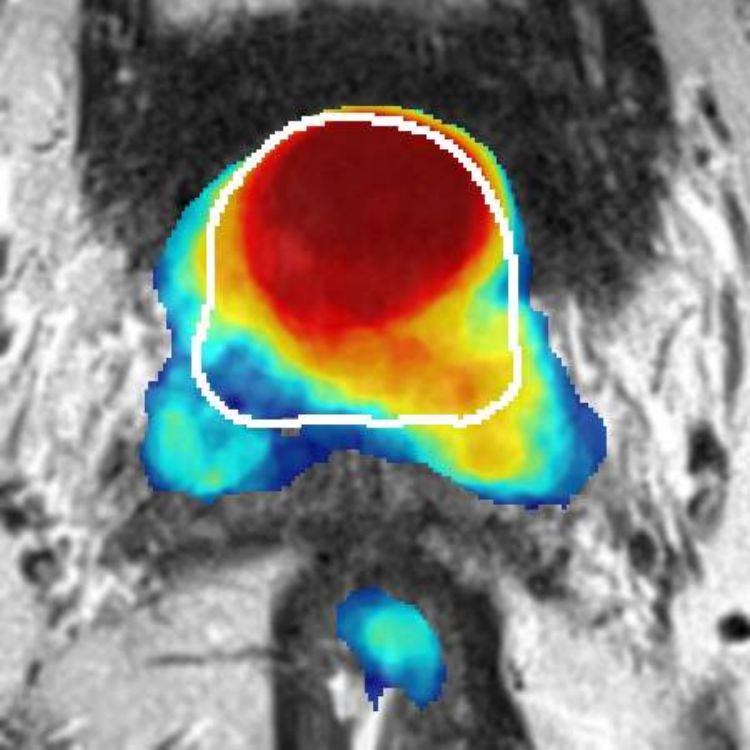}\\
	\multicolumn{8}{r}{\includegraphics[width=50mm]{figures/cb.pdf}}\\
\end{tabular}

\caption{
		Qualitative examples of 
		improvements in calibration and segmentation as a function of the number of models $M$ in the ensemble of models trained with Dice loss. 
		The overlaid probability maps show the results of inference for an ensemble of size M=2, M-20, and  M=50. White line shows the ground truth boundary of the structures.
}
\label{fig:n_ensembles_dice}
\end{figure*}

\begin{figure*}[h]
    \centering
    \setlength{\tabcolsep}{1pt}
    \begin{tabular}{llllllll}
            \parbox[t]{3mm}{\rotatebox[origin=l]{90}{\footnotesize{$~~~~~{M=2}$}}} &
            \includegraphics[width=20mm]{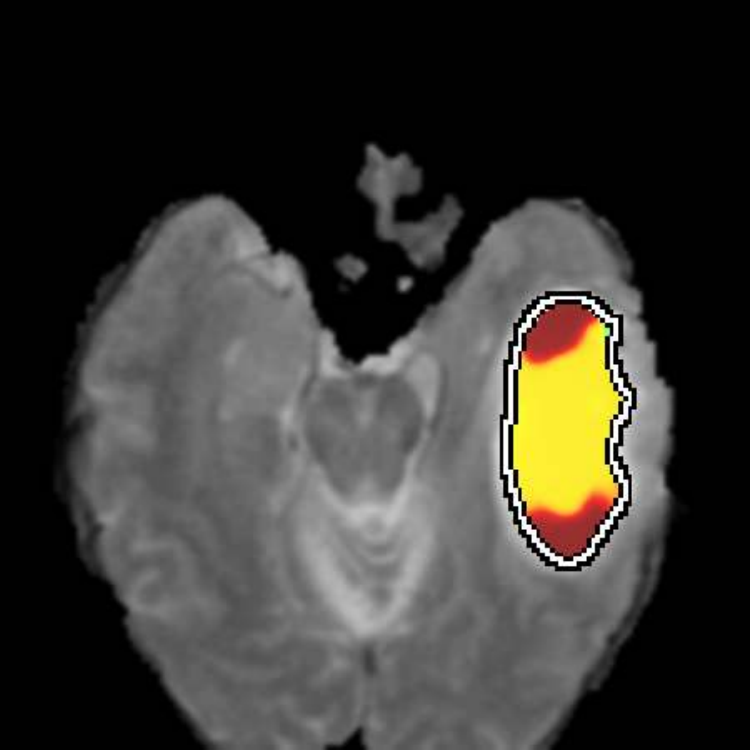}&
            \includegraphics[width=20mm]{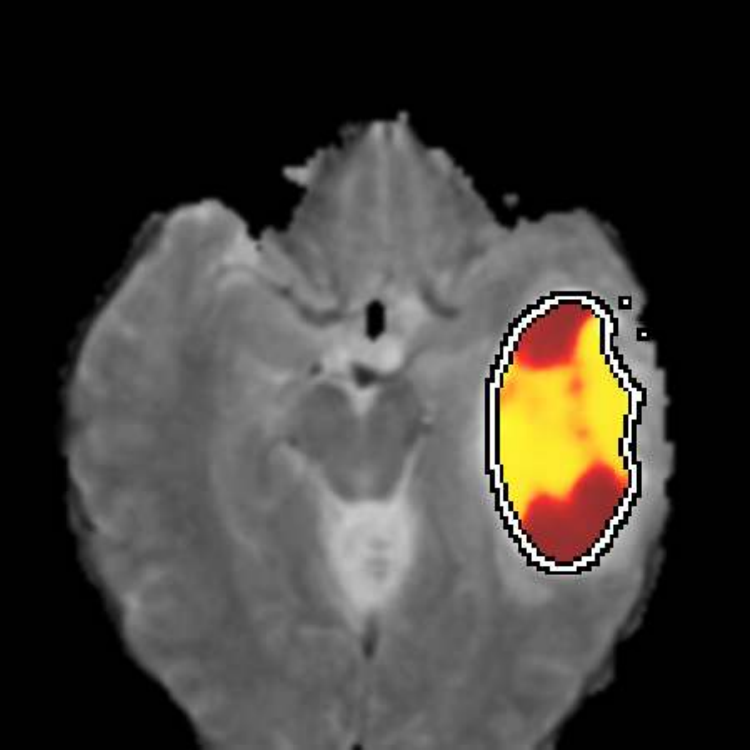}&
            \includegraphics[width=20mm]{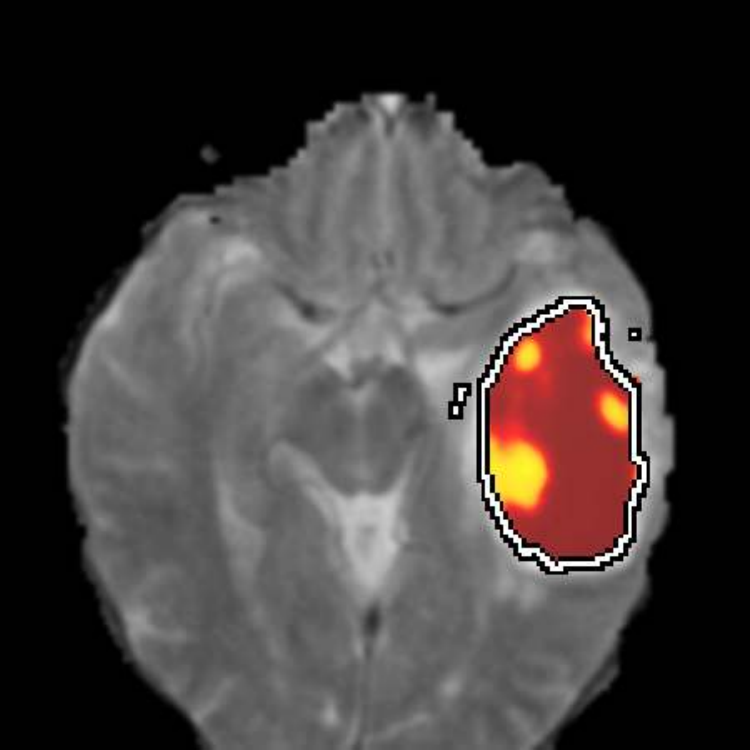}&
            \includegraphics[width=20mm]{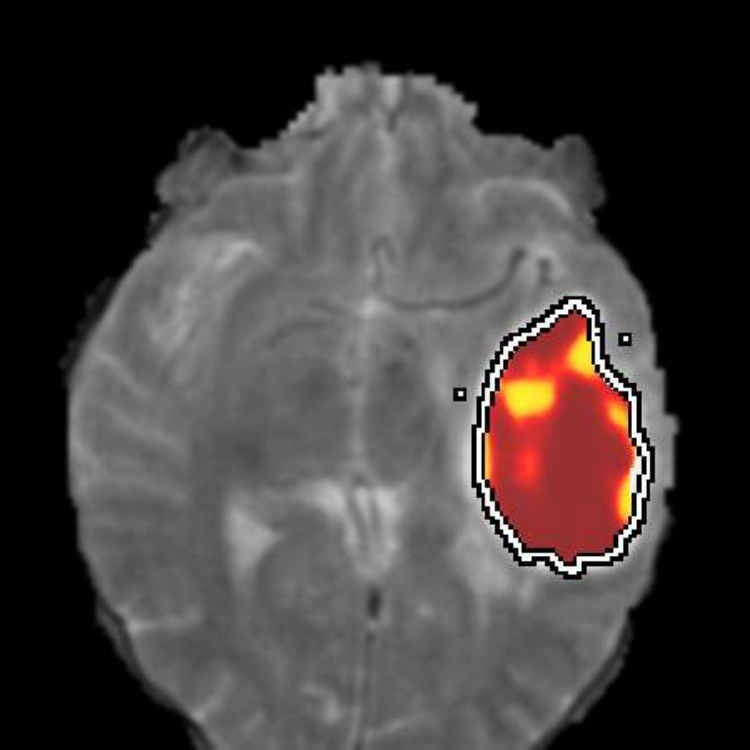}&
            \includegraphics[width=20mm]{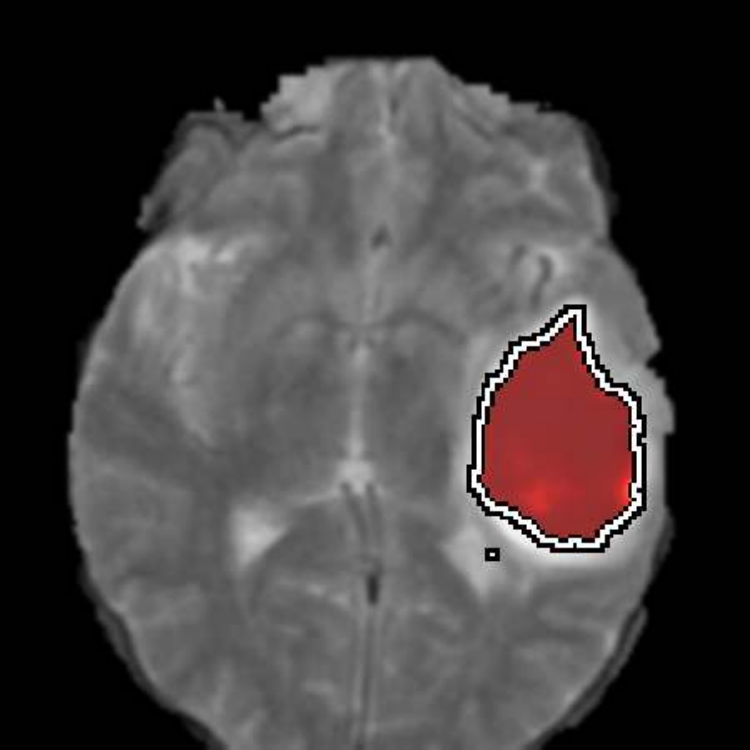}&
            \includegraphics[width=20mm]{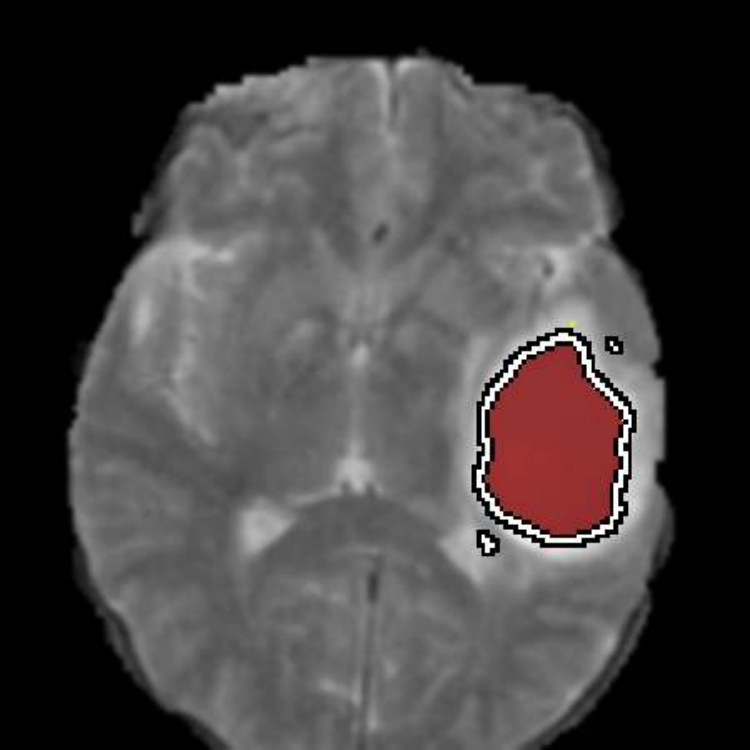}&
            \includegraphics[width=20mm]{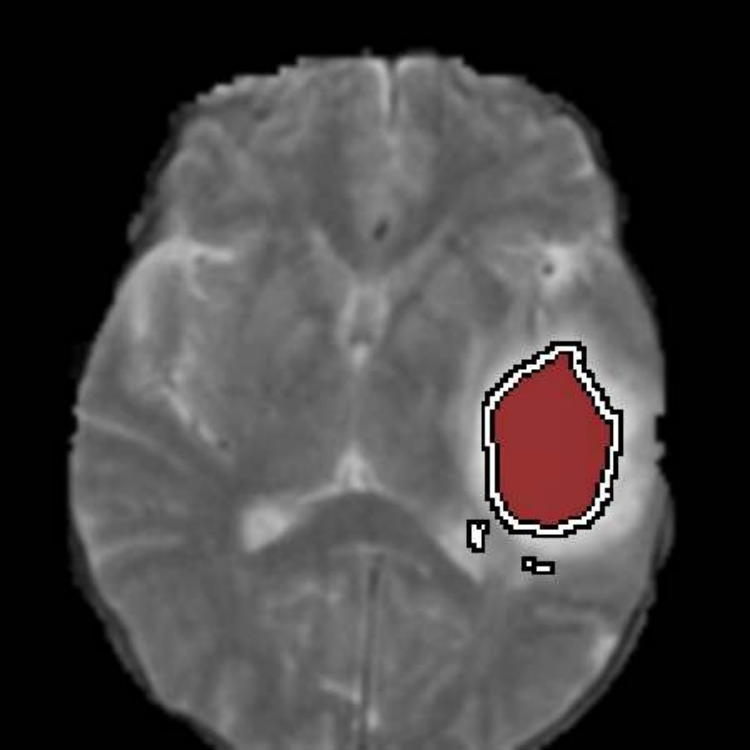}\\
            \parbox[t]{3mm}{\rotatebox[origin=l]{90}{\footnotesize{$~~~~{M=10}$}}} &
            \includegraphics[width=20mm]{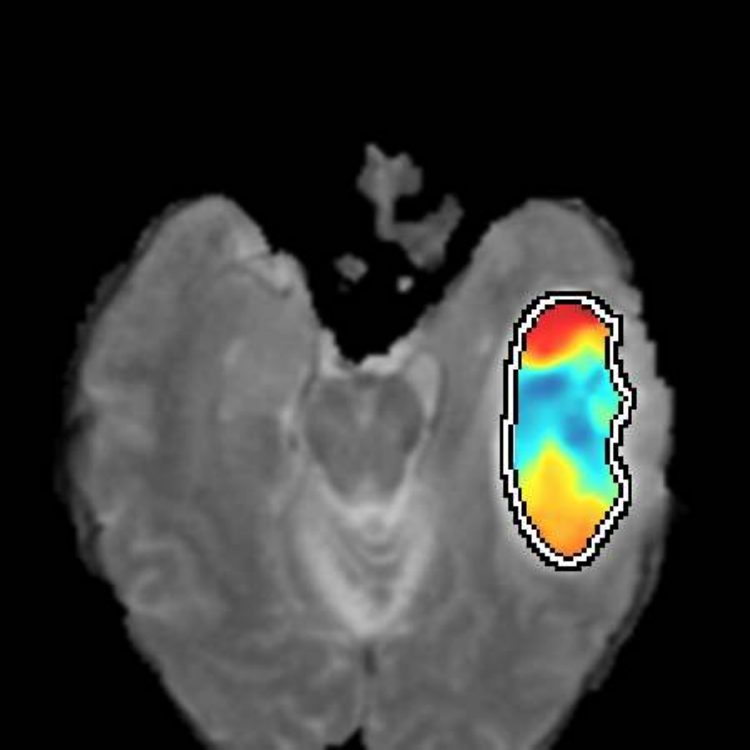}&
            \includegraphics[width=20mm]{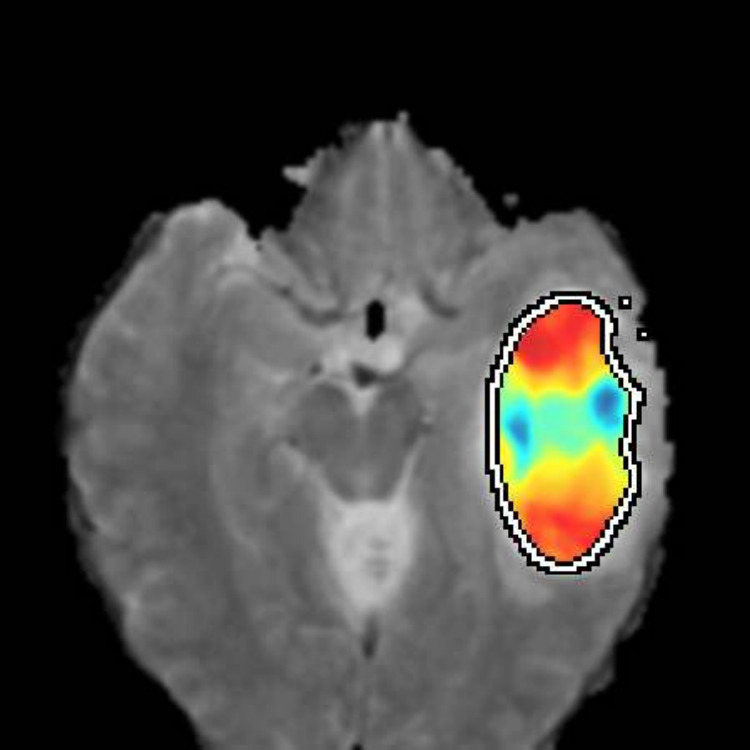}&
            \includegraphics[width=20mm]{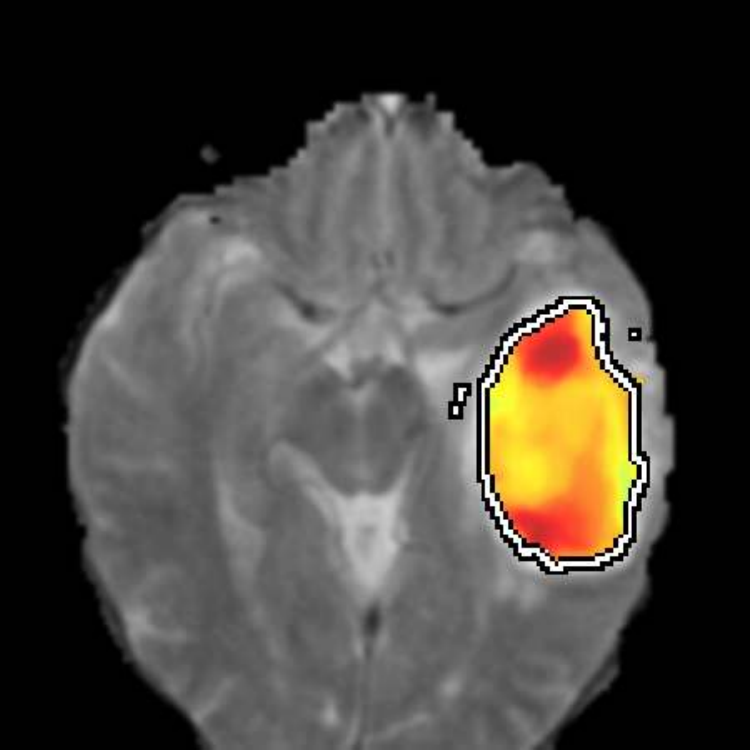}&
            \includegraphics[width=20mm]{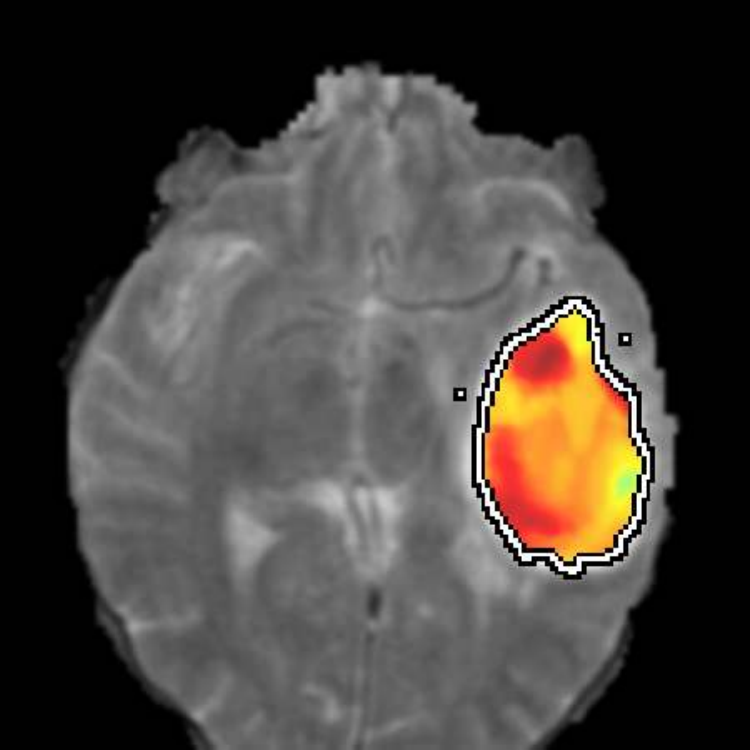}&
            \includegraphics[width=20mm]{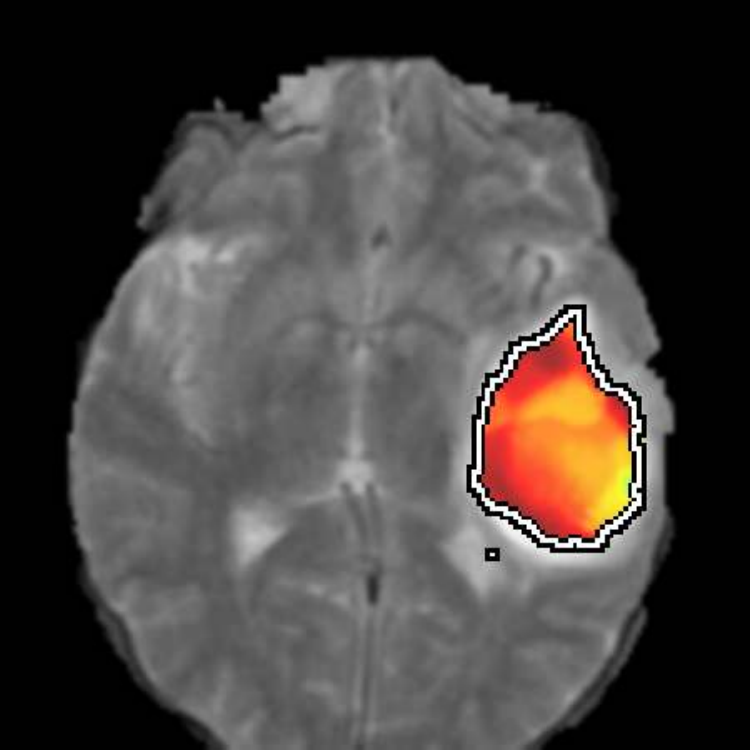}&
            \includegraphics[width=20mm]{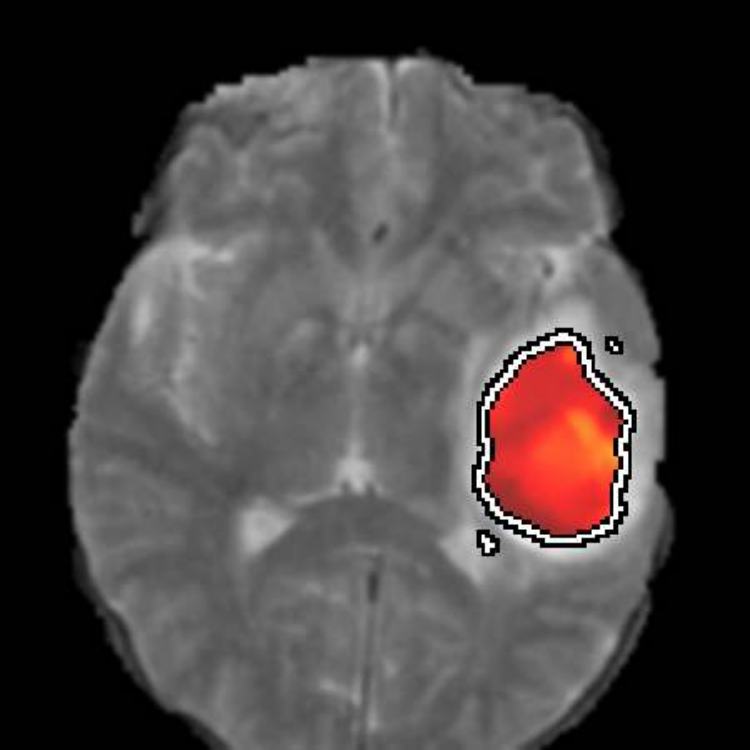}&
            \includegraphics[width=20mm]{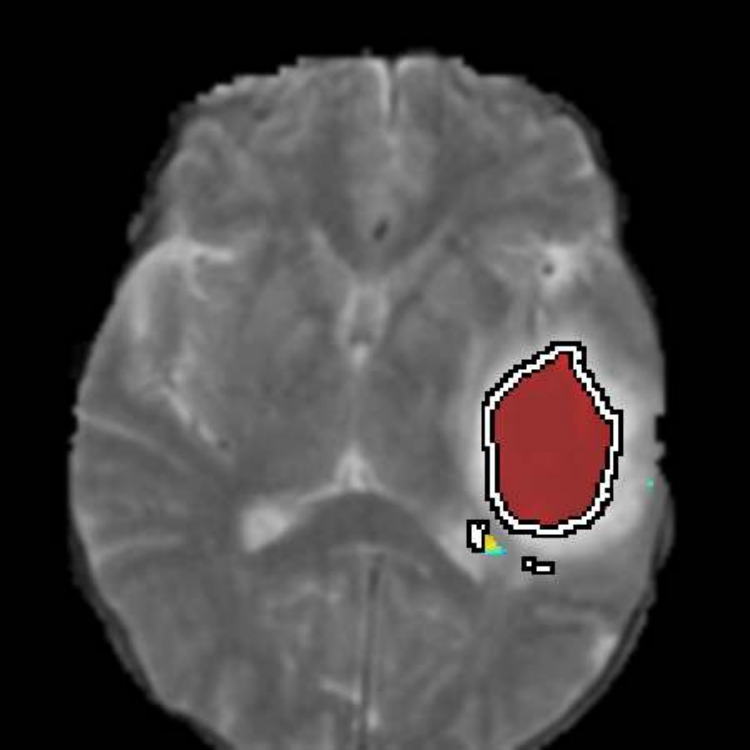}\\
            \parbox[t]{3mm}{\rotatebox[origin=l]{90}{\footnotesize{$~~~~{M=50}$}}} &
            \includegraphics[width=20mm]{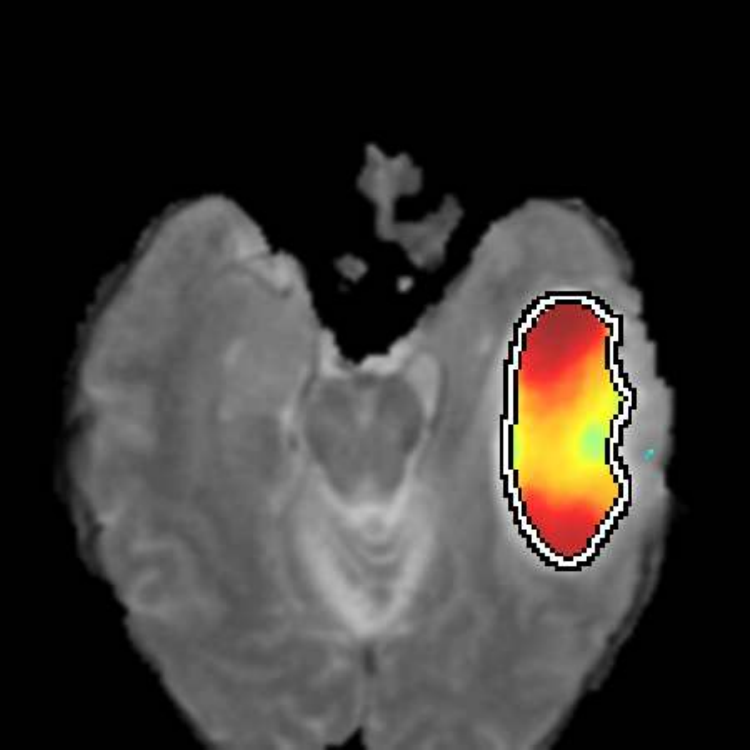}&
            \includegraphics[width=20mm]{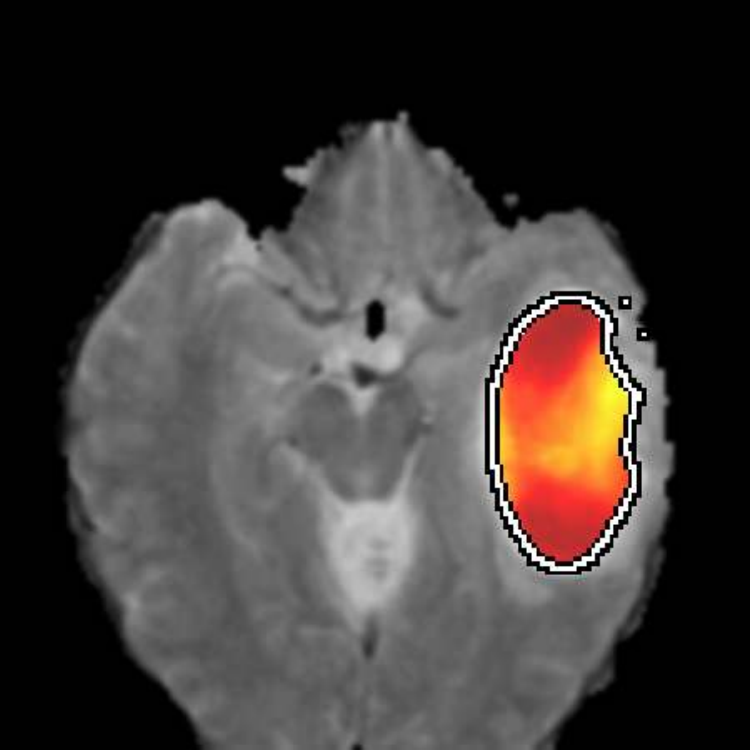}&
            \includegraphics[width=20mm]{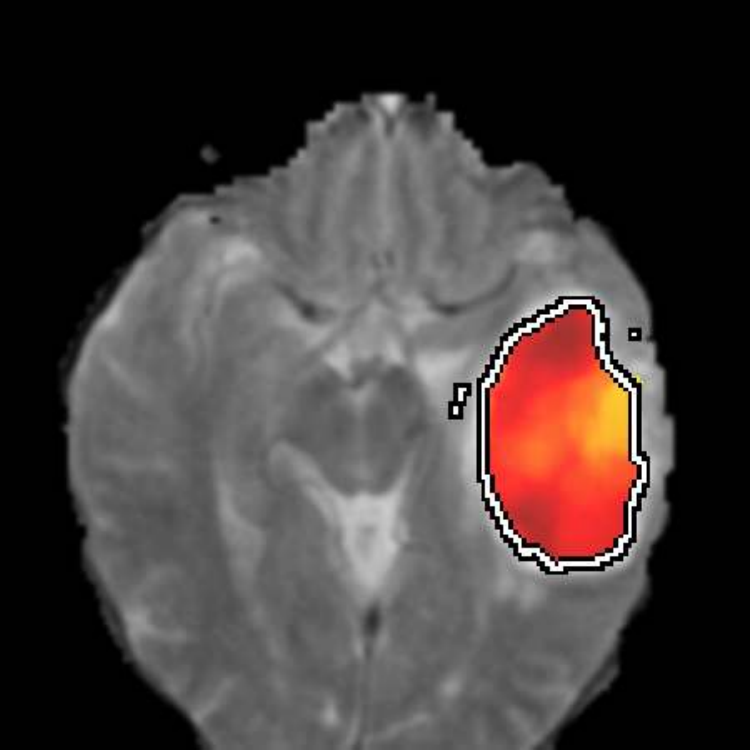}&
            \includegraphics[width=20mm]{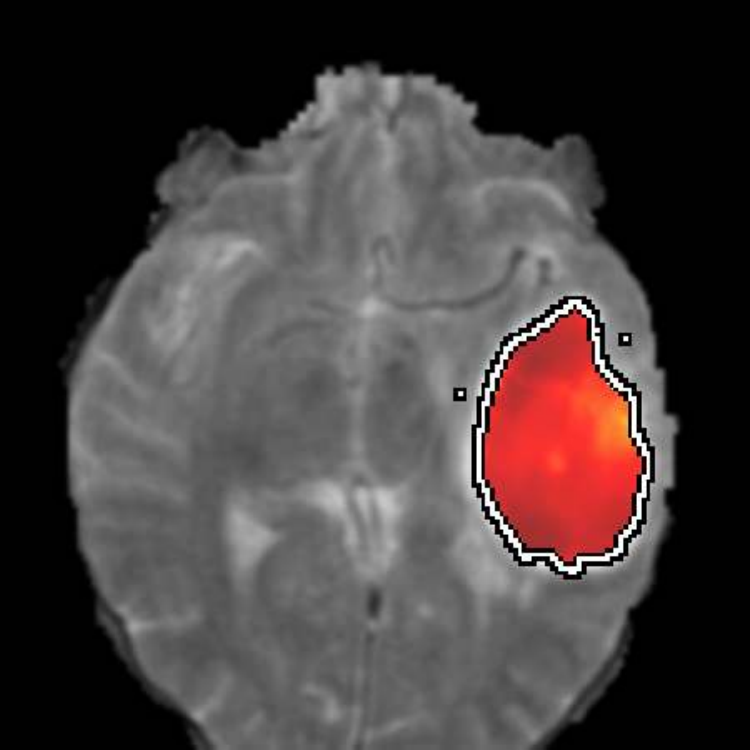}&
            \includegraphics[width=20mm]{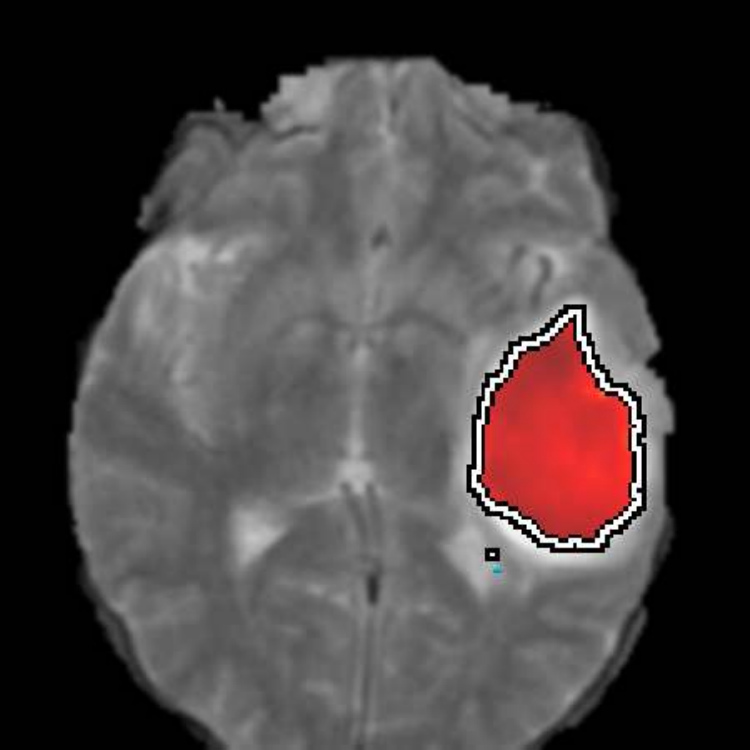}&
            \includegraphics[width=20mm]{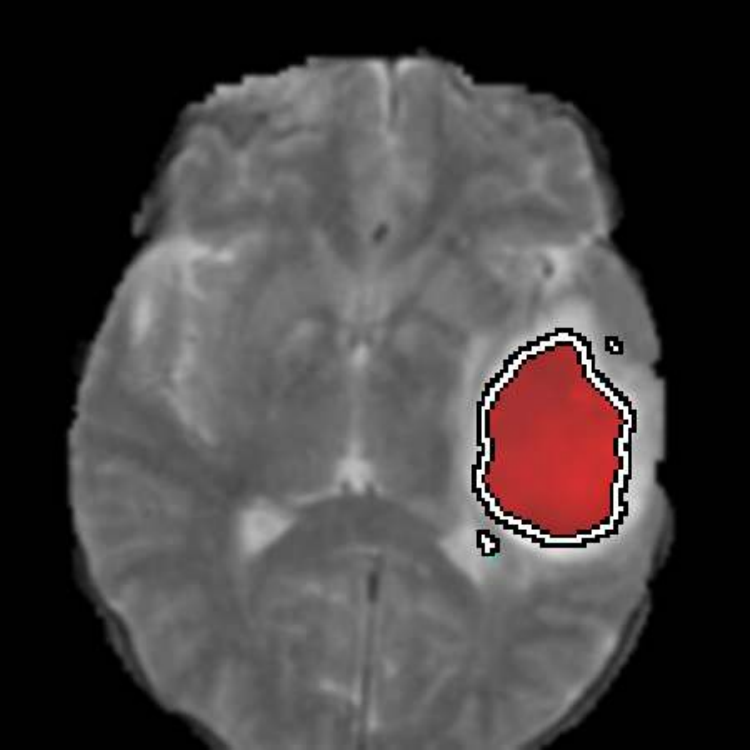}&
            \includegraphics[width=20mm]{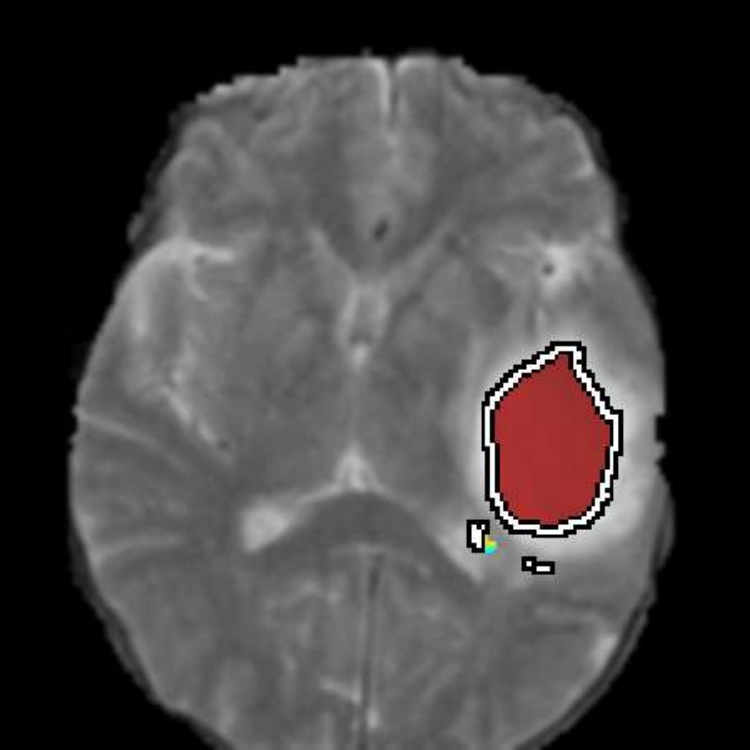}\\
    \end{tabular}
    
    \begin{tabular}{llllllll}
            \parbox[t]{3mm}{\rotatebox[origin=l]{90}{\footnotesize{$~~~~~{M=2}$}}} &
            \includegraphics[width=20mm]{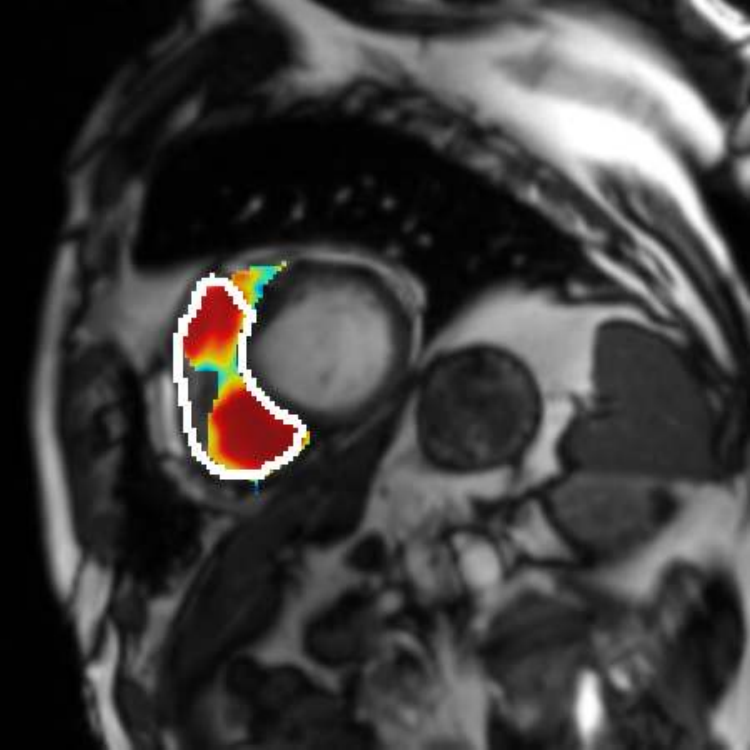}&
            \includegraphics[width=20mm]{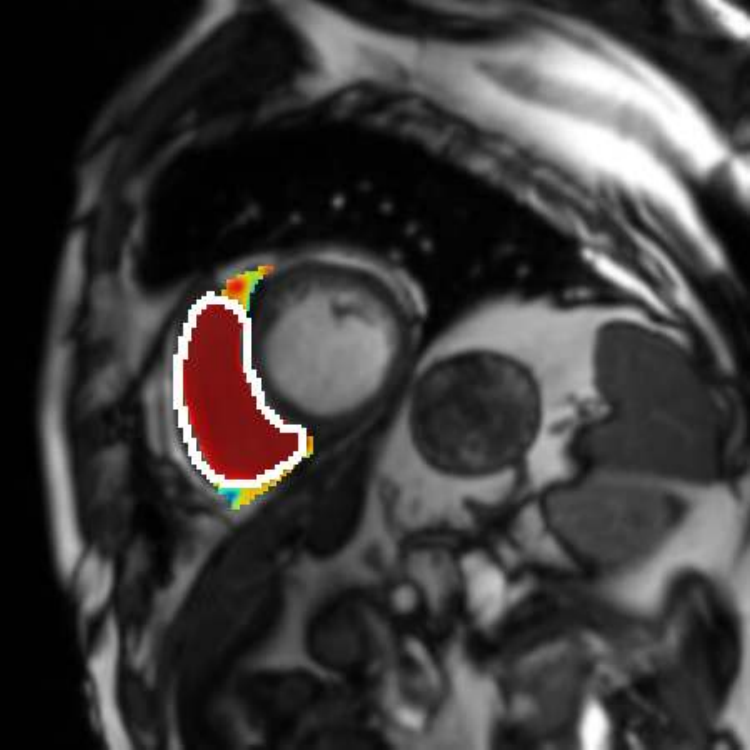}&
            \includegraphics[width=20mm]{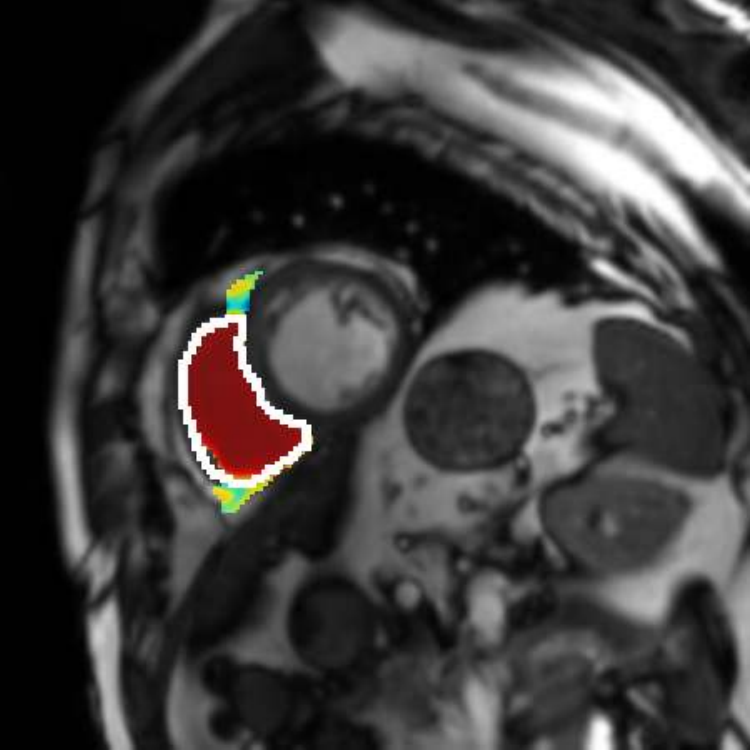}&
            \includegraphics[width=20mm]{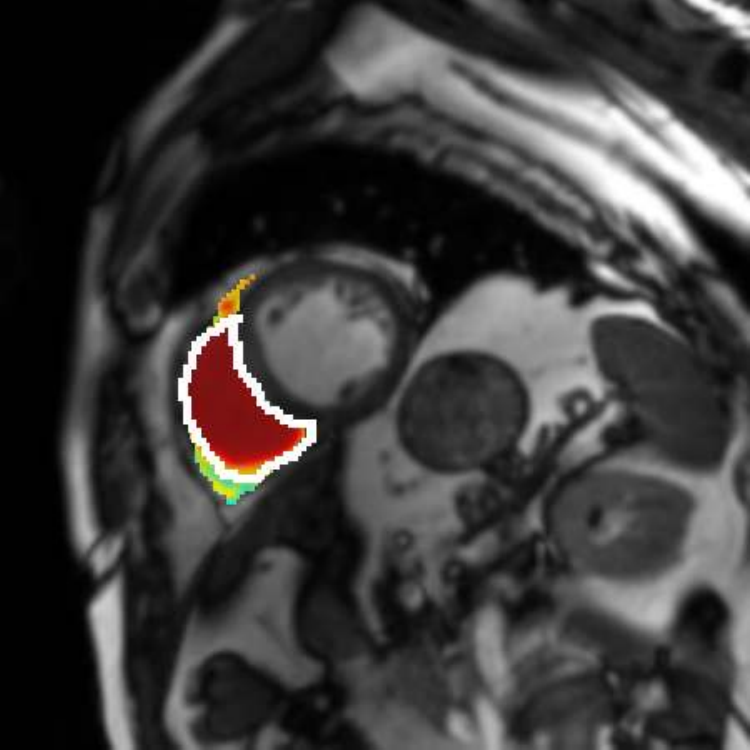}&
            \includegraphics[width=20mm]{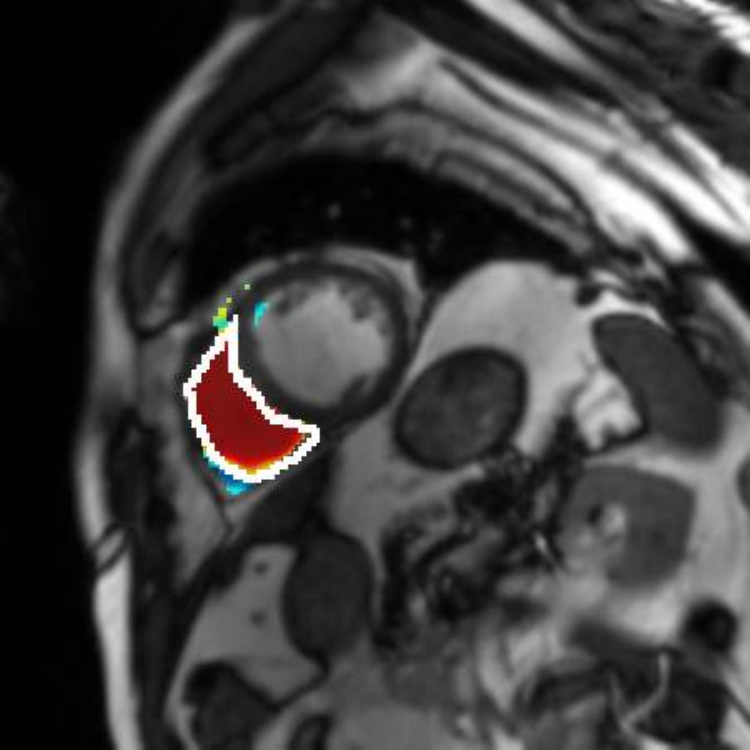}&
            \includegraphics[width=20mm]{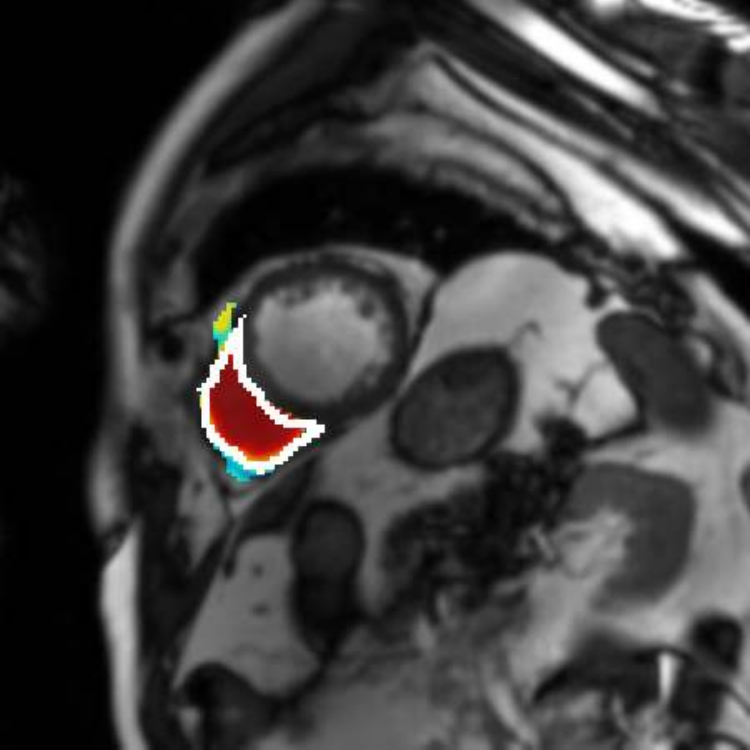}&
            \includegraphics[width=20mm]{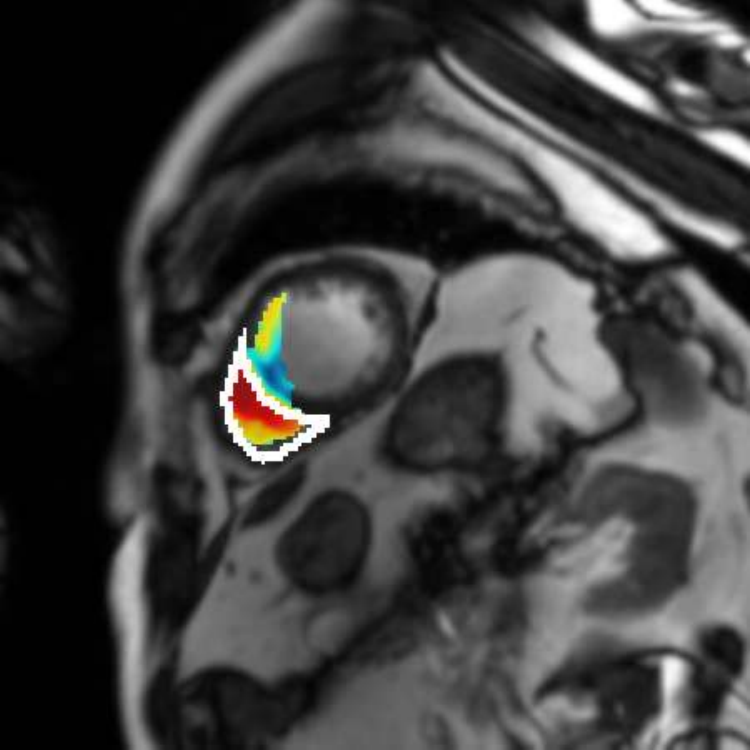}\\
            \parbox[t]{3mm}{\rotatebox[origin=l]{90}{\footnotesize{$~~~~{M=10}$}}} &
            \includegraphics[width=20mm]{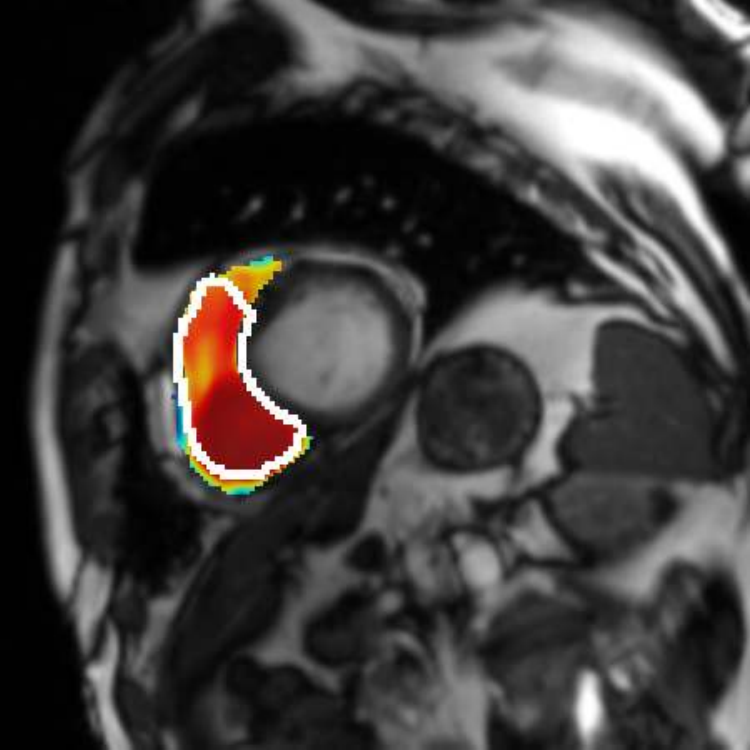}&
            \includegraphics[width=20mm]{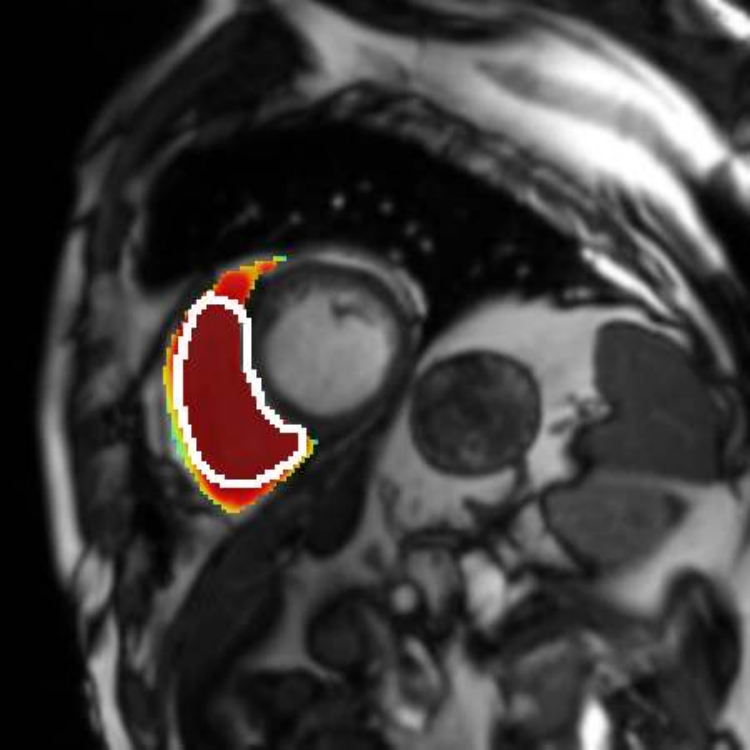}&
            \includegraphics[width=20mm]{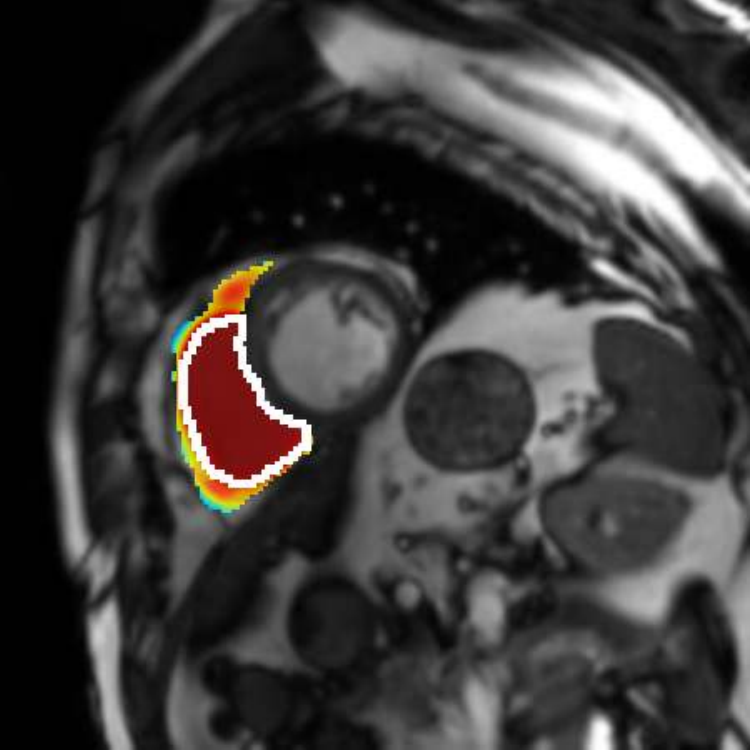}&
            \includegraphics[width=20mm]{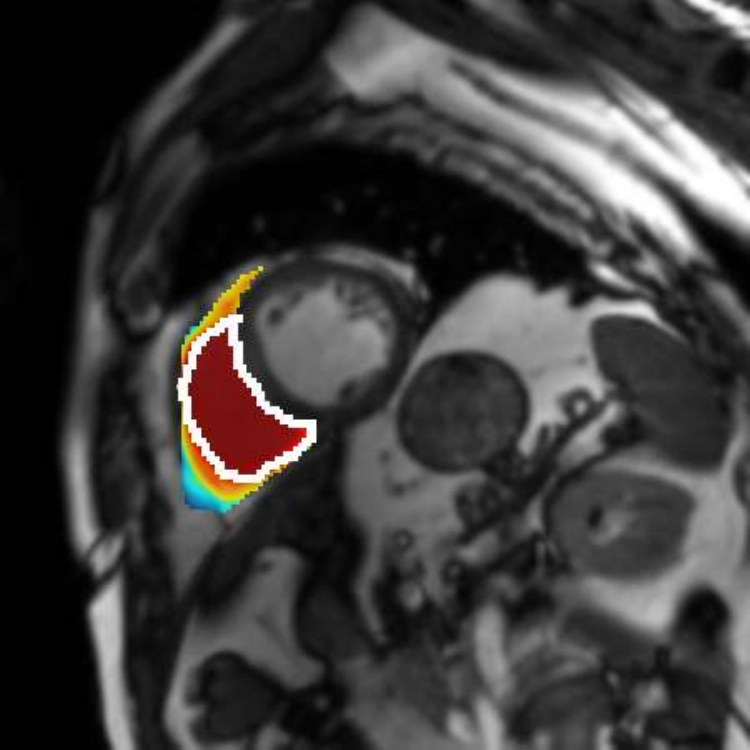}&
            \includegraphics[width=20mm]{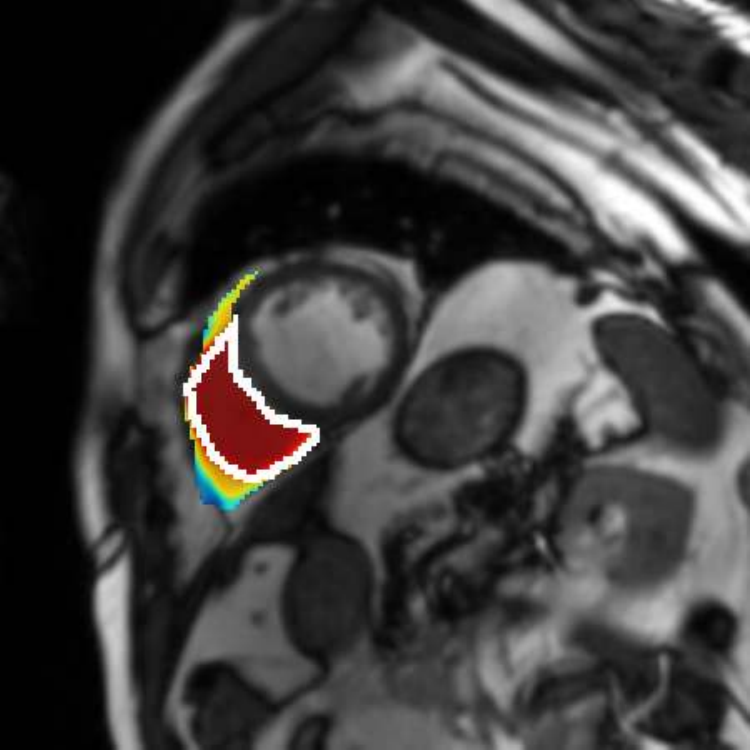}&
            \includegraphics[width=20mm]{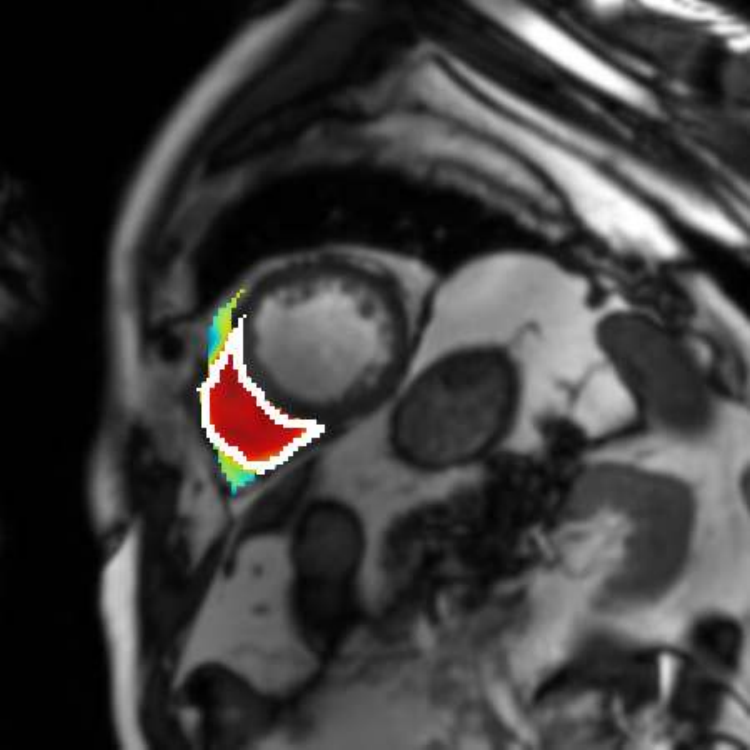}&
            \includegraphics[width=20mm]{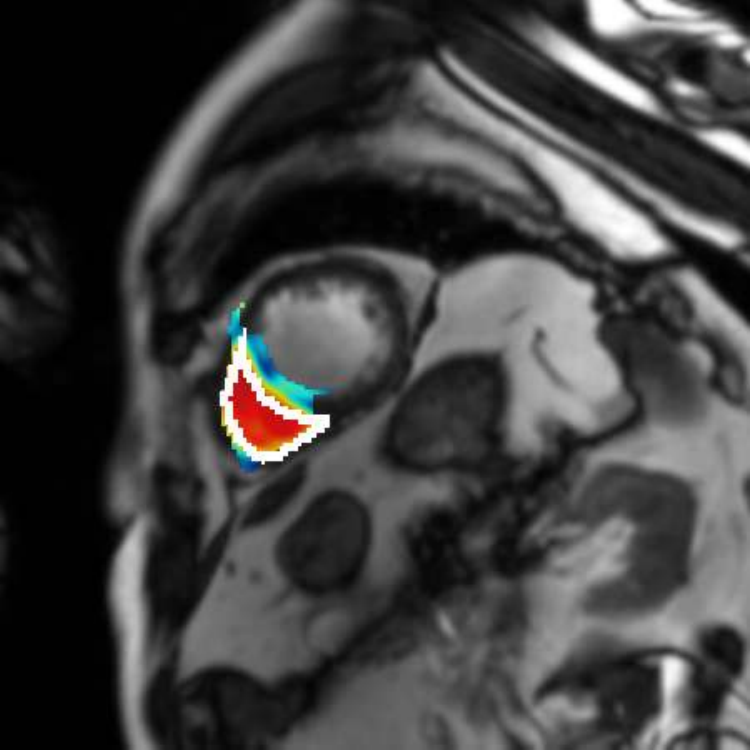}\\
            \parbox[t]{3mm}{\rotatebox[origin=l]{90}{\footnotesize{$~~~~{M=50}$}}} &
            \includegraphics[width=20mm]{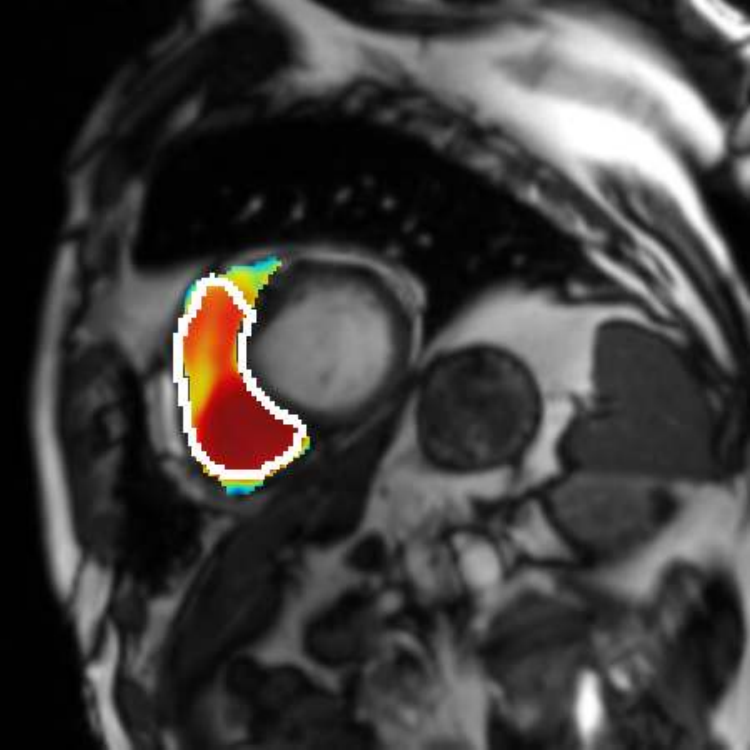}&
            \includegraphics[width=20mm]{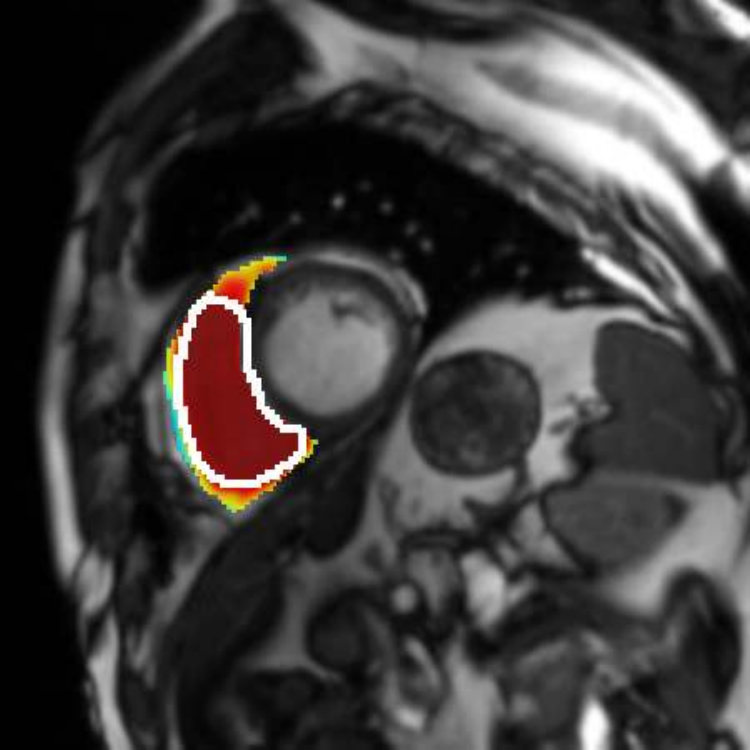}&
            \includegraphics[width=20mm]{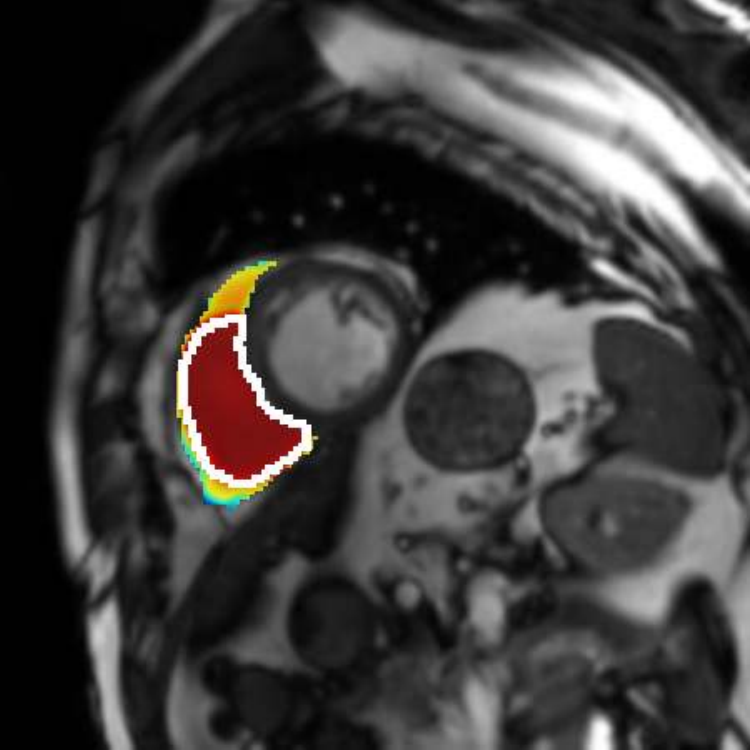}&
            \includegraphics[width=20mm]{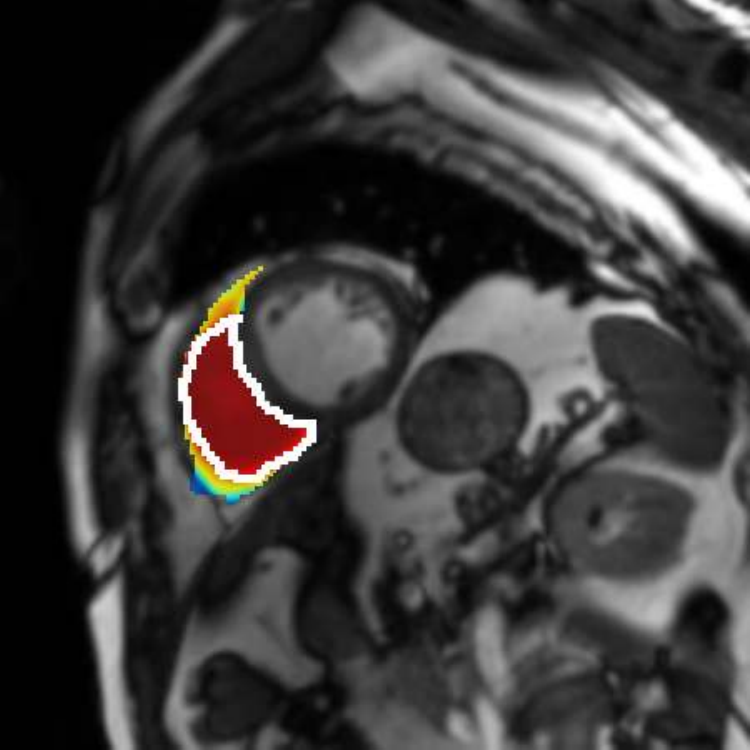}&
            \includegraphics[width=20mm]{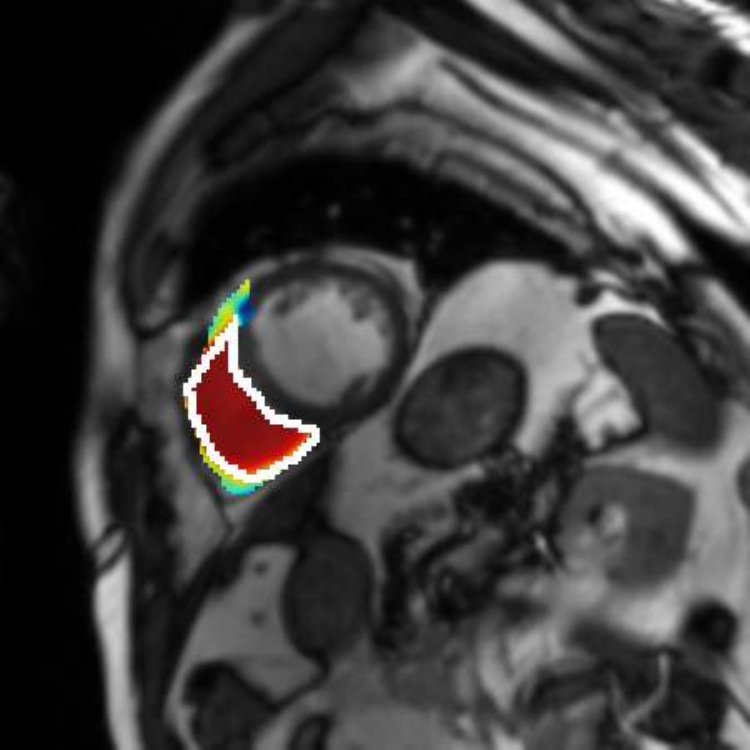}&
            \includegraphics[width=20mm]{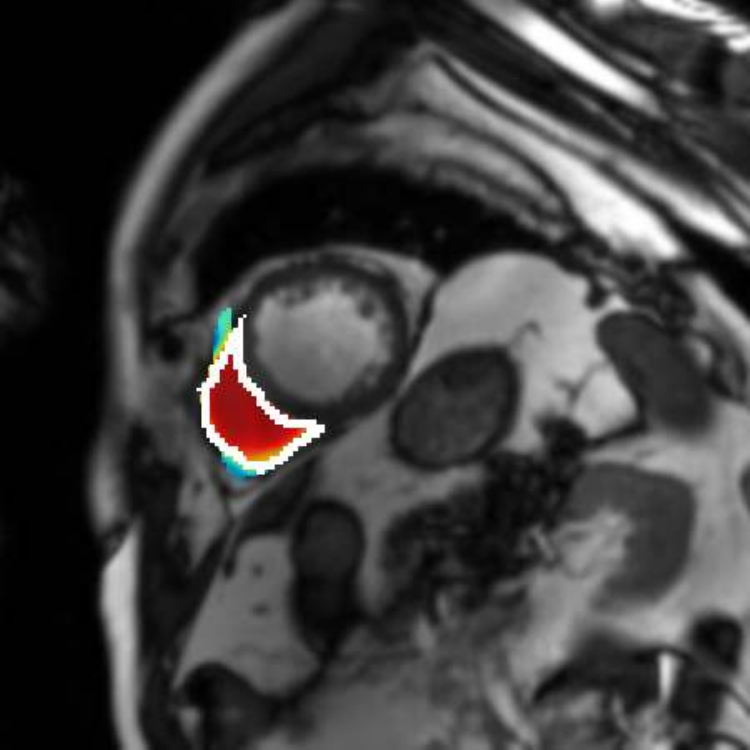}&
            \includegraphics[width=20mm]{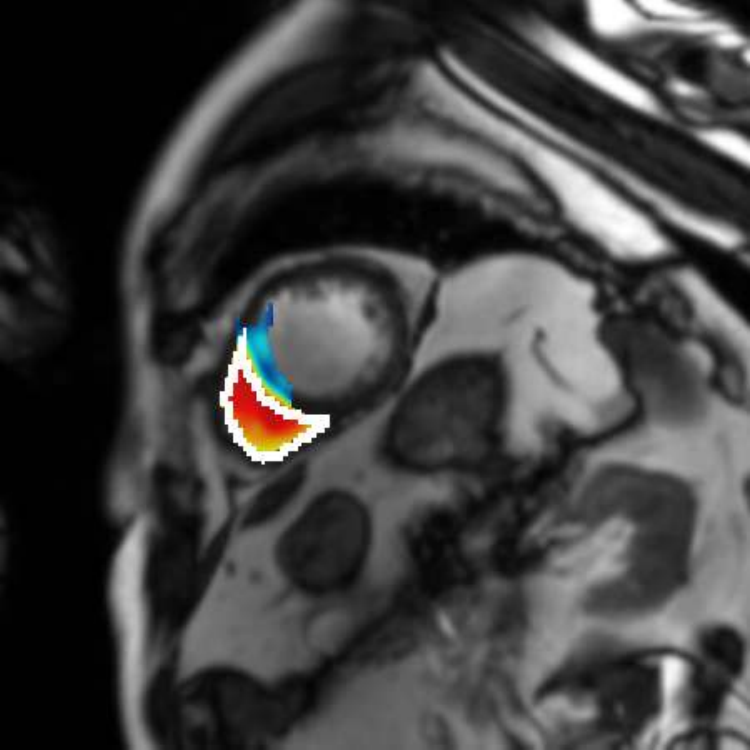}\\
    \end{tabular}
    
    \begin{tabular}{rlllllll}
            \parbox[t]{3mm}{\rotatebox[origin=l]{90}{\footnotesize{$~~~~~{M=2}$}}} &
            \includegraphics[width=20mm]{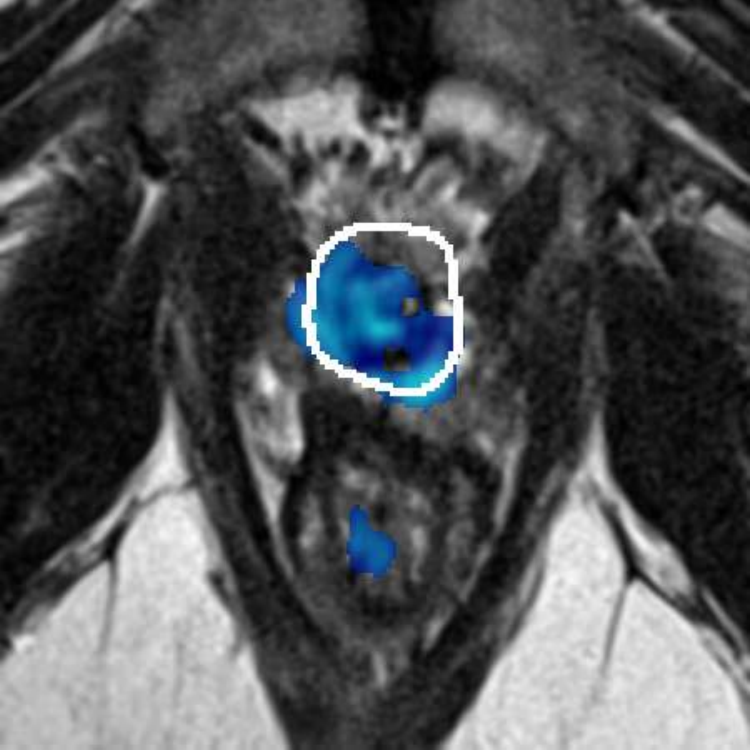}&
            \includegraphics[width=20mm]{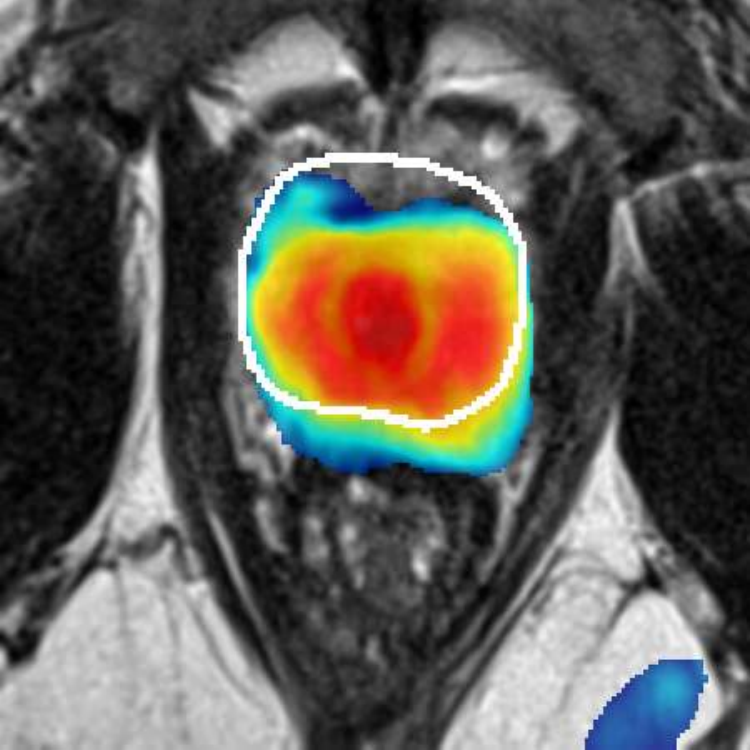}&
            \includegraphics[width=20mm]{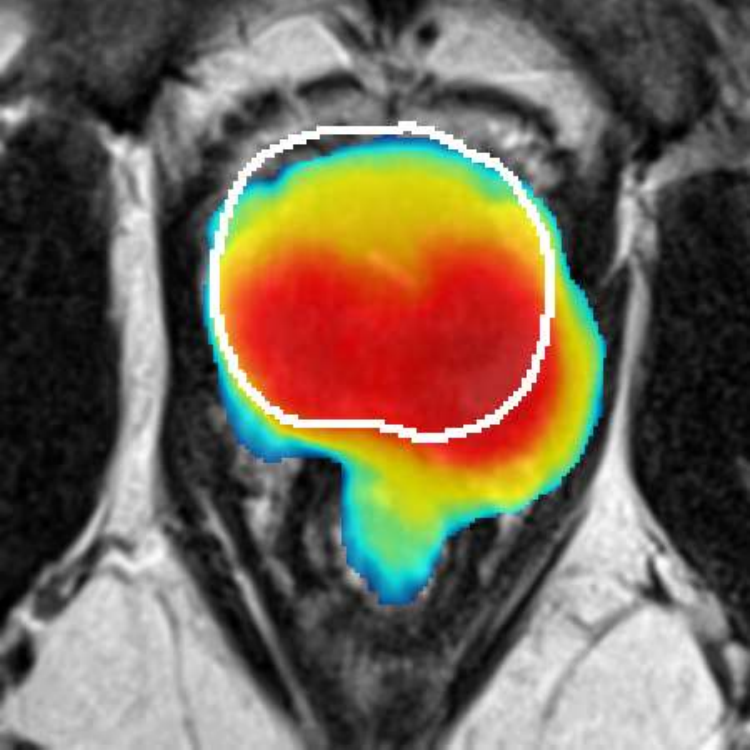}&
            \includegraphics[width=20mm]{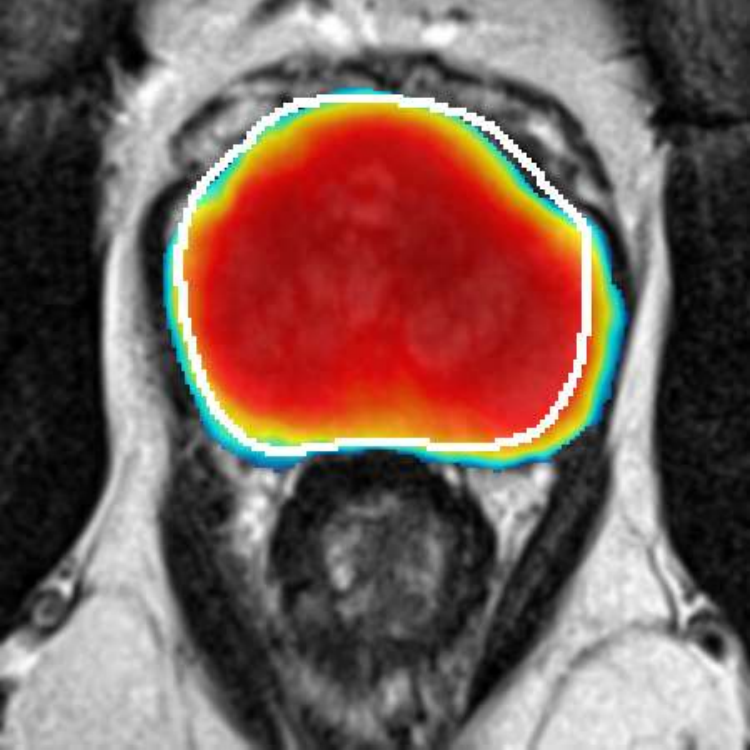}&
            \includegraphics[width=20mm]{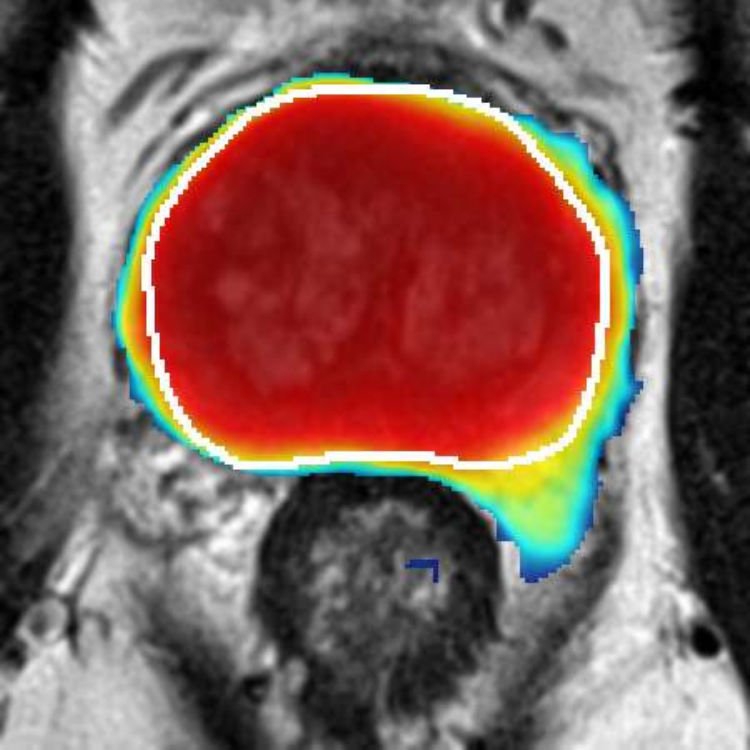}&
            \includegraphics[width=20mm]{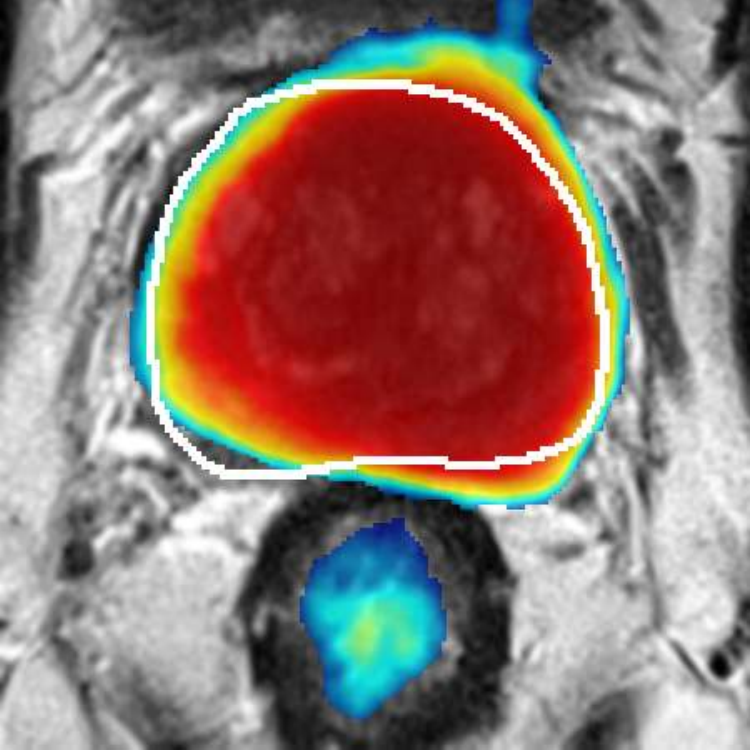}&
            \includegraphics[width=20mm]{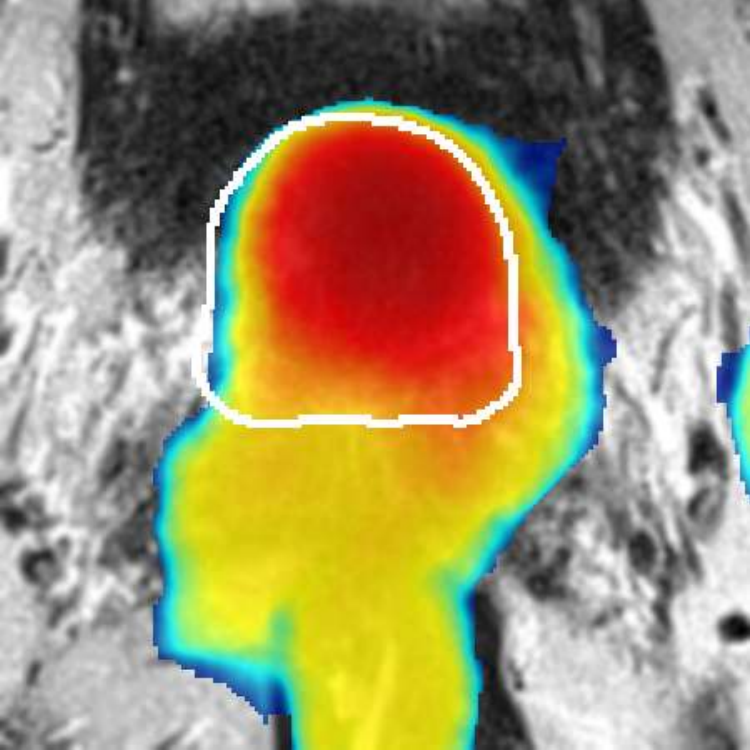}\\
            \parbox[t]{3mm}{\rotatebox[origin=l]{90}{\footnotesize{$~~~~{M=10}$}}} &
            \includegraphics[width=20mm]{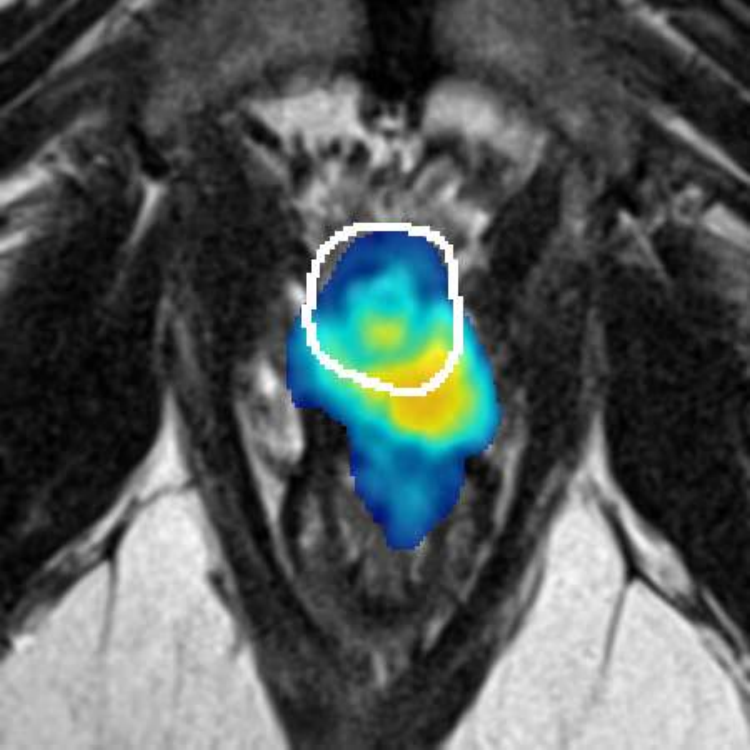}&
            \includegraphics[width=20mm]{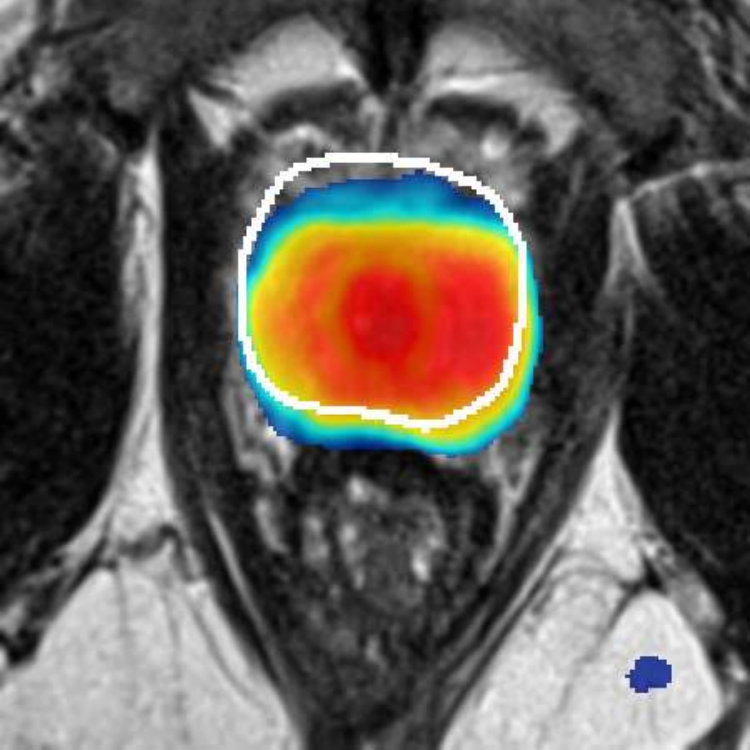}&
            \includegraphics[width=20mm]{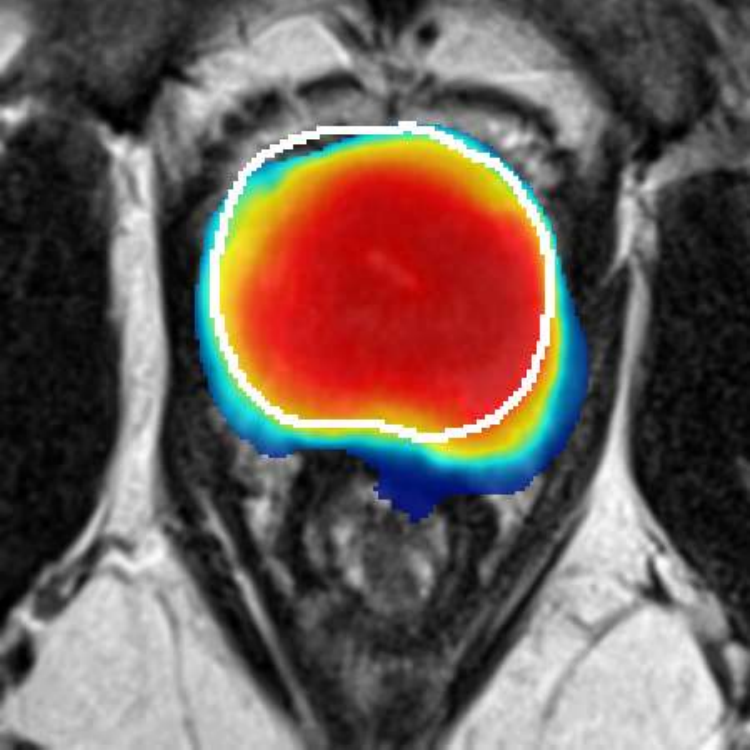}&
            \includegraphics[width=20mm]{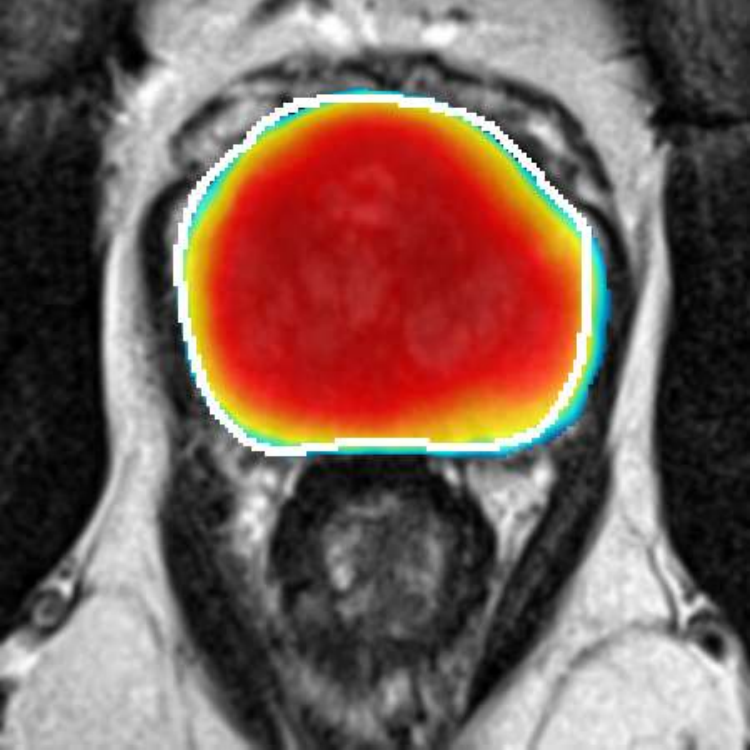}&
            \includegraphics[width=20mm]{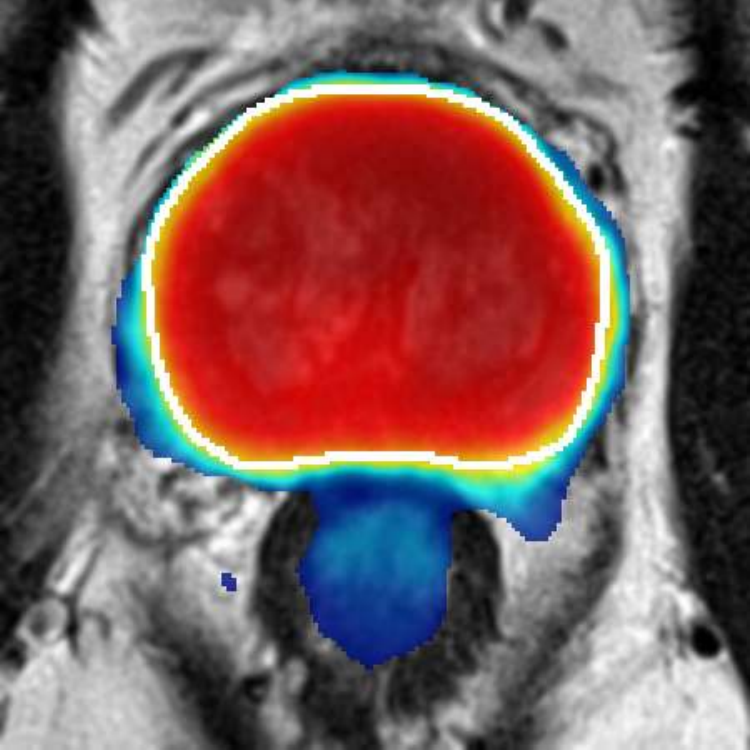}&
            \includegraphics[width=20mm]{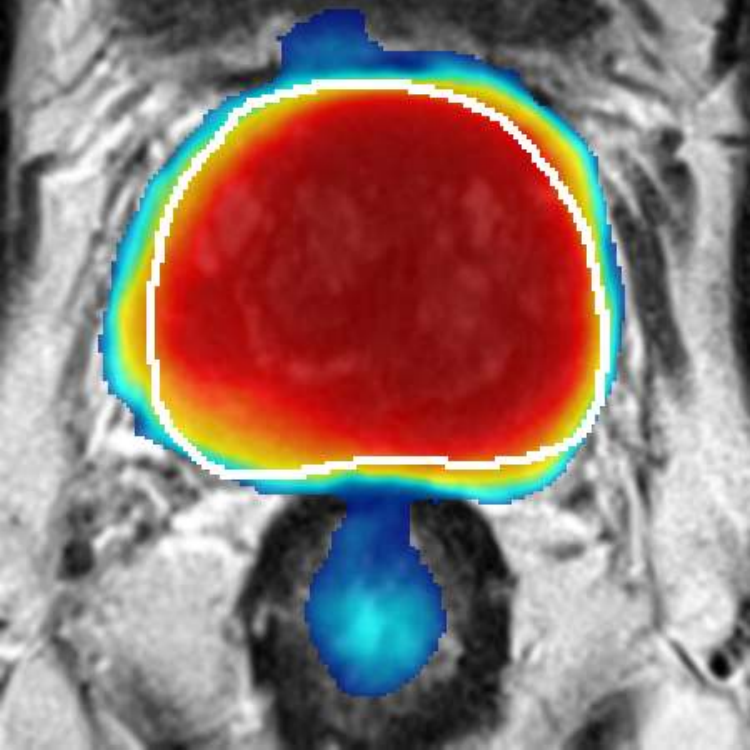}&
            \includegraphics[width=20mm]{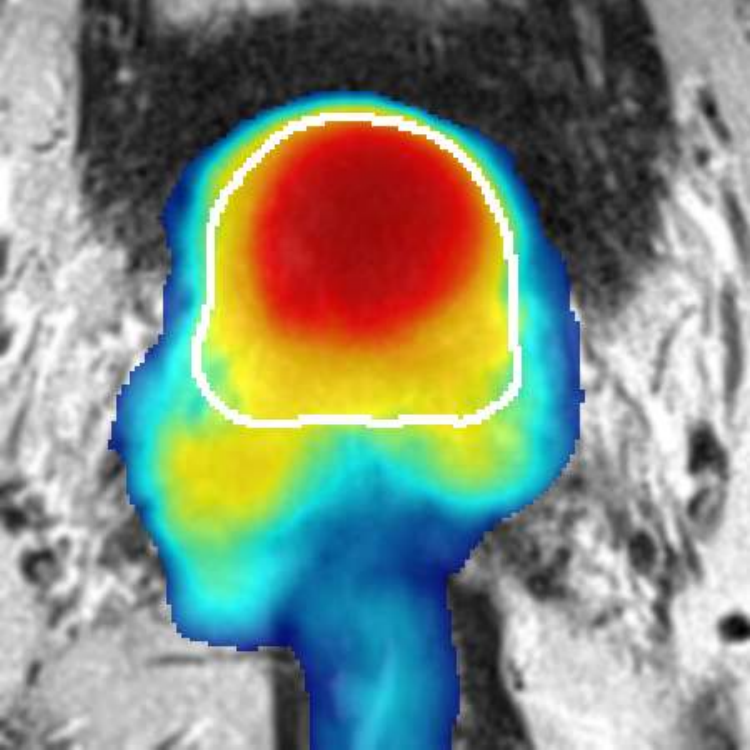}\\
            \parbox[t]{3mm}{\rotatebox[origin=l]{90}{\footnotesize{$~~~~{M=50}$}}} &
            \includegraphics[width=20mm]{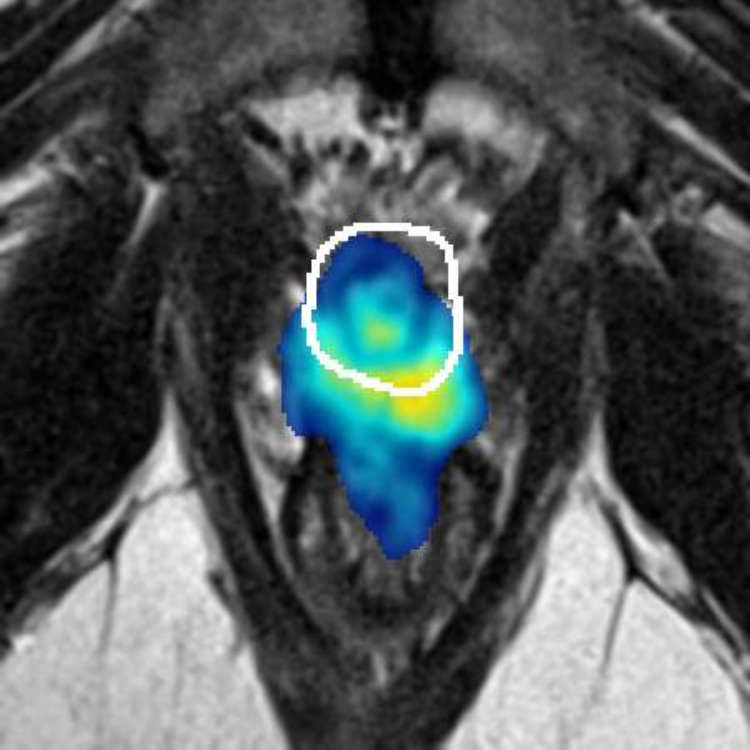}&
            \includegraphics[width=20mm]{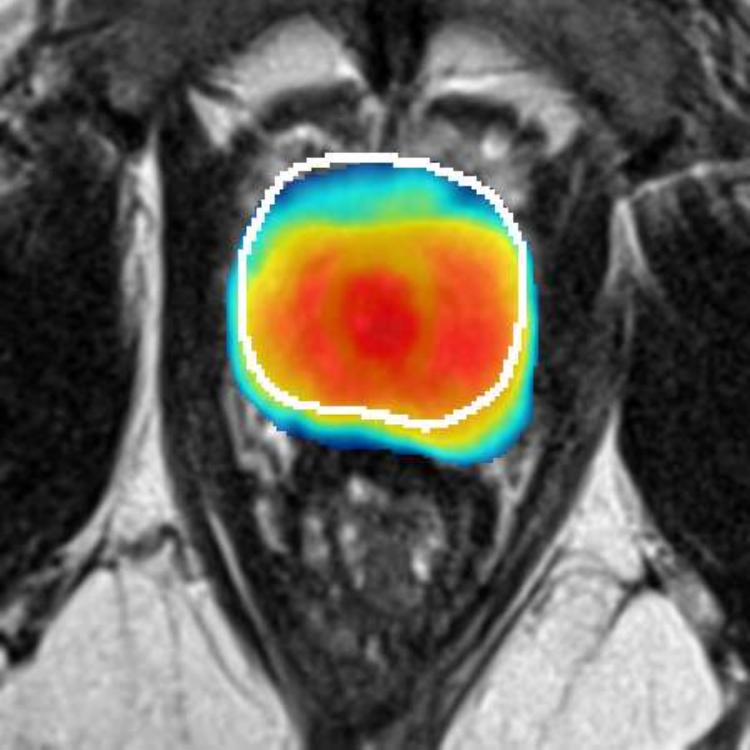}&
            \includegraphics[width=20mm]{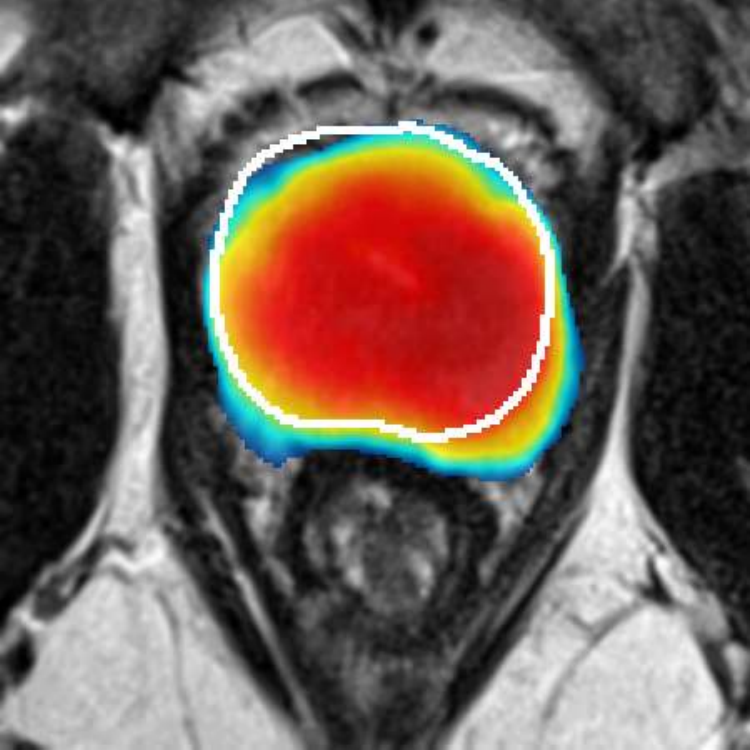}&
            \includegraphics[width=20mm]{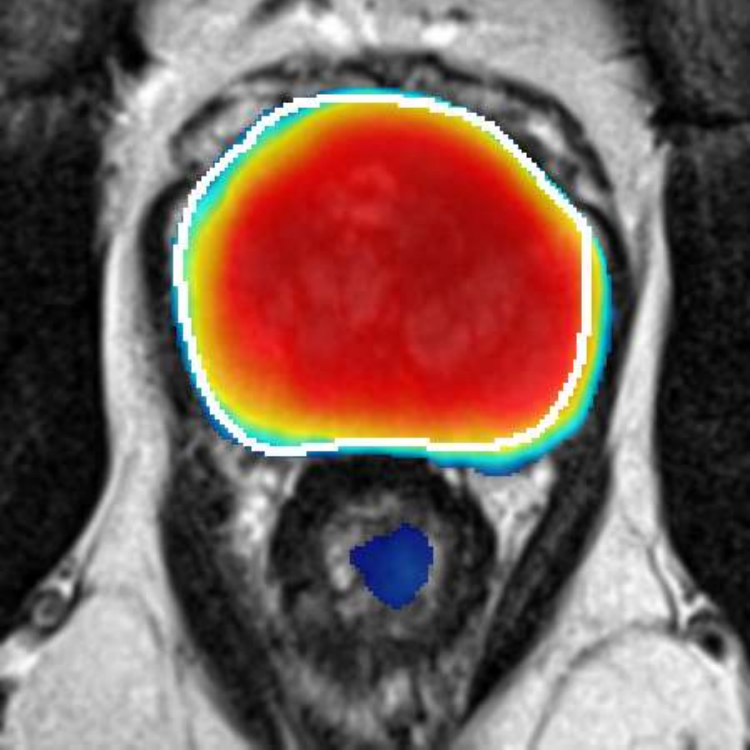}&
            \includegraphics[width=20mm]{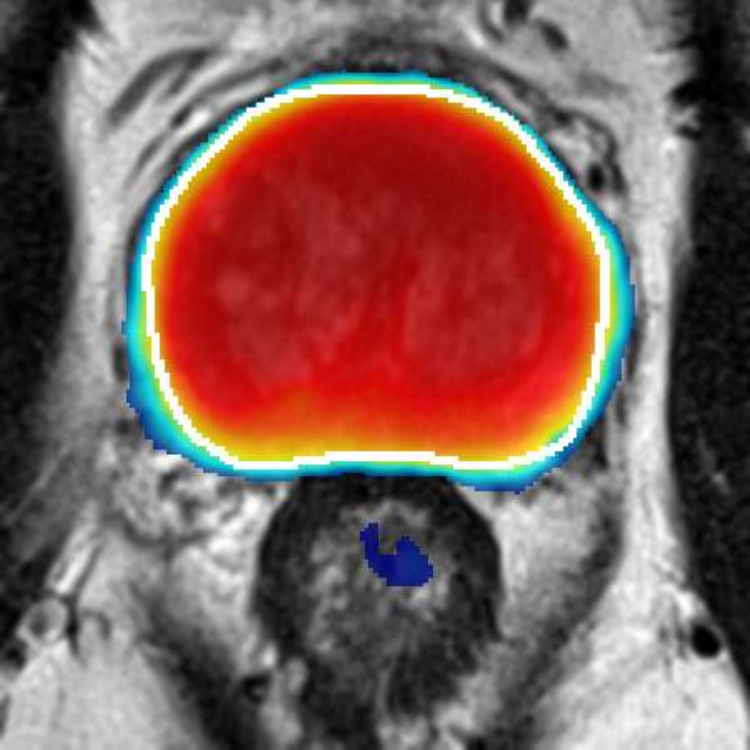}&
            \includegraphics[width=20mm]{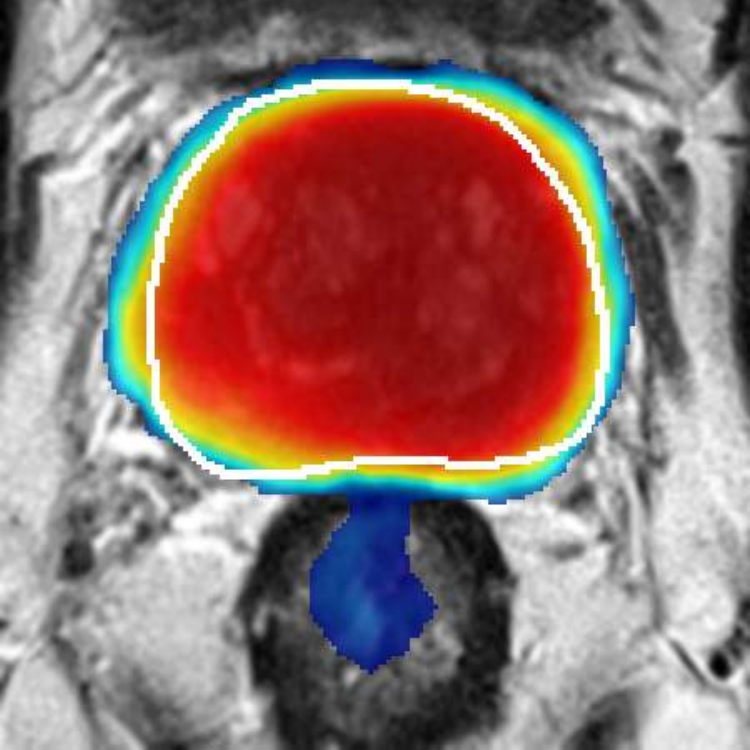}&
            \includegraphics[width=20mm]{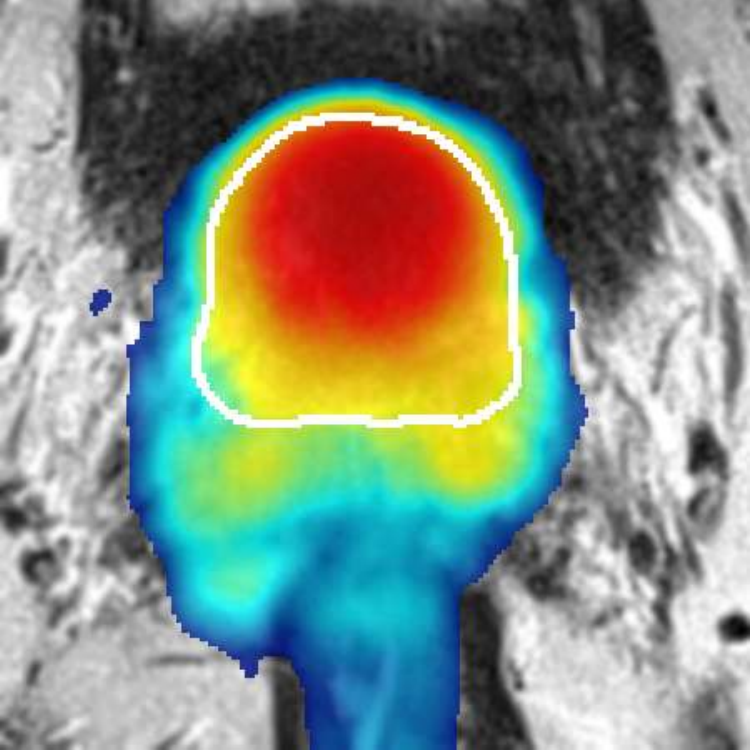}\\
            \multicolumn{8}{r}{\includegraphics[width=50mm]{figures/cb.pdf}}\\
    \end{tabular}

    \caption{
		Qualitative examples of 
		improvements in calibration and segmentation as a function of the number of models $M$ in the ensemble of models trained with cross entropy loss. 
		The overlaid probability maps show the results of inference for an ensemble of size M=2, M-20, and  M=50. White line shows the ground truth boundary of the structures.
	}
	\label{fig:n_ensembles_ce}
\end{figure*}

The three graphs in Figure \ref{fig:n_ensembles_with_ci} show the quantitative improvement in the calibration as a function of the number of models in the ensemble, M, with 0.95 CI. 
The images in Figure \ref{fig:n_ensembles_dice} and \ref{fig:n_ensembles_ce} qualitatively illustrate this calibration improvement by ensembling for models trained with Dice loss and CE loss, respectively.

\clearpage
\newpage
\section{Segment-level Predictive Uncertainty}
\begin{figure*}[h]
	\centering
	\setlength{\tabcolsep}{1pt}
	\begin{tabular}{ccccccccc}
		\includegraphics[height=18mm]{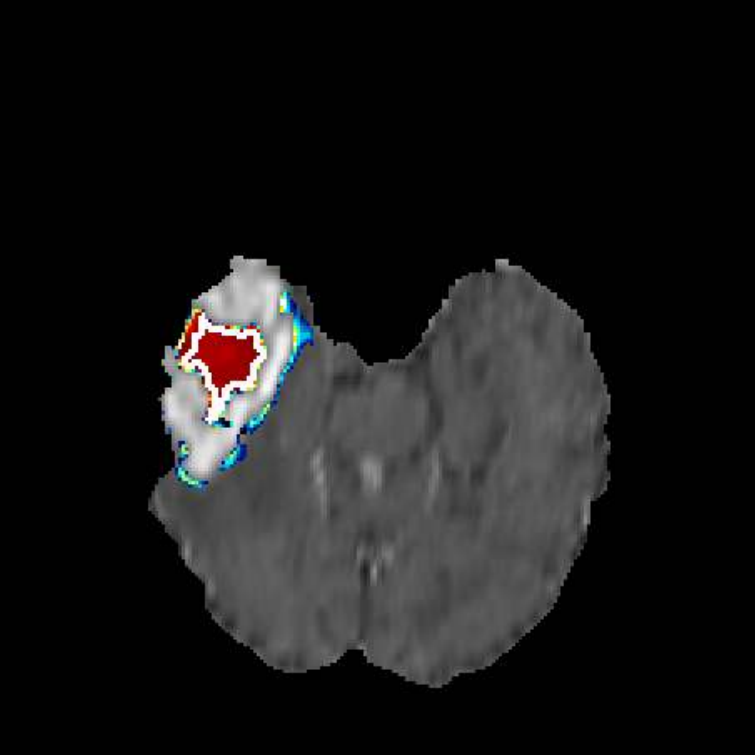} &
		\includegraphics[height=18mm]{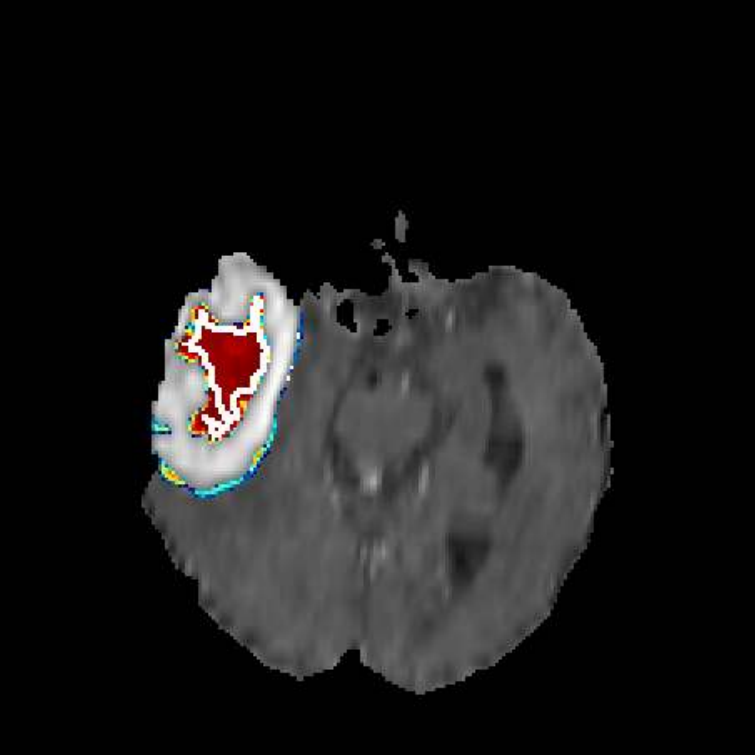} &
		\includegraphics[height=18mm]{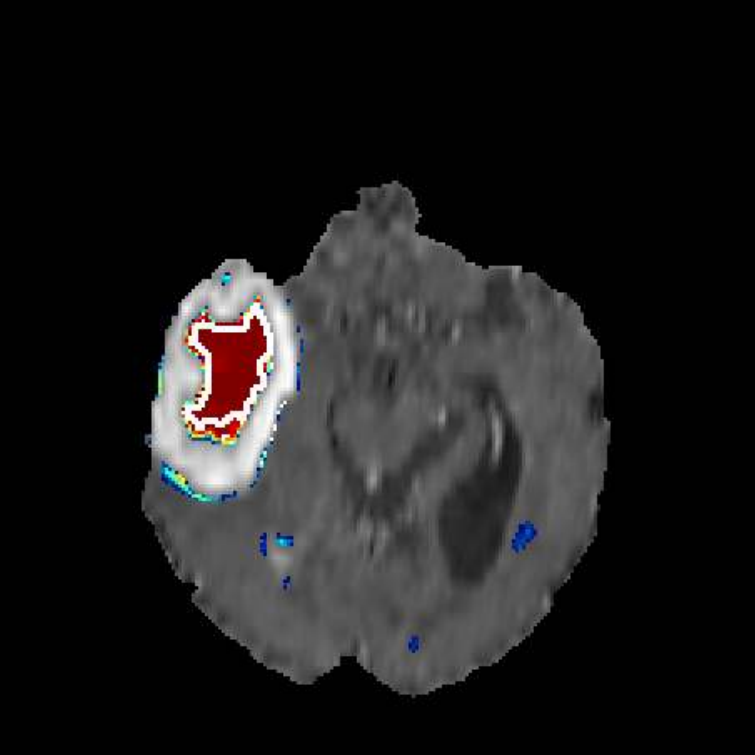} &
		\includegraphics[height=18mm]{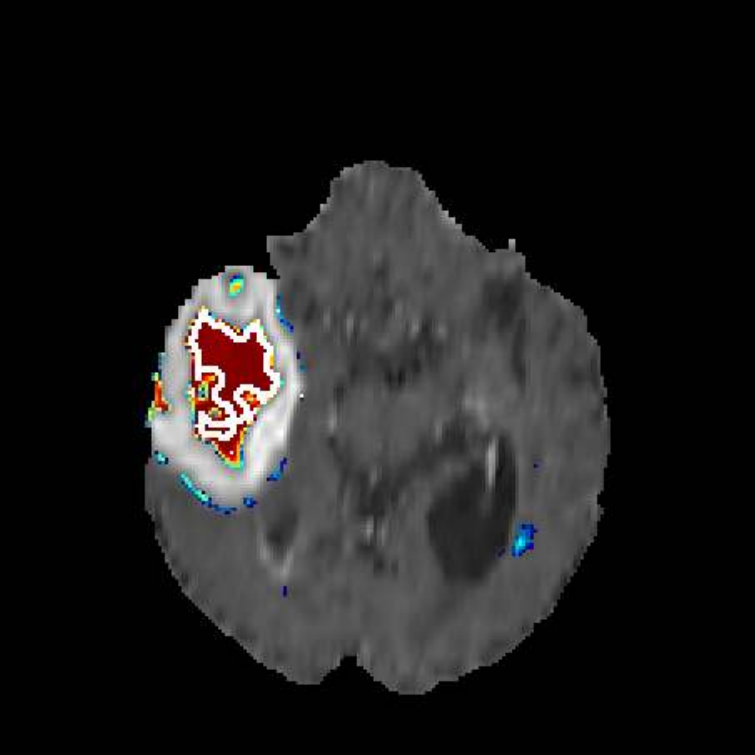} &
		\includegraphics[height=18mm]{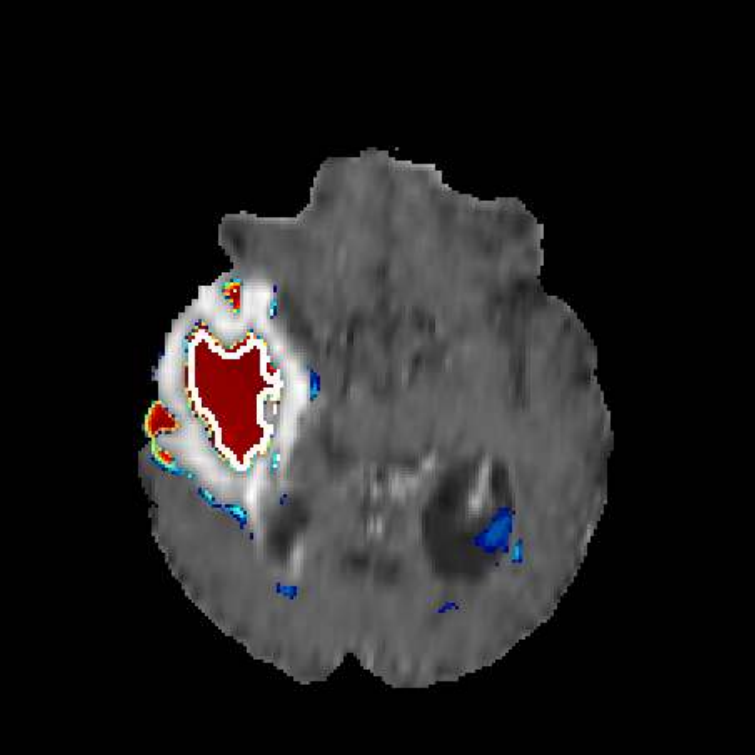} &
		\includegraphics[height=18mm]{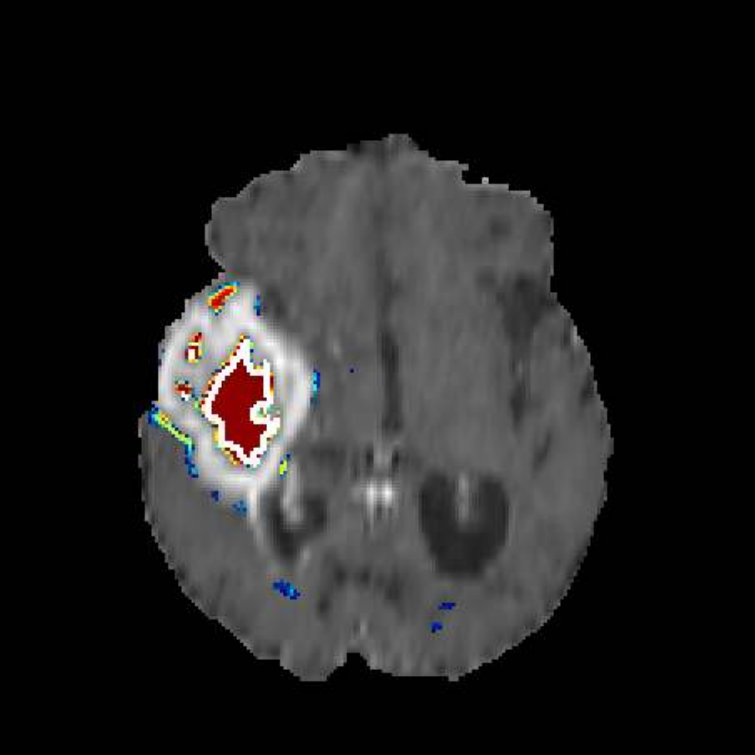} &
		\includegraphics[height=18mm]{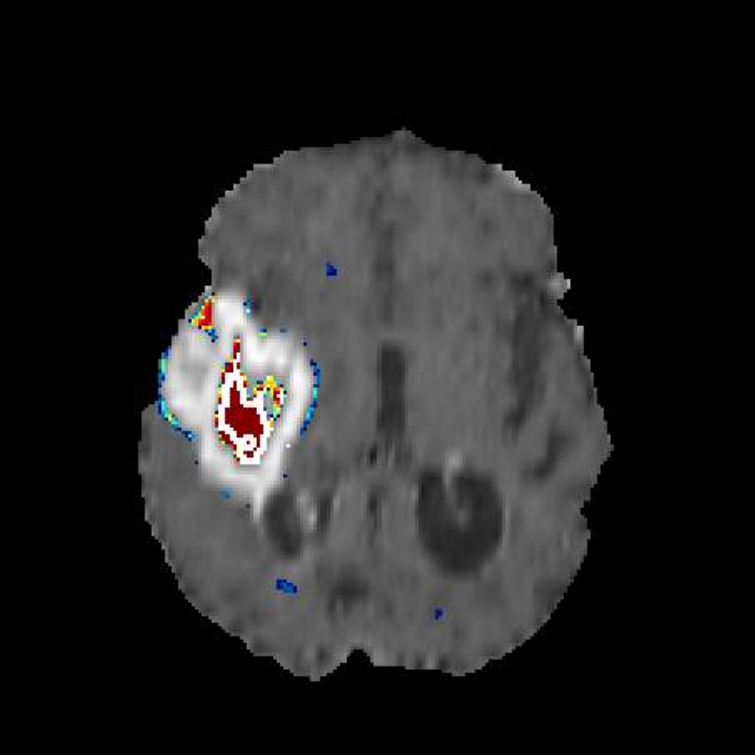} &
		\includegraphics[height=18mm]{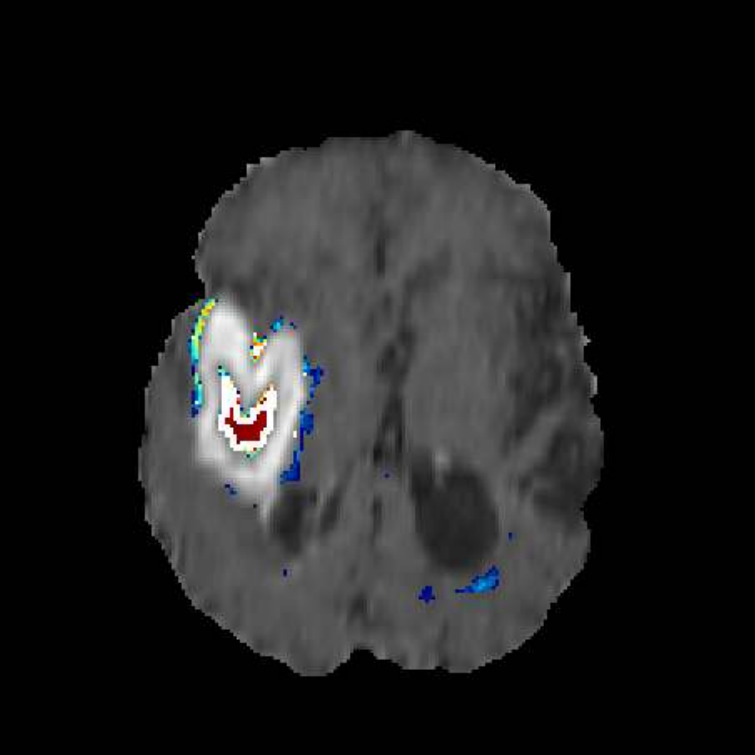} &
		\includegraphics[height=18mm]{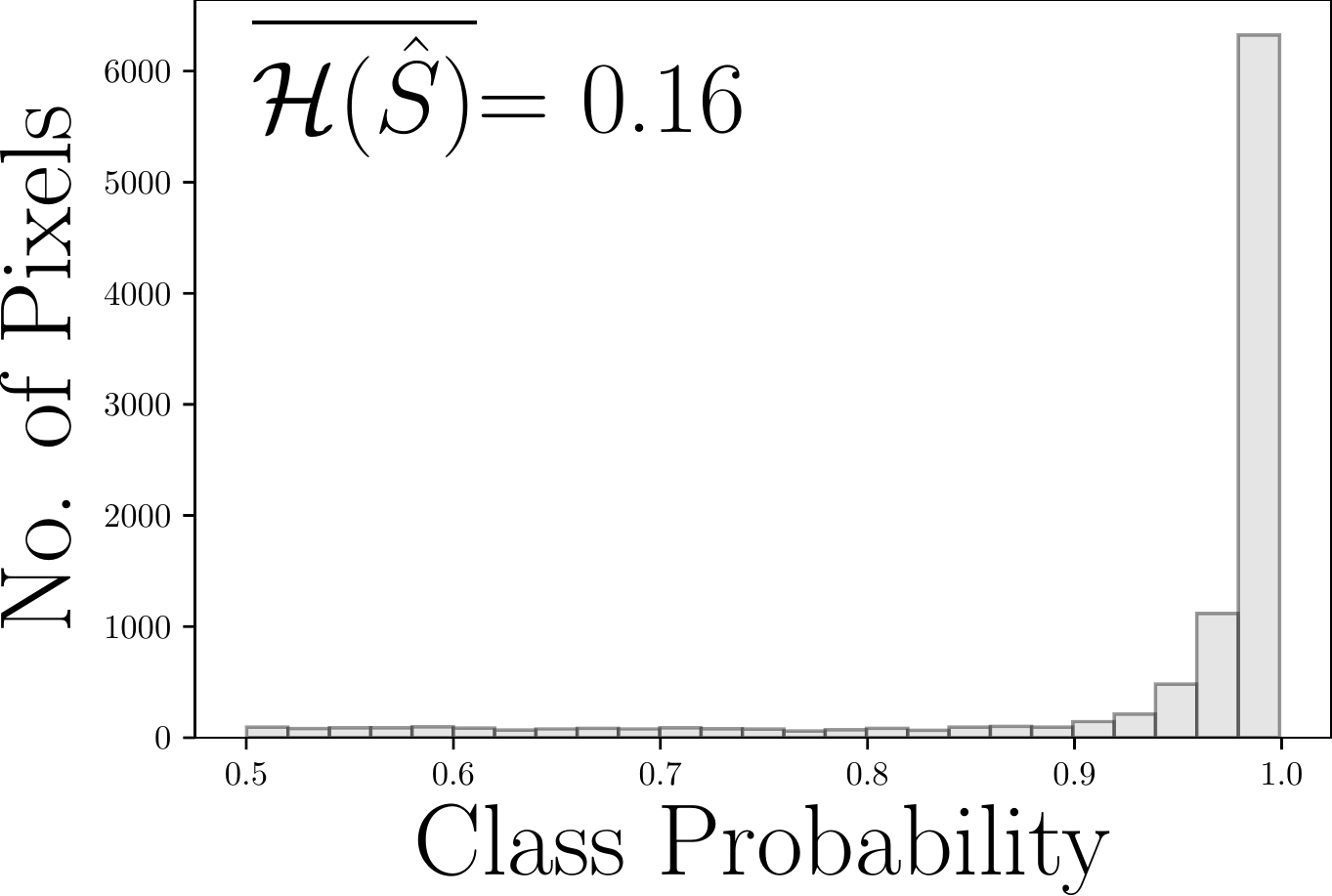} \\
		\includegraphics[height=18mm]{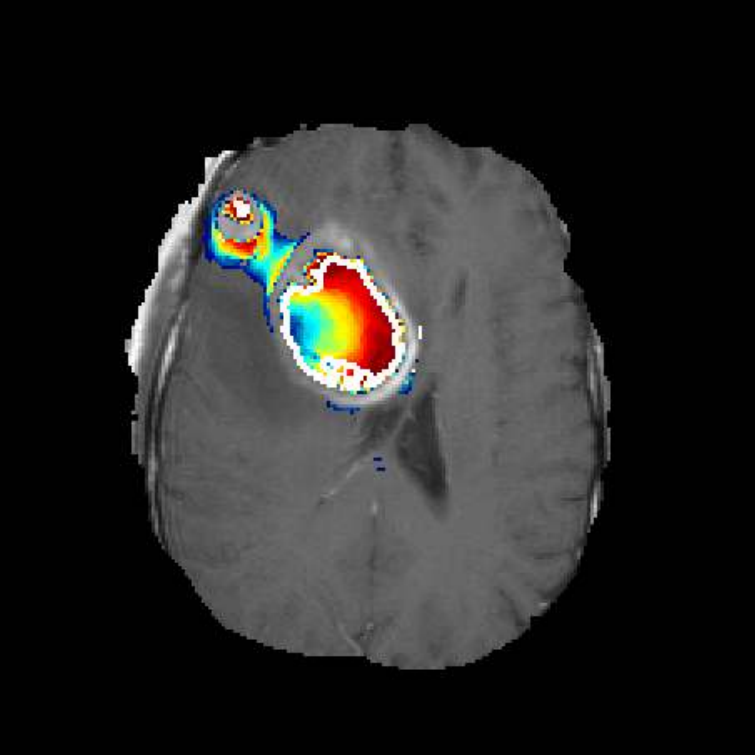} &
		\includegraphics[height=18mm]{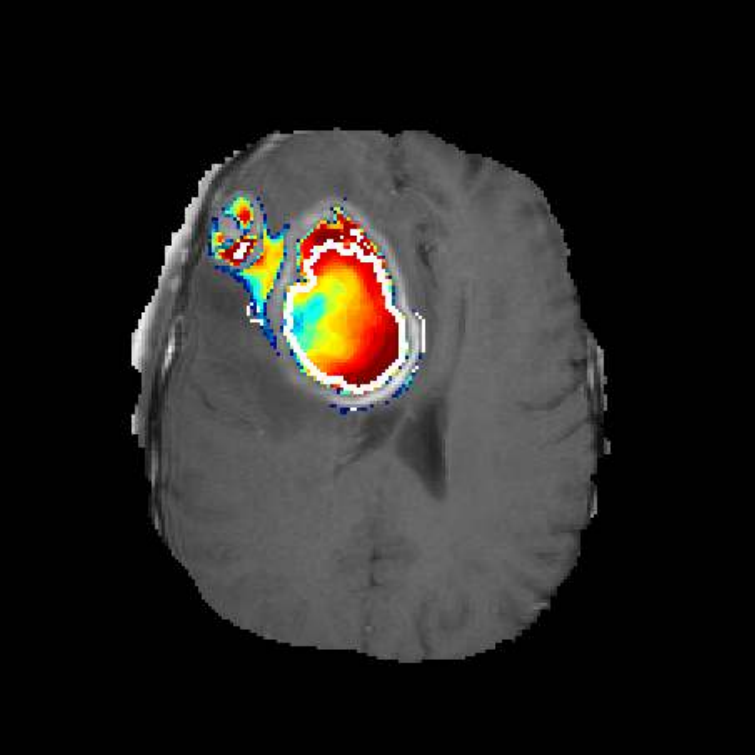} &
		\includegraphics[height=18mm]{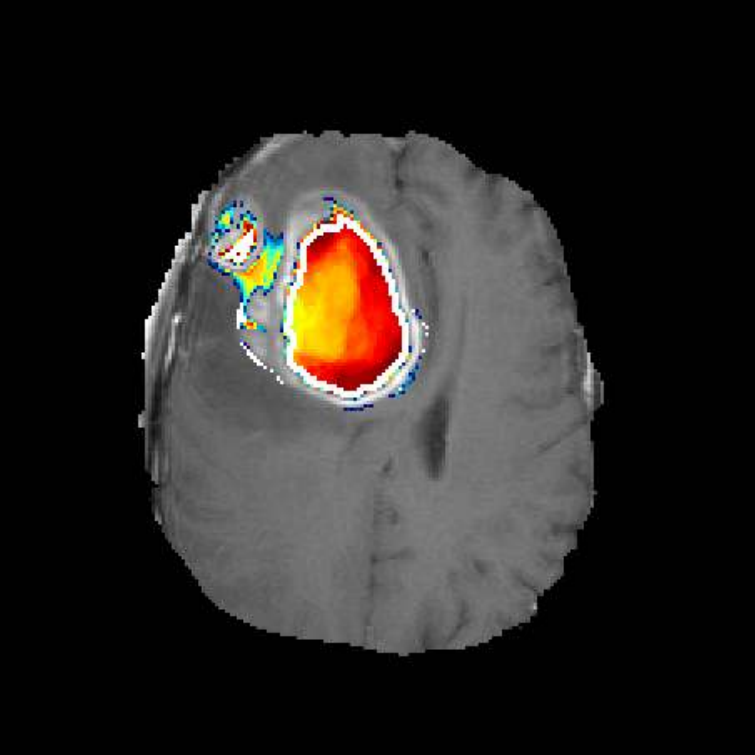} &
		\includegraphics[height=18mm]{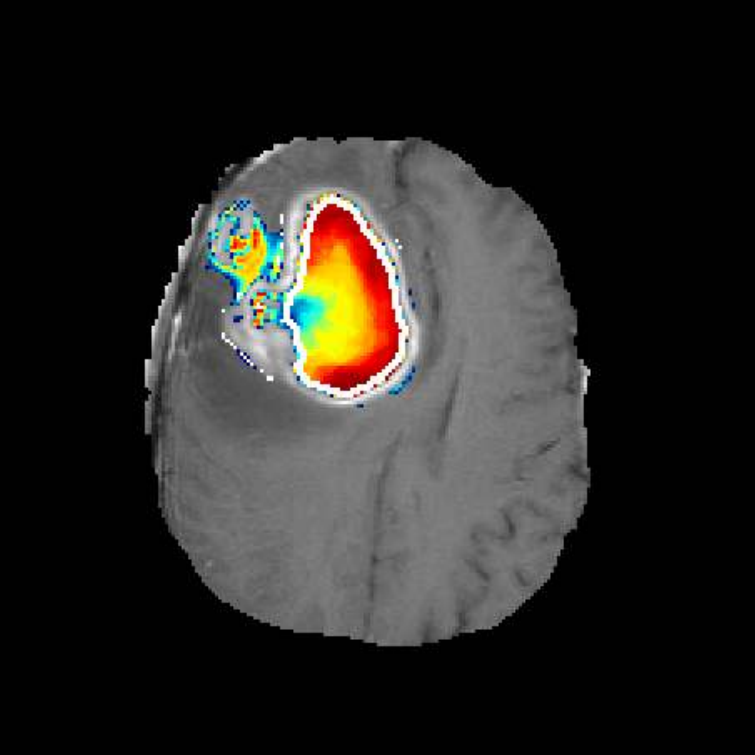} &
		\includegraphics[height=18mm]{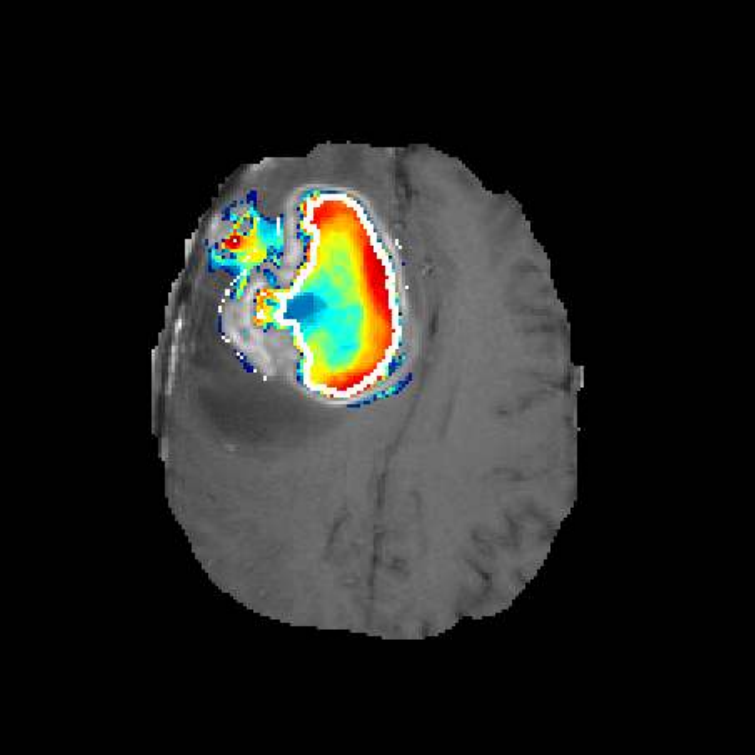} &
		\includegraphics[height=18mm]{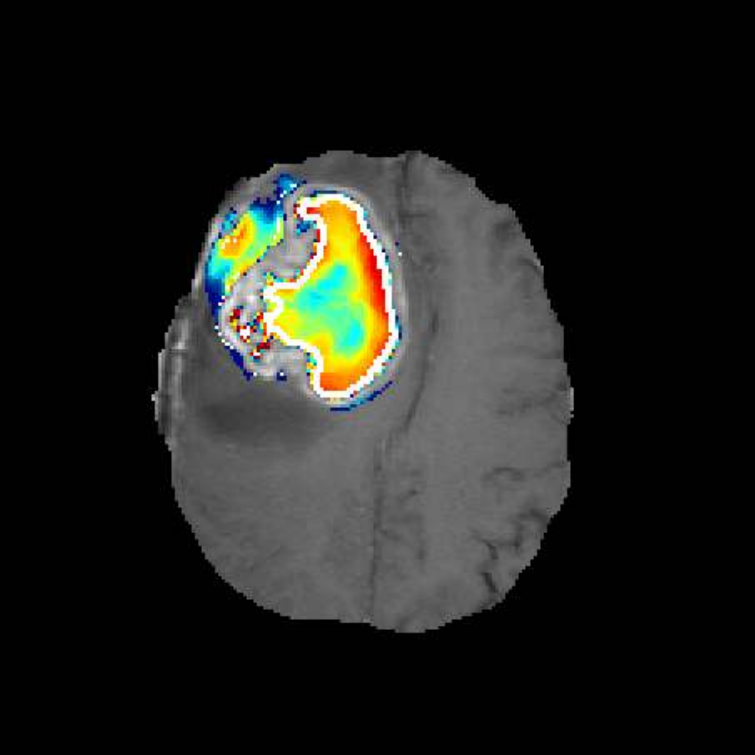} &
		\includegraphics[height=18mm]{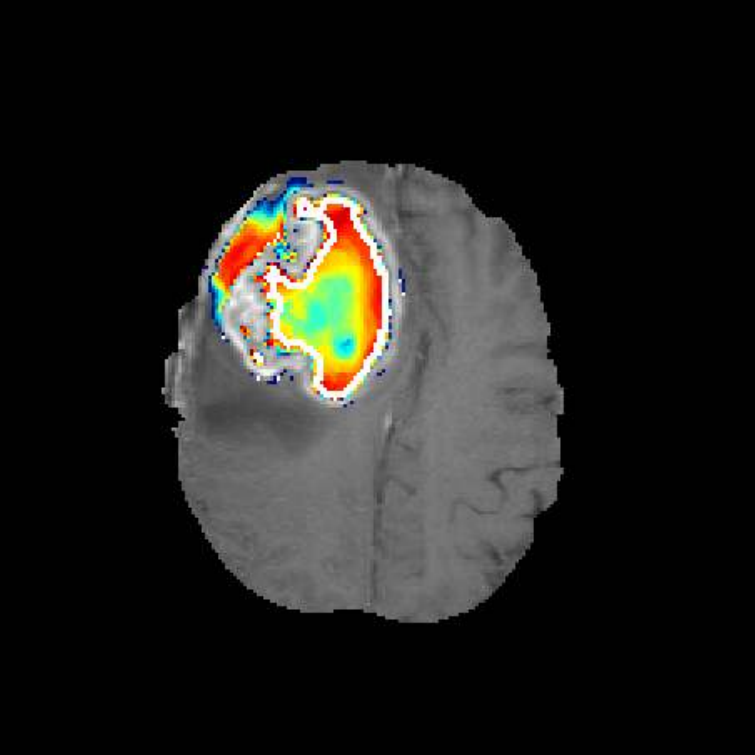} &
		\includegraphics[height=18mm]{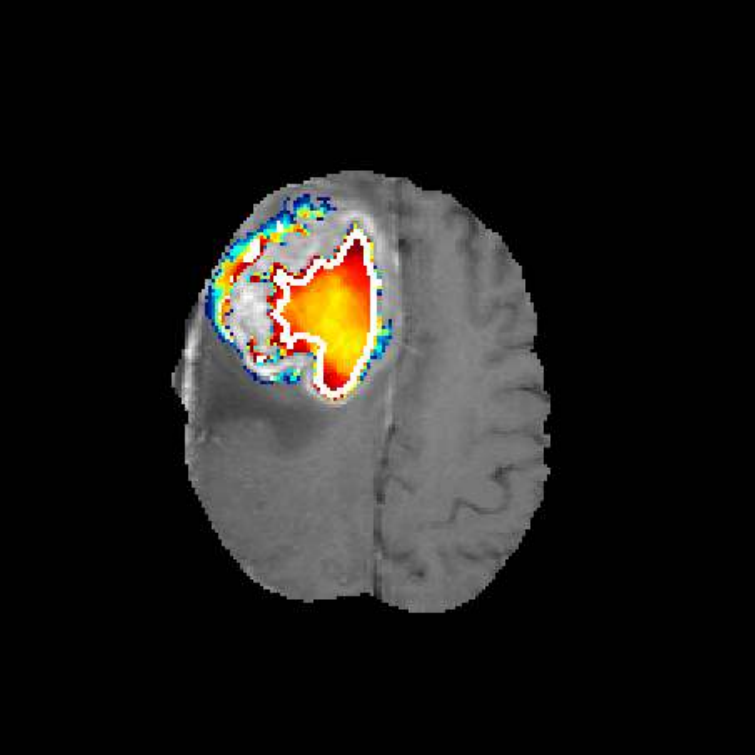} &
		\includegraphics[height=18mm]{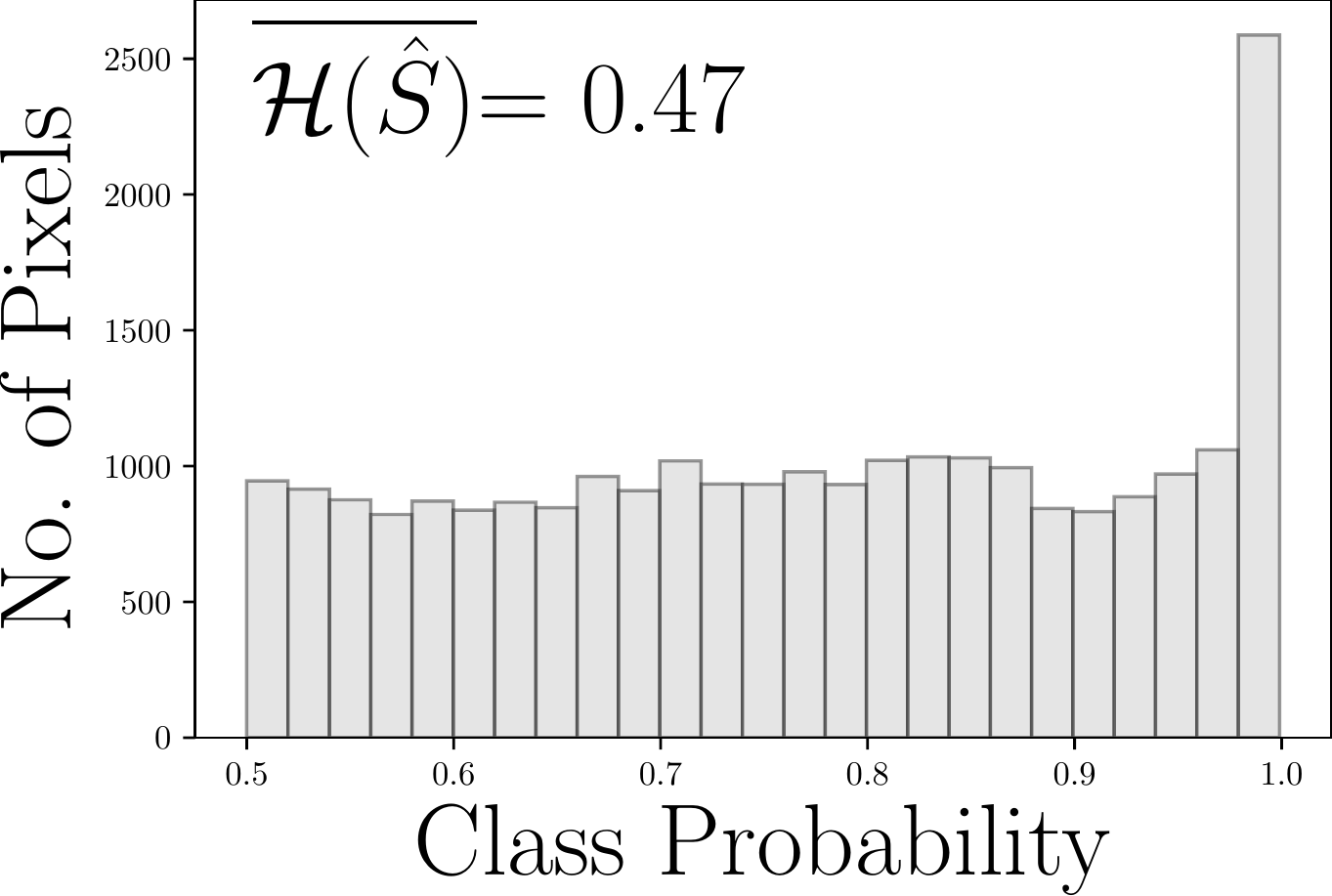} \\
	\end{tabular}
	\begin{tabular}{ccccccccc}
	\includegraphics[height=18mm]{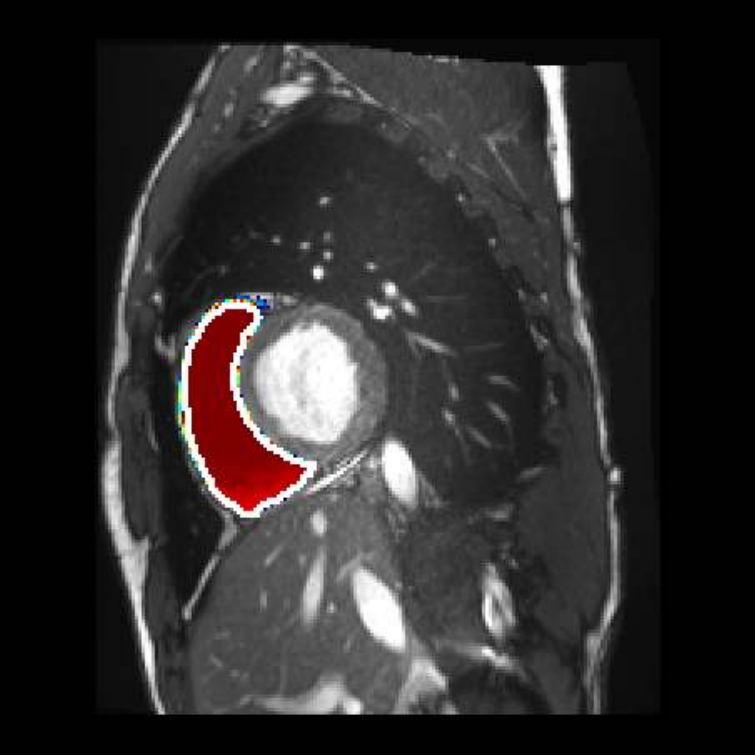} &
	\includegraphics[height=18mm]{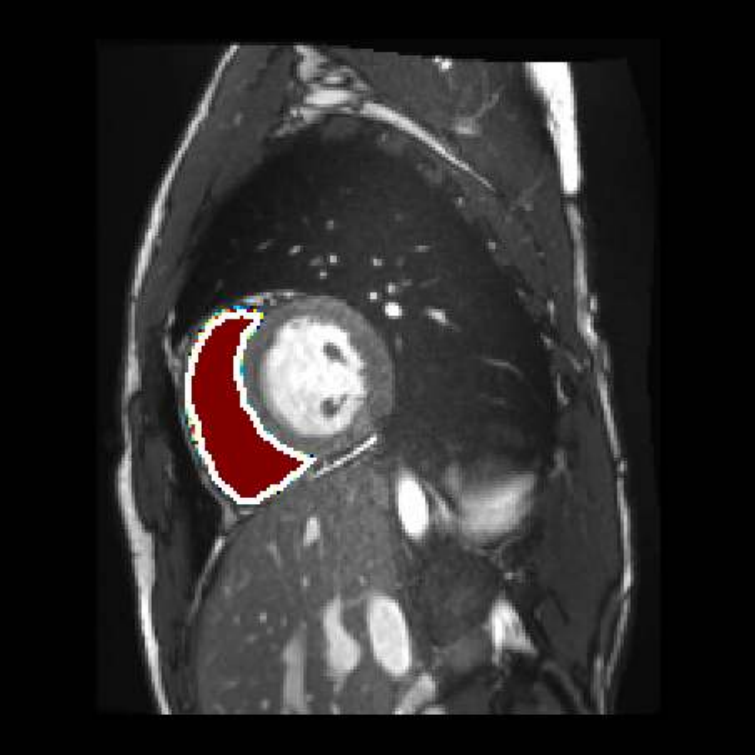} &
	\includegraphics[height=18mm]{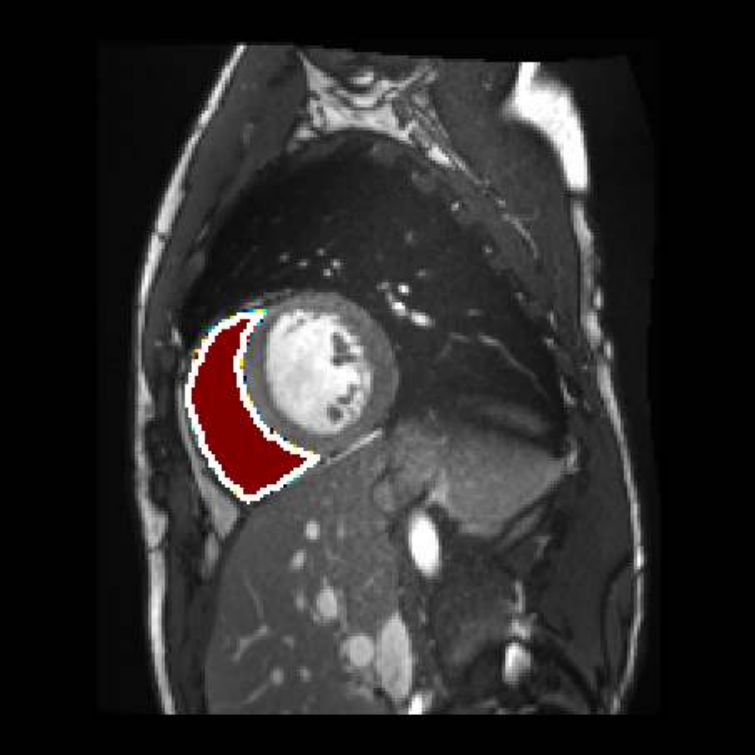} &
	\includegraphics[height=18mm]{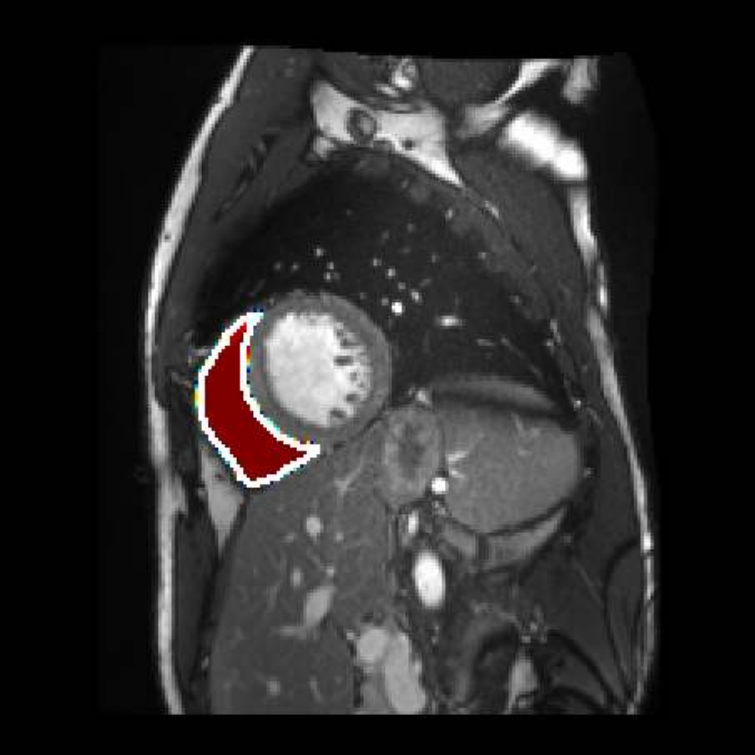} &
	\includegraphics[height=18mm]{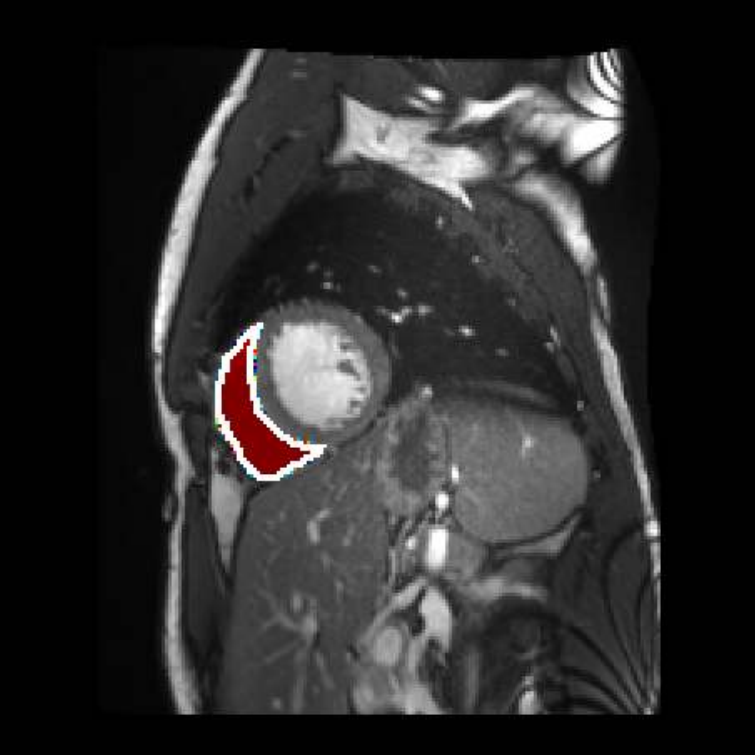} &
	\includegraphics[height=18mm]{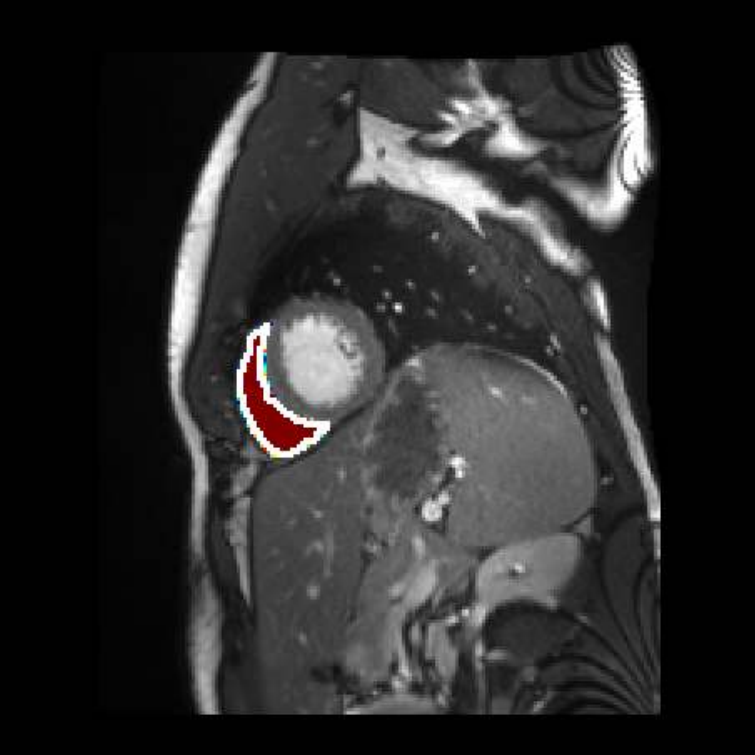} &
	\includegraphics[height=18mm]{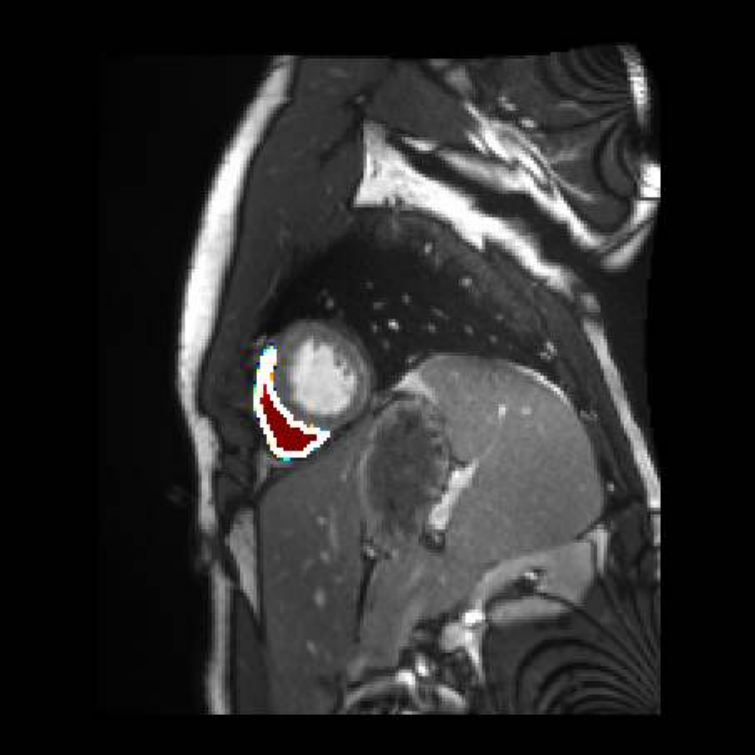} &
	\includegraphics[height=18mm]{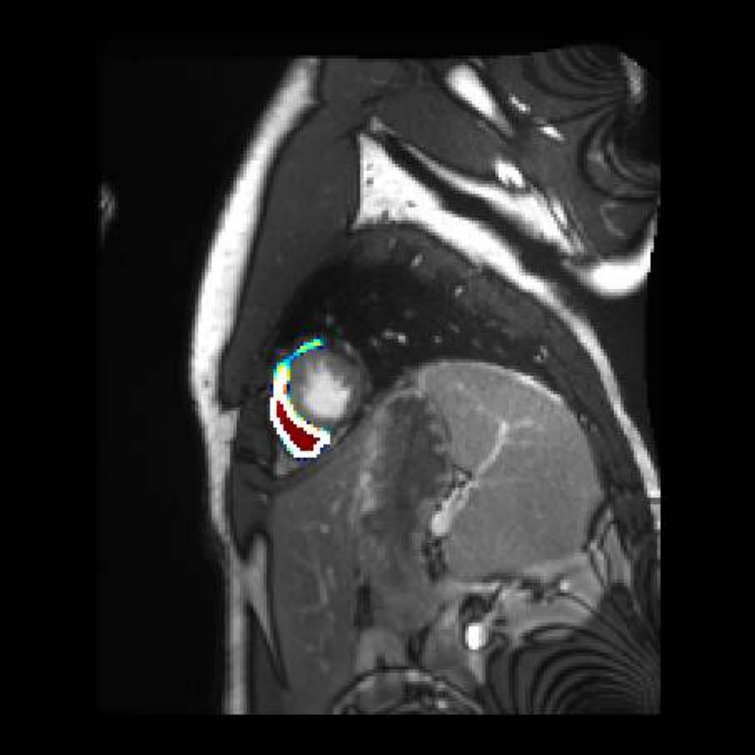} &
	\includegraphics[height=18mm]{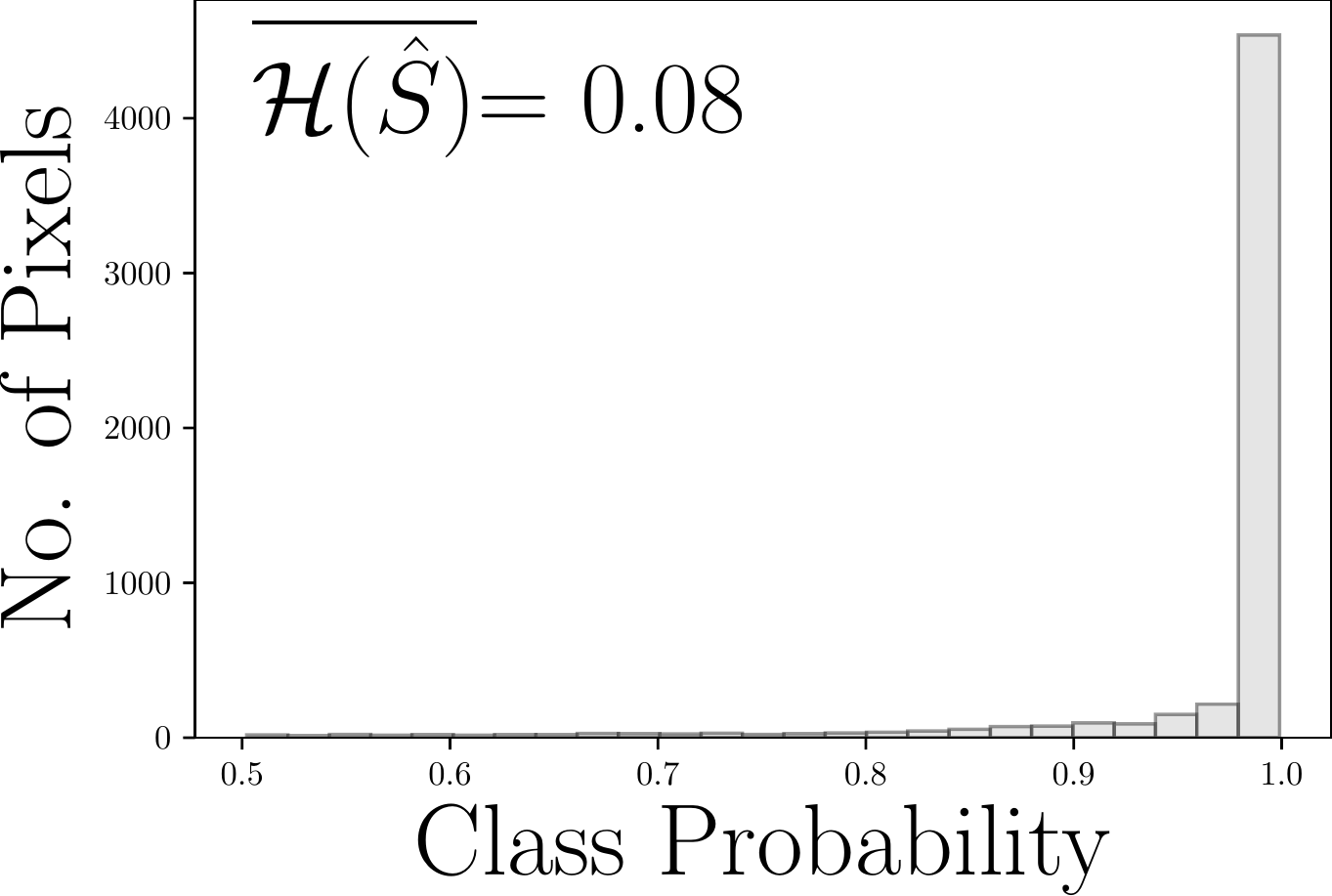} \\
	\includegraphics[height=18mm]{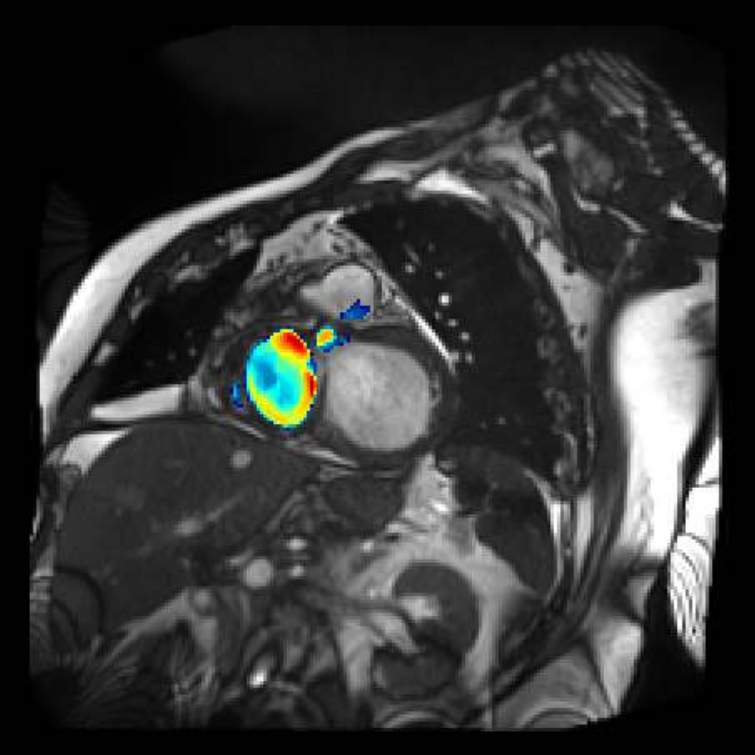} &
	\includegraphics[height=18mm]{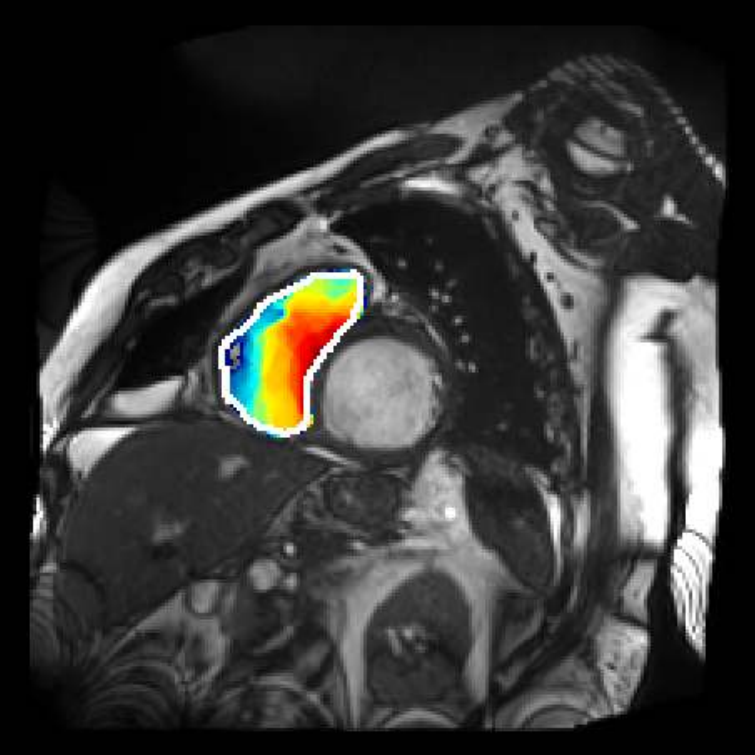} &
	\includegraphics[height=18mm]{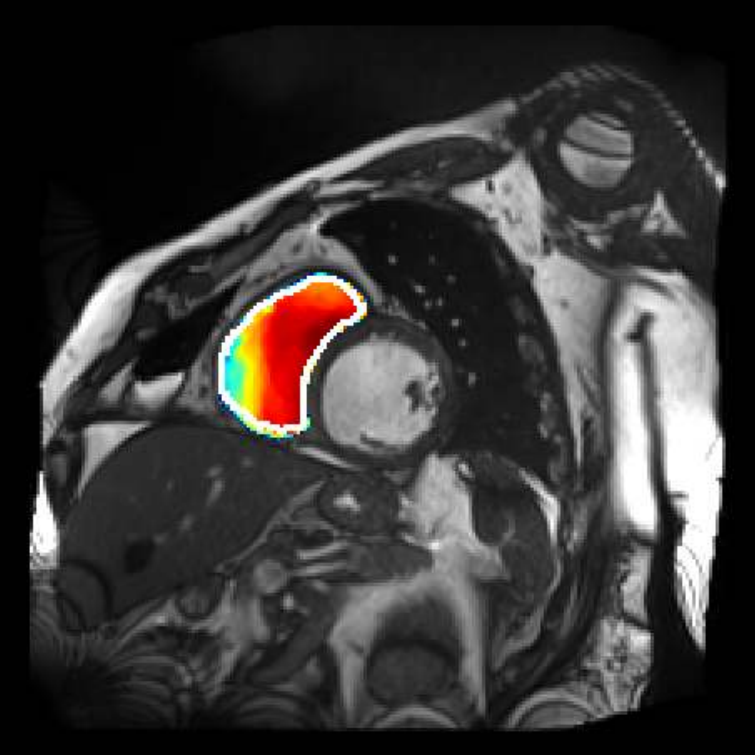} &
	\includegraphics[height=18mm]{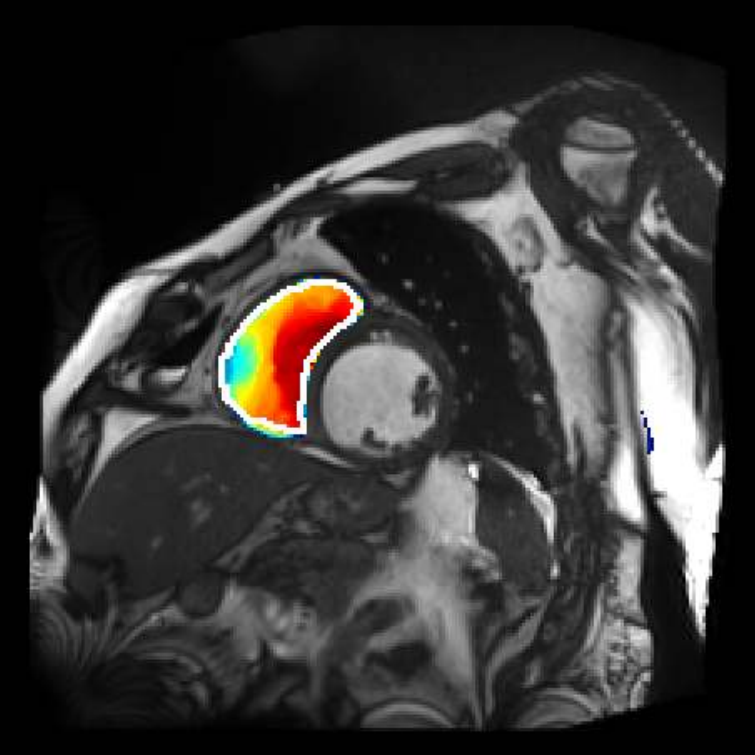} &
	\includegraphics[height=18mm]{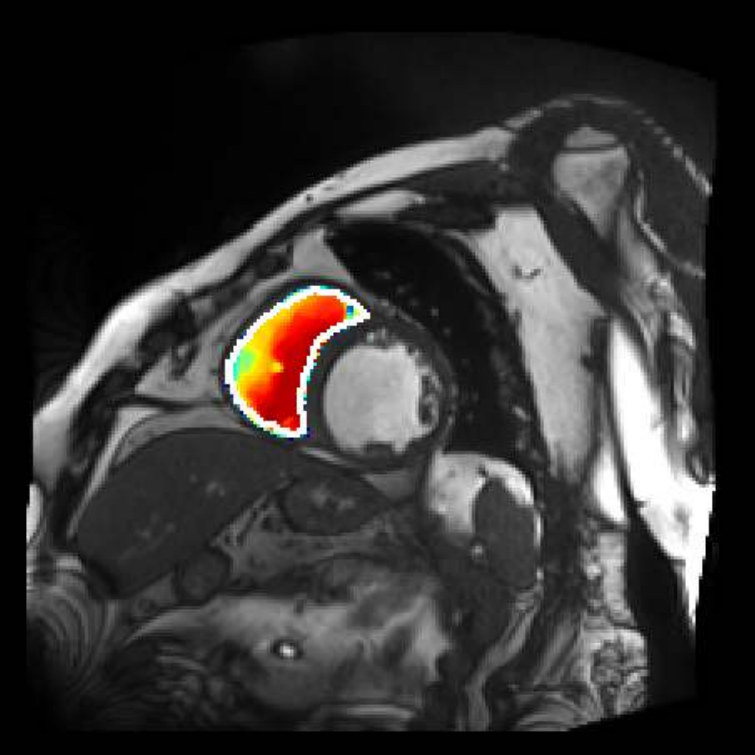} &
	\includegraphics[height=18mm]{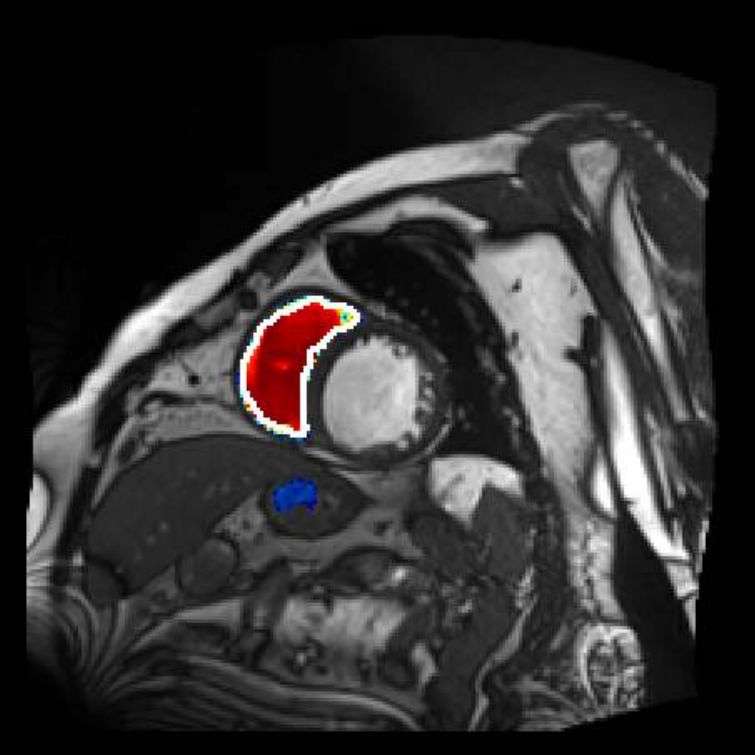} &
	\includegraphics[height=18mm]{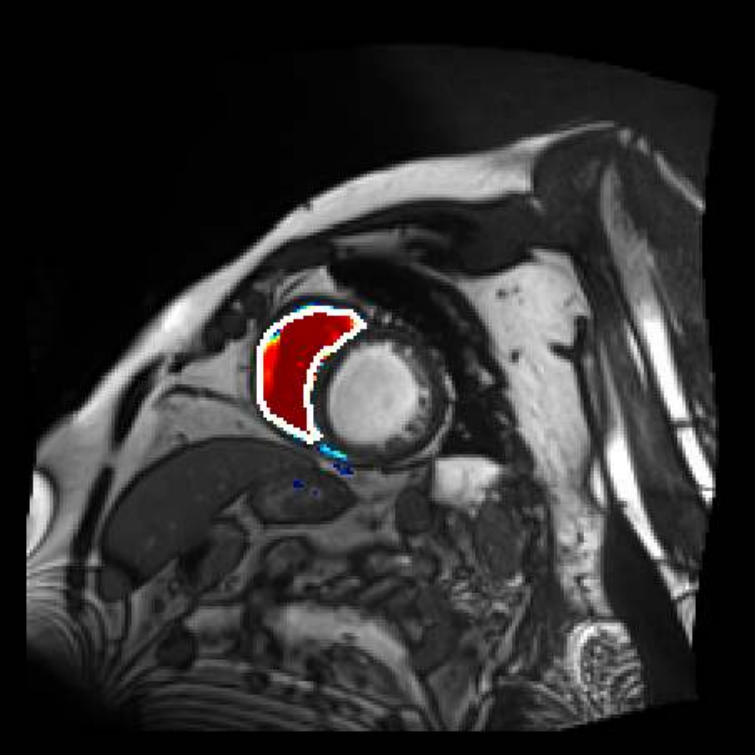} &
	\includegraphics[height=18mm]{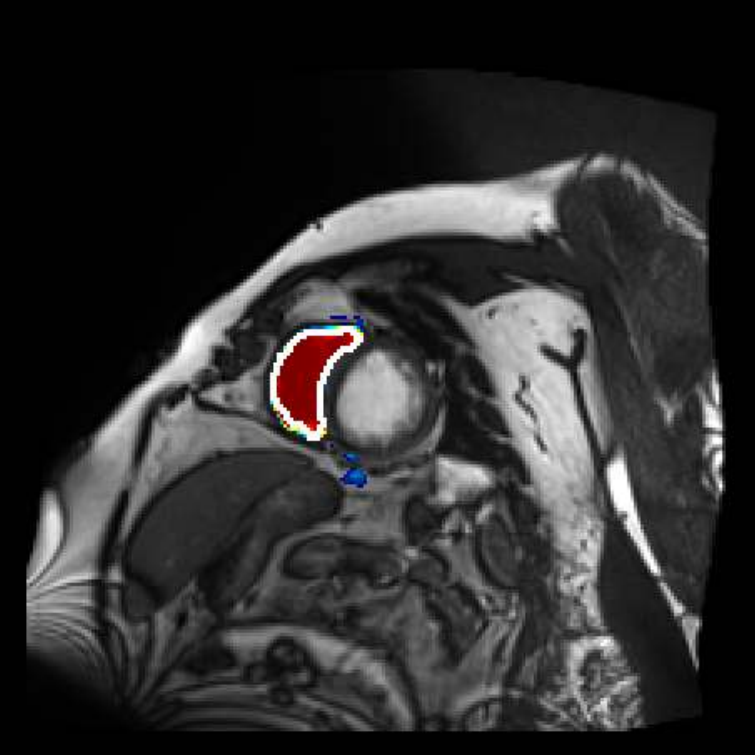} &
	\includegraphics[height=18mm]{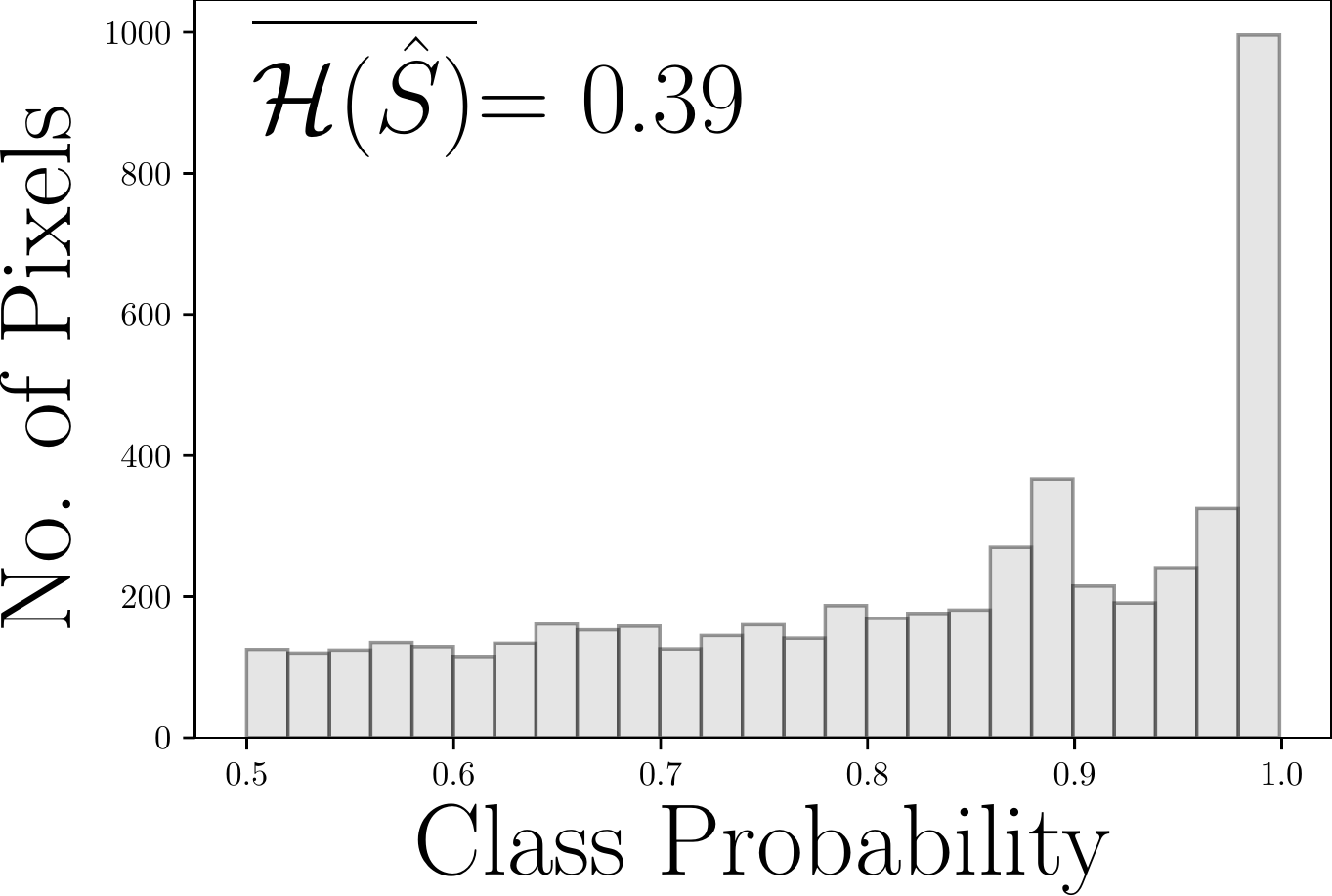} \\
\end{tabular}
	\caption{
	Examples from tumor segmentation and right ventricle segmentation tasks.
	}
	
	\label{fig:dice_vs_nll_more}
\end{figure*}

Figure \ref{fig:dice_vs_nll_more} shows examples of 
predictions with different levels of confidence for brain and heart application.

\begin{figure*}[h]
	\centering
	\setlength{\tabcolsep}{1pt}
	\begin{tabular}{ccc}
		\small{~~~Prostate Segmentation} &
		\small{~~~Brain Tumor Segmentation} &
		\small{~~~Cardiac Segmentation} \\
		\includegraphics[width=58mm]{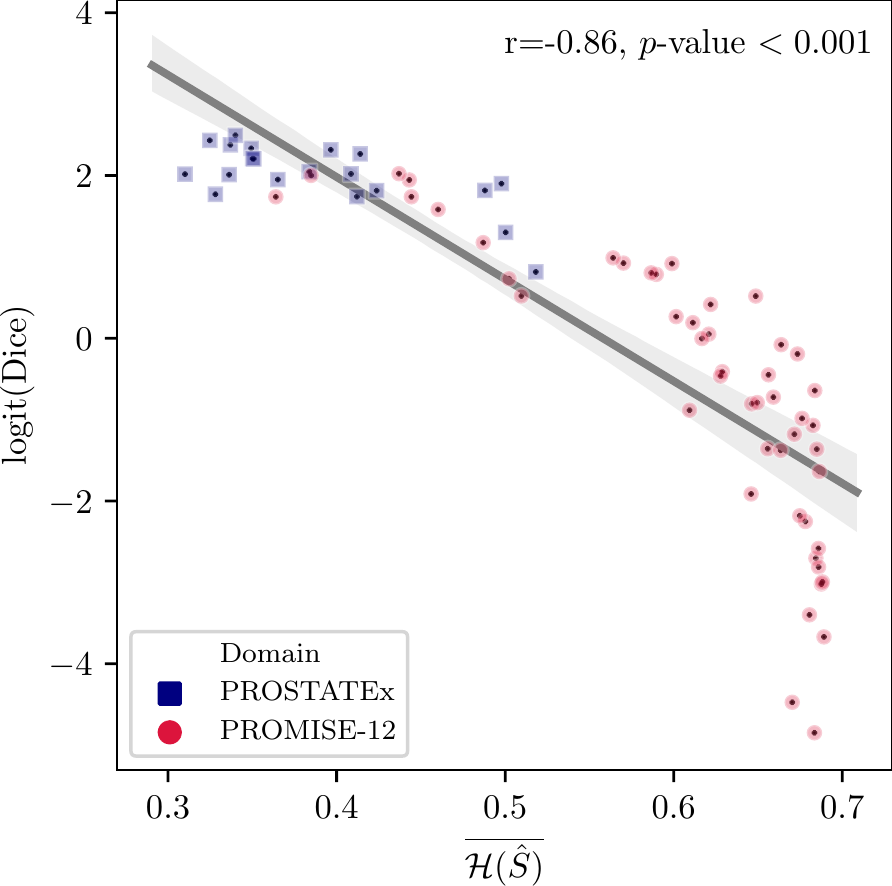} &
		\includegraphics[width=58mm]{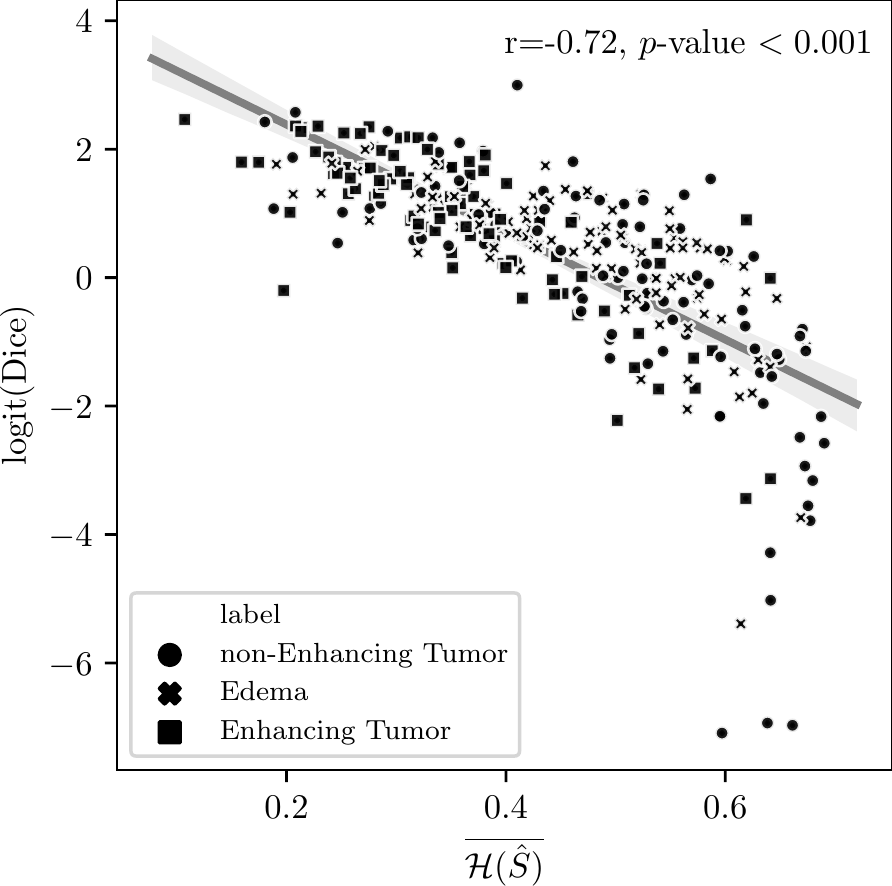} &
		\includegraphics[width=58mm]{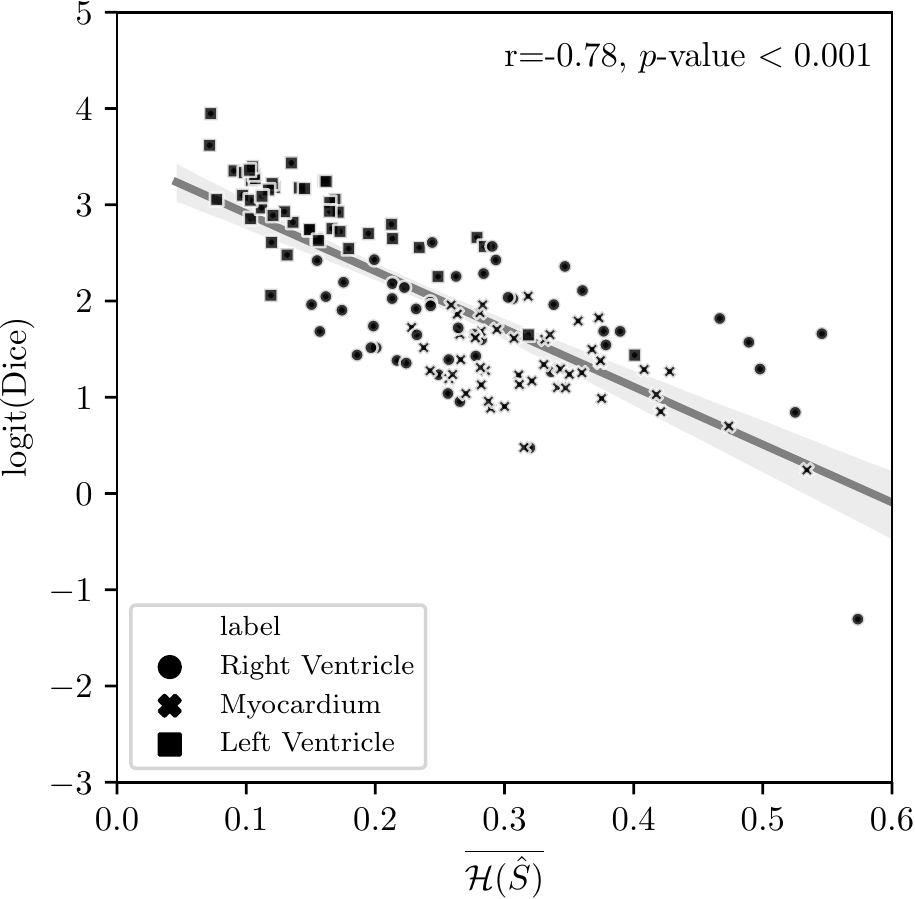} \\

	\end{tabular}
	\caption{
		Segment-level predictive uncertainty estimation for ensembles of models trained with cross entropy loss. Scatter plots and linear regression between Dice coefficient and average of entropy over the predicted segment.
	}
\label{fig:segment_level_uncertainty_ce}
\end{figure*}

Figure \ref{fig:segment_level_uncertainty_ce} provides scatter plots of Dice coefficient vs. the proposed segment-level predictive uncertainty metric,  $\overline{\mathcal{H(\hat{\mathcal{S}})}}$ (Equation \ref{eq:avg_entr}), for models trained with CE loss and calibrated with ensembling (M=50).
Similar to the charts in Figure \ref{fig:segment_level_uncertainty},
a reverse correlation holds between the average entropy over the segmented foreground
and the Dice scores.

\clearpage
\newpage
\section{3D FCN}

To test the generalizability of the proposed methods on 3D CNNs, we run limited experiments
on prostate gland segmentation.
Prostate images were resampled to resolution of $1 \times 1 \times 1$ mm and 
all the 3D volumes were then cropped at the center to create images of size $112 \times 112 \times 112$ pixels as the input size of the FCN.
Image intensities were normalized to be within the range of [0,1]. 
We constructed a 3D FCN with same number of kernels and depth as the 2D network described in Section \ref{sec:Methods}.
2D Convolutional, max pooling and upsampling layers were replaced 3D layers. 
Training parameters were kept the same except that the model was trained 10 times with cross entropy and 10 times with Dice loss, and 10 times with dropout layers for MC dropout experiments.
Table \ref{tab:3d_cnn} compares the calibration quality and segmentation performance of baselines and ensembles (M=50) trained with CE loss with those that were trained with Dice loss and those that were calibrated with MC dropout.

\begin{table}[h]
	\centering
	\caption{
		Observed average calibration quality and segmentation performance for 3D FCNs trained for prostate gland segmentation.} 
	\label{tab:3d_cnn}
		\begin{tabular}{l|ccc|cc}
			\toprule
			& \multicolumn{3}{c}{Calibration Quality (bounding boxes)}
			& \multicolumn{2}{c}{Segmentation Performance}\\
			
			\midrule
			Model & NLL & Brier  & ECE\%  
			& Dice Score & $95^{th}$ HD\\
		
			\midrule
			$\mathcal{L}_{CE}$& 
			
			0.24 & 
			0.15 & 
			8.25 & 
			0.84 & 
			9.25 \\ 
			
			MCDO ${\mathcal{L}_{CE}}$ &
						
			0.26 & 
			0.16 & 
			6.84 & 
			0.81 & 
			12.12 \\ 
	
			EN ${\mathcal{L}_{CE}}$ & 
			
			0.21 & 
			0.12 & 
			9.23 & 
			0.87 & 
			5.62 \\ 

			${\mathcal{L}_{DSC}}$& 
			
			0.40 & 
			0.13 & 
			5.32 & 
			0.89 & 
			6.06 \\ 

			MCDO ${\mathcal{L}_{DSC}}$ &
						
			0.41 & 
			0.11 & 
			5.27 & 
			0.88 & 
			7.47 \\ 
			
			EN ${\mathcal{L}_{DSC}}$& 
			
			0.19 & 
			0.08 & 
			2.82 & 
			0.90 & 
			4.61 \\ 
			
			\bottomrule
			\end{tabular}
\end{table}

\end{document}